\documentclass[a4paper,11pt]{article}
\usepackage[utf8]{inputenc}
\usepackage{amsfonts}
\usepackage{amsmath}
\usepackage{amssymb}
\usepackage{graphicx}
\usepackage[table]{xcolor}
\usepackage{todonotes}
\usepackage{multirow}
\usepackage{latexsym,graphics,amscd} 
\usepackage{bigints}
\usepackage{booktabs} 
\usepackage[ruled,vlined]{algorithm2e}
\usepackage{tabularx} 
\usepackage{threeparttable}
\usepackage{subfig}
\usepackage{lscape}
\usepackage{caption} 
\usepackage{mathtools}
\usepackage{array} 
\usepackage{setspace}
\usepackage{bbm}
\usepackage{easyReview}
\usepackage{makecell}
\usepackage[authoryear,round,comma]{natbib}
\usepackage{url}
\usepackage[final]{pdfpages}
\usepackage{hyperref}
\hypersetup{
    colorlinks=true,
    citecolor=blue,
    linkcolor=blue,
    filecolor=magenta,      
    urlcolor=blue,
}

\newcommand{\rset}{\mathbb{R}}%
\newcommand{\nset}{\mathbb{N}}%

\newcommand{\E}{\mathbb{E}}

\newcommand{\rmd}{\mathrm{d}} 

\newcommand{\tempo}[2]{{#1}\vert{#2}}
\renewcommand{\mod}{{\mskip 3mu}\textnormal{mod}{\mskip 3mu}}

\newcommand{\mf}{(\textnormal{\texttt{m}})}
\newcommand{\qf}{(\textnormal{\texttt{q}})}
\newcommand{\dsf}{(\textnormal{\texttt{6d}})}

\usepackage{geometry}
\addtocounter{footnote}{1}
\makeatletter
\@addtoreset{table}{bsection}
\def\thetable{\thesection.\@arabic\c@table}
\def\fps@table{h, t}
\@addtoreset{equation}{section}

\makeatother

\newtheorem{theorem}{Theorem}[section]
\newtheorem{definition}[theorem]{Definition}

\newtheorem{example}[theorem]{Example}

\newcommand{\cHL}{\cellcolor{black!15}}
\newcommand{\chl}{\cellcolor{black!8}}

\providecommand{\norm}[1]{\lVert#1\rVert}

\newcommand{\vertiii}[1]{{\left\vert\kern-0.25ex\left\vert\kern-0.25ex\left\vert #1 
    \right\vert\kern-0.25ex\right\vert\kern-0.25ex\right\vert}}
\newsavebox{\savepar}

\makeatletter
\makeatletter
\DeclareMathOperator*{\argmin}{arg\,min}

\usepackage{scalerel,stackengine}
\stackMath
\newcommand\reallywidehat[1]{%
\savestack{\tmpbox}{\stretchto{%
  \scaleto{%
    \scalerel*[\widthof{\ensuremath{#1}}]{\kern-.6pt\bigwedge\kern-.6pt}%
    {\rule[-\textheight/2]{1ex}{\textheight}}
  }{\textheight}%
}{0.5ex}}%
\stackon[1pt]{#1}{\tmpbox}%
}

\expandafter\def\expandafter\normalsize\expandafter{%
    \normalsize%
    \setlength\abovedisplayskip{6pt}%
    \setlength\belowdisplayskip{6pt}%
    \setlength\abovedisplayshortskip{-2pt}%
    \setlength\belowdisplayshortskip{2pt}%
}

\begin{document}

\title{\vspace{-25pt}\huge Reservoir Computing for Macroeconomic Forecasting\\ with Mixed Frequency Data}
\author{Giovanni Ballarin$^{1}$, Petros Dellaportas$^{2,3}$, Lyudmila Grigoryeva$^{4,5}$, \\ Marcel Hirt$^{8}$, Sophie van Huellen$^{6,7}$, Juan-Pablo Ortega$^{8}$}
\date{ \ }
\maketitle

\vspace{-20pt}

\begin{abstract}
	Macroeconomic forecasting has recently started embracing techniques that can deal with large-scale datasets and series with unequal release periods. MIxed-DAta Sampling (MIDAS) and Dynamic Factor Models (DFM) are the two main state-of-the-art approaches that allow modeling series with non-homogeneous frequencies. We introduce a new framework called the Multi-Frequency Echo State Network (MFESN) based on a relatively novel machine learning paradigm called reservoir computing. Echo State Networks (ESN) are recurrent neural networks formulated as nonlinear state-space systems with random state coefficients where only the observation map is subject to estimation. MFESNs are considerably more efficient than DFMs and allow for incorporating many series, as opposed to MIDAS models, which are prone to the curse of dimensionality. All methods are compared in extensive multistep forecasting exercises targeting US GDP growth. We find that our MFESN models achieve superior or comparable performance over MIDAS and DFMs at a much lower computational cost.
\end{abstract}

\bigskip



\vspace{1em}
{\footnotesize\noindent {\bf Acknowledgments:} The authors acknowledge financial support from UK Research and Innovation (grant number ES/V006347/1). GB thanks the hospitality and the generosity of the University of Warwick  without which this paper would not have happened. GB is grateful to the Center for Doctoral Studies in Economics (CDSE) at the University of Mannheim for providing travel funding through the 2020 CDSE Teaching Award. The authors acknowledge support by the state of Baden-Württemberg through bwHPC. The authors acknowledge the use of the UCL Myriad High Performance Computing Facility (Myriad@UCL), and associated support services, in completing this work. MH acknowledges that the majority of his work was done when he was affiliated with UCL and hence thanks the support of UCL in this project. The authors acknowledge the support of the Swiss National Science Foundation (grant number 200021\_175801/1). The authors also thank Thanos Moraitis and Alfonso Silva Ruiz for excellent research assistance and help in data collection and curation. }

\addtocounter{footnote}{1} \footnotetext{ 
	Department of Economics, University of Mannheim, L7, 3-5, Mannheim, 68131, Germany. {\texttt{Giovanni.Ballarin@gess.uni-mannheim.de} }}
\addtocounter{footnote}{1} \footnotetext{%
	Department of Statistical Science, UCL, Gower Str., London WC1E 6BT, UK. {\texttt{P.Dellaportas@ucl.ac.uk} }}
\addtocounter{footnote}{1} \footnotetext{%
	Department of Statistics, Athens University of Economics and Business, 10434 Athens, Greece. {\texttt{Petros@aueb.gr} }}
\addtocounter{footnote}{1} \footnotetext{%
	Faculty of Mathematics and Statistics, University of St. Gallen, Bodanstrasse 6, CH-9000 St.~Gallen, Switzerland. {\texttt{Lyudmila.Grigoryeva@unisg.ch} }}
\addtocounter{footnote}{1} \footnotetext{%
	Honorary Associate Professor, Department of Statistics, University of Warwick, Coventry CV4 7AL, UK. {\texttt{Lyudmila.Grigoryeva@warwick.ac.uk} }}
\addtocounter{footnote}{1} \footnotetext{%
	Global Development Institute (GDI), University of Manchester, Manchester M13 9PL, UK.
	{\texttt{Sophie.vanHuellen@manchester.ac.uk} }} 
\addtocounter{footnote}{1} \footnotetext{%
	Department of Economics, SOAS University of London, WC1H 0XG, London, UK. 
	{\texttt{Sv8@soas.ac.uk} }} 
\addtocounter{footnote}{1} \footnotetext{%
	Division of Mathematical Sciences, 
	Nanyang Technological University,
	21 Nanyang Link,
	Singapore 637371.
	\texttt{MarcelAndre.Hirt@ntu.edu.sg} (some of the work done while at UCL), {\texttt{Juan-Pablo.Ortega@ntu.edu.sg}}}
\makeatother

\newpage
\setcounter{footnote}{0} 

\newpage
\setstretch{1.1}

\section{Introduction}
\label{Introduction}

The availability of timely and accurate forecasts of key macroeconomic variables is of crucial importance to economic policymakers, businesses, and the banking sector alike. Fundamental macroeconomic figures, such as GDP growth, become available at low frequency with a considerable time lag and are subject to various rounds of revisions after their release. This is particularly problematic in a fast-changing and uncertain economic environment, as experienced during the Great Recession of 2007-2008 (\cite{Hindrayanto2016}) and the recent pandemic (\cite{Buell2021, Huber2021}). However, a large number of the potentially predictive financial market (and other macroeconomic) indicators are available at a daily or even higher frequency (\cite{Andreou2013}). The desire to utilize such high-frequency data for macroeconomic forecasting has led to the exploration of techniques that can deal with large-scale datasets and series with unequal release periods (see \cite{Borio2011, Borio2013,Morley2015}; we also refer the reader to \cite{Fuleky2020} for more details regarding high-dimensional data and to \cite{Armesto2010} and \cite{Banbura2013} for a review on mixed-frequency data). 

We contribute to the existing literature by proposing a new macroeconomic forecasting framework that utilizes high-dimensional and mixed-frequency input data, the Multi-Frequency Echo State Network (MFESN). The MFESN originates from a machine learning paradigm called Reservoir Computing (RC). RC is a family of learning models that take advantage of the information processing capabilities of complex dynamical systems~(see \cite{maass1, DynamicalSystemsMaass, Crutchfield2010}, and \cite{lukosevicius, tanaka:review} for reviews). Generally speaking, RC is a versatile class of recurrent neural network (RNN) models (see \cite{Salehinejad2017} for a detailed survey). Although conventional RNNs are well-suited for handling sequence data and dynamic problems, estimating their weights during the training phase is inherently difficult~(\cite{Pascanu2013a, Doya92}). Reservoir networks stand out due to the fact that their inner weights can be {\it randomly generated} and {\it fixed}, and only the output (readout) layer weights are subject to estimation (supervised training). Echo State Network (ESN) is one of the most popular instances of RC models with provable universality, generalization properties (see \cite{RC6, RC7, RC9, RC10, RC12, RC20}, and references therein for more details), and excellent performance in forecasting, classification, and learning of dynamical systems (see \cite{Hart2021,RC21}). While conventional RNNs have been adopted for macroeconomic forecasting in a few instances (see, for example, \cite{Paranhos2021}), to the best of our knowledge, we are the first to explore easily-trainable reservoir models in this context.  

Our main contribution is three-fold. First, inspired by the remarkable empirical success of ESNs in prediction tasks, we propose the so-called Multi-Frequency Echo State Network (MFESN) framework, which allows multistep forecasting of the target variable at lower or the same frequencies as those of the input series. Second, we introduce two different approaches to predicting within the MFESN framework, namely  {\it Single-Reservoir MFESN} (S-MFESN) and  {\it Multi-Reservoir MFESN} (M-MFESN). S-MFESN is determined by modifying the ESN architecture to accommodate input and target variables of mixed frequencies. In M-MFESN, several Echo State Networks are adopted to handle input time series, each ESN corresponding to a group of input variables quoted at one given frequency. Finally, our third contribution consists of an extensive empirical comparative analysis of the forecasting capability of the proposed approaches in a concrete task of predicting the quarterly U.S. output growth. We inspect the forecasting capabilities of the MFESN framework compared to two well-established benchmarks widely used in the macroeconomic literature and among practitioners and show its empirical superiority in several thoroughly conducted forecasting exercises. Moreover, as a bi-product, we propose a new data aggregation scheme that allows bridging these two standard forecasting approaches, which is not available in the literature.

In our empirical study, we evaluate the multistep forecasting performance of the MFESN framework targeting quarterly U.S. output growth (Gross Domestic Product (GDP) growth) and utilizing a small- and medium-sized set of monthly and daily financial and macroeconomic variables. We compare the MFESN approach against two state-of-the-art methods, MIDAS and DFM, known for their ability to incorporate data of heterogeneous frequencies and utilize high-dimensional data inputs. The MIxed DAta Sampling (MIDAS) model developed in \cite{Ghysels2004,Ghysels2007} has been adopted widely for macroeconomic forecasting with mixed-frequency data (see for instance \cite{Clements2008, Clements2009, Ghysels2009, Francis2011,  Monteforte2012, Galvao2010, Galvao2013, Andreou2013, Ghysels2016, Jardet2020}). However, MIDAS is prone to curse-of-dimensionality problems and performs poorly when the set of predictors is of even moderate size (\cite{Clements2009,Kostrov2021essays}) due to optimization-related issues. Recently, some attempts have been made in the literature to overcome these issues by employing variable selection techniques under some additional assumptions. For instance, \cite{Babii2022} proposes the MIDAS projection approach, which is more amenable to high-dimensional data environments under the assumption of sparsity. Even with these improvements, practical high-dimensional implementations of MIDAS remain challenging. This is in part caused by the ragged edges of the ``raw'' macroeconomic data, incomplete observations, and uneven sampling frequencies. The relative inflexibility of MIDAS regression lag specifications makes integrating daily and weekly data at true calendar frequencies (that is, without interpolation or aggregation) very complex. State-space models effectively mitigate these issues.

A strong state-of-the-art state-space competitor for our MFESN framework is the Dynamic Factor Model (DFM), which has been first introduced in \cite{geweke1977dynamic} and \cite{sargent1977business}. DFMs have become the standard workhorse for macroeconomic nowcasting and prediction (for more details, we refer the reader to \cite{Stock1996, Stock2002, stock2016dynamic, Giannone2008, Banbura2011, Chauvet2015, Hindrayanto2016}). Conventional DFMs for data of multiple sampling frequencies are linear state-space models with a latent low-frequency process of interest and high-dimensional input time series. Although their linear structure lends itself to inference with likelihood-based methods and Kalman filtering, using DFMs in the high-dimensional setting is limited by the associated computational effort. For Gaussian state-space models, some of these issues are proposed to be handled with a more compact matrix representation as in \cite{DELLEMONACHE201922}. Still, in the particular settings of nowcasting and forecasting of GDP growth, the computational complexity is one of the main reasons why DFMs are rarely used with daily input series, see \cite{Banbura2013} for a detailed review and \cite{Aruoba2009} for a mixed-frequency DFM wherein the latent factor process is updated daily, with the highest input frequency being weekly. We address these numerical difficulties using novel Python libraries for auto-differentiation and using GPUs for parallel computing, which allow the estimation of DFMs even in instances of high-frequency input observations.  Further, to adapt the DFM to mixed frequency tasks, we propose a new DFM aggregation scheme with Almon polynomial structure that bridges MIDAS and the DFM for our forecasting comparison. To our knowledge, we are the first to present this aggregation scheme which reduces the number of parameters subject to estimation. In contrast, previous DFM such as in \cite{mariano2003new, Banbura2011, Camacho2010a, Frale2011a} commonly assume a fixed aggregation scheme a-priori depending on whether the macroeconomic variable is a flow or stock variable.

To carry out a fair comparison of our MFESN framework with the state-of-the-art MIDAS and DFM models, we designed two model evaluation settings that differ regarding whether the financial crisis of 2007-2008 is included in the estimation period or not. In the first forecasting setting, all the competing models are estimated using the data from January 1st, 1990, until December 31st, 2007. Their performance in the forecasting into and after the financial crisis period is assessed. In the second evaluation setting, fitting is done with data largely encompassing the crisis period, again from January 1st, 1990 but now up to December 31st, 2011. In both cases, the forecasting (testing) period spans time up to the COVID-19 pandemic events, namely the fourth quarter of 2019.
Along with the two state-of-the-art DFM and MIDAS models, we use the unconditional mean of the sample as a baseline benchmark against the reservoir models. We find that our ESN-inspired models attain comparable or much better performance than DFMs at a much lower computational cost, even for a relatively long forecasting horizon of four quarters. Additionally, ESNs do not suffer from curse-of-dimensionality problems, which are known to be pervasive for MIDAS models and hence consistently outperform them in a number of forecasting exercises.

The remainder of the paper is structured as follows. Section~\ref{Reservoir models} presents reservoir models and discusses their advantages, as well as estimation, hyperparameter tuning, penalization and nonlinear multistep forecasting. In Section~\ref{Multi-Frequency Echo State Models}, we introduce the Multi-Frequency Echo State Network (MFESN) framework, propose the single-reservoir and multi-reservoir MFESN models, and spell out their defining features. Section~\ref{sec:empirical_study} contains the empirical study of the comparative GDP forecasting performance of MFESNs with respect to the set of benchmark models. We assess one-step and multistep forecasting results in several setups, with a small and a medium-sized set of regressors. We fit models with data before and after the 2007-08 financial crisis, and with different estimation windows. Section~\ref{Conclusions} concludes and discusses future research avenues and applications. Finally, the Appendix contains information regarding data sources, forecasting figures and formal details regarding our forecasting setups. The Supplementary Appendix gives detailed information on the implementation of all models, robustness checks and provides additional figures.

\paragraph{Code.} Our code, the data, and all results presented in the paper are made available in the GitHub repository at {\href{https://github.com/RCEconModelling/Reservoir-Computing-for-Macroeconomic-Modelling}{github.com/RCEconModelling/Reservoir-Computing-for-Macroeconomic-Modelling}}.


\subsection{Notation}
\label{Notation}
We use the symbol $\mathbb{N}$ (respectively, $\mathbb{N}^+$) to denote the set of natural numbers with the zero element included  (respectively, excluded). $\mathbb{Z}$ denotes the set of all integers. We use $\mathbb{R}$ (respectively, $\mathbb{R}_+$) to denote the set of all (respectively, positive excluding zero element) reals. We abbreviate the set $[n]=\{1, \ldots, n\}$, with $ n \in \mathbb{N}^+$.

\paragraph{Vector notation.} 
A column vector is denoted by a bold lowercase symbol like $\boldsymbol{r}$ and $\boldsymbol{r} ^\top $ indicates its transpose. Given a vector $\boldsymbol{v} \in \mathbb{R}  ^n $, we denote its entries by $v_i$, with $i \in \left\{ 1, \dots, n
\right\} $; we also write $\boldsymbol{v}=(v _i)_{i \in \left\{ 1, \dots, n\right\} }$. The symbols $\boldsymbol{i} _n, \boldsymbol{0} _n  \in \mathbb{R}  ^n $ stand for the vectors of length $n$ consisting of ones and of zeros, respectively. Additionally, given $n \in 
\mathbb{N}^+  $,   $\boldsymbol{e}_n ^{(i)} \in \mathbb{R}  ^n$, $i \in \left\{ 1, \dots, n\right\} $ denotes the 
canonical unit vector of length $n$ determined by $\boldsymbol{e} _n ^{(i)} = (\delta _{ij} )_{j\in \left\{ 1, \dots, n\right\} }$. For any $\boldsymbol{v} \in \mathbb{R}  ^{n} $,  $\|\boldsymbol{v}\|  $ denotes its Euclidean norm.

\paragraph{Matrix notation.}
We denote by $\mathbb{M}_{n ,  m }$ the space of real $n\times m$ matrices with $m, n \in \mathbb{N}^+ $. When $n=m$, we use the symbols $\mathbb{M}_n $ and $\mathbb{D}  _n $ to refer to the space of square and diagonal matrices of order 
$n$, respectively. Given a matrix $A \in \mathbb{M}  _{n , m} $, we denote its components by $A _{ij} $ and we write $A=(A_{ij})$, with $i \in \left\{ 1, \dots, n\right\} $, $j \in \left\{ 1, \dots m\right\} $.  The symbol $\mathbb{I}_n \in \mathbb{D}  _n $ denotes the identity matrix, and the symbol $\mathbb{O}_n$ stands for the zero matrix of dimension $n$. For any $A \in \mathbb{M}  _{n , m} $,  $\|A\| _2  $ denotes its matrix norm induced by the Euclidean norms in $\mathbb{R}^m $ and $\mathbb{R} ^n $, and  $\|A\| _2=\sigma_{{\rm max}}(A)$, with $\sigma_{{\rm max}}(A)$  the largest singular value of $A$. 
\paragraph{Input and target stochastic processes.} 
We fix a probability space $(\Omega,\mathcal{A},\mathbb{P})$ 
on which all random variables are defined. 
The input and target signals are modeled by discrete-time stochastic processes $\boldsymbol{ z} = (\boldsymbol{ z}_t)_{t \in \mathbb{Z}}$ and $\boldsymbol{ y} = (\boldsymbol{ y}_t)_{t \in \mathbb{Z}}$ taking values in $\mathbb{R}^K$ and $\mathbb{R}^J$, respectively. Moreover, we write $\boldsymbol{ z}(\omega)=(\boldsymbol{ z}_t( \omega))_{t \in \mathbb{Z}}$ and $\boldsymbol{ y}(\omega)=(\boldsymbol{ y}_t( \omega))_{t \in \mathbb{Z}}$ for each outcome $\omega \in \Omega$ to denote the realizations or sample paths of $\boldsymbol{ z}$ and $\boldsymbol{ y}$, respectively. Since  $\boldsymbol{ z}$ can be seen as a random sequence in $\mathbb{R}^K$, we write interchangeably $\boldsymbol{ z}:{\mathbb{Z}} \times \Omega \longrightarrow \mathbb{R}^K$ and $\boldsymbol{ z}: \Omega\longrightarrow (\mathbb{R}^K)^{\mathbb{Z}}$. 
The same applies to the analogous assignments involving $\boldsymbol{y}$.

\paragraph{Temporal notation.} 
Let $(u_t)_{t \in I}$, $u_t \in \rset$ be a (scalar) time series with $I$ some index set (in this paper it will always be discrete). Time series $(u _t)_{t \in I}$ will be denoted just as $(u _t)$ when the index set $I$ is specified by the context. We write $u_{s_1:s_2}=(u_t)_{t\in \{s_1, \ldots, s_2\}}$ for integers $s_1<s_2$ and time series $(u_t)$. To define the concept of the sampling frequency, we must introduce an additional series, call it $(v_s)_{s\in J} $. The time index $J$ is not the same as $I$. We assume that $u_t$ is sampled at the coarsest rate; equivalently, it has the \textit{lowest} sampling frequency, which we call in what follows the \textit{reference frequency}. In practice, this means that in the same window of time, $u_t$ will be observed at most as frequently as $v_s$. The case when the sampling frequency of $v_s$ is strictly higher than that of $u_t$ is of primary interest.

We assume that all sampling happens in instants that are evenly spaced in time. Series other than the reference one and with higher sampling frequencies are given an additional time index, the \textit{tempo index}, written $t, \tempo{\ast}{{\kappa}} \,$, where $\kappa$ is the \textit{frequency multiplier}. Our tempo notation assumes that low- and high-frequency series are sampled with temporal \textit{alignment}: this means that the reference time index $t$ and the tempo index $\tempo{\ast}{{\kappa}}$ have the following properties.

\begin{definition}
\label{Temporal notation}
	A reference time index $t \in \mathbb{N}$ and a tempo index $\tempo{\ast}{{\kappa}}$ for a given high-frequency ${\kappa} \in \mathbb{N}^+$ are such that the following relations hold
	\begin{description}
		\item[(i)] $t, \tempo{0}{{\kappa}} \equiv t $
		\item[(ii)] $t, \tempo{{\kappa}}{{\kappa}} \equiv t+1$
		\item[(iii)] $t, \tempo{s}{{\kappa}} \equiv t + \lfloor{s / {\kappa}}\rfloor, \tempo{(s\mod{\kappa})}{{\kappa}} \quad {\rm for} \enspace \forall s \in \mathbb{N}$
		\item[(iv)] $t, \tempo{-s}{{\kappa}} \equiv (t - 1) - \lfloor{s / {\kappa}}\rfloor, \tempo{{\kappa} - (s\mod{\kappa})}{{\kappa}} \quad {\rm for} \enspace \forall  s \in \mathbb{N}$,
	\end{description}
	where $\mod$ is the modulo operation and for any $x\in \mathbb{R}$ the floor operator $\lfloor x \rfloor$ outputs the greatest  $z \in \mathbb{N}$ such that $z\leq x$. 
\end{definition}

Since we can exchange ``frequency'' and ``frequency multiplier'' in the tempo notation, we will make no distinction between the two terms in what follows. 

\paragraph{Forecasting schemes.} 
The theoretical setup and design of the forecasting exercises conducted in this paper are carefully discussed in Appendix~\ref{Forecasting Schemes}. There, we formally distinguish between the so-called high-frequency and low-frequecy forecasting in the presence of mixed-frequency data. For more details regarding time series forecasting with economic data, we also refer the reader to \cite{Clements2008,Clements2009,Chen2010,Jardet2020} and references therein.



\section{Reservoir Models}
\label{Reservoir models}
In this section, we introduce {\it reservoir computing} models \citep{Jaeger04} for forecasting of stochastic time series of a single frequency. We focus on a family of RC systems called {\it Echo State Networks} (ESNs), which have been successfully applied to forecasting of deterministic dynamical systems \citep{pathak:chaos, Pathak:PRL, wikner2021using, arcomano2022hybrid}. In the following, we discuss the linear estimation of ESN model parameters, the hyperparameters tuning, the loss penalty selection, and how to carry out nonlinear forecasting. 

\subsection{Reservoir Models}
\label{subsec:reservoir_models}
Reservoir computing (RC) models are nonlinear state-space systems that, in the forecasting setting, are defined by the following equations:
\begin{align}
    \boldsymbol{x}_t & = F(\boldsymbol{x}_{t-1}, \boldsymbol{z}_t), \label{state_map}\\
    \boldsymbol{y}_{t+1} & = h_{\boldsymbol{\theta}}(\boldsymbol{x}_t) + \boldsymbol{\epsilon}_t, \label{output_map}
\end{align}
for all $t \in \mathbb{Z}$, where the {\it state map} $F: \rset^N \times \rset^K \to \rset^N$, $N, K \in \nset^+$ is called also the \textit{reservoir map}, and the \textit{observation map} $h_{\boldsymbol{\theta}}: \rset^N \to \rset^J$, $J\in \mathbb{N}^+$ is referred to as  the \textit{readout} layer, parametrized by $\boldsymbol{\theta} \in \Theta$.	
Sequences $(\boldsymbol{z}_t)_{t \in \mathbb{Z}}$, $\boldsymbol{z}_t \in \mathbb{R}^K$, and $(\boldsymbol{y}_t)_{t \in \mathbb{Z}}$, $\boldsymbol{y}_t \in \mathbb{R}^J$, stand for the \textit{input} and the \textit{output (target)} of the system, respectively, and $(\boldsymbol{x}_t)_{t \in \mathbb{Z}}$, $\boldsymbol{x}_t \in \mathbb{R}^N$, are the associated \textit{reservoir states}.
In \eqref{output_map}, $(\boldsymbol{\epsilon}_t)_{t \in \mathbb{Z}}$ are $J$-dimensional independent zero-mean innovations with variance $\sigma_\epsilon^2\mathbb{I}_J$ that are also independent of $\boldsymbol{x}_t$ across all $t$.
Importantly, many families of RC systems have been proven to have universal approximation properties for $L^p$-integrable stochastic processes \citep{RC8}, and estimation and generalization error bounds have been established in \cite{RC10, RC12}. 

In the case of an ESN model, the state and observation equations \eqref{state_map}-\eqref{output_map} are given by 
\begin{align}
    \boldsymbol{x}_t & = \alpha \boldsymbol{x}_{t-1} + (1-\alpha) \sigma(A \boldsymbol{x}_{t-1} + C \boldsymbol{z}_t + \boldsymbol{\zeta}) \label{eq:esn_model_states} \\
    \boldsymbol{y}_{t+1} & =\boldsymbol{a} +  W^\top \boldsymbol{x}_t + \boldsymbol{\epsilon}_t, \label{eq:esn_model_output}
\end{align}
where $A \in \mathbb{M}_{N}$ is the \textit{reservoir matrix}, $C \in \mathbb{M}_{N,K}$ is the \textit{input matrix},  $\boldsymbol{\zeta} \in \rset^N$ is the \textit{input shift}, $\alpha \in [0,1)$ is the \textit{leak rate} and $W \in \mathbb{M}_{N,J}$ are the \textit{readout coefficients}. The map $\sigma: \rset \to \rset$ is an activation function applied elementwise, which in what follows we take to be the hyperbolic tangent. We refer to $A$, $C$, $\boldsymbol{\zeta}$ as \textit{state parameters} that are randomly generated. Notice that if $A = 0$ and $\alpha = 0$ the state equation reduces to a nonlinear regression model with random coefficients (or a feedforward neural network with random weights) which is usually referred to as an {\it  Extreme Learning Machine} \citep{caoReviewNeuralNetworks2018, RC12}. 

\paragraph{Properties of ESN models.}
We focus on ESNs with the so-called {\it  echo state property (ESP)}, that is, when for any  $ \boldsymbol{z} \in (\mathbb{R}^K)^{\mathbb{Z}}$ there exists a unique $ \boldsymbol{y} \in (\mathbb{R}^J)^{\mathbb{Z}}$ such that \eqref{eq:esn_model_states}-\eqref{eq:esn_model_output} hold (see \cite{RC6, RC7, RC9} and references therein). One can require that the ESP holds only on the level of the state equation, that is for any input sequence $ \boldsymbol{z} \in (\mathbb{R}^K)^{\mathbb{Z}}$ there exists a unique state sequence $ \boldsymbol{x} \in (\mathbb{R}^N)^{\mathbb{Z}}$ such that \eqref{eq:esn_model_states} holds. The result in Corollary 3.2 in \citet{RC7}, which is also valid for the case of ESNs with the leak rate, shows that the sufficient condition of the ESP associated with \eqref{eq:esn_model_states} to hold is $\left\|A\right\|_2 L _\sigma<1 $ where $L _\sigma$ is the Lipschitz constant of the activation function $\sigma$ (in our setting, $L _{tanh}=1$). This sufficient ESP condition has been extensively studied in the ESN literature; see \cite{jaeger2001, Jaeger04, Buehner:ESN, zhang:echo, Yildiz2012, Wainrib2016, Manjunath:Jaeger} for more details. The result in Corollary 3.2 in \citet{RC7} also shows that this condition implies the so-called {\it fading memory property} \citep{Boyd1985}, which from the practical point of view means that the impact of initial $\boldsymbol{x}_0$ is negligible for sufficiently long samples. 

In the stochastic setting, part (i) of Proposition 4.2 in \citet{RC16} proves that the condition $\left\|A\right\|_2<1$  guarantees variance stationarity of the states associated with variance stationary inputs. Moreover, \cite{RC27} show that this condition implies the so-called stochastic state contractivity ensuring a stochastic analog of the ESP. Notably, violations of $\left\|A\right\|_2 <1$  do not have detrimental implications for the performance of ESNs in various learning tasks, as reported in multiple empirical studies.

\paragraph{Computational advantages of ESNs.}
We emphasize that the core computational advantage of ESNs is that state parameters $A$, $C$, and $\boldsymbol{\zeta}$ are randomly sampled and need not be estimated. Additionally, since observation equation \eqref{eq:esn_model_output} is linear in $\boldsymbol{x}_t$, coefficients $W$ can be estimated via (penalized) least squares regression, as we explain in the following subsection. The choice of properties of state parameters determines memory properties and forecasting performance of linear \citep{RC23} and nonlinear ESNs \citep{RC15} as we discuss in Section~\ref{subsec:Hyperparameter Tuning}.

\subsection{Estimation}
\label{Estimation}
We now discuss in detail the estimation of coefficients $W$ in \eqref{eq:esn_model_output}. Let a sample $(\boldsymbol{z}_t, \boldsymbol{y}_t)_{t=1}^T$ of input and target pairs be available. Given an initial state $\boldsymbol{x}_0$, the reservoir states can be computed iteratively according to state equation \eqref{eq:esn_model_states} as:
\begin{equation*}
	\boldsymbol{x}_1 
	= \alpha \boldsymbol{x}_{0} + (1-\alpha) \sigma(A \boldsymbol{x}_{0} + C \boldsymbol{z}_1 + \boldsymbol{\zeta}) , 
	\quad \ldots, \quad
	\boldsymbol{x}_T 
	= \alpha \boldsymbol{x}_{T-1} + (1-\alpha) \sigma(A \boldsymbol{x}_{T-1} + C \boldsymbol{z}_T + \boldsymbol{\zeta}) .
\end{equation*}
Collect the states and the targets into the state and the observation matrices, respectively, as
\begin{equation*}
	X = (\boldsymbol{x}_1, \boldsymbol{x}_2, \ldots, \boldsymbol{x}_{T-1})^\top \in \mathbb{M}_{T-1,N}, 
	\qquad 
	Y = (\boldsymbol{y}_2, \boldsymbol{y}_3, \ldots, \boldsymbol{y}_T)^\top \in \mathbb{M}_{T-1,J} .
\end{equation*}
Consider the ridge regression estimator for $W$ given by
\begin{equation}\label{eq:ridge_regression}
	\widehat{W}_\lambda 
	\: := \: 
	\argmin_{W \in \mathbb{R}^N} \sum_{t=1}^{T-1} \left\lVert \boldsymbol{y}_{t+1} - {W}^\top \boldsymbol{x}_t \right\rVert_2^2 + \lambda \lVert {W} \rVert_2^2
	\: = \: 
	\left( X^\top X + \lambda ((T-1)\, \mathbb{I}_N) \right)^{-1} X^\top Y ,
\end{equation}
where $\lambda \in \mathbb{R}_+$ is the ridge penalty strength. When $\lambda \to 0$, the estimator $\widehat{W}	_\lambda$ converges to the minimum-norm least squares solution \citep{ishwaranGeometryPropertiesGeneralized2014b}. 
In applications, ridge regression is the most commonly used estimation method applied to ESNs, as it provides a straightforward regularization scheme both when $N < T$ and $N \geq T$. This is especially important since in practice the ESN state dimension is often chosen to be $10^3$--$10^4$ (see for example \cite{pathak:chaos}). Additionally, a virtue of the ridge regression problem is the fact that the associated objective function is convex and, hence, it can be efficiently solved using stochastic gradient descent even when ${\rm min}\{N, T\}$ is large and one decides against the closed-form solution~\eqref{eq:ridge_regression}. Finally, as mentioned in the properties of reservoir systems in Subsection~\ref{subsec:reservoir_models}, we notice that in the presence of the fading memory property, the estimation does not depend significantly on the choice of $\boldsymbol{x}_0$ as sample size $T$ increases. 

We refer to \eqref{eq:ridge_regression} as the \textit{fixed-parameter} estimator. In our empirical analyses, we also implement \textit{expanding} and \textit{rolling window} estimation strategies which update $\widehat{W}_\lambda$ as new observations become available (we refer the reader to Supplementary Appendix \ref{sec:estimation_setup} for details). In the rest of the paper, for brevity, we use $\widehat{W}$ to denote the ridge estimator of coefficients $W$ assuming that the appropriate choice of the penalty strength $\lambda$ is made for each concrete situation.

\subsubsection{Hyperparameter Tuning}
\label{subsec:Hyperparameter Tuning}

As discussed in Subsection~\ref{subsec:reservoir_models}, the performance of ESNs depends on the choice of randomly drawn state parameters $A$, $C$, $\boldsymbol{\zeta}$. Much work has been put into determining optimal specifications (see for example \cite{Rodan2011, Goudarzi2016, farkas:bosak:2016, GHLO2014_capacity,  RC3, RC15}). We construct these parameters by first sampling $\widetilde{A}$, $\widetilde{C}$ and $\widetilde{\boldsymbol{\zeta}}$ from appropriately chosen laws. Then, we normalize each element of the tuple such that
\begin{equation}
\label{normalizing}
	\overline{A} = {\widetilde{A}}/{\rho(\widetilde{A})}
	, \quad
	\overline{C} =  {\widetilde{C}}/{\norm{\widetilde{C}}}
	, \quad
	\overline{\boldsymbol{\zeta}} = {\widetilde{\boldsymbol{\zeta}}}/{\norm{\widetilde{\boldsymbol{\zeta}}}}
	,
\end{equation}
where $\rho(\widetilde{A})$ denotes the spectral radius of $\widetilde{A}$. As discussed in the properties of reservoir systems in Subsection~\ref{subsec:reservoir_models}, the sufficient condition of the ESP is $\|A\|_2<1$. By this normalizing choice, we allow for some more flexibility in terms of marginal violations of the non-sharp ESP constraint. Finally, defining $A= \rho \overline{A}$, $C=\gamma \overline{C}$, and $\boldsymbol{\zeta}=\omega \overline{\boldsymbol{\zeta}}$, we can rewrite state equation \eqref{eq:esn_model_states} as
\begin{equation}\label{eq:esn_hyperparams}
	\boldsymbol{x}_t =  \alpha \boldsymbol{x}_{t-1} + (1-\alpha) \sigma(\rho \overline{A} \boldsymbol{x}_{t-1} + \gamma \overline{C} \boldsymbol{z}_t + \omega \overline{\boldsymbol{\zeta}}) .
\end{equation}
We refer to tuple $\boldsymbol{\varphi}:=(\alpha, \rho, \gamma, \omega)$ as the \textit{hyperparameters} of the ESN. Specifically, $\alpha \in [0,1)$ is the leak rate and $\rho \in \mathbb{R}_+$ is called the \textit{spectral radius} of the reservoir matrix, $\gamma \in \rset_+$ is the \textit{input scaling}, and $\omega \in \rset_+$ is the \textit{shift scaling}. 
The choice of the hyperparameters determines the properties of the state map. For simplicity, in Section~\ref{sec:empirical_study}, we choose the hyperparameters based on the empirical ESN literature. In Supplementary Appendix \ref{Hyperparameter Tuning}, we also propose a general though more computationally intensive procedure to select hyperparameters in a data-driven way that could be interesting to practitioners.

\subsubsection{Penalty Selection}

To apply ridge estimator \eqref{eq:ridge_regression}, it is necessary to first select a penalty $\lambda$. Cross-validation (CV) is a common selection procedure for regularization strength in penalized methods such as ridge, LASSO, and Elastic Net. CV techniques have also been applied in the time series context \citep{kockPenalizedTimeSeries2020,Ballarin2023a} with their validity established in \cite{bergmeir2018}. 

In our empirical study, to account for temporal dependence, we use a sequential CV strategy with ten validation folds.
More precisely, we reserve the last 50 observations for validation and all other previous data points for training. The first fold consists of the first five observations out of the validation set, and the model is fitted using all training data. The following validation fold comprises the next five subsequent validation observations while the training set is expanded by five data points (from the previous fold). This procedure is repeated ten times and the CV loss is the average of the one-step-ahead forecast MSE on each fold.
In expanding or rolling window setups, we rerun the CV penalty selection to ensure that estimated ESN coefficients do not induce oversmoothing. We refer the reader to Supplementary Appendix~\ref{sec:cross_validation} for additional details.

\subsection{Relation to Nonparametric Regression}

Together with hyperparameters and penalty strength selection, the choice of the state dimension $N$ is a key ingredient of an ESN model. A large state space generally implies better approximation bounds \citep{RC12, RC24}. Although it is customary in the empirical literature to take $N$ as large as possible \citep{Lukosevicius2012}, some recent literature discusses both the statistical risk bounds and the approximation-risk trade-off bounds for various RC families (see \cite{RC10} and \cite{RC24} for details). Under simplified assumptions that $\alpha = 0$ and $\rho = 0$ in \eqref{eq:esn_hyperparams}, ESNs have a natural connection to random-weights neural networks \citep{caoReviewNeuralNetworks2018} and random projection regression \citep{maillardLinearRegressionRandom2012}, and are thus comparable to nonparametric sieve methods. If the data were independently sampled, known results on sieve estimation would require that at most $N/T = o(1)$ up to logarithmic factors for consistency \citep{belloniNewAsymptoticTheory2015a}. \cite{chenOptimalUniformConvergence2015} have extended this result to $\beta$-mixing data with B-spline and wavelet sieves.  
Sieve rates appear to suggest that choosing $N = O(T)$ in echo state networks could lead to nontrivial forecasting bias owing to poor approximation properties. Unfortunately, this comparison relies on neglecting the dynamic component of the ESN model, and as such it is only qualitative. It is, therefore, an important topic for future research.
	
A different but related problem is the potential degradation of forecasting performance when a model is at the interpolation threshold in the overparametrized regime, $N \geq T$. Ridge regression is also commonly applied to address generalization concerns in statistical learning (see \cite{hastie2009elements}). Recent work has studied more in-depth the link between regularization and generalization: \cite{hastieSurprisesHighdimensionalRidgeless2022} show that ``ridgeless", that is interpolation, solutions can be optimal in some scenarios. However, in our empirical evaluations in Section~\ref{sec:empirical_study}, cross-validation consistently selects non-zero ridge penalties, confirming that ridge penalization plays an important role in ESN forecasting performance.

\subsection{ESN Forecasting}\label{subsec:esn_forecasting}

We are primarily interested in using ESN models to construct conditional forecasts of target variables. Given that the conditional mean is the best mean square error estimator for $h$-step-ahead target $\boldsymbol{y}_{t+h}$, $h \geq 1$, our main focus is approximating
\begin{equation*}
     \widehat{\boldsymbol{y}}_{t+h|t} := \E\left[\boldsymbol{y}_{t+h} | \boldsymbol{x}_{0:t}, \boldsymbol{z}_{0:t}\right].
\end{equation*}
The case $h=1$ is trivial, since the ESN model is estimated by regressing $\boldsymbol{y}_{t+1}$ on state $\boldsymbol{x}_t$, and thus we can set $\widetilde{\boldsymbol{y}}_{t+1|t} = \widehat{W}^\top \boldsymbol{x}_t $. However, when $h > 1$ the nonlinear state dynamics precludes a direct computation of the conditional mean. This is in contrast to linear models like VARMAs or DFMs, where the assumption of linearity implies that conditional expectations reduce to simple matrix-vector operations. In particular, linear models are such that the variance (and any other higher-order moments) of the noise term do not impact the conditional mean forecast. 

Let $p_{{\theta}}(\boldsymbol{x}_t \vert \boldsymbol{x}_{t-1}, \boldsymbol{z}_t)$ and $g_{\theta}(\boldsymbol{y}_{t+1} \vert \boldsymbol{x}_t)$ be the state transition  and observation densities, respectively. Then, for $h > 1$,
\begin{equation}\label{eq:esn_condexp_integral}
    \widehat{\boldsymbol{y}}_{t+h|t} = 
    \int \boldsymbol{y}_{t+h} \; g_{{\theta}}(\boldsymbol{y}_{t+h} \vert \boldsymbol{x}_{t+h-1}) \prod_{j=1}^{h-1} p_{{\theta}}(\boldsymbol{x}_{t+j} \vert \boldsymbol{x}_{t+j-1}, \boldsymbol{z}_{t+j}) 
    \nu(\boldsymbol{z}_{t+j} \vert \boldsymbol{x}_{t+j-1}) \rmd \boldsymbol{z}_{t+j} \rmd \boldsymbol{x}_{t+j} \rmd \boldsymbol{y}_{t+h}
    ,
    %
\end{equation}
where $\nu(\boldsymbol{z}_{t+j} \vert \boldsymbol{x}_{t+j-1})$ is the conditional density of inputs. Here, we introduce the additional assumption that $\boldsymbol{x}_{t+j-1}$ is sufficient to condition on past states and inputs, that is 
\begin{equation}
    \nu(\boldsymbol{z}_{t+j} \vert \boldsymbol{x}_{t+j-1}) \equiv \nu(\boldsymbol{z}_{t+j} \vert \boldsymbol{x}_{0:t+j-1}, \boldsymbol{z}_{0:t+j-1}).
\label{eq:state_conditioning}
\end{equation}
Some elements in the expectation integral are not directly available. Specifically, while an ESN explicitly models both $p_{{\theta}}(\boldsymbol{x}_t \vert \boldsymbol{x}_{t-1}, \boldsymbol{z}_t)$ and $g_{{\theta}}(\boldsymbol{y}_{t+1} \vert \boldsymbol{x}_t)$, the density $\nu(\boldsymbol{z}_{t+j} \vert \boldsymbol{x}_{t+j-1})$ is unavailable.

In the remaining part of this subsection, we present a novel ESN-based approach to forecasting the target variable. Our idea is to enrich the ESN model with an auxiliary observation equation for the input covariates. As we demonstrate in Section~\ref{sec:empirical_study}, our proposed method shows superior performance with respect to the standard state-of-the-art benchmarks.

\subsubsection{Multi-step Forecasting of Targets via Iterative Forecasting of Inputs} 

In general, we are interested in constructing forecasts of target variables that are not the same as the model inputs. To do so, we resolve the issue of the intractability of \eqref{eq:esn_condexp_integral} while simultaneously capitalizing on the available results using ESNs in the forecasting of dynamical systems. More explicitly, we add to the ESN specification \eqref{eq:esn_model_states}-\eqref{eq:esn_model_output}  an equation that allows sidestepping modeling the density $\nu$ directly, thus making the computation of $\widehat{\boldsymbol{y}}_{t+h|t}$ feasible even when $h > 1$. 

Consider the ESN where the reservoir states $(\boldsymbol{x}_t)_{t \in \mathbb{Z}}$ follow \eqref{eq:esn_model_states}, while the target sequence is the same as the input sequence $(\boldsymbol{z}_t)_{t \in \mathbb{Z}}$, 
\begin{align}
    \boldsymbol{x}_t & = \alpha \boldsymbol{x}_{t-1} + (1-\alpha) \sigma(A \boldsymbol{x}_{t-1} + C \boldsymbol{z}_t + \boldsymbol{\zeta}) \label{eq:autonomous_esn_states} \\
    \boldsymbol{z}_{t+1} & = \mathcal{W}^\top \boldsymbol{x}_t + \boldsymbol{u}_{t+1}. \label{eq:autonomous_esn_obs}
\end{align}
Here, we use symbol $\mathcal{W}$ for the output coefficients to separate this case from the general ESN equations \eqref{eq:esn_model_states}-\eqref{eq:esn_model_output}. In \eqref{eq:autonomous_esn_obs}, $(\boldsymbol{u}_t)_{t \in \mathbb{Z}}$ are $K$-dimensional independent zero-mean innovations with variance $\sigma_u^2\mathbb{I}_K$ that are also independent of $\boldsymbol{x}_t$ across all $t$. 

In this case, the reservoir map $F(\boldsymbol{x}_{t-1}, \boldsymbol{z}_{t})$ in \eqref{state_map} is determined by \eqref{eq:autonomous_esn_states}, and it is possible to re-feed the forecasted variables back into the state equation as inputs. This yields the following state recursion:
\begin{equation*}
    \boldsymbol{x}_t = F(\boldsymbol{x}_{t-1}, \mathcal{W}^\top \boldsymbol{x}_{t-1} + \boldsymbol{u}_{t}) =: G_{\theta}(\boldsymbol{x}_{t-1}, \boldsymbol{u}_{t}),
\end{equation*}
where the subscript $\theta$ denotes the dependence on the model coefficients. In the reservoir computing literature, regimes, where the ESN state equation is iteratively fed with the model outputs, are called ``autonomous'' \citep{RC15}. They are widely and successfully utilized for the prediction of deterministic dynamical systems. Indeed, in those instances, provided that the ridge estimate $\widehat{\mathcal{W}}$ is available from data according to Subsection~\ref{Estimation},  the $h > 1$ steps autonomous state iteration is given by 
\begin{equation*}
     {F}^{*}_{\theta}(\boldsymbol{x}_{t}) := \alpha \boldsymbol{x}_{t} + (1-\alpha) \sigma((A + C \widehat{\mathcal{W}}^\top) \boldsymbol{x}_t + \boldsymbol{\zeta}) 
\end{equation*}
and
\begin{equation*}
    \boldsymbol{x}_{t+h} = \underbrace{{F}^{*}_{\theta} \circ {F}^{*}_{\theta} \circ \cdots \circ {F}^{*}_{\theta} }_{h \textnormal{ times}} \; (\boldsymbol{x}_t ).
\end{equation*}
Hence one can directly obtain the $h$-steps ahead predictions of the input time series as  $\boldsymbol{z}_{t+h} = \widehat{\mathcal{W}}^\top \boldsymbol{x}_{t+h-1}$.

In the case of stochastic target variables, assuming \eqref{eq:state_conditioning}, we notice that for the conditional forecast of the states, it holds that 
\begin{equation}\label{eq:esn_condexp_nu_int}
     \widehat{\boldsymbol{x}}_{t+1|t} = \E\left[\boldsymbol{x}_{t+1} | \boldsymbol{x}_{0:t}, \boldsymbol{z}_{0:t}\right] = \int \boldsymbol{x}_t \, p_{{\theta}}(\boldsymbol{x}_t \vert \boldsymbol{x}_{t-1}, \boldsymbol{z}_t) \nu(\boldsymbol{z}_t \vert \boldsymbol{x}_{t-1}) \rmd \boldsymbol{z}_t
    =
    \int G_{{\theta}}(\boldsymbol{x}_{t-1}, \boldsymbol{u}_t) \phi(\boldsymbol{u}_t) \rmd \boldsymbol{u}_t,
\end{equation}
where density $\phi$ of $\boldsymbol{u}_t$ is, again, unavailable. Note that, even under the assumption $\boldsymbol{u}_t \sim \mathcal{N}(\boldsymbol{0}, \Sigma_{\boldsymbol{u}})$, which is standard in the filtering literature, the presence of nonlinear map $G_{{\theta}}$ makes the computation of the forecasts of $\boldsymbol{z}_{t+h}$ a non-straightforward exercise. Nevertheless, this forecast construction can be readily used when one is interested exclusively in predicting the time series $\boldsymbol{z}_t$.

Whenever the final goal of the exercise is forecasting some other explained variable $\boldsymbol{y}_{t+h}$ $h$-steps ahead, additional issues arise. In this case, one needs to compute the conditional expectation in \eqref{eq:esn_condexp_integral} which is intractable even under Gaussian assumptions on the innovations. One option is to apply particle filtering techniques such as bootstrap sampling or sequential importance sampling (SIS) to evaluate the expectation \citep{doucet2001sequential}. We emphasize that the state dimension is usually chosen to be large, and hence implementing filtering techniques requires some care.

Our approach is to avoid dealing with the nonlinear densities involved in \eqref{eq:esn_condexp_integral} with the help of \eqref{eq:esn_condexp_nu_int} and, instead, to reduce the computation of the conditional expectation $\widehat{\boldsymbol{y}}_{t+h|t}$ to a composition of functions. By the linearity of observation equation \eqref{eq:esn_model_output} and the assumption of independence in the zero-mean noise $\boldsymbol{\epsilon}_{t+h}$, we write
\begin{align}
    \widehat{\boldsymbol{y}}_{t+h|t} 
    & = W^\top \widehat{\boldsymbol{x}}_{t+h-1|t} =
    \bigintssss W^\top \boldsymbol{x}_{t+h-1} \prod_{j=1}^{h-1} p_{{\theta}}(\boldsymbol{x}_{t+j} \vert \boldsymbol{x}_{t+j-1}, \boldsymbol{z}_{t+j}) 
    \nu(\boldsymbol{z}_{t+j} \vert \boldsymbol{x}_{t+j-1}) \rmd \boldsymbol{x}_{t+j} \rmd \boldsymbol{z}_{t+j} \nonumber
\end{align}
and use the approximation
\begin{align}
\widehat{\boldsymbol{y}}_{t+h|t}  \approx \widetilde{\boldsymbol{y}}_{t+h} = W^\top \underbrace{{F}^{*}_{{\theta}} \circ {F}^{*}_{{\theta}} \circ \cdots \circ {F}^{*}_{{\theta}} }_{h-1 \textnormal{ times}} \; (\boldsymbol{x}_{t}), \label{eq:esn_forecast_integral_approx}
\end{align}
which originates from
\begin{equation}\label{eq:esn_condexp_state_auto_approx}
     \widehat{\boldsymbol{x}}_{t|t-1} = \int G_{{\theta}}(\boldsymbol{x}_{t-1}, \boldsymbol{u}_t) \phi(\boldsymbol{u}_t) \rmd \boldsymbol{u}_t
    \approx
    G_{{\theta}}(\boldsymbol{x}_{t-1}, \E[\boldsymbol{u}_t])
    = 
    F( \boldsymbol{x}_{t-1}, \mathcal{W}^\top \boldsymbol{x}_{t-1} )
    \equiv
    {F}^{*}_{{\theta}}( \boldsymbol{x}_{t-1} ),
\end{equation}
where $\boldsymbol{u}_t$ is assumed to be zero-mean. The validity of \eqref{eq:esn_condexp_state_auto_approx} itself requires implicit assumptions on the nature of the distribution of $\boldsymbol{u}_t$, but here we want to keep the analysis of $\widehat{\boldsymbol{y}}_{t+h|t}$ to a minimum, and just use the insights from the dynamical systems ESN literature. We are hence not delving deeper into alternative approaches to estimate forecasts or, more generally, to compute conditional expectations of ESN models with stochastic inputs.

\section{Multi-Frequency Echo State Models}
\label{Multi-Frequency Echo State Models}

In this subsection, we construct a broad class of ESN models that can accommodate input and target time series sampled at distinct sampling frequencies. We call this family of reservoir models the \textit{Multi-Frequency Echo State Networks} (MFESNs). The state-space structure of MFESNs is naturally amenable to the setting of time series with mixed frequencies. Additionally, the prediction strategy discussed in Section~\ref{subsec:esn_forecasting} is straightforward to extend to MFESNs.

We present two groups of MFESN architectures. The first family is based on a single echo state network architecture and we call these models \textit{Single-Reservoir Multi-Frequency Echo State Networks} (S-MFESNs). The second group, referred to as \textit{Multi-Reservoir Multi-Frequency Echo State Networks} (M-MFESNs), allows for as many state equations as the number of distinct sampling frequencies present in the input data. 

\subsection{Single-Reservoir MFESN}\label{subsec:Single-reservoir MFESN}

Recall that, in the temporal notation of Definition~\ref{Temporal notation}, we reserve $t$ to be the reference time index, which is also used for the target variable, and all other frequencies will be measured with respect to the reference frequency.

Consider $L$ collections of different time series. We assume that the $l$th collection, $l\in [L]$, consists of $n_l$ time series that are sampled at a common frequency $\kappa_l$ and contain observations $(\boldsymbol{z}^{(l)}_{t,\tempo{s}{\kappa_l}})_{t,s}$ with $\boldsymbol{z}^{(l)}_{t,\tempo{s}{\kappa_l}} \in \rset^{n_l}$ for all $t \in \mathbb{Z}$ and $s \in \{0, \ldots, \kappa_l-1\}$. Let $\kappa_{\max} = \max_l \kappa_l$ be the highest sampling frequency among the $L$ time series groups and let $q_l := \kappa_{\max} / \kappa_l$ indicate how low each $\kappa_l$ sampling frequency is with respect to $\kappa_{\max}$. We can now stack together and repeat the observations in a way that is consistent with the high-frequency index by defining
\begin{equation*}
    \boldsymbol{z}_{t,\tempo{s}{\kappa_{\max}}} := 
    \left( \boldsymbol{z}^{(1)\,\top}_{t,\tempo{\lfloor s / q_1 \rfloor}{\kappa_1}} , 
    \: \boldsymbol{z}^{(2)\,\top}_{t,\tempo{\lfloor s / q_2 \rfloor}{\kappa_2}} , 
    \: \ldots , 
    \: \boldsymbol{z}^{(L)\,\top}_{t,\tempo{ \lfloor s / q_L \rfloor}{\kappa_L}} \right)^\top\in \mathbb{R}^{\sum_{l=1}^{L}n_l}, \enspace s\in \{0, \ldots, \kappa_{\max}-1\},
\end{equation*}
where for all $l\in [L]$, $\boldsymbol{z}^{(l)}_{0,\tempo{0}{\kappa_l}} = \mathbf{0}_{n_l}$.
Thus, it is possible to write a single high-frequency ESN as 
\begin{align}
    \boldsymbol{x}_{t,\tempo{s}{\kappa_{\max}}} &= \alpha \boldsymbol{x}_{t,\tempo{s-1}{\kappa_{\max}}} + (1-\alpha) \sigma(A \boldsymbol{x}_{t,\tempo{s-1}{\kappa_{\max}}} + C \boldsymbol{z}_{t,\tempo{s}{\kappa_{\max}}} + \boldsymbol{\zeta}),\label{single ESN state}\\
    \boldsymbol{z}_{t,\tempo{s+1}{\kappa_{\max}}} & = \mathcal{W}^\top \boldsymbol{x}_{t,\tempo{s}{\kappa_{\max}}} + \boldsymbol{u}_{t,\tempo{s+1}{\kappa_{\max}}},\label{single ESN input obs}
\end{align}
where $\mathcal{W}\in \mathbb{M}_{N, \sum_{l=1}^{L}n_l}$ and $s >0$.
We term this class of MFESN models the \textit{Single-Reservoir Multi-Frequency ESNs} (S-MFESNs).

Notice that equations \eqref{single ESN state}-\eqref{single ESN input obs} of the S-MFESN model prescribe the dynamics at the highest frequency, $\kappa_{\max}$.
In order to forecast a lower frequency target, we map high-frequency states $\boldsymbol{x}_{t,\tempo{s}{\kappa_{\max}}}$ to low-frequency targets $\boldsymbol{y}_{t+1} \in \mathbb{R}^J$ by introducing a \textit{state alignment} scheme.
An \textit{aligned} S-MFESN uses the most recent state with respect to the reference time index $t$ to construct the forecast. More precisely, the state equation of an S-MFESN is iterated $\kappa_{\max}$ times until the state $\boldsymbol{x}_{t-1,\tempo{\kappa_{\max}}{\kappa_{\max}}} = \boldsymbol{x}_{t,\tempo{0}{\kappa_{\max}}}$ is obtained and then target $\boldsymbol{y}_{t+1}$ is forecast with observation equation
\begin{equation}
\label{single ESN obs}
    \boldsymbol{y}_{t+1} = W^\top \boldsymbol{x}_{t,\tempo{0}{\kappa_{\max}}} + \boldsymbol{\epsilon}_{t+1}, \quad W\in \mathbb{M}_{N,J}.
\end{equation}
\paragraph{Estimation of aligned S-MFESN.}
Both coefficient matrices  ${W}$ and ${\mathcal{W}}$ can be estimated as explained in Subsection~\ref{Estimation} under appropriate choices of corresponding penalty strengths. In particular, in order to obtain $\widehat{\mathcal{W}}$, the state and the observation matrices in \eqref{eq:ridge_regression} are given by 
\begin{align*}
	X_{\kappa_{\max}} &= (\boldsymbol{x}_{1,\tempo{0}{\kappa_{\max}}}, \ldots, \boldsymbol{x}_{1,\tempo{\kappa_{\max}-1}{\kappa_{\max}}}, \ldots, \boldsymbol{x}_{T-1,\tempo{0}{\kappa_{\max}}}, \ldots, \boldsymbol{x}_{T-1,\tempo{\kappa_{\max}-1}{\kappa_{\max}}})^\top \in \mathbb{M}_{(T-1)\kappa_{\max}-1,N},\\
 	Y_{\kappa_{\max}} &= (\boldsymbol{z}_{1,\tempo{1}{\kappa_{\max}}}, \ldots, \boldsymbol{z}_{1,\tempo{\kappa_{\max}}{\kappa_{\max}}}, \ldots, \boldsymbol{z}_{T-1,\tempo{1}{\kappa_{\max}}}, \ldots, \boldsymbol{z}_{T-1,\tempo{\kappa_{\max}}{\kappa_{\max}}})^\top \in \mathbb{M}_{(T-1)\kappa_{\max}-1,\sum_{l=1}^{L}n_l},
\end{align*}
while
\begin{align*}
 	X &= \left(\boldsymbol{x}_{1,\tempo{0}{\kappa_{\max}}}, \boldsymbol{x}_{2,\tempo{0}{\kappa_{\max}}}, \ldots, \boldsymbol{x}_{T-1,\tempo{0}{\kappa_{\max}}}\right)^\top \in \mathbb{M}_{T-1,N},\\
 	Y &= \left(\boldsymbol{y}_{2}, \ldots, \boldsymbol{y}_{T}\right)^\top \in \mathbb{M}_{T-1,J},
\end{align*}
are used for the estimation of $\widehat{W}$. We note that the state equation \eqref{single ESN state} of S-MFESN can be initialized by $\boldsymbol{x}_{0,\tempo{0}{\kappa_{\max}}}$, which under the fading memory property is inconsequential for long enough samples (see the discussion in Subsection~\ref{Reservoir models}).

\paragraph{Forecasting with aligned S-MFESN.}

Let $\widehat{W}$ and $\widehat{\mathcal{W}}$ be the sample estimates of the readout matrices as explained above. The fitted  high-frequency autonomous state transition map associated with \eqref{single ESN state} is given by 
\begin{equation}
\label{single ESN state map}
    F_{\kappa_{\max}}(\boldsymbol{x}_{t,\tempo{s-1}{\kappa_{\max}}}) := \alpha \boldsymbol{x}_{t,\tempo{s-1}{\kappa_{\max}}} + (1-\alpha) \sigma\left( (A + C \widehat{\mathcal{W}}^\top) \boldsymbol{x}_{t,\tempo{s-1}{\kappa_{\max}}} + \boldsymbol{\zeta} \right),
\end{equation}
which, composed with itself exactly $\kappa_{\max}$ times, yields the target-frequency-aligned autonomous state transition map
\begin{equation}
\label{single ESN state aligned}
    F(\boldsymbol{x}_{t,\tempo{0}{\kappa_{\max}}}) := \, \underbrace{F_{\kappa_{\max}} \circ F_{\kappa_{\max}} \circ \cdots \circ F_{\kappa_{\max}}}_{\kappa_{\max} \textnormal{ times}}\, ( \boldsymbol{x}_{t, \tempo{0}{\kappa_{\max}}} ).
\end{equation}
Finally, from \eqref{eq:esn_forecast_integral_approx} the $h$-steps ahead low-frequency forecasts, $h\in \mathbb{N}$, can be computed as
\begin{align}
\label{single ESN forecast}
    \widetilde{y}_{T+h \vert T} & = \widehat{W}^\top \big( \underbrace{F_{ } \circ F \circ \cdots \circ F}_{h-1 \textnormal{ times}}\, ( \boldsymbol{x}_{T, \tempo{0}{\kappa_{\max}}} ) \big).
\end{align}

\begin{figure}[!t]
	\centering
	\includegraphics[width=0.97\linewidth]{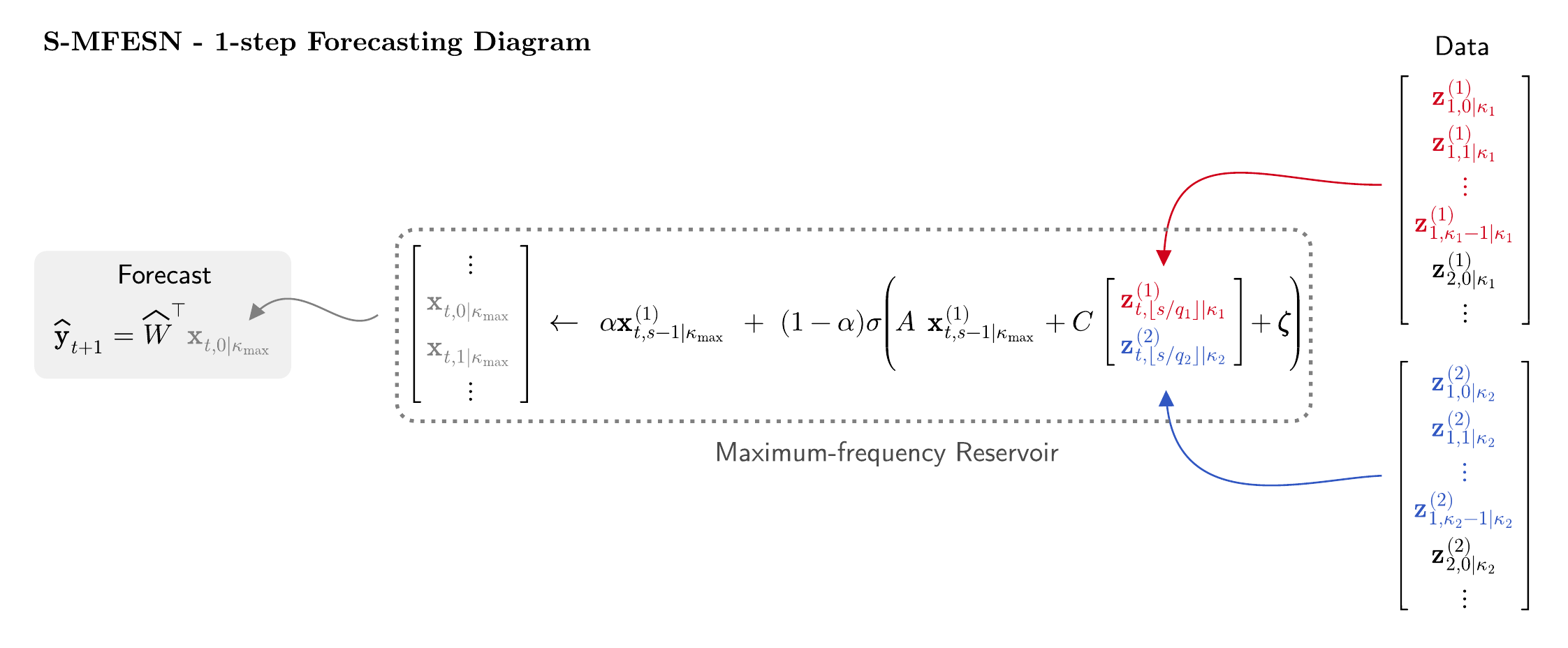}
	\caption{Scheme of a Single-Reservoir MFESN (S-MFESN) model combining input data sampled at two frequencies with state alignment and estimation for one-step ahead forecasting of the target series.}
	\label{fig:single_reservoir_sfesn_diagram}
\end{figure}

Figure~\ref{fig:single_reservoir_sfesn_diagram} gives a graphical diagram of the 1-step forecasting procedure for an S-MFESN. Additionally, Figure \ref{fig:single_reservoir_sfesn_multistep_diagram} in Supplementary Appendix \ref{subsec:additional_figures} provides a similar diagram for the case of multistep forecasts.

The following example illustrates this proposed forecasting strategy for the case of quarterly GDP forecasting using monthly and daily series inputs.
\begin{example}
\normalfont
\label{ex:s-mfesn}

Suppose that we wish to use an aligned {S-MFESN} model to forecast a quarterly one-dimensional target $(y_t)$ using $n_{(\textnormal{\texttt{m}})}$ monthly and $n_{(\textnormal{\texttt{d}})}$ daily series,  $(\boldsymbol{z}^{(\textnormal{\texttt{m}})}_{t,\tempo{s}{\kappa_1}})$ and   $(\boldsymbol{z}^{(\textnormal{\texttt{d}})}_{t,\tempo{s}{\kappa_2}})$, respectively.
We adopt the assumption that daily data is released 24 days over each calendar month and hence $\kappa_1 = 3$, $\kappa_2 = 72$ and $\kappa_{\max} = 72$, while $q_1 = 24$ and $q_2=1$. Let $t, \tempo{\ast}{72}$ be the temporal index with a quarterly reference frequency. The input vector for the S-MFESN state equation consistent with the daily frequency is given by
\begin{equation*}
    \boldsymbol{z}^{(\textnormal{\texttt{m,d}})}_{t, \tempo{s}{72}} := (
        {\boldsymbol{z}^{(\textnormal{\texttt{m}})}_{t,\tempo{\lfloor s / 24 \rfloor}{3}}}^\top,\:
        {\boldsymbol{z}^{(\textnormal{\texttt{d}})}_{t, \tempo{s}{72}}}^\top
    )^\top\in \mathbb{R}^{n_{(\textnormal{\texttt{m}})}+n_{(\textnormal{\texttt{d}})}} \enspace {\rm with}\enspace \boldsymbol{z}^{(\textnormal{\texttt{d}})}_{0,\tempo{0}{3}} = \mathbf{0}_{n_{(\textnormal{\texttt{d}})}}\enspace {\rm and}\enspace\boldsymbol{z}^{(\textnormal{\texttt{m}})}_{0,\tempo{0}{24}} = \mathbf{0}_{n_{(\textnormal{\texttt{m}})}}.
\end{equation*}
The complete S-MFESN model with the state space dimension $N$ can be written as:
\begin{align}
    \boldsymbol{x}^{(\textnormal{\texttt{m,d}})}_{t,\tempo{s}{72}} & = \alpha \boldsymbol{x}^{(\textnormal{\texttt{m,d}})}_{t,\tempo{s-1}{72}} + (1-\alpha) \sigma(A \boldsymbol{x}^{(\textnormal{\texttt{m,d}})}_{t,\tempo{s-1}{72}} + C \boldsymbol{z}^{(\textnormal{\texttt{m,d}})}_{t,\tempo{s}{72}} + \boldsymbol{\zeta}), \label{eq:s-mfesn_state} \\
    \boldsymbol{z}^{(\textnormal{\texttt{m,d}})}_{t,\tempo{s+1}{72}} & = \mathcal{W}^\top \boldsymbol{x}^{(\textnormal{\texttt{m,d}})}_{t,\tempo{s}{72}} + \boldsymbol{u}_{t,\tempo{s+1}{72}}, \label{eq:s-mfesn_input_obs} \\
    y_{t+1} & = W^\top \boldsymbol{x}^{(\textnormal{\texttt{m,d}})}_{t,\tempo{0}{\kappa_{\max}}} + \boldsymbol{\epsilon}_{t+1}, \label{eq:s-mfesn_target_obs}
\end{align}
where the state equations \eqref{eq:s-mfesn_state}-\eqref{eq:s-mfesn_input_obs} are run in their own maximum frequency temporal index $s>0$, and only the states $\boldsymbol{x}_{t-1,\tempo{\kappa_{\max}}{\kappa_{\max}}} = \boldsymbol{x}_{t,\tempo{0}{\kappa_{\max}}}$ are used in the observation equation \eqref{eq:s-mfesn_target_obs}.
Provided the input-target pairs sample of length $T$, the coefficient matrices $\mathcal{W}\in \mathbb{M}_{N, n_{(\textnormal{\texttt{m}})}+n_{(\textnormal{\texttt{d}})}}$ in \eqref{eq:s-mfesn_input_obs} and $W\in \mathbb{R}^{N}$ in \eqref{eq:s-mfesn_target_obs} can be estimated via ridge regression as explained above.

From \eqref{single ESN state map} the high-frequency autonomous state transition map is given by 
\begin{equation*}
    F_{72}^{(\textnormal{\texttt{m,d}})}(\boldsymbol{x}^{(\textnormal{\texttt{m,d}})}_{t,\tempo{s-1}{72}}) := \alpha \boldsymbol{x}^{(\textnormal{\texttt{m,d}})}_{t,\tempo{s-1}{72}} + (1-\alpha) \sigma\left( (A + C \widehat{\mathcal{W}}^\top) \boldsymbol{x}^{(\textnormal{\texttt{m,d}})}_{t,\tempo{s-1}{72}} + \boldsymbol{\zeta} \right),
\end{equation*}
which, composed with itself exactly $72$ times, by \eqref{single ESN state aligned} yields the target-frequency-aligned autonomous state transition map
\begin{equation*}
    F^{(\textnormal{\texttt{m,d}})}(\boldsymbol{x}^{(\textnormal{\texttt{m,d}})}_{t,\tempo{0}{72}}) := \, \underbrace{F_{72}^{(\textnormal{\texttt{m,d}})} \circ F_{72}^{(\textnormal{\texttt{m,d}})} \cdots \circ F_{72}^{(\textnormal{\texttt{m,d}})}}_{72 \textnormal{ times}}\, ( \boldsymbol{x}^{(\textnormal{\texttt{m,d}})}_{t, \tempo{0}{72}} ).
\end{equation*}
By applying $F^{(\textnormal{\texttt{m,d}})}$ to state $\boldsymbol{x}^{(\textnormal{\texttt{m,d}})}_{t,\tempo{0}{72}}$ we iterate the S-MFESN forward in time to provide an estimate for $\boldsymbol{x}^{(\textnormal{\texttt{m,d}})}_{t+1,\tempo{0}{72}}$, which can then be linearly projected using $\widehat{W}$ to yield a forecast for $y_{t+2}$. For the target variable, as well as forecasts, we do not use our temporal notation for the sake of compactness and clarity of exposition. Finally, the quarterly forecasts for $h\in \mathbb{N}$ can be computed using \eqref{single ESN forecast} as
\begin{align*}
    \widetilde{y}_{T+h \vert T} & = \widehat{W}^\top \big( \underbrace{F^{(\textnormal{\texttt{m,d}})}_{\phantom{.}} \circ F^{(\textnormal{\texttt{m,d}})} \circ \cdots \circ F^{(\textnormal{\texttt{m,d}})}}_{h-1 \textnormal{ times}}\, ( \boldsymbol{x}^{(\textnormal{\texttt{m,d}})}_{T, \tempo{0}{72}} ) \big).
\end{align*}
\end{example}

\subsection{Multi-Reservoir MFESN}\label{subsec:Multi-reservoir MFESN}

Constructing an MFESN with a single reservoir is not necessarily the most effective modeling strategy. Having more than one reservoir allows more flexible modeling of state dynamics for different subsets of input variables sampled at common frequencies. For example, suppose quarterly and monthly data are used as regressors. Our presentation is general enough to accommodate other types of partitioning of series into the corresponding reservoir models. We leave it to future research to test other approaches based, for instance, on markets or data types as done in \cite{Huellen2020}.

Assume again $L$ groups of series with input observations $(\boldsymbol{z}^{(l)}_{t,\tempo{s}{\kappa_l}})_{t,s}$ with $\boldsymbol{z}^{(l)}_{t,\tempo{s}{\kappa_l}} \in \rset^{n_l}$, $l\in [L]$, for all $t \in \mathbb{Z}$ and $s \in \{0, \ldots, \kappa_l-1\}$ sampled at common frequencies $\{\kappa_1, \ldots, \kappa_L\}$, respectively. 
For each of the $L$ groups of input series we define the corresponding ESN model as
\begin{align}
    \boldsymbol{x}^{(l)}_{t,\tempo{s}{\kappa_l}} & = \alpha_l \boldsymbol{x}^{(l)}_{t,\tempo{s-1}{\kappa_l}} + (1-\alpha_l) \sigma(A_l \boldsymbol{x}^{(l)}_{t,\tempo{s-1}{\kappa_l}} + C_l \boldsymbol{z}^{(l)}_{t,\tempo{s}{\kappa_l}} + \boldsymbol{\zeta}_l),\label{mult ESN state}\\
    \boldsymbol{z}^{(l)}_{t,\tempo{s+1}{\kappa_l}} & = \mathcal{W}_{l}^\top \boldsymbol{x}^{(l)}_{t,\tempo{s}{\kappa_l}} + \boldsymbol{u}^{(l)}_{t,\tempo{s+1}{\kappa_l}}, \label{mult ESN input obs} \quad l \in [L],
\end{align}
with $s>0$, $\mathcal{W}_{l}\in \mathbb{M}_{N_l, n_l}$ with $N_l$ the dimension of the state space. Notice that the time index $s$ is different for each $l$ according to our temporal notation introduced in Definition~\ref{Temporal notation} and each state equation runs at its own frequency $\kappa_l$. The dimensions $\{N_1, N_2, \ldots, N_L\}$ of the state spaces can be chosen for the $L$ reservoir models individually. Additionally, multiple reservoirs have the associated hyperparameter tuples $\{\boldsymbol{\varphi}_1, \ldots, \boldsymbol{\varphi}_L\}$ to be tuned. This requires some care whenever one wants to optimize all hyperparameters jointly. Since there are $L$ reservoir state equations, we call this class of MFESN models \textit{Multi-Reservoir Multi-Frequency ESN} (M-MFESN).

Similar to S-MFESN, all $L$ state equations are iterated each $\kappa_l$ times respectively until the states $\boldsymbol{x}^{(l)}_{t-1,\tempo{\kappa_l}{\kappa_l}}=\boldsymbol{x}^{(l)}_{t,\tempo{0}{\kappa_l}}$ are obtained. The \textit{aligned} M-MFESN observation equation is given by
\begin{align}
\label{mult ESN obs}
    \boldsymbol{y}_{t+1} = W^\top \boldsymbol{x}_{t,L} + \boldsymbol{\epsilon}_{t+1}, \enspace {\rm with} \enspace \boldsymbol{x}_{t,L} = \left(\begin{array}{c}
              \boldsymbol{x}^{(1)}_{t,\tempo{0}{\kappa_1}} \\
              \vdots \\
              \boldsymbol{x}^{(L)}_{t,\tempo{0}{\kappa_L}}
         \end{array}\right) \in \mathbb{R}^{\sum_{l=1}^L N_l}, \enspace W\in \mathbb{M}_{\sum_{l=1}^L N_l, J}.
\end{align}
\paragraph{Estimation of aligned M-MFESN.}
The coefficient matrices  ${W}_l$, $l\in [L]$, and ${\mathcal{W}}$ can be estimated similarly to the case of S-MFESN. The state and observation matrices for the estimation of $\widehat{\mathcal{W}}_l$, $l\in [L]$, in \eqref{eq:ridge_regression} are constructed as 
\begin{align*}
	X^{(l)} &= (\boldsymbol{x}^{(l)}_{1,\tempo{0}{\kappa_{l}}}, \ldots, \boldsymbol{x}^{(l)}_{1,\tempo{\kappa_{l}-1}{\kappa_{l}}}, \ldots, \boldsymbol{x}^{(l)}_{T-1,\tempo{0}{\kappa_{l}}}, \ldots, \boldsymbol{x}^{(l)}_{T-1,\tempo{\kappa_{l}-1}{\kappa_{l}}})^\top \in \mathbb{M}_{(T-1)\kappa_{l}-1,N_l},\\
 	Y^{(l)} &= (\boldsymbol{z}^{(l)}_{1,\tempo{1}{\kappa_{l}}}, \ldots, \boldsymbol{z}^{(l)}_{1,\tempo{\kappa_{l}}{\kappa_{l}}}, \ldots, \boldsymbol{z}^{(l)}_{T-1,\tempo{1}{\kappa_{l}}}, \ldots, \boldsymbol{z}^{(l)}_{T-1,\tempo{\kappa_{l}}{\kappa_{l}}})^\top \in \mathbb{M}_{(T-1)\kappa_{l}-1,n_l},
\end{align*}
while with the notation as in \eqref{mult ESN obs}
\begin{align*}
 	X &= \left(\boldsymbol{x}_{1,L}, 
               \boldsymbol{x}_{2,L}, 
         \ldots, 
               \boldsymbol{x}_{T-1,L}
         \right)^\top \in \mathbb{M}_{T-1,\sum_{l=1}^L N_l},\\
 	Y &= \left(\boldsymbol{y}_{2}, \ldots, \boldsymbol{y}_{T}\right)^\top \in \mathbb{M}_{T-1,J},
\end{align*}
are used for the estimation of $\widehat{W}$. Again, the state equations \eqref{mult ESN state} of M-MFESN can be started with $\boldsymbol{x}^{(l)}_{0,\tempo{0}{\kappa_{l}}}=\mathbf{0}_{N_l}$ (see Subsection~\ref{Reservoir models} for more details).

\paragraph{Forecasting with aligned M-MFESN.}

Let $\widehat{W}$ and $\widehat{\mathcal{W}}_{l}$, $l\in [L]$, be the sample estimates of the readout matrices. For any $l\in[L]$ the $\kappa_l$-frequency autonomous state transition map is given by 
\begin{equation}
\label{mult ESN l state}
    F^{(l)}_{\kappa_l}(\boldsymbol{x}^{(l)}_{t,\tempo{s-1}{\kappa_{l}}}) := \alpha_l \boldsymbol{x}^{(l)}_{t,\tempo{s-1}{\kappa_{l}}} + (1-\alpha_l) \sigma\left( (A_l + C_l \widehat{\mathcal{W}}_l^\top) \boldsymbol{x}^{(l)}_{t,\tempo{s-1}{\kappa_{l}}} + \boldsymbol{\zeta}_l \right).
\end{equation}
The target-frequency-aligned autonomous state transition map associated with each frequency $l$ is hence defined as
\begin{equation}
\label{mult ESN state aligned}
    F^{(l)}(\boldsymbol{x}_{t,\tempo{0}{\kappa_{l}}}) := \, \underbrace{F^{(l)}_{\kappa_l} \circ F^{(l)}_{\kappa_l} \circ \cdots \circ F^{(l)}_{\kappa_l}}_{\kappa_{l} \textnormal{ times}}\, ( \boldsymbol{x}^{(l)}_{t, \tempo{0}{\kappa_{l}}} ).
\end{equation}
Finally, from \eqref{eq:esn_forecast_integral_approx} the $h$-steps ahead forecasts can be computed as
\begin{align}
\label{mult ESN forecast}
    \widetilde{y}_{T+h \vert T} & = \widehat{W}^\top \left(\begin{array}{c}
              \underbrace{ F^{(1)} \circ  F^{(1)} \circ \cdots \circ  F^{(1)}}_{h-1 \textnormal{ times}}\, ( \boldsymbol{x}^{(1)}_{T,\tempo{0}{\kappa_1}} ) \\
              \vdots \\
              \underbrace{ F^{(L)} \circ F^{(L)} \circ \cdots \circ F^{(L)}}_{h-1 \textnormal{ times}}\, ( \boldsymbol{x}^{(L)}_{T,\tempo{0}{\kappa_L}} )
         \end{array}\right). 
\end{align}

\begin{figure}[!t]
    \centering
    \includegraphics[width=0.97\linewidth]{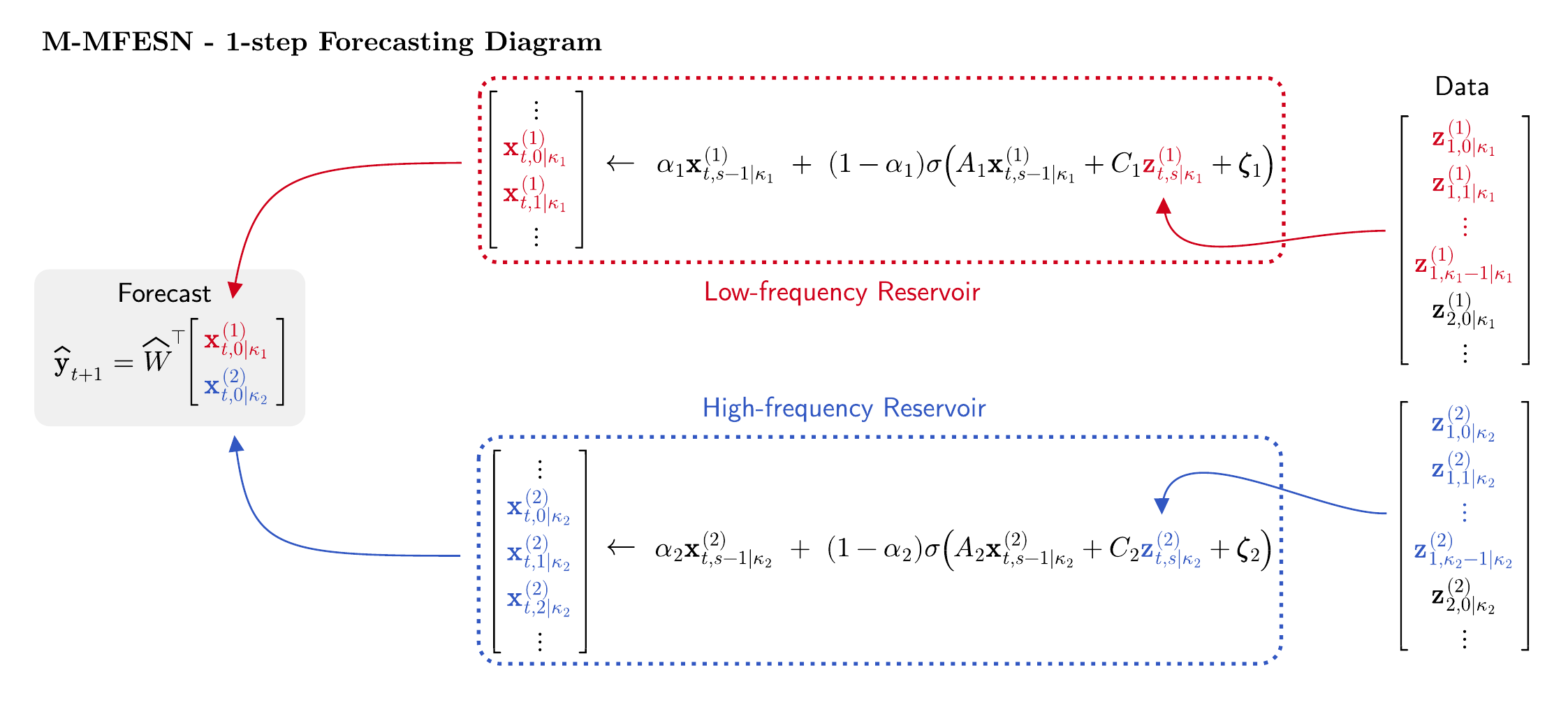}
    \caption{Scheme of a Multi-Reservoir MFESN (M-MFESN) model combining input data sampled at two frequencies with state alignment and estimation for one-step ahead forecasting of the target series.}
    \label{fig:multi_reservoir_mfesn_diagram}
\end{figure}

In Figure~\ref{fig:multi_reservoir_mfesn_diagram} we provide a diagram for the case of 1-step ahead forecasting with an aligned M-MFESN involving regressors of only two frequencies. Figure~\ref{fig:multi_reservoir_mfesn_multistep_diagram} in Supplementary Appendix \ref{subsec:additional_figures} provides a similar diagram for the case of multistep forecasting.

\begin{example}
\label{ex:m-mfesn}
\normalfont

Similar to Example \ref{ex:s-mfesn}, we aim to forecast a quarterly target with monthly and daily series, but this time we use an M-MFESN model. We have to define two independent state equations, one for monthly and one for daily series; in the observation equations, two states must be aligned temporally and stacked to form the full set of regressors. The data consists again of quarterly $(y_t)$, $n_{(\textnormal{\texttt{m}})}$ monthly series $(\boldsymbol{z}^{(\textnormal{\texttt{m}})}_{t,\tempo{s}{3}})$ and  $n_{(\textnormal{\texttt{d}})}$ daily series $(\boldsymbol{z}^{(\textnormal{\texttt{d}})}_{t,\tempo{s}{72}})$.

The aligned M-MFESN model with two reservoirs of dimensions $N_{(\textnormal{\texttt{m}})}$ and $N_{(\textnormal{\texttt{d}})}$, respectively, is given by
\begin{align}
    \boldsymbol{x}^{(\textnormal{\texttt{m}})}_{t,\tempo{s}{3}} & = \alpha_1 \boldsymbol{x}^{(\textnormal{\texttt{m}})}_{t,\tempo{s-1}{3}} + (1-\alpha_1) \sigma(A_1 \boldsymbol{x}^{(\textnormal{\texttt{m}})}_{t,\tempo{s-1}{3}} + C_1 \boldsymbol{z}^{(\textnormal{\texttt{m}})}_{t,\tempo{s}{3}} + \boldsymbol{\zeta}_1) \label{eq:m-mfesn_state_1}, \\
    \boldsymbol{z}^{(\textnormal{\texttt{m}})}_{t,\tempo{s+1}{3}} & = \mathcal{W}_{(\textnormal{\texttt{m}})}^\top \boldsymbol{x}^{(\textnormal{\texttt{m}})}_{t,\tempo{s}{3}} + \boldsymbol{u}^{(\textnormal{\texttt{m}})}_{t,\tempo{s+1}{3}} \label{eq:m-mfesn_input_obs_1}, \\[3pt]
    \boldsymbol{x}^{(\textnormal{\texttt{d}})}_{t,\tempo{s}{72}} & = \alpha_2 \boldsymbol{x}^{(\textnormal{\texttt{d}})}_{t,\tempo{s-1}{72}} + (1-\alpha_2) \sigma(A_2 \boldsymbol{x}^{(\textnormal{\texttt{d}})}_{t,\tempo{s-1}{72}} + C_2 \boldsymbol{z}^{(\textnormal{\texttt{d}})}_{t,\tempo{s}{72}} + \boldsymbol{\zeta}_2) \label{eq:m-mfesn_state_2}, \\
    \boldsymbol{z}^{(\textnormal{\texttt{d}})}_{t,\tempo{s+1}{72}} & = \mathcal{W}_{(\textnormal{\texttt{d}})}^\top \boldsymbol{x}^{(\textnormal{\texttt{d}})}_{t,\tempo{s}{72}} + \boldsymbol{u}^{(\textnormal{\texttt{d}})}_{t,\tempo{s+1}{72}} \label{eq:m-mfesn_input_obs_2}, \\[3pt]
    y_{t+1} & = W^\top \left(\begin{array}{c}
              \boldsymbol{x}^{(\textnormal{\texttt{m}})}_{t,\tempo{0}{3}} \\[3pt]
              \boldsymbol{x}^{(\textnormal{\texttt{d}})}_{t,\tempo{0}{72}}
         \end{array}\right) + \boldsymbol{\epsilon}_{t+1}, \label{eq:m-mfesn_target_obs}
\end{align}
where $s>0$, $\mathcal{W}_{(\textnormal{\texttt{m}})} \in \mathbb{M}_{N_{(\textnormal{\texttt{m}})},n_{(\textnormal{\texttt{m}})}}$, $\mathcal{W}_{(\textnormal{\texttt{d}})} \in \mathbb{M}_{N_{(\textnormal{\texttt{d}})},n_{(\textnormal{\texttt{d}})}}$ and $W\in \mathbb{R}^{N_{(\textnormal{\texttt{m}})}+N_{(\textnormal{\texttt{d}})}}$.
Here, the monthly reservoir $(\boldsymbol{x}^{(\textnormal{\texttt{m}})}_{t,\tempo{s}{3}})$ has the temporal index of frequency $3$, while the daily reservoir $(\boldsymbol{x}^{(\textnormal{\texttt{d}})}_{t,\tempo{s}{72}})$ of $72$; the high-frequency index $s$ is different for the two models. Notice that in an M-MFESN model it is necessary to introduce $2$ additional observation equations for the states, that is \eqref{eq:m-mfesn_input_obs_1} and \eqref{eq:m-mfesn_input_obs_2}). Notice that the state equations are iterated each $\kappa_l$ times to collect the states to be aligned in the observation equation \eqref{eq:m-mfesn_target_obs}. Again, the sample-based estimates of coefficient matrices $\widehat{\mathcal{W}}_{(\textnormal{\texttt{m}})}$, $\widehat{\mathcal{W}}_{(\textnormal{\texttt{d}})}$ and $\widehat{W}$ in \eqref{eq:m-mfesn_input_obs_1}, \eqref{eq:m-mfesn_state_2}, and in \eqref{eq:m-mfesn_target_obs}, respectively, can be obtained via the ridge regression as discussed above.

Exactly as in Example~\ref{ex:s-mfesn}, using \eqref{mult ESN l state} we can introduce high-frequency autonomous state maps $F_{3}^{(\textnormal{\texttt{m}})}$ and $F_{72}^{(\textnormal{\texttt{d}})}$ as
\begin{align*}
    F_{3}^{(\textnormal{\texttt{m}})}(\boldsymbol{x}^{(\textnormal{\texttt{m}})}_{t,\tempo{s-1}{3}}) & := \alpha_1 \boldsymbol{x}^{(\textnormal{\texttt{m}})}_{t,\tempo{s-1}{3}} + (1-\alpha_1) \sigma\left( (A_1 + C_1 \widehat{\mathcal{W}}_{(\textnormal{\texttt{m}})}^\top) \boldsymbol{x}^{(\textnormal{\texttt{m}})}_{t,\tempo{s-1}{3}} + \boldsymbol{\zeta}_1 \right), \\
    F_{72}^{(\textnormal{\texttt{d}})}(\boldsymbol{x}^{(\textnormal{\texttt{d}})}_{t,\tempo{s-1}{72}}) & := \alpha_2 \boldsymbol{x}^{(\textnormal{\texttt{d}})}_{t,\tempo{s-1}{72}} + (1-\alpha_2) \sigma\left( (A_2 + C _2\widehat{\mathcal{W}}_{(\textnormal{\texttt{d}})}^\top) \boldsymbol{x}^{(\textnormal{\texttt{d}})}_{t,\tempo{s-1}{72}} + \boldsymbol{\zeta}_2 \right),
\end{align*}
as well as their target-frequency aligned counterparts $F^{(\textnormal{\texttt{m}})}$ and $F^{(\textnormal{\texttt{d}})}$, by \eqref{mult ESN state aligned}, as
\begin{align*}
    F^{(\textnormal{\texttt{m}})}(\boldsymbol{x}^{(\textnormal{\texttt{m}})}_{t,\tempo{0}{3}}) &:= \, \underbrace{F_{3}^{(\textnormal{\texttt{m}})} \circ F_{3}^{(\textnormal{\texttt{m}})}  \circ F_{3}^{(\textnormal{\texttt{m}})}}_{3 \textnormal{ times}}\, ( \boldsymbol{x}^{(\textnormal{\texttt{m}})}_{t, \tempo{0}{3}} ),\\
    F^{(\textnormal{\texttt{d}})}(\boldsymbol{x}^{(\textnormal{\texttt{d}})}_{t,\tempo{0}{72}}) &:= \, \underbrace{F_{72}^{(\textnormal{\texttt{d}})} \circ F_{72}^{(\textnormal{\texttt{d}})} \circ \cdots \circ F_{72}^{(\textnormal{\texttt{d}})}}_{72 \textnormal{ times}}\, ( \boldsymbol{x}^{(\textnormal{\texttt{d}})}_{t, \tempo{0}{72}} ).
\end{align*}
The $h$-step ahead forecasts can be computed using the approximation in \eqref{mult ESN forecast} as 
\begin{align*}
    \widetilde{y}_{T+h \vert T} & = \widehat{W}^\top \left(\begin{array}{c}
              \underbrace{ F^{(\textnormal{\texttt{m}})}_{\phantom{.}} \circ F^{(\textnormal{\texttt{m}})} \circ \cdots \circ F^{(\textnormal{\texttt{m}})} }_{h-1 \textnormal{ times}}\, ( \boldsymbol{x}^{(\textnormal{\texttt{m}})}_{T,\tempo{0}{3}} ) \\[13pt]
              \underbrace{ F^{(\textnormal{\texttt{d}})}_{\phantom{.}} \circ F^{(\textnormal{\texttt{d}})} \circ \cdots \circ F^{(\textnormal{\texttt{d}})} }_{h-1 \textnormal{ times}}\, ( \boldsymbol{x}^{(\textnormal{\texttt{d}})}_{T,\tempo{0}{72}} )
         \end{array}\right). 
\end{align*}
In this case, it is important to note that while both $F^{(\textnormal{\texttt{m}})}$ and $F^{(\textnormal{\texttt{d}})}$ are composed $h-1$ times at step $h$, the underlying number of autonomous reservoir iterations is different for the monthly and daily reservoirs, namely $3$ and $72$, and depends on their own frequencies. This also suggests that one should take into account the different time dynamics when, for example, tuning M-MFESN hyperparameters $\boldsymbol{\varphi}^{(\textnormal{\texttt{m}})}$ and $\boldsymbol{\varphi}^{(\textnormal{\texttt{d}})}$, as proposed in Subsection~\ref{Hyperparameter Tuning}.
\end{example}

\section{Empirical Study}
\label{sec:empirical_study}

In this section, we compare the forecasting performance of our proposed MFESN to state-of-the-art benchmarks. We use a combination of macroeconomic and financial data sampled at low and high-frequency intervals, respectively. Our empirical exercises encompass several setups, with a small and a medium-sized set of regressors, fitting models with data before and after the 2007-08 crisis, and with fixed, rolling, and expanding estimation windows.

\subsection{Data}
\label{subsec:data}
Two sets of predictors of different sizes are compiled: Small-MD with 9 predictors and Medium-MD with 33 predictors in monthly and daily frequency. The reference frequency is quarterly: this is the frequency at which the target variable, US GDP growth, is available. Seasonally adjusted quarterly and monthly data is obtained from the Federal Reserve Bank of St. Louis Monthly (FRED-MD) and Quarterly (FRED-QD) Databases for Macroeconomic Research (see \cite{McCracken2016,McCracken2020} for detail). Daily data is obtained from Refinitiv Datastream, a subscription-based data service. All data is the last revised vintage data. The macroeconomic target and predictors, their transformations, and availability are provided in full detail in Table \ref{tab:dataMD} in Appendix~\ref{subsec:data_table}.

The selection of predictors follows the seminal work by \cite{Stock1996,Stock2006} in which the FRED-MD and FRED-QD data are proposed. Variations of their dataset have been used profusely in the literature (for example, see \cite{Boivin2005,Marcellino2006,Hatzius2010}). Indicators from ten macroeconomic and financial categories are considered: (1) output and income, (2) labor market, (3) housing, (4) orders and inventories, (5) price indices, (6) money and credit, (7) interest rates, (8) exchange rates, (9) equity, and (10) derivatives. The latter five categories represent financial market conditions and are sourced at daily frequency. The exception is interest rates, which move relatively slowly and enter as monthly aggregates, available in the FRED-MD data. We refer to this dataset as Medium-MD. A subset of predictors is selected for the Small-MD dataset by choosing variables that have been identified as leading indicators in the empirical literature \citep{Ingenito1996, Clements2008, Andreou2013, marsilli:thesis, clement:midas, Carriero2019, Jardet2020}. Data availability is an additional criterion, and predictors unavailable before 1990 are not considered. This excludes the VIX volatility index, which has been identified as a leading indicator in some studies, for example in \cite{Andreou2013,Jardet2020}.

We follow instructions by \cite{McCracken2016,McCracken2020} on pre-processing macroeconomic predictors before they are used as input for forecasting. These are mainly differenced for detrending. We further transform financial predictors to capture market disequilibrium and volatility. Disequilibrium indicators, such as interest rate spreads, have been found to be more relevant for macroeconomic prediction than routine changes captured by differencing (see \cite{Borio2002,Gramlich2010,Qin2022}). In addition to disequilibrium indicators, realized stock market volatility has been found to improve macroeconomic predictions \citep{Chauvet2015}. In the absence of intraday trading data from the 1990s onward, which prevents us from utilizing conventional daily realized volatility indicators, we extract volatility indicators from daily price series by fitting a GARCH(1,1) by \citet{bollerslev:garch}.\footnote{We include a control \texttt{scale = 1} to ensure convergence of the optimization algorithm and only include a constant mean term in the return process for simplicity.} In addition to volatility of stock and commodity prices, term structure indicators are used. The term structure is forward-looking, capturing information about future demand and supply, and has been found to be a leading predictor of GDP growth (see for example \citet{Hong2012,Kang2020}). 

The data spans the period January 1st, 1990 to December 31st, 2019.\footnote{In the Small-MD dataset experiments we make a small variation and instead include data starting from 1st January 1975, but \textit{only} for the initial CV selection of ridge penalties for MFESN models. Our aim is to make sure that at least for the fixed window estimation strategy -- where $\lambda$ is cross-validated once and only one $\widehat{W}$ is estimated -- the ridge estimator is robust. In practice, when we compare to expanding and rolling window estimators, where $\lambda$ is re-selected at each window, we find that extending the initial CV data window has little impact on out-of-sample performance.} We are interested in evaluating model performance under two stylized settings. First, a researcher fits all models up until the Great Recession, including data from Q1 1990 to Q4 2007. Second, fitting is done with data largely encompassing the crisis period, again from Q1 1990 but now up to Q4 2011. In both cases, the testing sample ranges from the next GDP growth observation after fitting up to Q4 2019. All exercises exclude the global COVID-19 economic depression, as we consider it as an extreme, unpredictable event that induces significant structural changes in the underlying macroeconomic dynamics.\footnote{In the macroeconomic literature this falls under the category of ``natural disaster'' events, and should not be na\"ively modeled together with previous observations. In this section, we therefore avoid dealing with post-COVID-19 macroeconomic data altogether.}

To avoid having to handle the many edge cases that daily data in its ``raw'' calendar releases involves, we use an interpolation approach. We set \textit{ex ante} the number of working days in \textit{any} month to be exactly 24: given that in forecasting the most recent information sets are more relevant, when interpolating daily data over months with less than calendar 24 observations, we linearly interpolate the ``missing'' data starting from a months' beginning (using the previous months' last observation). The choice of 24 as a daily frequency is transparent by noting that this is the closest number to actual commonly observed data releases, whilst also being a multiple of both 4 (approximate number of weeks per month) and 6 (upper bound on the number of working days per week).

\subsection{Models}

In this section, we present the set of models that we use throughout our empirical exercises. For a general overview, Table \ref{table:models_empirical} summarizes all models, including hyperparameters. In our analysis, we compare the competing models based on several performance measures, which we introduce in Supplementary Appendix~\ref{Performance measures}. 

\begin{table}[!t]
	\centering
	\setlength\tabcolsep{6pt}
    \setlength\extrarowheight{0pt}
	\linespread{1}\selectfont\centering
	\renewcommand{\arraystretch}{1.5}
	%
	%
	\begin{tabular}{m{0.15\textwidth}<{\centering}m{0.45\textwidth}m{0.32\textwidth}}
	   \hline
	     Model Name & Description & Specification \\
	   \hline 
	     Mean &
	        Unconditional mean of target series over estimation sample. & 
	        None \\[4pt]
	       AR(1) &
	        Autoregressive model of target series estimated using OLS. & 
	        None \\[8pt]
	     MIDAS &
	        Almon-weighted MIDAS regression, linear (unconstrained) autoregressive component. &
	        \shortstack[l]{
	            Autoregressive lags: 3 \\
	            Monthly freq. lags: 9 \\
	            Daily freq. lags: 30 
	        } \\[15pt]
	     DFM [A] & 
	        Stock aggregation, VAR(1) factor process. & 
	        \shortstack[l]{ 
                Factors: 5 for Small-MD \\
                \phantom{Factors:} 10 for Medium-MD 
            } \\[10pt]
	     DFM [B] &
	        Almon aggregation, VAR(1) factor process & 
	        \shortstack[l]{
                Factors: 5 for Small-MD \\ 
                \phantom{Factors:} 10 for Medium-MD 
            } \\[10pt]
	     singleESN [A] &
	        \shortstack[l]{
	            S-MFESN model: \\ 
	            $\quad$ Sparse-normal $\widetilde{A}$, sparse-uniform $\widetilde{C}$, $\widetilde{\boldsymbol{\zeta}} = \mathbf{0}$. \\
	            $\quad$ Isotropic ridge regression fit.\vspace{10pt}
	        } & \shortstack[l]{
	            Reservoir dim: 30 \\
	            Sparsity: $33.3\%$ \\
	            $\rho = 0.5$, $\gamma = 1$, $\alpha = 0.1$ 
	        } \\[20pt]
	     singleESN [B] &
	        \shortstack[l]{
	            S-MFESN model: \\ 
	            $\quad$ Sparse-normal $\widetilde{A}$, sparse-uniform $\widetilde{C}$, $\widetilde{\boldsymbol{\zeta}} = \mathbf{0}$. \\
	            $\quad$ Isotropic ridge regression fit.
	        } & \shortstack[l]{
	            Reservoir dim: 120 \\
	            Sparsity: $8.3\%$ \\
	            $\rho = 0.5$, $\gamma = 1$, $\alpha = 0.1$ 
	        } \\[20pt]
	     multiESN [A] &
	        \shortstack[l]{
	            M-MFESN model: \\ 
	            $\quad$ Monthly and daily frequency reservoirs. \\
	            $\quad$ Sparse-normal $\widetilde{A}_1$, $\widetilde{A}_2$,\\
	            $\quad$ sparse-uniform $\widetilde{C}_1$, $\widetilde{C}_2$,  $\widetilde{\boldsymbol{\zeta}}_1 = \mathbf{0}$, $\widetilde{\boldsymbol{\zeta}}_2 = \mathbf{0}$. \\
	            $\quad$ Isotropic ridge regression fit.\vspace{10pt}
	        } & \shortstack[l]{
	            Reservoir dims: M=100, D=20 \\
	            Sparsity: M=$10\%$, D=$50\%$ \\
	            M: $\rho = 0.5$, $\gamma = 1.5$, $\alpha = 0$ \\
	            D: $\rho = 0.5$, $\gamma = 0.5$, $\alpha = 0.1$ 
	        } \\[20pt]
	     multiESN [B] &
	        \shortstack[l]{
	            M-MFESN model: \\ 
	            $\quad$ Monthly and daily frequency reservoirs. \\
	            $\quad$ Sparse-normal $\widetilde{A}_1$, $\widetilde{A}_2$, \\
	            $\quad$ sparse-uniform $\widetilde{C}_1$, $\widetilde{C}_2$, $\widetilde{\boldsymbol{\zeta}}_1 = \mathbf{0}$, $\widetilde{\boldsymbol{\zeta}}_2 = \mathbf{0}$. \\	            $\quad$ Isotropic ridge regression fit.\vspace{10pt}
	        } & \shortstack[l]{
	            Reservoir dims: M=100, D=20 \\
	            Sparsity: M=$10\%$, D=$50\%$ \\
	            M: $\rho = 0.08$, $\gamma = 0.25$, $\alpha = 0.3$ \\
	            D: $\rho = 0.01$, $\gamma = 0.01$, $\alpha = 0.99$ 
	        } \\[15pt]
	    \hline
	\end{tabular}
	\caption{Table of models used in applied forecasting exercises. MFESN hyperparameters are defined with respect to normalized state parameters c.f. \eqref{normalizing}.}
	\label{table:models_empirical}
	\renewcommand{\arraystretch}{1}
\end{table}

\subsubsection{Benchmarks}
\label{Benchmarks}
\paragraph{Unconditional mean.}
We use the unconditional mean of the sample used for fitting as a baseline benchmark. For GDP growth forecasting, there is evidence that the unconditional mean produces forecasts that are competitive with linear models such as VARs in terms of mean square forecasting errors (MSFE), even at relatively short horizons \citep{aroraNonlinearNonparametricModeling2013}. It is therefore an important reference for the performance of all other models and we report relative MSFE with respect to the unconditional mean in the tables below.

\paragraph{AR(1) model.}
 A simple autoregressive process of order one on the target variable is included as a benchmark model.\footnote{Suggested by an anonymous referee.} This is also a common benchmark in the literature, as AR(1) models are often able to capture key dynamics and produce meaningful forecasts for macroeconomic variables \citep{Stock2002,Bai2008}. We emphasize that since AR(1) model is fit to the series of quarterly GDP targets and does not use any additional information, its forecasts are identical for both the Small-MD and Medium-MD samples.

\paragraph{MIxed DAta Sampling (MIDAS).}
The first mixed-frequency model benchmark is given by a MIDAS model \citep{Ghysels2004, Ghysels2007}. Our dynamic MIDAS specification includes autoregressive lags of the target series and uses an Almon weighting scheme. As shown in \cite{Bai2010}, exponential Almon MIDAS regressions are related to dynamic factor models, which we also consider as benchmarks. The MIDAS model includes three lags of quarterly GDP target variable, and 30 daily and 9 monthly lags for all daily and monthly series, respectively.  This model prescription allows for some parsimony as the Almon polynomial weighing reduces the number of daily and monthly lag coefficients. 

A thorough description of our MIDAS implementation can be found in Supplementary Appendix~\ref{subsec_midas}. To make optimization more efficient, we use explicit expressions for MIDAS loss gradients as in \cite{Kostrov2021essays}. The MIDAS estimation can be hard to perform in practice due to the complexity of nonlinear optimization. First, exponential weighting schemes might require computing floating-point numbers that exceed numerical precision. Therefore, it is a better choice to start the gradient descent close to the origin of the parameter space. Second, even with this choice of starting points, one may encounter issues with optimization results since the Almon-scheme MIDAS loss can have a large number of distinct local minima. In Supplementary Appendix \ref{sec:midas_robustness} we document, using a simple replication experiment, that even small changes in the initial conditions can result in different local minima picked by the numerical optimization algorithm.\footnote{We set the initial coefficient values to zero in all empirical exercises.} These important robustness issues are present even when using closed-form gradients and multi-start optimization routines for the MIDAS models. The computational issues become more pronounced as the number of MIDAS parameters increases unless a careful model/variable selection step is performed. We, therefore, do not include any MIDAS model specifications in the Medium-MD setup. 

\paragraph{Dynamic Factor Model (DFM).}
The dynamic factor model framework has been extensively applied in macroeconometrics, starting with \cite{geweke1977dynamic} and \cite{sargent1977business}. A DFM specification assumes that predictable dynamics of a large set of time series can be explained by a small number of factors with an autoregressive dependence (see for example \cite{forni2005generalized, doz2011two, stock2016dynamic}). We generalize the standard two-frequency DFM modeling setup \citep{mariano2003new,banbura2014maximum} to a flexible mixed-frequency DFM that encompasses any number of data frequencies. Moreover, we derive a novel weighting scheme that effectively links the MIDAS and DFM approaches. For a detailed discussion of our factor model setup, we refer the reader to Supplementary Appendix \ref{subsec_dfm}. Two distinct DFM specifications are used. The first one termed DFM [A] uses the standard linear aggregation scheme, as provided in Example \ref{ex:stock}, while the second is a variation that implements an Almon weighting scheme as presented in Example \ref{ex:almon} (we name it DFM [B]). The latter is similar to a MIDAS-type aggregation scheme \citep{Marcellino2010}: the factor structure effectively mitigates the parameter proliferation.  

A key choice for a DFM model is the dimension of the factor process. While a number of methods have been developed over the years to systematically derive the number of factors (see, for example, the review of \cite{stock2016dynamic}), commonly used macroeconomic panels feature a number of challenges, such as weak factors \citep{onatskiAsymptoticsPrincipalComponents2012}. Moreover, as mentioned in Supplementary Appendix~\ref{sec:DFM_implementation}, factor number selection in the mixed-frequency setting has not been sufficiently addressed in the literature. To sidestep these issues, we construct both DFM models with 5 unobserved factors for Small-MD and 10 for Medium-MD, respectively, and assume that they follow a VAR(1) process.

One extant issue with integrating daily data is its very high release frequency compared to monthly and especially quarterly releases: computationally this can be extremely taxing, which might be one of the reasons why to our knowledge we are {\it the first to provide DFM forecasts that include daily data}. Our solution is to reduce aggregate daily data every 6 days by averaging, thus leaving 4 observations per month. This eases the computational burden to estimate coefficients and latent states considerably (12 versus 72 daily observations per quarter).

\subsubsection{Multi-Frequency ESNs}

The first set of ESNs we propose is given by two S-MFESN models, based on Example \ref{ex:s-mfesn}. One model uses a reservoir of 30 neurons (we call it singleESN [A]); the other has a larger reservoir of dimension 120 (named singleESN [B]). The sparsity degree of state parameters for both models is set to be $10/N$, where $N$ is the reservoir size. Both MFESNs share the same hyperparameters, $\rho = 0.5$, $\gamma = 1$, $\alpha = 0.1$ (see \eqref{eq:esn_hyperparams}). These values have not been tuned but are presumed credible given other ESN implementations in the literature. To make a fair comparison with DFMs, we fit the S-MFESN models using 6-day-averaged daily data. Note here that for MFESN models the computational gains of averaging are negligible, and are most apparent when tuning the ridge penalty via cross-validation.

Our second set of proposed models consists of two M-MFESNs according to Example \ref{ex:m-mfesn}. Both models have two reservoirs, one for each data frequency -- monthly and daily -- with 100 and 20 neurons, respectively; sparsity degrees are again adjusted to be $10/N$, where $N$ is the reservoir state dimension. The first M-MFESN has hyperparameters that are hand-selected among reasonable values: we note that the monthly-frequency reservoir has no state leak and a larger input scaling, while the daily frequency reservoir features smaller scaling than usual (to avoid compressing high volatility events with the activation function) and the same leak rate as in S-MFESN models (we call this specification multiESN~[A]). For the second M-MFESN, we change hyperparameters more radically: we aim to set up a model that has a very high input memory \citep{RC23}, and that also features long-term smoothing of states. Note that here input scaling values are small, spectral radii are an order of magnitude smaller than in previous models, and leak rates are large (we term this model multiESN~[B]).

\subsection{Results}
\label{subsec:results}
We start by commenting on the computational efficiency of competing models and report execution times in seconds in Table~\ref{tab:execution_times}. Firstly, DFM models appear to be the most computationally effortful models among all specifications. For the Small-MD dataset, the simplest MFESN models, that is, singleESN [A] and [B], have execution times which are at most $3.5$ times higher than the MIDAS model, while still being at least $15.6$ times computationally cheaper than any of the DFM models. The more resource-demanding models MFESN, multiESN [A] and [B], are nevertheless at least $2.6$ times faster to run than the best DFM model (DFM [A]). When moving to the Medium-MD dataset, where the MIDAS model is not, as explained earlier, a feasible choice, the most inefficient MFESN model (singleESN [B]) still outperforms the best DFM model, DFM [A], by $8.4$ times, while the same holds for multiESN [A] model versus DFM [A] model by $2.7$ times. We can conclude that our proposed MFESN architectures provide an attractive and computationally efficient framework for GDP forecasting in the multifrequency framework which is feasible for computations on low-cost machine configurations available to practitioners.

\begin{table}[!t]
    \centering
    \setlength\extrarowheight{2pt}
	\linespread{1.2}\selectfont\centering
    Execution Time (Seconds) for Model Estimation \\[5pt]
    \begin{tabular*}{\textwidth}{@{\extracolsep{\fill}}*{10}{c}}
        \toprule
        &  &  &  & \multicolumn{2}{c}{DFM} & \multicolumn{2}{c}{singleESN} & \multicolumn{2}{c}{multiESN} \\
		\cline{5-6} \cline{7-8} \cline{9-10} 
        Dataset & Mean & AR(1) & MIDAS & [A] & [B] & [A] & [B] & [A] & [B] \\
        \midrule
        Small-MD & 0.1 & 0.7 & 1.3 & $40.5$ & $85.5$ & $2.6$ & $4.5$ & $15.3$ & $14.6$ \\
        Medium-MD & 0.1 & 0.8 & $-$ & $48.0$ & $226.5$ & $2.5$ & $5.7$ & $17.7$ & $14.7$ \\
        \bottomrule
    \end{tabular*}
    \caption{\small Execution time in seconds for model estimation measured over a single run on a quad-core computer. MFESN models timing includes ridge penalty cross-validation. MIDAS estimation time refers to optimization from a single initial value. DFM models were estimated on a single-core server and times are adjusted by a factor of $1/4$ for comparison.  }
    \label{tab:execution_times}
\end{table}

Competing forecasts are compared using the Model Confidence Set (MCS) test derived in \cite{Hansen2011}. One should note that due to the intrinsic nature of data availability of macroeconomic time series and panels, our sample sizes are modest. This implies that the small sample sensitivities of the MCS test need to be taken into account when evaluating our comparisons. Recent analyses of the finite sample properties of the MCS methodology have shown that it requires signal-to-noise ratios which are unattainable in most empirical settings, an issue that undermines its applicability \citep{Aparicio2018}.  Given this fact, we also conduct pairwise model comparison tests with the Modified Diebold-Mariano (MDM) test for predictive accuracy \citep{Diebold2002,Harvey1997}. 

As we also provide multiple-steps-ahead forecasts, we test for the best subset of models uniformly across all horizons using the Uniform Multi-Horizon MCS (uMCS) test proposed by \cite{quaedvliegMultiHorizonForecastComparison2021}. Since there is relatively little systematic knowledge regarding the power properties of the uMCS test in small samples, our inclusion of this procedure is meant as a statistical counterpoint to simple relative forecasting error comparisons, which provide limited information about the significance of performance differences. We provide more details on our implementation of the test in Supplementary Appendix~\ref{uMCS}.
Finally, we do not report uMCS test outcomes for the expanding window setup, as \cite{quaedvliegMultiHorizonForecastComparison2021} argues that in such context the test is invalid.


\subsubsection{Small Dataset}

We begin our discussion of the Small-MD forecasting results by reviewing Table \ref{table:1sa_GDP_smallMD}. For both sample setups (2007 and 2011) and all three estimation strategies (fixed, expanding, and rolling windows) we provide relative MSFE metrics, with the unconditional mean being used as a reference. Plots of each of the model's forecasts are given in Figures~\ref{fig:1sa_forecasts_2007_GDP_smallMD} and \ref{fig:1sa_forecasts_2011_GDP_smallMD} in Appendix~\ref{subsec:forecast_figures}; additional plots for cumulative SFE, cumulative RMSFEs and other metrics can be found in Supplementary Appendix~\ref{subsec:additional_figures}.

\begin{table}[!t]
	\centering
	\setlength\tabcolsep{3.3pt}
    \setlength\extrarowheight{2pt}
	\linespread{1}\selectfont\centering
	1-Step-ahead GDP Forecasting - Small-MD Dataset \\[5pt]
	\begin{tabular*}{\textwidth}{@{}*{13}{c}}
		\toprule
		& \multicolumn{4}{c}{Fixed Parameters} & \multicolumn{4}{c}{Expanding Window} & \multicolumn{4}{c}{Rolling Window} \\
		\cline{2-5} \cline{6-9} \cline{10-13}
		Model & \multicolumn{2}{c}{2007} & \multicolumn{2}{c}{2011} & \multicolumn{2}{c}{2007} & \multicolumn{2}{c}{2011} & \multicolumn{2}{c}{2007} & \multicolumn{2}{c}{2011} \\
	    \cline{2-3} \cline{4-5} \cline{6-7} \cline{8-9} \cline{10-11} \cline{12-13}
		 & MSFE & MCS & MSFE & MCS & MSFE & MCS & MSFE & MCS & MSFE & MCS & MSFE & MCS \\
		\midrule
          Mean & 1.000 & * & 1.000 & ** & 1.000 & ** & 1.000 & ** & 1.000 & ** & 1.000 & ** \\ 
          AR(1) & 0.758 & * & 1.230 & ** & 0.789 & ** & 1.226 & ** & 0.775 & ** & 1.209 & ** \\ 
          MIDAS & \cHL\textbf{0.533} & \cHL** & 1.300 &  & 0.596 & ** & 1.129 & * & 0.709 & ** & 1.170 & * \\ 
          DFM [A] & 0.799 & * & 1.337 &  & 0.980 & * & 1.320 &  & 0.919 & * & 1.226 &  \\ 
          DFM [B] & 0.885 &  & 1.221 & ** & 0.982 & * & 1.022 & ** & 0.948 &  & 1.028 & ** \\ 
          singleESN [A] & 0.721 & ** & 1.015 & ** & 0.597 & ** & \chl0.867 & \chl** & \cHL\textbf{0.529} & \cHL** & \chl0.863 & \chl** \\ 
          singleESN [B] & 0.758 & * & \cHL\textbf{0.921} & \cHL** & 0.602 & ** & \cHL\textbf{0.844} & \cHL** & 0.561 & ** & 0.930 & ** \\ 
          multiESN [A] & 0.802 & * & 1.250 &  & 0.635 & ** & 0.874 & ** & 0.621 & ** & \cHL\textbf{0.859} & \cHL** \\ 
          multiESN [B] & \chl0.590 & \chl** & \chl0.969 & \chl** & \cHL\textbf{0.552} & \cHL** & 0.895 & ** & \chl0.530 & \chl** & 0.921 & ** \\ 
		\bottomrule
	\end{tabular*}
	\caption{\small Relative MSFE and Model Confidence Set (MCS) comparison between models in 1-step-ahead forecasting exercises. Unconditional mean MSFE is used as a reference. MCS columns show inclusion among best models: $\ast$ indicates inclusion at 90\%, $\ast\ast$ indicates inclusion at 75\% confidence.}
	\label{table:1sa_GDP_smallMD}
\end{table}
 
The overall finding is that MFESN models perform excellent, and, when we exclude the 2007 fixed parameters setup, they perform the best. It is easy to see from Figure \ref{fig:1sa_forecasts_2007_GDP_smallMD} (a) why the 2007 fixed window estimation case is different from other cases: the 2008 Financial Crisis induced a deep drop in quarter-to-quarter GDP growth that was in stark contrast with previous business cycle fluctuations. By keeping model parameters fixed, and using only information from 1990 to 2007 -- periods where systematic fluctuations are small -- DFM and MFESN models are fit to produce smooth, low-volatility forecasts. MIDAS, on the other hand, yields an exponential smoothing which can be more responsive to changes in monthly and daily series. From Figure \ref{fig:1sa_forecasts_2007_GDP_smallMD} (b) and (c) it is possible to see that expanding and rolling window estimation resolves this weakness of state-space models. At the same time, the AR(1) model outperforms the unconditional mean only in the 2007 sample with fixed parameters, losing to the MIDAS model in all but one scenario.

Table \ref{table:1sa_GDP_smallMD} shows that MFESN models always perform better than the mean in terms of MSFE, something which no other model class achieves across all setups. In both expanding and rolling window setups they also always outperform the AR(1) model. Furthermore, at least one MFESN model for each subclass (single or multi-reservoir) is always included in the model confidence set at the highest confidence level. We remind again that the MCS test of \cite{Hansen2011} might be distorted due to the modest sample sizes considered, even more so in the 2011 test sample. To complement the MCS, we provide graphical tables for pairwise Modified Diebold-Mariano (MDM) tests, with 10\% level rejections highlighted in Figure \ref{fig:1sa_GDP_smallMD_diebold_mariano}, Supplementary Appendix \ref{subsec:additional_figures}. The MDM tests broadly agree with the results of Table \ref{table:1sa_GDP_smallMD}, although they do not account for multiple testing, and therefore cannot be interpreted as yielding subsets of the most accurate forecasting models in a statistical sense.

For multiple-steps-ahead forecasts, relative RMSFE and uMCS are reported in Tables \ref{table:msa_2007_GDP_smallMD} and \ref{table:msa_2011_GDP_smallMD}: we constrain our exercise to $h \in \{ 1, \ldots, 8 \}$ steps, since we are interested in GDP growth forecasts within 2 years. Note that for $h = 1$ our results are similar, but do not reduce to the one-step-ahead results. To make correct multistep RMSFE evaluations and execute the uMCS procedure one must select $h$ different vectors of residuals of the same length: this implies that residuals at the end of the forecasting sample must be trimmed off to compute short-term multistep RMSFEs that are comparable to the long-term ones.  Generally, we notice that MIDAS, as well as S-MFESNs, provide the worst-performing multistep forecasts, with RMSFEs considerably exceeding the unconditional mean baseline after horizon 1. Figures \ref{fig:msa_2007_GDP_smallMD} and \ref{fig:msa_2011_GDP_smallMD} in Appendix~\ref{subsec:forecast_figures} reproduce the RMSFE numbers of the aforementioned tables graphically. 

\begin{table}[!p]
\vspace{5em}
\centering
	\setlength\tabcolsep{6.5pt}
    \setlength\extrarowheight{2pt}
	\linespread{1}\selectfont\centering
	Multistep-ahead GDP Forecasting - Small-MD Dataset - 2007 Sample \\[5pt]
	\begin{tabular*}{\textwidth}{@{}*{11}{c}}
  \toprule
    Setup & Model & \multicolumn{8}{c}{Horizon} & uMCS \\
    \cline{3-10}
    & & 1 & 2 & 3 & 4 & 5 & 6 & 7 & 8 & \\ 
  \midrule
      FIX & Mean & 1.000 & 1.000 & 1.000 & 1.000 & 1.000 & 1.000 & 1.000 & 1.000 &  \\ 
      FIX & AR(1) & 0.870 & \cHL\textbf{0.950} & \cHL\textbf{0.982} & \chl0.991 & 0.992 & 0.991 & 0.992 & 0.992 & ** \\ 
      FIX & MIDAS & \chl0.823 & 1.672 & 2.737 & 1.816 & 2.213 & 2.791 & 1.888 & 1.921 &  \\ 
      FIX & DFM [A] & 0.890 & 0.969 & 1.014 & 1.077 & 1.341 & 1.701 & 2.001 & 2.180 & * \\ 
      FIX & DFM [B] & 0.937 & 1.069 & 1.202 & 1.344 & 1.799 & 2.310 & 2.638 & 2.801 &  \\ 
      FIX & singleESN [A] & 0.852 & 0.994 & 0.995 & 0.995 & 0.993 & 0.991 & 0.991 & 0.991 & * \\ 
      FIX & singleESN [B] & 0.871 & 0.986 & 0.989 & \cHL\textbf{0.989} & \cHL\textbf{0.985} & \cHL\textbf{0.981} & \cHL\textbf{0.981} & \cHL\textbf{0.981} & ** \\ 
      FIX & multiESN [A] & 0.898 & 0.980 & 0.990 & \chl0.991 & \chl0.988 & \chl0.985 & \chl0.985 & \chl0.985 & ** \\ 
      FIX & multiESN [B] & \cHL\textbf{0.767} & \chl0.954 & \chl0.983 & \chl0.991 & 0.991 & 0.990 & 0.991 & 0.991 & ** \\  
  \midrule
      EW & Mean & 1.000 & 1.000 & 1.000 & 1.000 & 1.000 & 1.000 & \cHL\textbf{1.000} & \cHL\textbf{1.000} & - \\ 
      EW & AR(1) & 0.887 & 0.922 & 0.951 & \chl0.962 & \chl0.957 & \cHL\textbf{0.981} & \chl1.001 & \chl1.008 & - \\ 
      EW & MIDAS & 0.814 & 1.283 & 1.518 & 1.596 & 1.697 & 1.391 & 1.951 & 1.800 & - \\ 
      EW & DFM [A] & 0.985 & 1.109 & 1.123 & 1.114 & 1.217 & 1.226 & 1.241 & 1.539 & - \\ 
      EW & DFM [B] & 0.989 & 1.082 & 1.149 & 1.199 & 1.315 & 1.412 & 1.373 & 1.425 & - \\ 
      EW & singleESN [A] & \chl0.771 & 1.260 & 1.485 & 1.564 & 2.070 & 2.728 & 2.550 & 2.834 & - \\ 
      EW & singleESN [B] & 0.772 & 1.031 & 1.135 & 1.319 & 1.831 & 2.279 & 2.449 & 2.556 & - \\ 
      EW & multiESN [A] & 0.792 & \chl0.897 & \chl0.941 & 0.976 & 1.015 & 1.240 & 1.377 & 1.227 & - \\ 
      EW & multiESN [B] & \cHL\textbf{0.740} & \cHL\textbf{0.853} & \cHL\textbf{0.894} & \cHL\textbf{0.911} & \cHL\textbf{0.873} & \chl0.993 & 1.020 & 1.020 & - \\  
  \midrule
      RW & Mean & 1.000 & 1.000 & 1.000 & 1.000 & 1.000 & 1.000 & 1.000 & 1.000 & * \\ 
      RW & AR(1) & 0.898 & \chl0.943 & \chl0.968 & \chl0.974 & \chl0.963 & \cHL\textbf{0.968} & \chl0.970 & \cHL\textbf{0.962} & ** \\ 
      RW & MIDAS & 0.933 & 1.438 & 1.642 & 1.993 & 1.794 & 1.661 & 1.816 & 1.973 & * \\ 
      RW & DFM [A] & 0.931 & 1.017 & 1.033 & 1.020 & 1.024 & \chl1.003 & \cHL\textbf{0.918} & \chl1.062 & * \\ 
      RW & DFM [B] & 0.942 & 0.973 & 0.970 & 1.045 & 1.059 & 1.203 & 1.225 & 1.263 & * \\ 
      RW & singleESN [A] & \cHL\textbf{0.714} & 1.320 & 1.693 & 1.972 & 2.733 & 3.669 & 3.391 & 3.719 & * \\ 
      RW & singleESN [B] & 0.737 & 1.100 & 1.248 & 1.667 & 2.327 & 2.765 & 2.842 & 2.792 & * \\ 
      RW & multiESN [A] & 0.773 & 0.972 & 1.053 & 1.111 & 1.187 & 1.293 & 1.505 & 1.131 & * \\ 
      RW & multiESN [B] & \chl0.716 & \cHL\textbf{0.895} & \cHL\textbf{0.916} & \cHL\textbf{0.926} & \cHL\textbf{0.890} & 1.041 & 1.102 & 1.105 & ** \\ 
   \bottomrule
\end{tabular*}
    \caption{Relative RMSFE and Uniform Multi-Horizon Model Confidence Set (uMCS) comparison between models in multiple-steps-ahead forecasting exercises. Unconditional mean RMSFE used as reference. FIX: Fixed parameters, EW: Expanding window, and RW: Rolling window. uMCS columns show inclusion among best models: $\ast$ indicates inclusion at 90\% confidence, $\ast\ast$ indicates inclusion at 75\% confidence.}
	\label{table:msa_2007_GDP_smallMD}
\end{table}

\begin{table}[!p]
\vspace{5em}
\centering
	\setlength\tabcolsep{6.5pt}
    \setlength\extrarowheight{2pt}
	\linespread{1}\selectfont\centering
	Multistep-ahead GDP Forecasting - Small-MD Dataset - 2011 Sample \\[5pt]
	\begin{tabular*}{\textwidth}{@{}*{11}{c}}
  \toprule
    Setup & Model & \multicolumn{8}{c}{Horizon} & uMCS \\
    \cline{3-10}
    & & 1 & 2 & 3 & 4 & 5 & 6 & 7 & 8 & \\ 
  \midrule
      FIX & Mean & 1.000 & \chl1.000 & 1.000 & \chl1.000 & \cHL\textbf{1.000} & 1.000 & \chl1.000 & \chl1.000 & ** \\ 
      FIX & AR(1) & 1.119 & 1.031 & 1.008 & 1.001 & \chl1.001 & \chl0.999 & \cHL\textbf{0.999} & \cHL\textbf{0.998} & * \\ 
      FIX & MIDAS & 1.090 & 1.721 & 1.793 & 2.203 & 2.363 & 1.997 & 2.846 & 2.328 &  \\ 
      FIX & DFM [A] & 1.112 & 1.051 & 0.999 & 1.079 & 1.084 & 1.025 & 1.020 & 1.061 & * \\ 
      FIX & DFM [B] & 1.058 & \cHL\textbf{0.945} & \cHL\textbf{0.916} & 1.003 & 1.012 & \cHL\textbf{0.970} & 1.038 & 1.033 & ** \\ 
      FIX & singleESN [A] & \chl0.978 & 1.705 & 2.561 & 2.704 & 3.314 & 3.151 & 2.999 & 3.316 &  \\ 
      FIX & singleESN [B] & \cHL\textbf{0.930} & 1.095 & 1.885 & 2.356 & 2.650 & 2.704 & 2.880 & 2.844 & ** \\ 
      FIX & multiESN [A] & 1.059 & 1.148 & 1.262 & 1.312 & 1.339 & 1.409 & 1.424 & 1.162 &  \\ 
      FIX & multiESN [B] & 0.981 & 1.007 & \chl0.985 & \cHL\textbf{0.994} & 1.008 & \chl0.999 & \cHL\textbf{0.999} & \cHL\textbf{0.998} & ** \\ 
  \midrule
      EW & Mean & 1.000 & 1.000 & 1.000 & 1.000 & 1.000 & 1.000 & \chl1.000 & 1.000 & - \\ 
      EW & AR(1) & 1.117 & 1.033 & 1.011 & 1.002 & 1.007 & 1.003 & 1.004 & 1.003 & - \\ 
      EW & MIDAS & 1.005 & 1.382 & 1.339 & 1.354 & 1.609 & 1.444 & 1.803 & 1.263 & - \\ 
      EW & DFM [A] & 1.144 & 1.132 & 1.057 & 1.093 & 1.076 & 1.067 & 1.038 & 1.016 & - \\ 
      EW & DFM [B] & 0.985 & \cHL\textbf{0.940} & \cHL\textbf{0.918} & 0.995 & 1.010 & \cHL\textbf{0.980} & 1.050 & \cHL\textbf{0.971} & - \\ 
      EW & singleESN [A] & 0.935 & 1.645 & 2.184 & 1.929 & 2.388 & 1.959 & 1.810 & 2.266 & - \\ 
      EW & singleESN [B] & \cHL\textbf{0.911} & 1.092 & 1.101 & 1.529 & 2.195 & 1.843 & 1.847 & 2.060 & - \\ 
      EW & multiESN [A] & \chl0.922 & \chl0.965 & 1.089 & \chl0.978 & \cHL\textbf{0.977} & 1.043 & 1.278 & \chl0.995 & - \\ 
      EW & multiESN [B] & 0.944 & 0.992 & \chl0.978 & \cHL\textbf{0.977} & \chl0.991 & \chl0.985 & \cHL\textbf{0.990} & 0.996 & - \\  
  \midrule
      RW & Mean & 1.000 & \chl1.000 & 1.000 & \chl1.000 & 1.000 & 1.000 & 1.000 & 1.000 &  \\ 
      RW & AR(1) & 1.080 & \chl1.000 & 0.984 & \cHL\textbf{0.989} & \chl0.982 & 0.976 & \cHL\textbf{0.963} & 0.968 &  \\ 
      RW & MIDAS & 1.051 & 1.303 & 1.310 & 1.674 & 1.762 & 1.467 & 1.643 & 1.463 &  \\ 
      RW & DFM [A] & 1.061 & 1.033 & 1.012 & 1.088 & 1.077 & 1.015 & 1.040 & 1.069 &  \\ 
      RW & DFM [B] & 0.947 & \cHL\textbf{0.893} & \cHL\textbf{0.901} & 1.009 & 1.040 & \cHL\textbf{0.966} & 1.030 & \cHL\textbf{0.949} & ** \\ 
      RW & singleESN [A] & \chl0.919 & 1.788 & 2.359 & 2.483 & 2.981 & 2.401 & 2.234 & 2.690 &  \\ 
      RW & singleESN [B] & 0.944 & 1.132 & 1.214 & 1.762 & 2.608 & 2.552 & 2.517 & 2.541 &  \\ 
      RW & multiESN [A] & \cHL\textbf{0.896} & 1.047 & 1.222 & 1.124 & 1.122 & 1.410 & 1.666 & 1.316 &  \\ 
      RW & multiESN [B] & 0.940 & 1.003 & \chl0.969 & \cHL\textbf{0.989} & \cHL\textbf{0.979} & \chl0.972 & \chl0.967 & \chl0.961 & ** \\ 
   \bottomrule
\end{tabular*}
    \caption{Relative RMSFE and Uniform Multi-Horizon Model Confidence Set (uMCS) comparison between models in multiple-steps-ahead forecasting exercises. Unconditional mean RMSFE used as reference. FIX: Fixed parameters, EW: Expanding window, and RW: Rolling window. uMCS columns show inclusion among best models: $\ast$ indicates inclusion at 90\%, $\ast\ast$ indicates inclusion at 75\% confidence.}
	\label{table:msa_2011_GDP_smallMD}
\end{table}

For MIDAS, we have already discussed how the existence of multiple loss minima can generate numerical instabilities. Model re-fitting at each horizon can amplify this problem, as the loss landscape itself changes as new observations are added to the fitting sample. We provide more discussions in Supplementary Appendix~\ref{sec:midas_robustness}.
In the case of S-MFESN models, the reason is structural: we have discussed how in our framework multistep MFESN forecasting entails iterating the state map, which can have multiple attraction (stable) points. If the hyperparameters and estimated full model $\widehat{W}$s jointly do not define a contraction, the limit of the multistep forecast does not have to be the estimated MFSEN model intercept. However, Figures \ref{fig:msa_2007_GDP_smallMD} and \ref{fig:msa_2011_GDP_smallMD} show that our M-MFESN models, multiESN [A] and multiESN [B], both perform on par or better than DFM models even after horizon $h = 4$. For example, in the 2007 expanding and rolling window experiments, multiESN [B] is able to outperform both DFMs and an unconditional mean forecast by meaningful margins for forecasts up to a year into the future.

\subsubsection{Medium Dataset}

\begin{table}[t!]
	\centering
	\setlength\tabcolsep{3.3pt}
    \setlength\extrarowheight{2pt}
	\linespread{1}\selectfont\centering
	1-Step-ahead GDP Forecasting - Medium-MD Dataset \\[5pt]
	\begin{tabular*}{\textwidth}{@{}*{13}{c}}
		\toprule
		& \multicolumn{4}{c}{Fixed Parameters} & \multicolumn{4}{c}{Expanding Window} & \multicolumn{4}{c}{Rolling Window} \\
		\cline{2-5} \cline{6-9} \cline{10-13}
		Model & \multicolumn{2}{c}{2007} & \multicolumn{2}{c}{2011} & \multicolumn{2}{c}{2007} & \multicolumn{2}{c}{2011} & \multicolumn{2}{c}{2007} & \multicolumn{2}{c}{2011} \\
	    \cline{2-3} \cline{4-5} \cline{6-7} \cline{8-9} \cline{10-11} \cline{12-13}
		 & MSFE & MCS & MSFE & MCS & MSFE & MCS & MSFE & MCS & MSFE & MCS & MSFE & MCS \\
		\midrule
          Mean & 1.000 &  & \chl1.000 & \chl** & 1.000 & ** & 1.000 & ** & 1.000 & ** & 1.000 & ** \\ 
          AR(1) & \chl0.758 & \chl* & 1.230 & ** & 0.789 & ** & 1.226 & ** & 0.775 & * & 1.209 & ** \\ 
          DFM [A] & 0.841 & * & 1.325 & * & 0.682 & ** & 1.272 & ** & 0.747 & * & 1.517 & ** \\ 
          DFM [B] & 1.118 & * & 1.408 & ** & 0.821 & * & 1.117 & ** & 0.926 &  & 1.186 & ** \\ 
          singleESN [A] & 0.967 & * & 1.717 & * & 0.775 & ** & 1.072 & ** & 0.791 & * & 1.493 & * \\ 
          singleESN [B] & 0.826 & * & 1.278 & ** & 0.655 & ** & 1.028 & ** & 0.561 & ** & 0.944 & ** \\ 
          multiESN [A] & 0.901 & * & 1.080 & ** & \chl0.618 & \chl** & \chl0.913 & \chl** & \chl0.556 & \chl** & \chl0.884 & \chl** \\ 
          multiESN [B] & \cHL\textbf{0.682} & \cHL** & \cHL\textbf{0.748} & \cHL** & \cHL\textbf{0.587} & \cHL** & \cHL\textbf{0.774} & \cHL** & \cHL\textbf{0.547} & \cHL** & \cHL\textbf{0.728} & \cHL** \\ 
		\bottomrule
	\end{tabular*}
	\caption{\small Relative MSFE and Model Confidence Set (MCS) comparison between models in 1-step-ahead forecasting exercises. Unconditional mean MSFE is used as a reference. MCS columns show inclusion among best models: $\ast$ indicates inclusion at 90\% confidence, $\ast\ast$ indicates inclusion at 75\% confidence.}
	\label{table:1sa_GDP_mediumMD}
\end{table}

We now present the results for the Medium-MD dataset, which includes more than 30 regressors and many high-frequency daily series. The same metrics as in the previous subsection are used for this dataset to evaluate the relative performance of different methods. 

The main difference in our empirical exercises is that now we a priori exclude MIDAS from the set of forecasting methods as explained in detail in Subsection~\ref{Benchmarks}. Table \ref{table:1sa_GDP_mediumMD} showcases the relative performance of DFM and MFESN models in the Medium-MD forecast setup. We find that the MFESN model multiESN [B] performs best in all setups, particularly under fixed parameters, where MCS testing reveals that it is the only model included at a 75\% confidence level. Of course, for the MCS results we must again take into account the relatively small sample size, which could distort the selection of best model subsets. MDM  tests of Figure \ref{fig:1sa_GDP_mediumMD_diebold_mariano} in Supplementary Appendix \ref{subsec:additional_figures} largely agree with the MCS results: in the fixed parameter setup any pairwise comparison of an alternative model against MFESN multiESN [B] is rejected in favor of the latter. 
A visual inspection of one-step-ahead forecasts in Figures \ref{fig:1sa_forecasts_2007_GDP_mediumMD} and \ref{fig:1sa_forecasts_2011_GDP_mediumMD} in Appendix~\ref{subsec:forecast_figures} also shows that DFM models estimated over the Medium-MD datasets produce forecasts with larger variability than MFESN methods, which is likely the key driver of the difference in performance.

The multistep-ahead experiments are run as for the Small-MD dataset, with a maximum horizon of 8 quarters. Tables \ref{table:msa_2007_GDP_mediumMD} and \ref{table:msa_2011_GDP_mediumMD} present the relative RMSFE performance of multistep forecasts for all models, and we use Figures \ref{fig:msa_2007_GDP_mediumMD} and \ref{fig:msa_2011_GDP_mediumMD} of RMSFEs as references for our discussion, available in Appendix~\ref{subsec:forecast_figures}. What can be seen visually -- and is also reproduced in the Tables -- is that multi-reservoir MFESN models and DFM model [A] have the best performance up to 4 quarters ahead; overall, taking into account also the longer term, expanding or rolling window estimation of model multiESN [B] yields the best forecasting results in the 2007 sample setup. The post-crisis 2011 sample setup makes comparison harder, as DFM and M-MFESN models largely produce results in line with the unconditional sample mean. This evaluation is confirmed by uMCS tests, consistently with the multistep results obtained with the Small-MD dataset.

\begin{table}[!p]
\vspace{5em}
\centering
	\setlength\tabcolsep{6.5pt}
    \setlength\extrarowheight{2pt}
	\linespread{1}\selectfont\centering
	Multistep-ahead GDP Forecasting - Medium-MD Dataset - 2007 Sample \\[5pt]
	\begin{tabular*}{\textwidth}{@{}*{11}{c}}
  \toprule
    Setup & Model & \multicolumn{8}{c}{Horizon} & uMCS \\
    \cline{3-10}
    & & 1 & 2 & 3 & 4 & 5 & 6 & 7 & 8 & \\ 
  \midrule
      FIX & Mean & 1.000 & 1.000 & 1.000 & 1.000 & 1.000 & 1.000 & 1.000 & 1.000 & * \\ 
      FIX & AR(1) & \chl0.870 & \chl0.950 & 0.982 & 0.991 & 0.992 & 0.991 & 0.992 & 0.992 &  \\ 
      FIX & DFM [A] & 0.914 & \cHL\textbf{0.947} & \cHL\textbf{0.955} & 0.988 & 1.015 & 1.027 & 1.034 & 0.995 & ** \\ 
      FIX & DFM [B] & 1.046 & 1.204 & 1.293 & 1.341 & 1.649 & 1.984 & 2.101 & 2.070 & * \\ 
      FIX & singleESN [A] & 0.985 & 0.995 & 0.995 & 0.995 & 0.994 & 0.992 & 0.992 & 0.992 & * \\ 
      FIX & singleESN [B] & 0.912 & 0.985 & 0.985 & \cHL\textbf{0.985} & \cHL\textbf{0.980} & \cHL\textbf{0.976} & \cHL\textbf{0.976} & \cHL\textbf{0.976} & * \\ 
      FIX & multiESN [A] & 0.950 & 0.993 & 0.994 & 0.994 & 0.992 & 0.990 & 0.990 & 0.990 & * \\ 
      FIX & multiESN [B] & \cHL\textbf{0.826} & 0.972 & \chl0.988 & \chl0.990 & \chl0.989 & \chl0.986 & \chl0.985 & \chl0.985 & * \\ 
  \midrule
      EW & Mean & 1.000 & 1.000 & 1.000 & 1.000 & 1.000 & 1.000 & \cHL\textbf{1.000} & \cHL\textbf{1.000} & - \\ 
      EW & AR(1) & 0.887 & 0.922 & \chl0.951 & \chl0.962 & \chl0.957 & \chl0.981 & \chl1.001 & \chl1.008 & - \\ 
      EW & DFM [A] & 0.805 & \chl0.916 & 0.978 & 1.038 & 1.077 & 1.126 & 1.077 & 1.073 & - \\ 
      EW & DFM [B] & 0.893 & 1.134 & 1.418 & 1.567 & 2.238 & 2.964 & 3.375 & 3.629 & - \\ 
      EW & singleESN [A] & 0.879 & 1.125 & 1.305 & 1.442 & 1.860 & 2.166 & 2.361 & 2.443 & - \\ 
      EW & singleESN [B] & 0.802 & 1.174 & 1.439 & 1.744 & 2.305 & 2.869 & 2.935 & 3.167 & - \\ 
      EW & multiESN [A] & \chl0.780 & 0.935 & 1.012 & 1.005 & 1.093 & 1.337 & 1.328 & 1.313 & - \\ 
      EW & multiESN [B] & \cHL\textbf{0.760} & \cHL\textbf{0.874} & \cHL\textbf{0.911} & \cHL\textbf{0.891} & \cHL\textbf{0.863} & \cHL\textbf{0.971} & 1.030 & 1.051 & - \\   
  \midrule
      RW & Mean & 1.000 & 1.000 & 1.000 & 1.000 & 1.000 & 1.000 & \chl1.000 & \chl1.000 &  \\ 
      RW & AR(1) & 0.898 & 0.943 & 0.968 & 0.974 & \chl0.963 & \cHL\textbf{0.968} & \cHL\textbf{0.970} & \cHL\textbf{0.962} &  \\ 
      RW & DFM [A] & 0.837 & \chl0.913 & \chl0.924 & \chl0.954 & 1.012 & 0.997 & 1.018 & 1.005 &  \\ 
      RW & DFM [B] & 0.932 & 1.116 & 1.232 & 1.414 & 1.952 & 2.704 & 3.183 & 3.294 &  \\ 
      RW & singleESN [A] & 0.873 & 1.274 & 1.530 & 1.652 & 2.095 & 2.575 & 2.786 & 3.014 &  \\ 
      RW & singleESN [B] & \chl0.732 & 1.190 & 1.490 & 1.712 & 2.218 & 2.861 & 2.967 & 3.094 &  \\ 
      RW & multiESN [A] & \chl0.732 & 0.914 & 0.960 & 1.011 & 1.202 & 1.618 & 1.683 & 1.572 &  \\ 
      RW & multiESN [B] & \cHL\textbf{0.731} & \cHL\textbf{0.871} & \cHL\textbf{0.875} & \cHL\textbf{0.844} & \cHL\textbf{0.771} & \chl0.971 & 1.014 & 1.014 & ** \\
   \bottomrule
\end{tabular*}
    \caption{Relative RMSFE and Uniform Multi-Horizon Model Confidence Set (uMCS) comparison between models in multiple-steps-ahead forecasting exercises. Unconditional mean RMSFE used as reference. FIX: Fixed parameters, EW: Expanding window, and RW: Rolling window. uMCS columns show inclusion among best models: $\ast$ indicates inclusion at 90\% confidence, $\ast\ast$ indicates inclusion at 75\% confidence.}
	\label{table:msa_2007_GDP_mediumMD}
\end{table}

\begin{table}[!p]
\vspace{5em}
\centering
	\setlength\tabcolsep{6.5pt}
    \setlength\extrarowheight{2pt}
	\linespread{1}\selectfont\centering
	Multistep-ahead GDP Forecasting - Medium-MD Dataset - 2011 Sample \\[5pt]
	\begin{tabular*}{\textwidth}{@{}*{11}{c}}
  \toprule
    Setup & Model & \multicolumn{8}{c}{Horizon} & uMCS \\
    \cline{3-10}
    & & 1 & 2 & 3 & 4 & 5 & 6 & 7 & 8 & \\ 
  \midrule
      FIX & Mean & 1.000 & 1.000 & 1.000 & \chl1.000 & \cHL\textbf{1.000} & 1.000 & \chl1.000 & \chl1.000 & * \\ 
      FIX & AR(1) & 1.119 & 1.031 & 1.008 & 1.001 & \chl1.001 & \chl0.999 & \cHL\textbf{0.999} & \cHL\textbf{0.998} & ** \\ 
      FIX & DFM [A] & 1.126 & 0.987 & \chl0.962 & 1.054 & 1.031 & \cHL\textbf{0.988} & 1.001 & 1.002 & ** \\ 
      FIX & DFM [B] & 1.149 & 0.987 & \cHL\textbf{0.885} & 1.064 & 1.142 & 1.134 & 1.273 & 1.296 &  \\ 
      FIX & singleESN [A] & 1.283 & 1.921 & 2.527 & 3.038 & 3.285 & 3.154 & 3.193 & 3.655 &  \\ 
      FIX & singleESN [B] & 1.059 & 1.523 & 1.918 & 2.417 & 2.812 & 2.683 & 2.703 & 2.970 &  \\ 
      FIX & multiESN [A] & 1.011 & 1.061 & 1.434 & 1.477 & 1.748 & 2.030 & 2.023 & 1.994 &  \\ 
      FIX & multiESN [B] & \cHL\textbf{0.841} & \cHL\textbf{0.945} & 0.997 & \cHL\textbf{0.978} & 1.004 & 1.015 & 1.013 & 1.014 & ** \\  
  \midrule
      EW & Mean & 1.000 & 1.000 & \chl1.000 & \chl1.000 & \cHL\textbf{1.000} & \chl1.000 & \chl1.000 & \chl1.000 & - \\ 
      EW & AR(1) & 1.117 & 1.033 & 1.011 & 1.002 & 1.007 & 1.003 & 1.004 & 1.003 & - \\ 
      EW & DFM [A] & 1.092 & \chl0.942 & \cHL\textbf{0.944} & 1.049 & 1.026 & \cHL\textbf{0.994} & \cHL\textbf{0.996} & \cHL\textbf{0.999} & - \\ 
      EW & DFM [B] & 0.971 & 1.046 & 1.031 & 1.114 & 1.238 & 1.116 & 1.223 & 1.310 & - \\ 
      EW & singleESN [A] & 1.039 & 1.451 & 1.980 & 2.385 & 2.699 & 2.353 & 2.506 & 2.608 & - \\ 
      EW & singleESN [B] & 0.992 & 1.828 & 2.465 & 3.072 & 3.547 & 3.357 & 3.368 & 3.610 & - \\ 
      EW & multiESN [A] & \chl0.934 & 1.014 & 1.391 & 1.252 & 1.371 & 1.369 & 1.228 & 1.279 & - \\ 
      EW & multiESN [B] & \cHL\textbf{0.857} & \cHL\textbf{0.931} & 1.003 & \cHL\textbf{0.973} & \chl1.002 & 1.009 & 1.025 & 1.029 & - \\ 
  \midrule
      RW & Mean & 1.000 & 1.000 & 1.000 & \chl1.000 & \chl1.000 & 1.000 & 1.000 & 1.000 & ** \\ 
      RW & AR(1) & 1.080 & 1.000 & \chl0.984 & \cHL\textbf{0.989} & \cHL\textbf{0.982} & \cHL\textbf{0.976} & \cHL\textbf{0.963} & \cHL\textbf{0.968} & ** \\ 
      RW & DFM [A] & 1.113 & 0.982 & \cHL\textbf{0.927} & 1.038 & 1.030 & 0.997 & 1.016 & 1.028 & * \\ 
      RW & DFM [B] & 0.881 & 0.996 & 1.021 & 1.098 & 1.150 & 1.114 & 1.114 & 1.212 & ** \\ 
      RW & singleESN [A] & 1.193 & 2.267 & 3.265 & 3.580 & 4.090 & 3.790 & 4.015 & 4.562 &  \\ 
      RW & singleESN [B] & 0.927 & 1.933 & 2.612 & 3.265 & 3.753 & 3.567 & 3.556 & 3.792 &  \\ 
      RW & multiESN [A] & 0.900 & 1.049 & 1.500 & 1.465 & 1.789 & 1.707 & 1.505 & 1.462 &  \\ 
      RW & multiESN [B] & \cHL\textbf{0.816} & \cHL\textbf{0.916} & 0.977 & 1.009 & \cHL\textbf{0.982} & \chl0.988 & \chl0.974 & \chl0.981 & ** \\ 
   \bottomrule
\end{tabular*}
    \caption{Relative RMSFE and Uniform Multi-Horizon Model Confidence Set (uMCS) comparison between models in multiple-steps-ahead forecasting exercises. Unconditional mean RMSFE used as reference. FIX: Fixed parameters, EW: Expanding window, and RW: Rolling window. uMCS columns show inclusion among best models: $\ast$ indicates inclusion at 90\% confidence, $\ast\ast$ indicates inclusion at 75\% confidence.}
	\label{table:msa_2011_GDP_mediumMD}
\end{table}

\section{Conclusions}
\label{Conclusions}

Macroeconomic forecasting -- especially long-term forecasting of macroeconomic aggregates -- is a topic of crucial importance for institutional policymakers, private companies, and economic researchers. Given the modern-day availability of ``big data'' resources, methods capable of integrating heterogeneous data sources are increasingly sought to provide more precise and robust forecasts. 

This paper presents a new methodological framework inspired by the Reservoir Computing literature to deal with data sampled at multiple frequencies and with multiple-step-ahead forecasts. We have then taken Echo State Networks -- a type of RC models -- and formally extended them to allow the modeling of data with multiple release frequencies. Our discussion encompasses model fitting, hyperparameter tuning, and forecast computation. As a result, we provide two classes of models, single- and multiple reservoir multi-frequency ESNs, that can be effectively applied to our empirical setup: forecasting US GDP growth using monthly and daily data series. Along with the unconditional mean and AR(1) model, we considered two well-known methods, MIDAS and DFMs, as the current benchmarks available in the literature. In our applications, we find that MFESN models are computationally more efficient and easier to implement than DFMs and MIDAS, respectively, and perform better than or as well as benchmarks in terms of MSFE. These improvements are statistically significant in a number of setups, as shown by our MCS and MDM tests. Thus, we argue that our machine learning-based methodology can be a useful addition to the toolbox of contemporary macroeconomic forecasters.

Lastly, we wish to highlight the many potential areas of research that we believe would be interesting to explore in the future. We have not discussed the role of the distribution from which we sample the entries of the reservoir matrices. While it is known that these can have significant effects on the forecasting capacity of an ESN model, the literature lacks definitive theoretical results (even for dynamical systems applications) or systematic studies with stochastic inputs and targets. The hyperparameter tuning routine we have developed neither allows separating individual hyperparameters nor does it tackle the identification problem. Moreover, we assume that the ridge regression penalty strength, $\lambda$, is tuned \textit{ex ante}: it would be interesting and desirable to understand if it is possible to jointly tune $\lambda$ and $\boldsymbol{\varphi}$, or rather if one can fully separate their selection. In our preliminary experiments, we have noticed that the roles of the ridge penalty and the input scaling, for example, cannot be trivially disentangled -- thus prompting the $\psi$-form normalization. Model selection for the dimension of MFESN models is another question that would be key to exploring and designing more efficient and effective ESN models, especially when dealing with multiple frequencies and reservoirs. Finally, practitioners may be interested in identifying the combination of frequencies in the regressor series that would lead to the most accurate GDP forecasts produced by MFESN models.

\newpage

{\footnotesize
\bibliography{GOLibrary.bib}
}


\appendix

\newpage

\begin{center}\huge
    Appendix
\end{center}

\section{Data Table}
\label{subsec:data_table}

\begin{table}[!h]
 \renewcommand{\arraystretch}{0.8}
 \begin{center}
  \begin{threeparttable}
   \caption{Variables, Frequencies and Transformations for Small and Medium \label{tab:dataMD}}
     \begin{small}
      \begin{tabular*}{\textwidth}{l @{\extracolsep{\fill}} lllllll}
       \toprule
       S & M & Start Date & T & Code & Name & Description \\  
        \midrule
        \multicolumn{7}{l}{Quarterly} \\
        \midrule
        X & X & 31/03/1959 & 5 & GDPC1 & Y & Real Gross Domestic Product\\
        \midrule
        \multicolumn{7}{l}{Monthly} \\
        \midrule
        X & X & 30/01/1959 & 5 & INDPRO & XM1 & Industrial Production Index\\
        X & X & 30/01/1959 & 5 & PAYEMS & XM4 & Payroll All Employees: Total nonfarm \\
        X & X & 30/01/1959 & 4 & HOUST & XM5 & Housing Starts: Total New Privately Owned \\
        X & X & 30/01/1959 & 5 & RETAILx & XM7 & Retail and Food Services Sales\\
        X & X & 31/01/1973 & 5 & TWEXMMTH & XM11 & Nominal effective exchange rate US \\
        X & X & 30/01/1959 & 2 & FEDFUNDS & XM12 & Effective Federal Funds Rate\\
        X & X & 30/01/1959 & 1 & BAAFFM & XM14 & Moody’s Baa Corporate Bond Minus FEDFUNDS \\
        X & X & 30/01/1959 & 1 & COMPAPFFx & XM15 & 3-Month Commercial Paper Minus FEDFUNDS \\
        & X & 30/01/1959 & 2 & CUMFNS & XM2 & Capacity Utilization: Manufacturing \\
        & X & 30/01/1959 & 2 & UNRATE & XM3 & Civilian Unemployment Rate \\
        & X & 30/01/1959 & 5 & DPCERA3M086SBEA & XM6 & Real personal consumption expenditures \\
        & X & 30/01/1959 & 5 & AMDMNOx & XM8 & New Orders for Durable Goods \\
        & X & 31/01/1978 & 2 & UMCSENTx & XM9 & Consumer Sentiment Index \\
        & X & 30/01/1959 & 6 & WPSFD49207 & XM10 & PPI: Finished Goods \\
        & X & 30/01/1959 & 1 & AAAFFM & XM13 & Moody’s Aaa Corporate Bond Minus FEDFUNDS \\
        & X & 30/01/1959 & 1 & TB3SMFFM  & XM16 & 3-Month Treasury C Minus FEDFUNDS \\
        & X & 30/01/1959 & 1 & T10YFFM & XM17 & 10-Year Treasury C Minus FEDFUNDS \\
        & X & 30/01/1959 & 2 & GS1 & XM18 & 1-Year Treasury Rate \\
        & X & 30/01/1959 & 2 & GS10 & XM19 & 10-Year Treasury Rate \\
        & X & 30/01/1959 & 1 & GS10-TB3MS & XM20 & 10-Year Treasury Rate - 3-Month Treasury Bill \\
        \midrule
        \multicolumn{7}{l}{Daily} \\
        \midrule
        X & X & 30/01/1959 & 8 & DJINDUS & XD3 & DJ Industrial price index \\
        & X & 31/12/1963 & 8 & S\&PCOMP & XD1 & S\&P500 price index \\
        & X & 01/05/1982 & 1 & ISPCS00-S\&PCOMP\textsuperscript{\dag} & XD2 & S\&P500 basis spread \\
        & X & 11/09/1989 & 8 & SP5EIND & XD4 & S\&P Industrial price index \\
        & X & 31/12/1969 & 8 & GSCITOT & XD5 & Spot commodity price index \\
        & X & 10/01/1983 & 8 & CRUDOIL & XD6 & Spot price oil \\
        & X & 02/01/1979 & 8 & GOLDHAR & XD7 & Spot price gold \\
        & X & 30/03/1982 & 8 & WHEATSF & XD8 & Spot price wheat \\
        & X & 01/11/1983 & 8 & COCOAIC,COCINUS\textsuperscript{\ddag} & XD9 & Spot price cocoa \\
        & X & 30/03/1983 & 1 & NCLC.03-NCLC.01 & XD10 & Futures price oil term structure \\
        & X & 30/10/1978 & 1 & NGCC.03-NGCC.01 & XD11 & Futures price gold term structure\\
        & X & 02/01/1975 & 1 & CWFC.03-CWFC.01 & XD12 & Futures price wheat term structure\\
        & X & 02/01/1973 & 1 & NCCC.03-NCCC.01 & XD13 & Futures price cocoa term structure\\
       \bottomrule
     \end{tabular*}
      \end{small}
    \begin{tablenotes}[para,flushleft]
      \small
      Notes: S and M stand for small and medium datasets, respectively. An `X' indicates selection into the dataset. `Start Date' is the date for which the series is first available (before data transformations). Following \cite{McCracken2016,McCracken2020}, the transformation codes in column `T' indicate with D for difference and log for natural logarithm 1: none, 2: D, 3: DD, 4: Log, 5: Dlog, 6: DDlog, 7: percentage change, 8: GARCH volatility. `Codes' are the codes in the FRED-QD and FRED-MD datasets for quarterly and monthly data and Datastream mnemonic for the remaining frequencies. Missing values due to public holidays are interpolated by averaging over the previous five observations. \textsuperscript{\ddag}Available until 20/09/2021. \textsuperscript{\ddag}Average before 29/12/2017, COCINUS mean adjusted thereafter.     
    \end{tablenotes}
  \end{threeparttable}
 \end{center}
 \renewcommand{\arraystretch}{1}
\end{table}

\newpage

\section{Forecasting Figures}
\label{subsec:forecast_figures}

\begin{figure}[!h]
    \centering
    \caption{1-Step-ahead GDP Forecasting -- 2007 Sample -- Small-MD Dataset}
    \label{fig:1sa_forecasts_2007_GDP_smallMD}
    \vspace{1em}
    \subfloat[Fixed]{%
      \includegraphics[width= 1\linewidth]{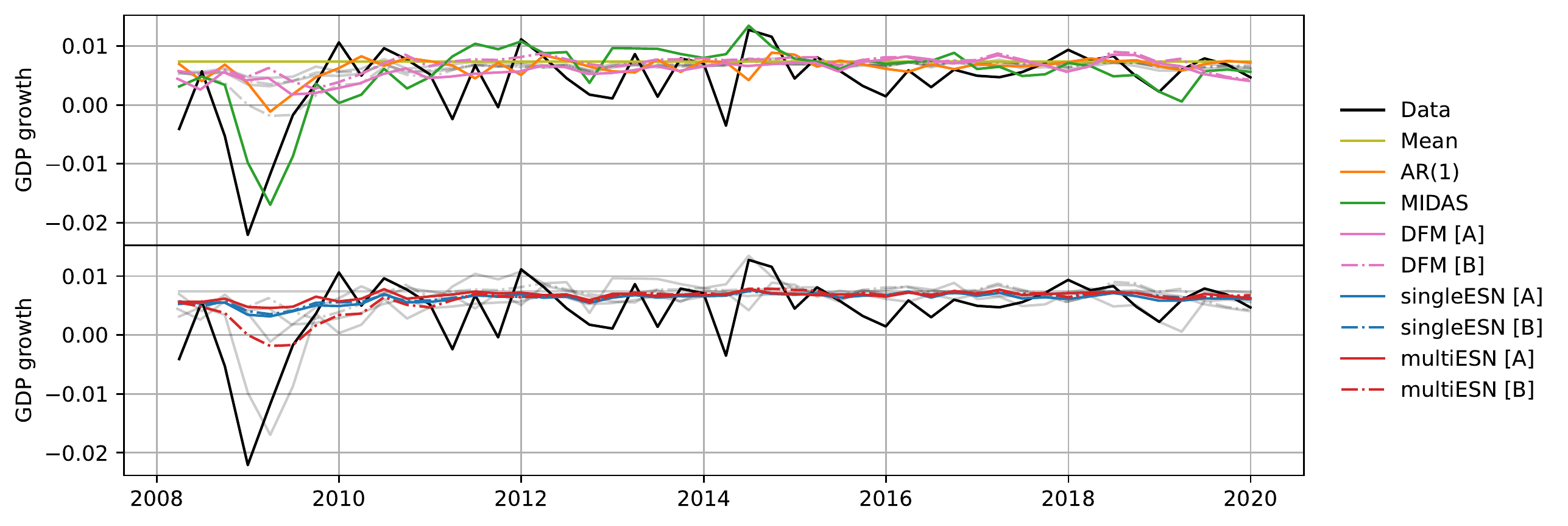}
    }
    \vspace{2em}
    \\
    \subfloat[Expanding]{%
      \includegraphics[width= 1\linewidth]{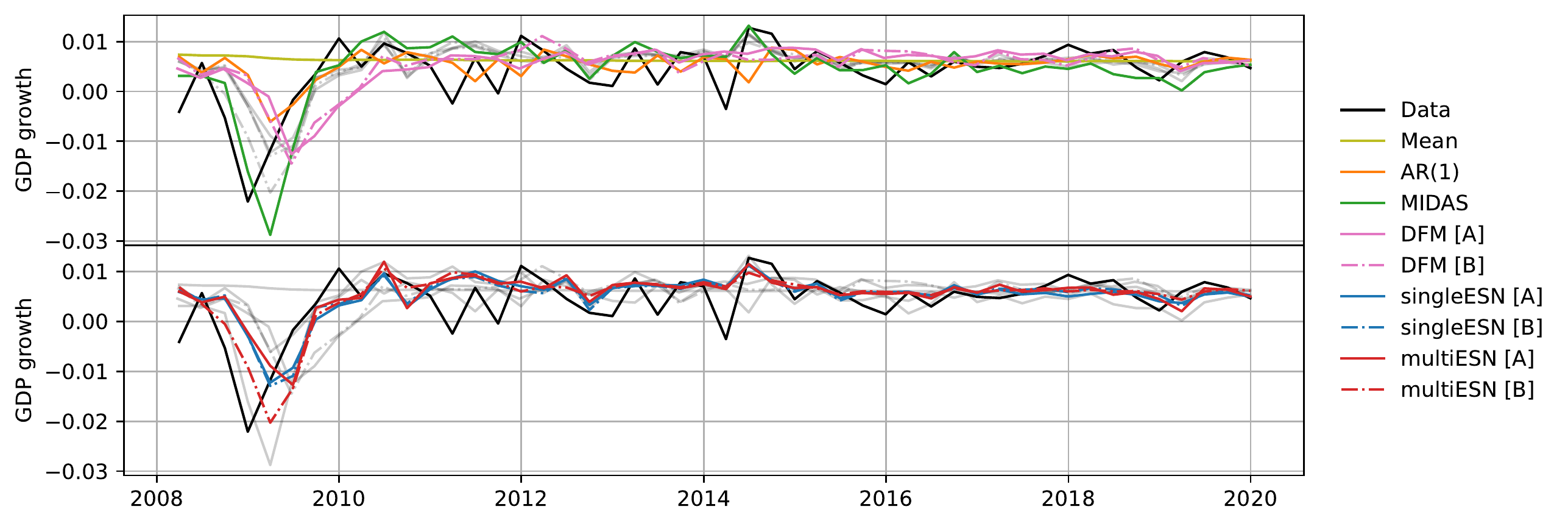}
    }
    \vspace{2em}
    \\
    \subfloat[Rolling]{%
      \includegraphics[width= 1\linewidth]{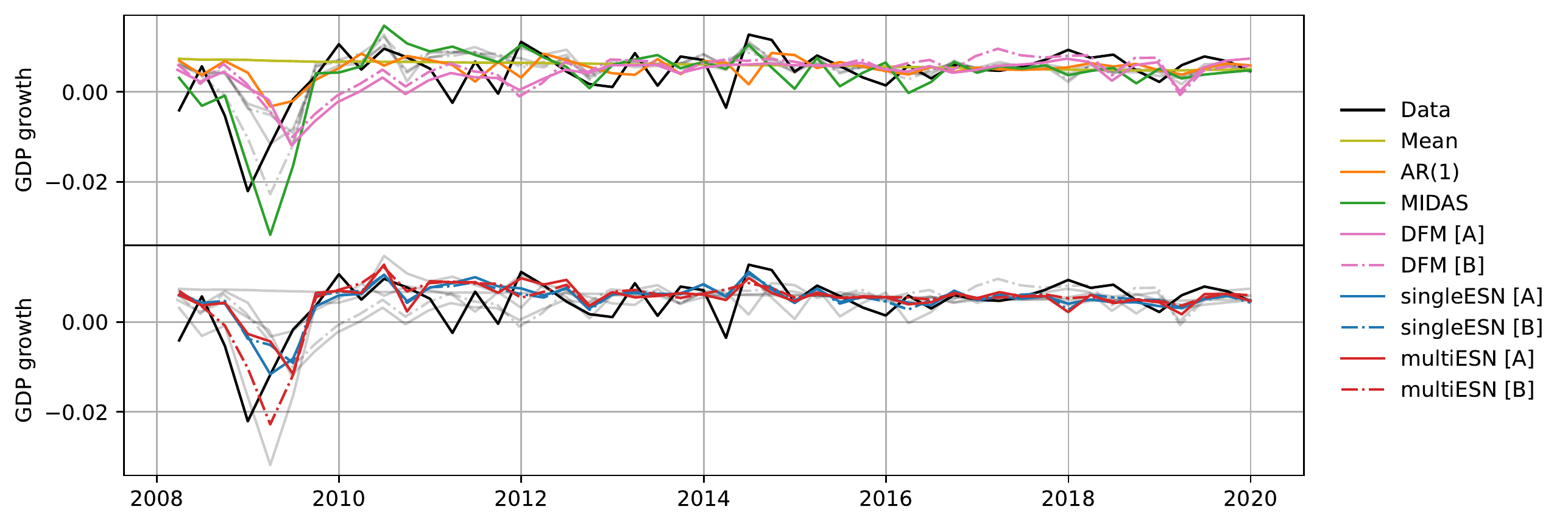}
    }
 \end{figure}
 
 \begin{figure}[!p]
    \centering
    \caption{1-Step-ahead GDP Forecasting -- 2011 Sample -- Small-MD Dataset}
    \label{fig:1sa_forecasts_2011_GDP_smallMD}
    \vspace{1em}
    \subfloat[Fixed]{%
      \includegraphics[width= 1\linewidth]{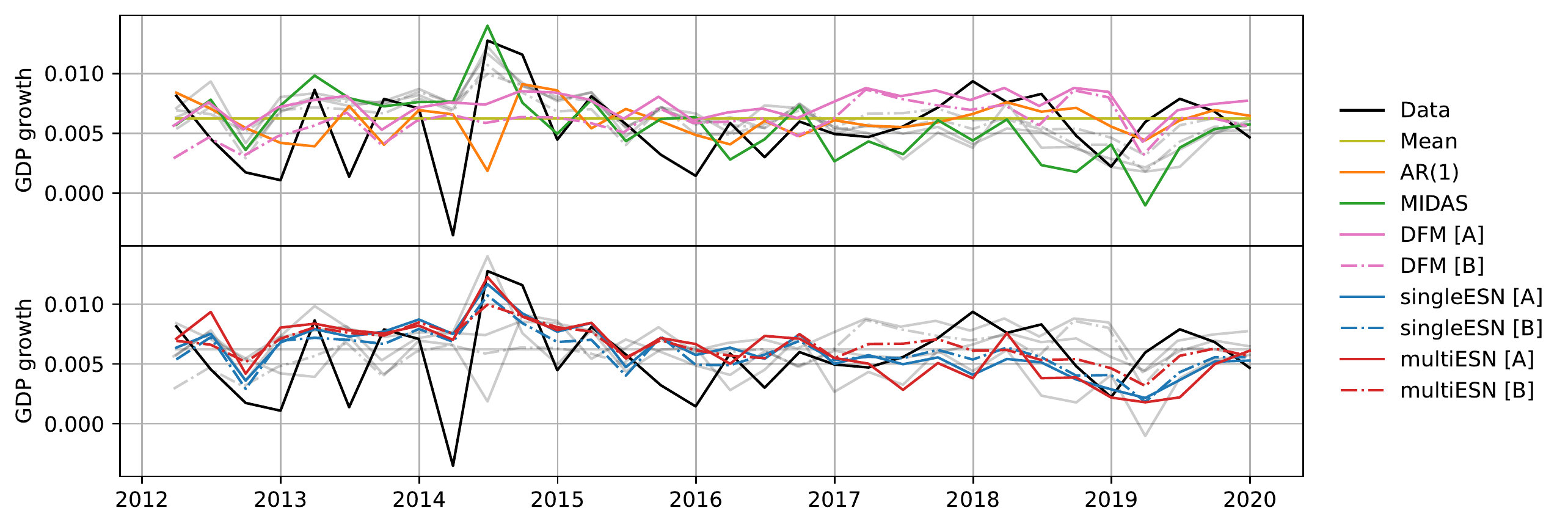}
    }
    \vspace{2em}
    \\
    \subfloat[Expanding]{%
      \includegraphics[width= 1\linewidth]{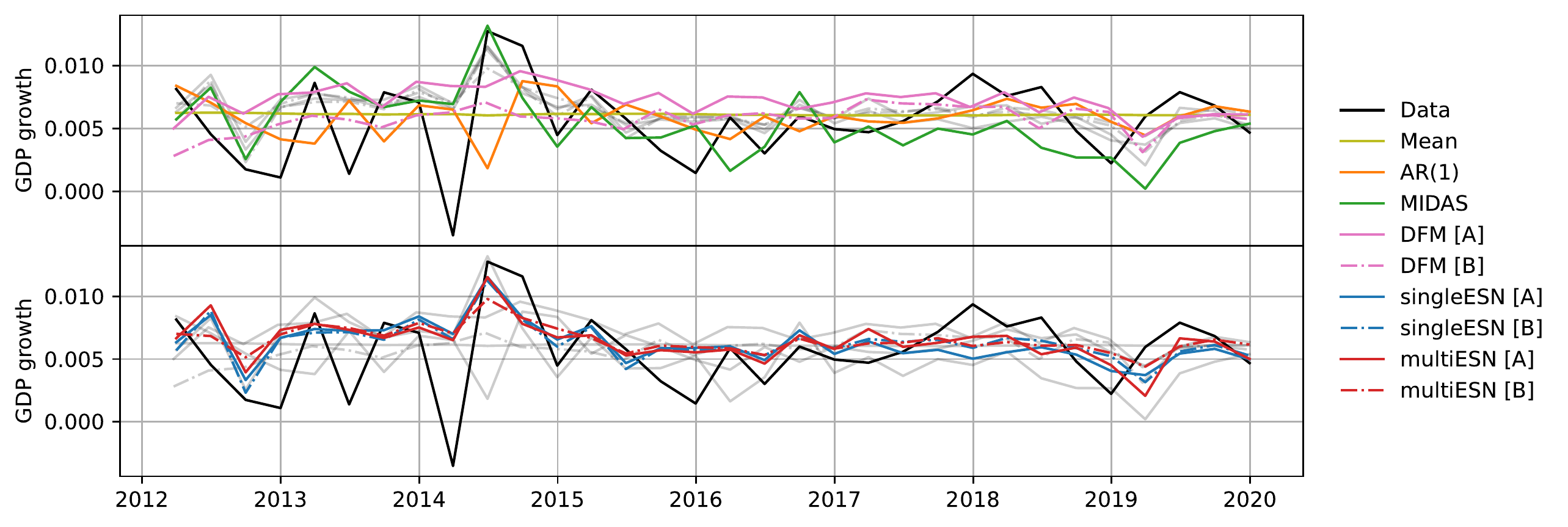}
    }
    \vspace{2em}
    \\
    \subfloat[Rolling]{%
      \includegraphics[width= 1\linewidth]{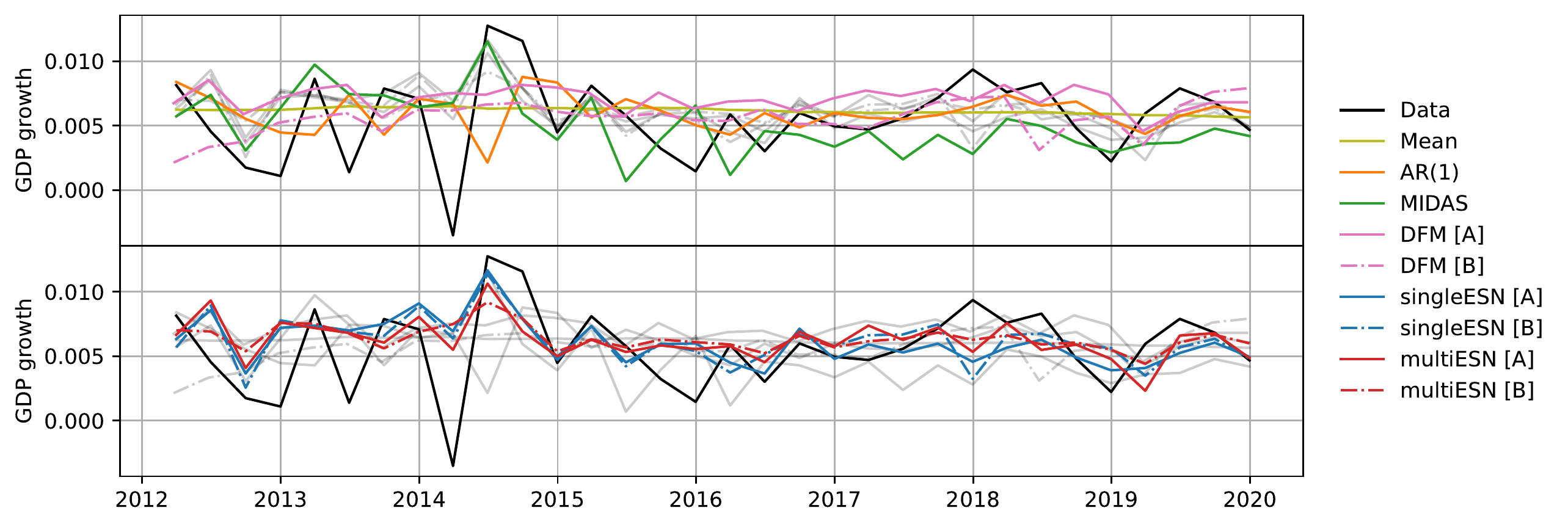}
    }
 \end{figure}
 
 \begin{figure}[!p]
    \centering
    \caption{Multistep-ahead GDP Forecasting, RMSFE -- 2007 Sample -- Small-MD Dataset}
    \label{fig:msa_2007_GDP_smallMD}
    \vspace{1em}
    \subfloat[Fixed]{%
      \includegraphics[width=1\linewidth]{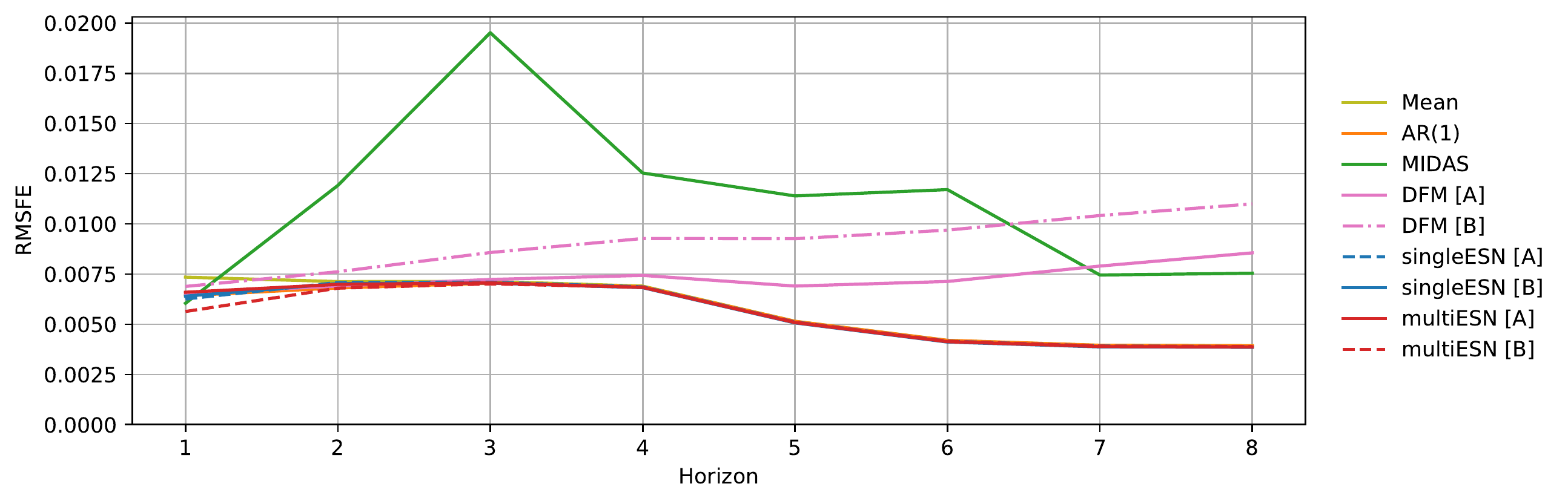}
    }
    \vspace{2em}
    \\
    \subfloat[Expanding]{%
      \includegraphics[width=1\linewidth]{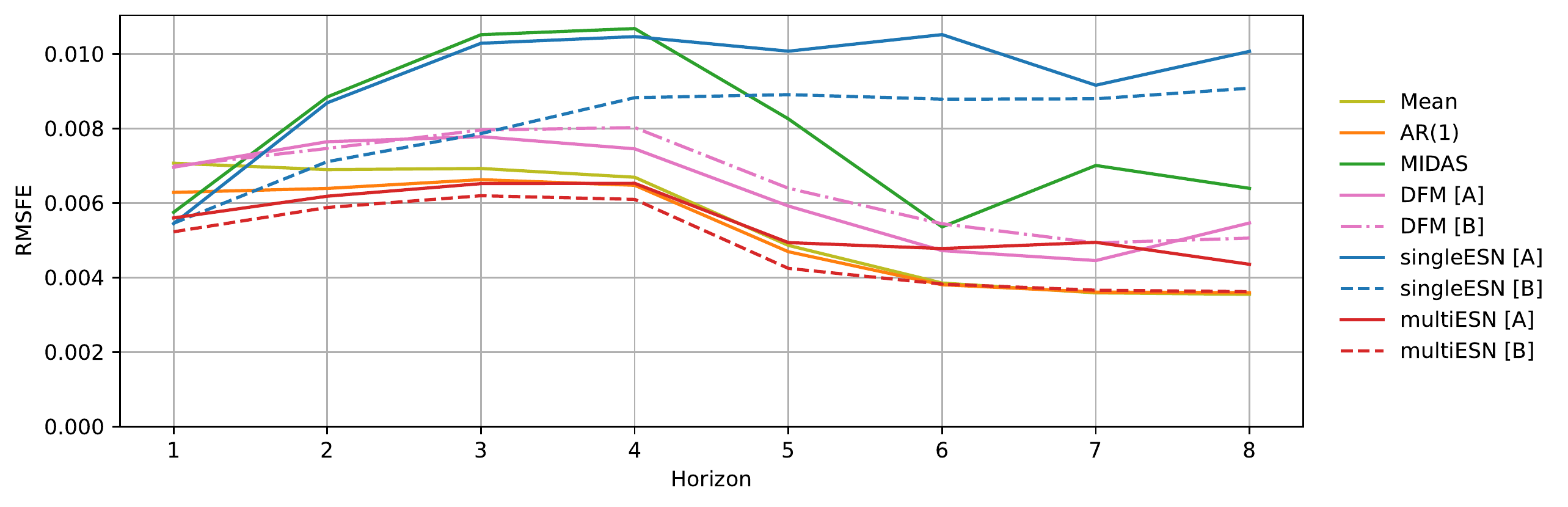}
    }
    \vspace{2em}
    \\
    \subfloat[Rolling]{%
      \includegraphics[width=1\linewidth]{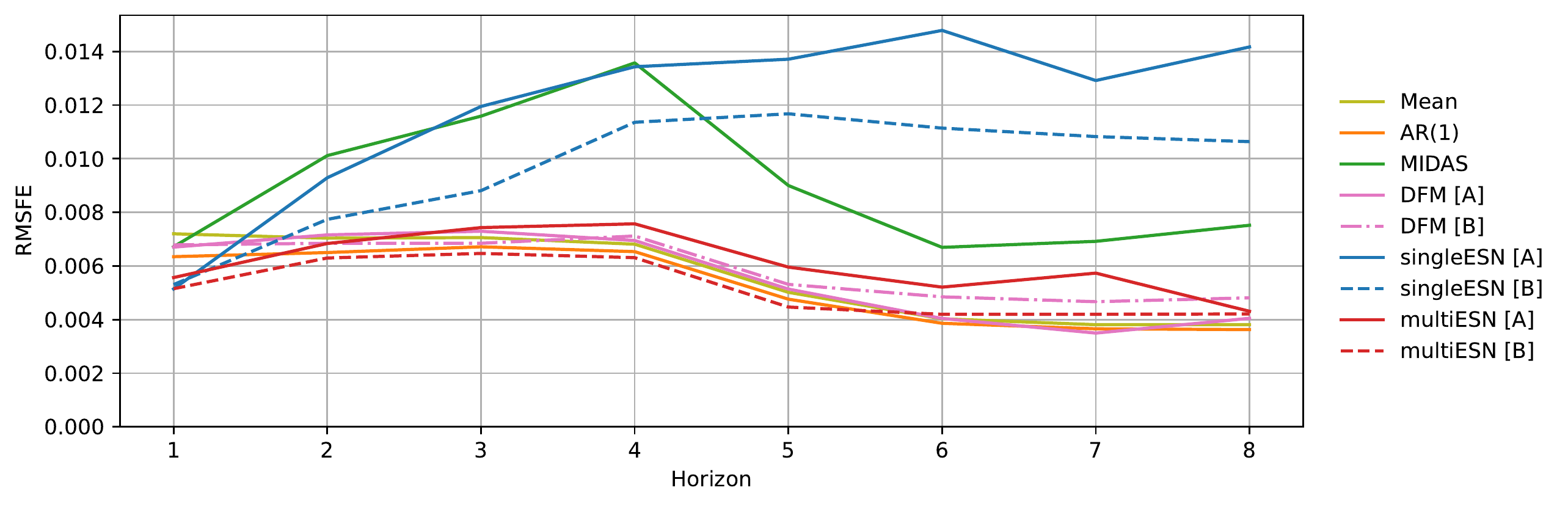}
    }
 \end{figure}
 
  \begin{figure}[!p]
    \centering
    \caption{Multistep-ahead GDP Forecasting, RMSFE -- 2011 Sample -- Small-MD Dataset}
    \label{fig:msa_2011_GDP_smallMD}
    \vspace{1em}
    %
    \subfloat[Fixed]{%
      \includegraphics[width=1\linewidth]{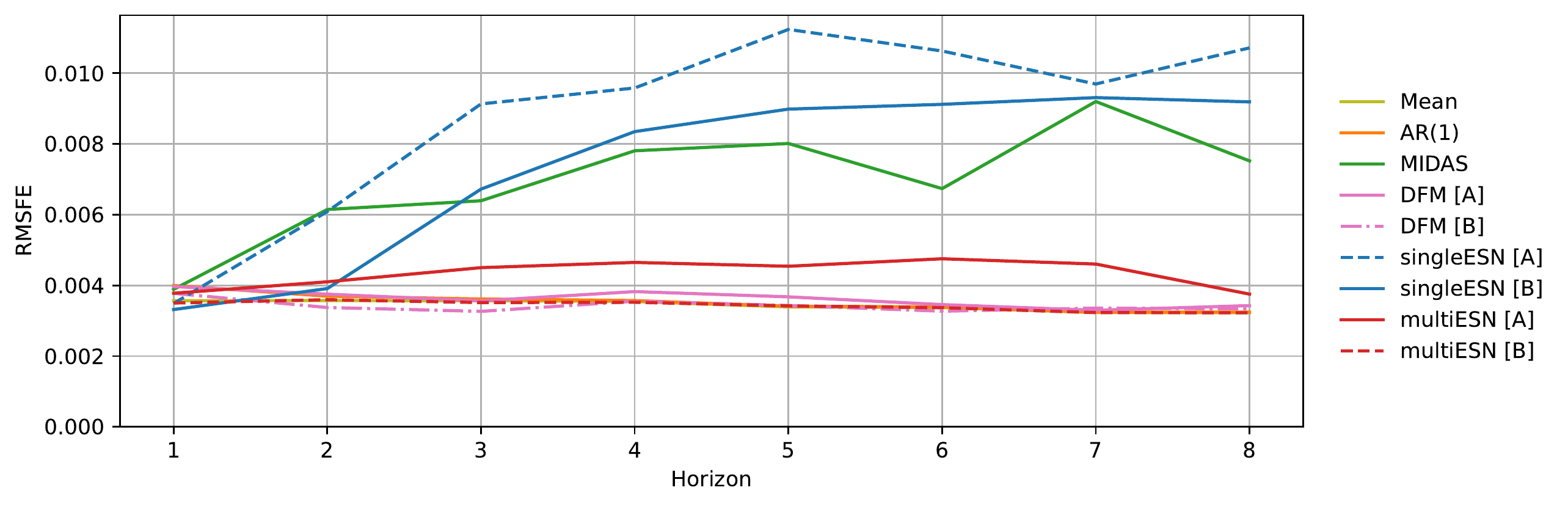}
    }
    \vspace{2em}
    \\
    \subfloat[Expanding]{%
      \includegraphics[width=1\linewidth]{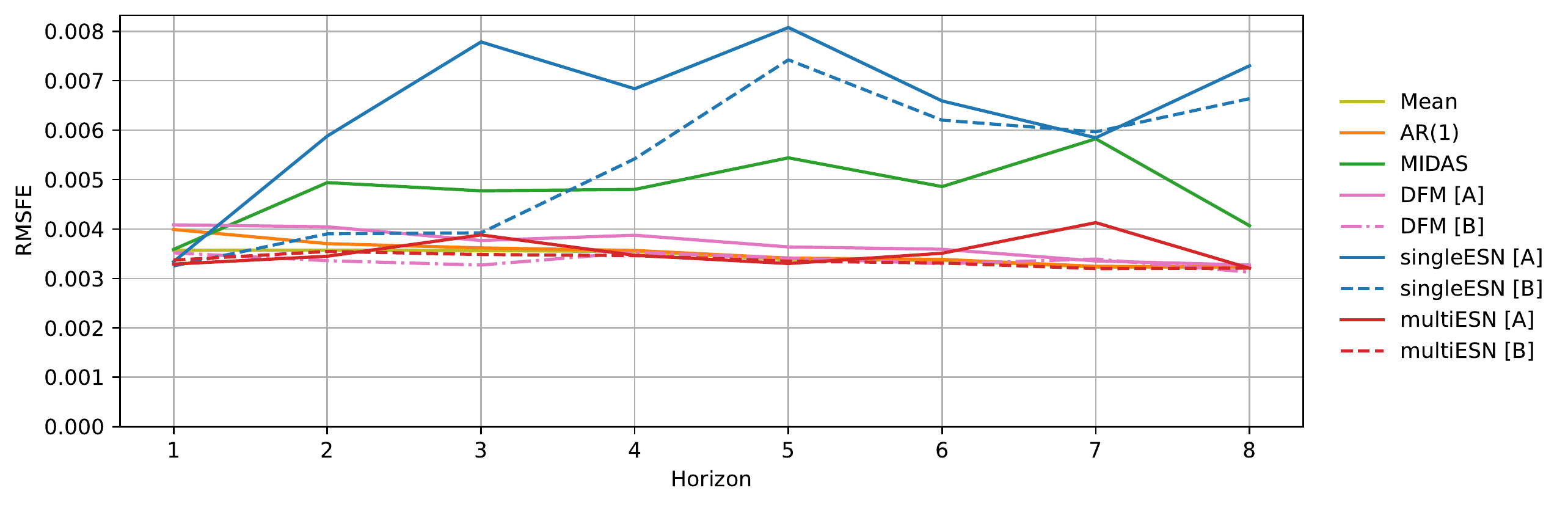}
    }
    \vspace{2em}
    \\
    \subfloat[Rolling]{%
      \includegraphics[width=1\linewidth]{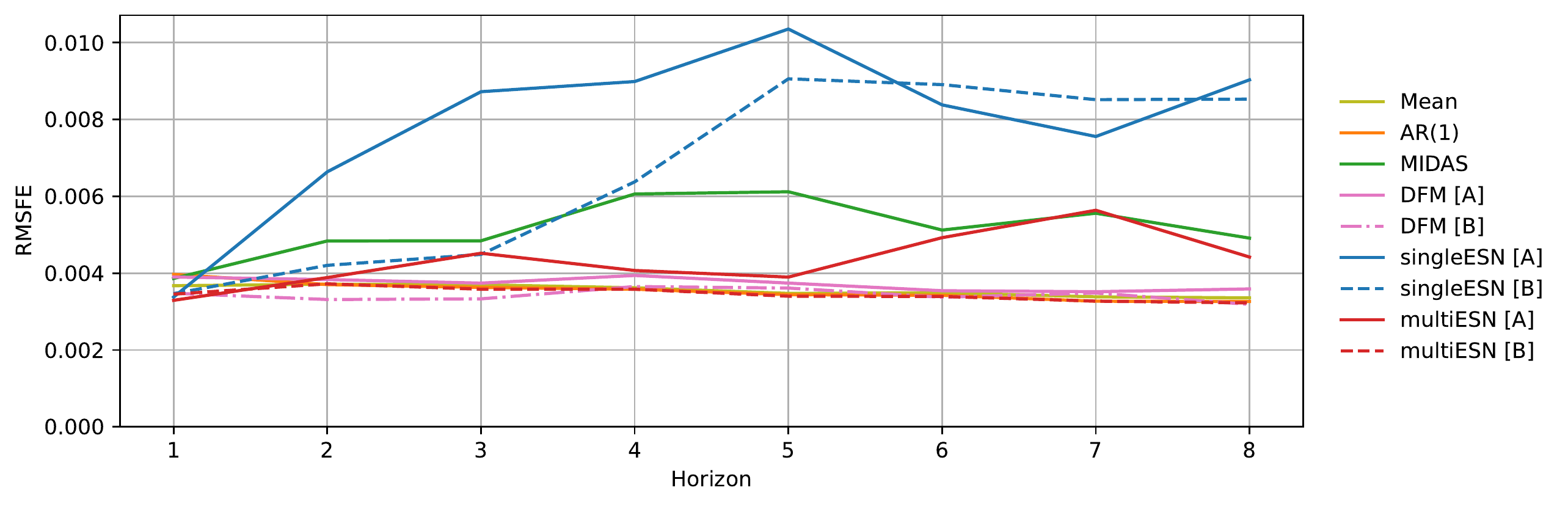}
    }
 \end{figure} 
 
 \begin{figure}[!p]
    \centering
    \caption{1-Step-ahead GDP Forecasting -- 2007 Sample -- Medium-MD Dataset}
    \label{fig:1sa_forecasts_2007_GDP_mediumMD}
    \vspace{1em}
    \subfloat[Fixed]{%
      \includegraphics[width=1\linewidth]{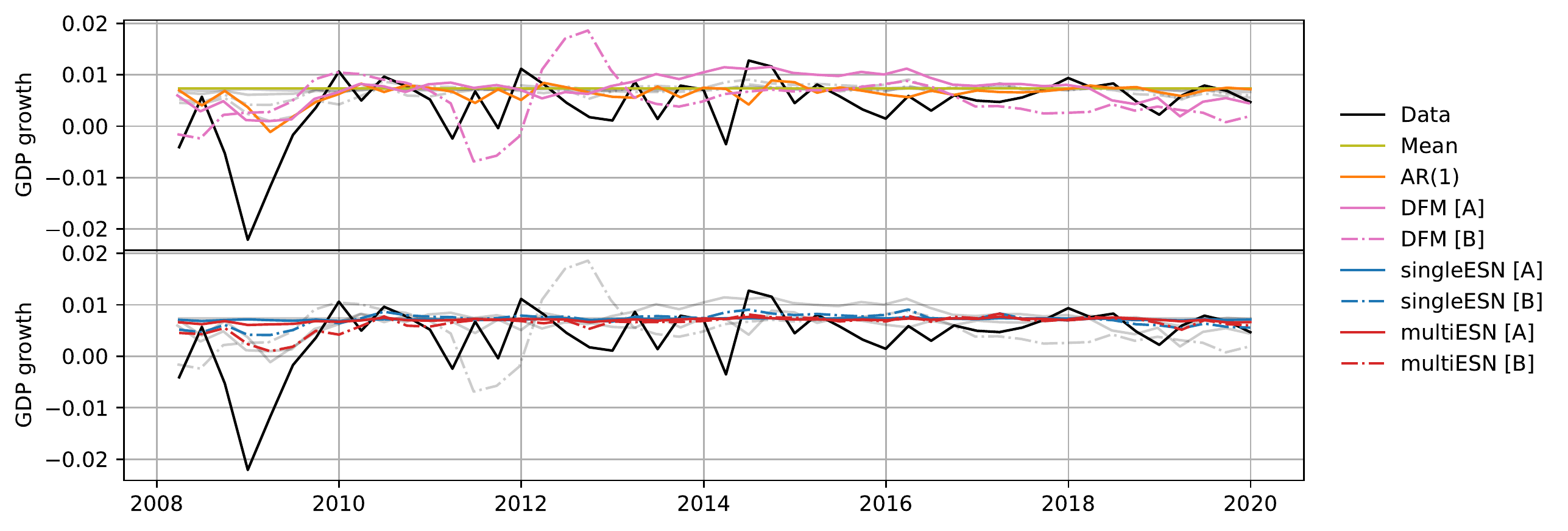}
    }
    \vspace{2em}
    \\
    \subfloat[Expanding]{%
      \includegraphics[width=1\linewidth]{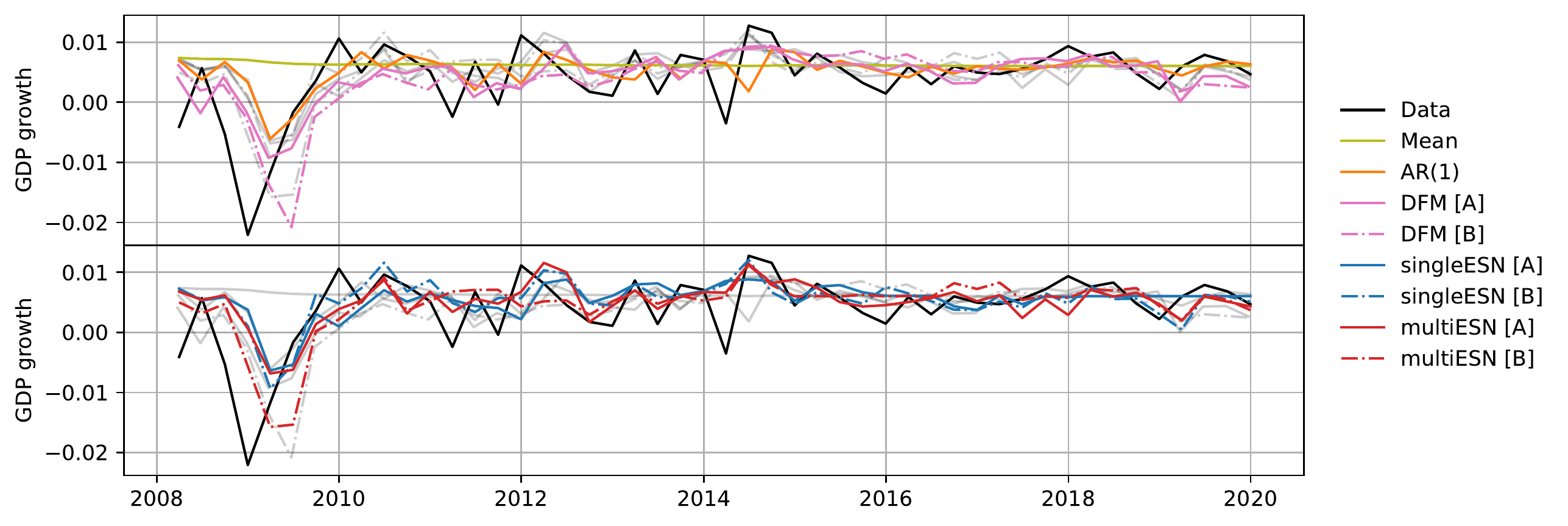}
    }
    \vspace{2em}
    \\
    \subfloat[Rolling]{%
      \includegraphics[width=1\linewidth]{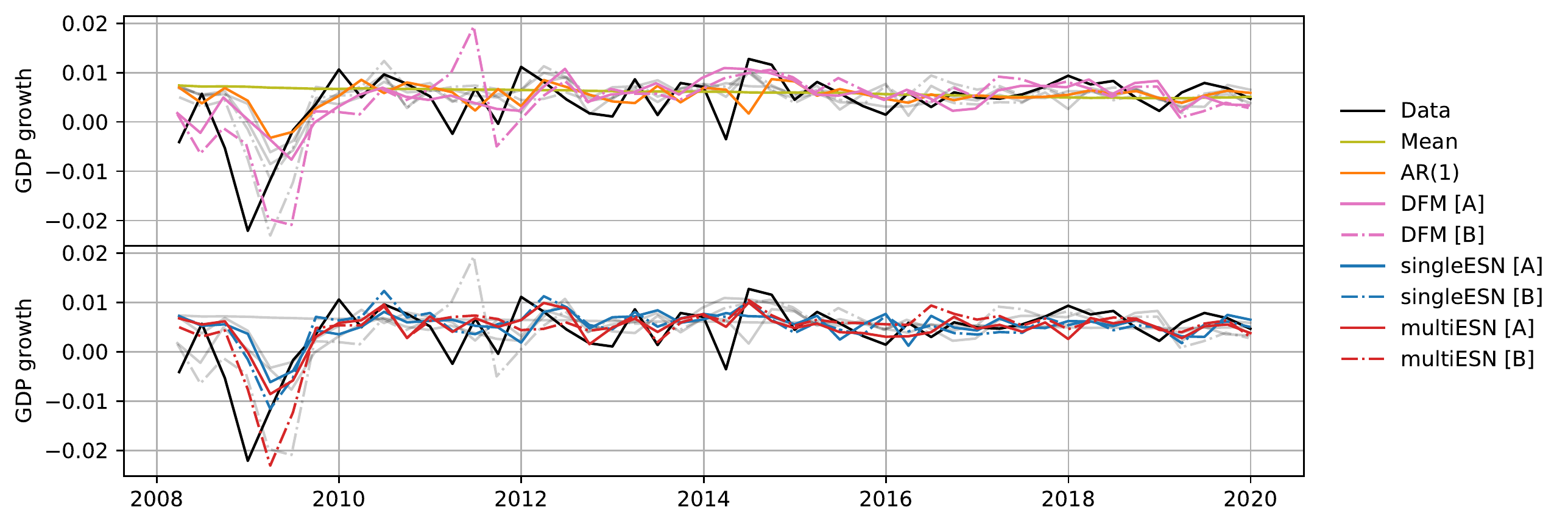}
    }
 \end{figure}
 
 \begin{figure}[!p]
    \centering
    \caption{1-Step-ahead GDP Forecasting -- 2011 Sample -- Medium-MD Dataset}
    \label{fig:1sa_forecasts_2011_GDP_mediumMD}
    \vspace{1em}
    \subfloat[Fixed]{%
      \includegraphics[width=1\linewidth]{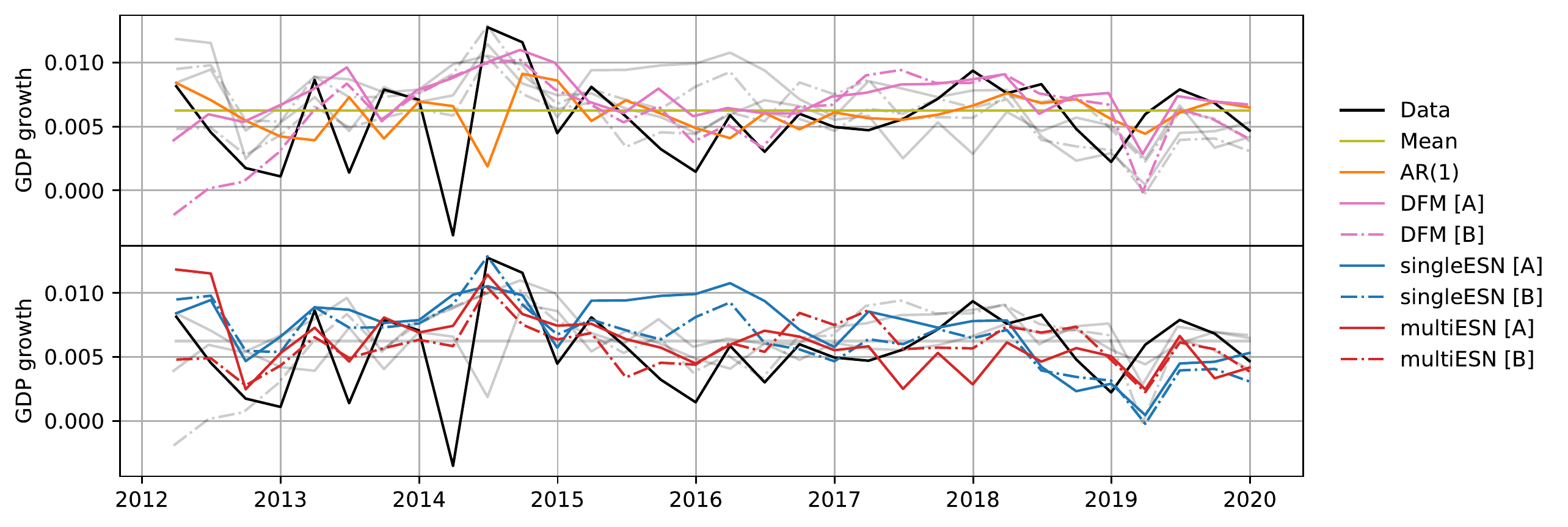}
    }
    \vspace{2em}
    \\
    \subfloat[Expanding]{%
      \includegraphics[width=1\linewidth]{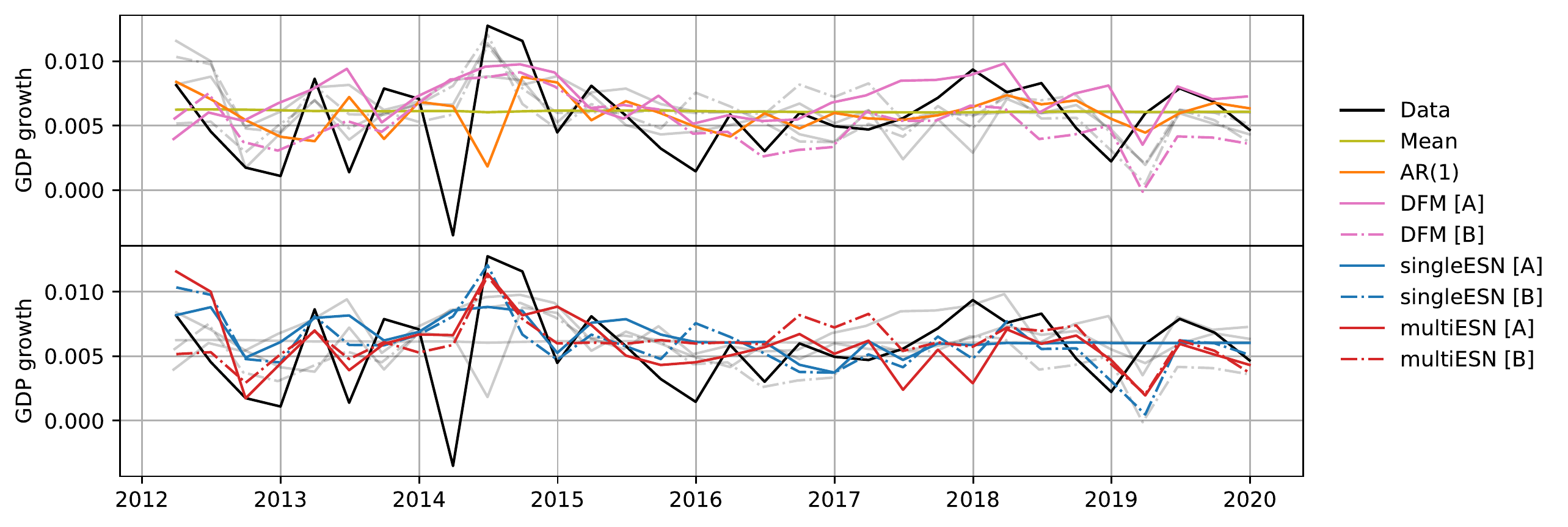}
    }
    \vspace{2em}
    \\
    \subfloat[Rolling]{%
      \includegraphics[width=1\linewidth]{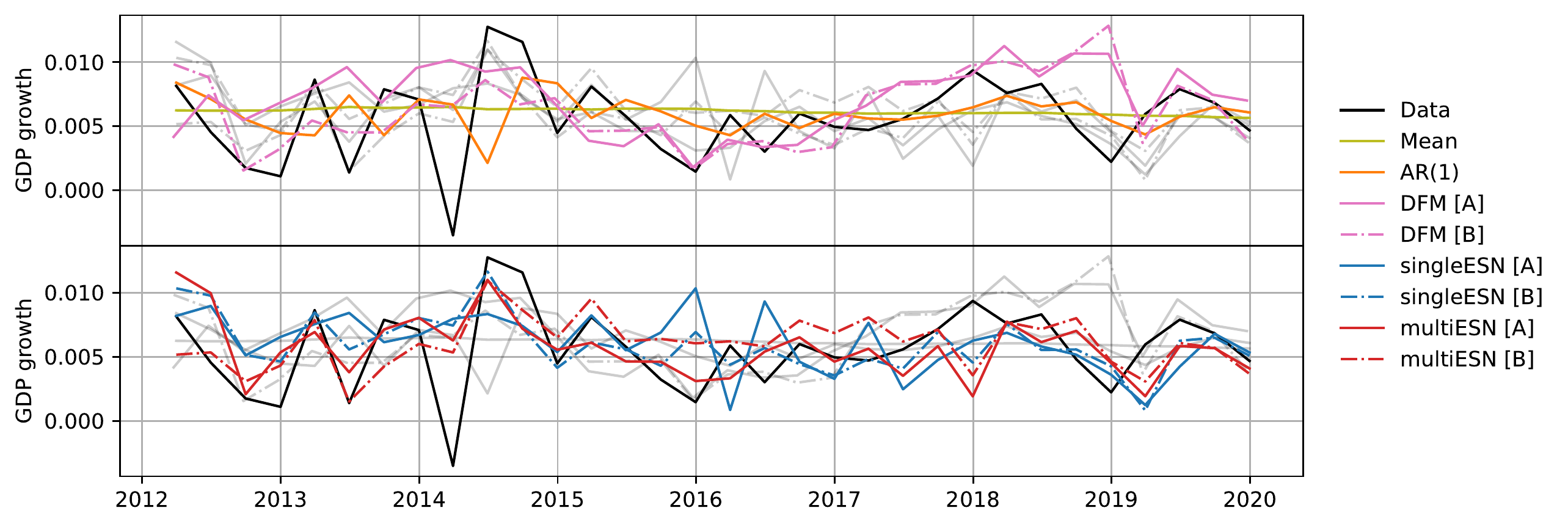}
    }
 \end{figure}

 \begin{figure}[!p]
    \centering
    \caption{Multistep-ahead GDP Forecasting, RMSFE -- 2007 Sample -- Medium-MD Dataset}
    \label{fig:msa_2007_GDP_mediumMD}
    \vspace{1em}
    \subfloat[Fixed]{%
      \includegraphics[width=1\linewidth]{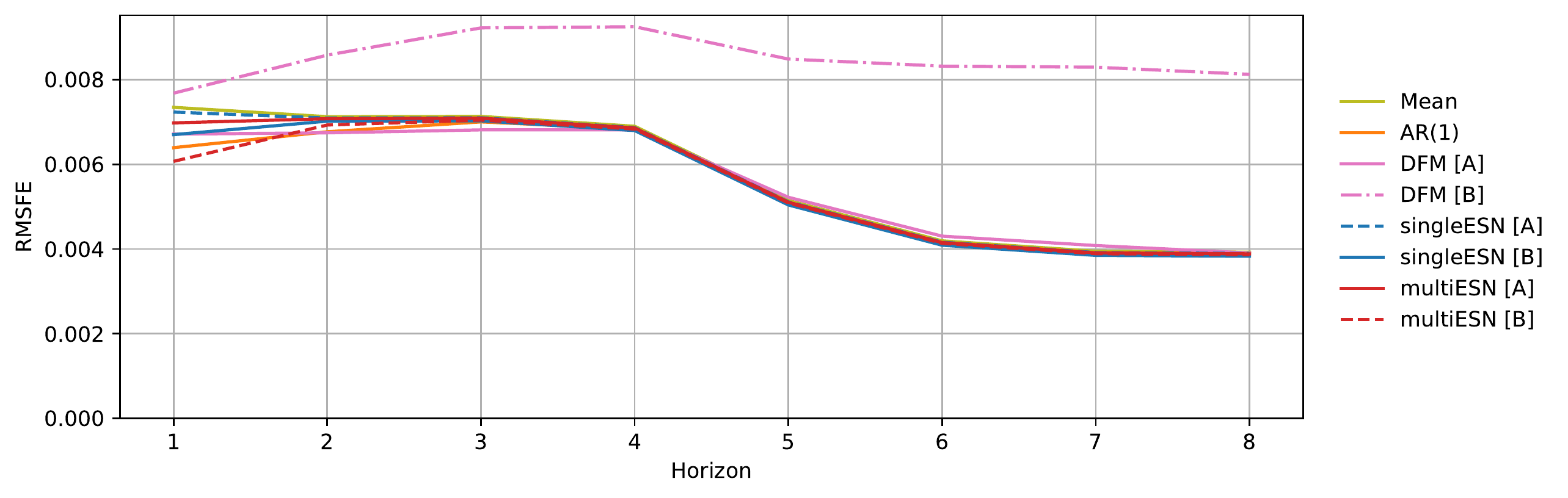}
    }
    \vspace{2em}
    \\
    \subfloat[Expanding]{%
      \includegraphics[width=1\linewidth]{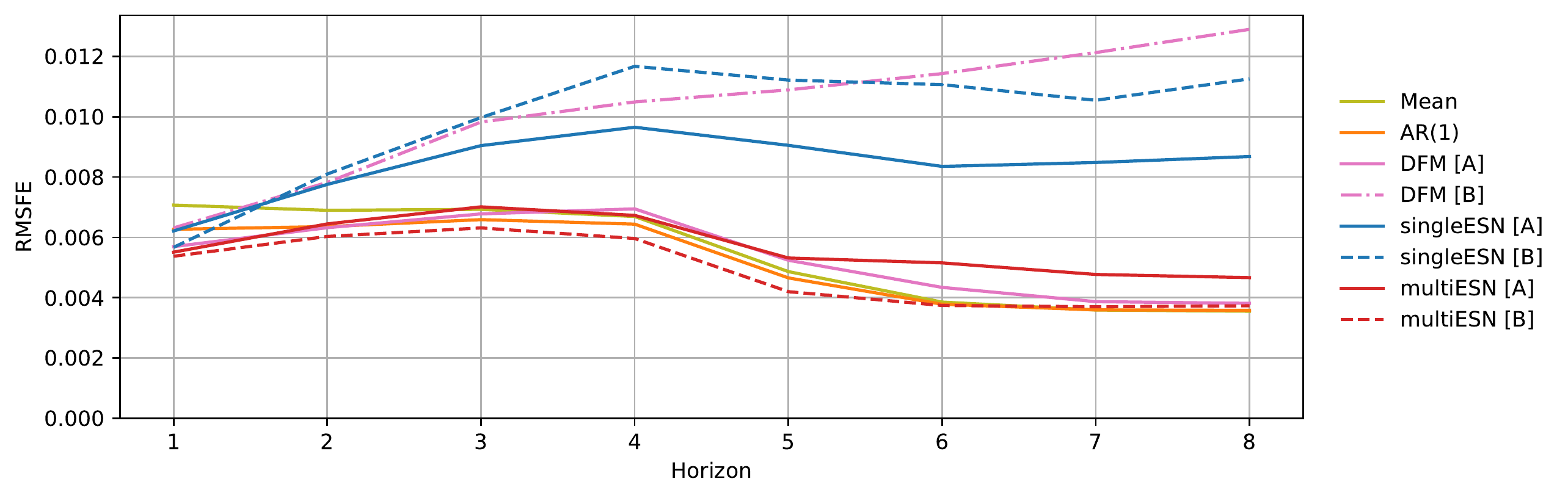}
    }
    \vspace{2em}
    \\
    \subfloat[Rolling]{%
      \includegraphics[width=1\linewidth]{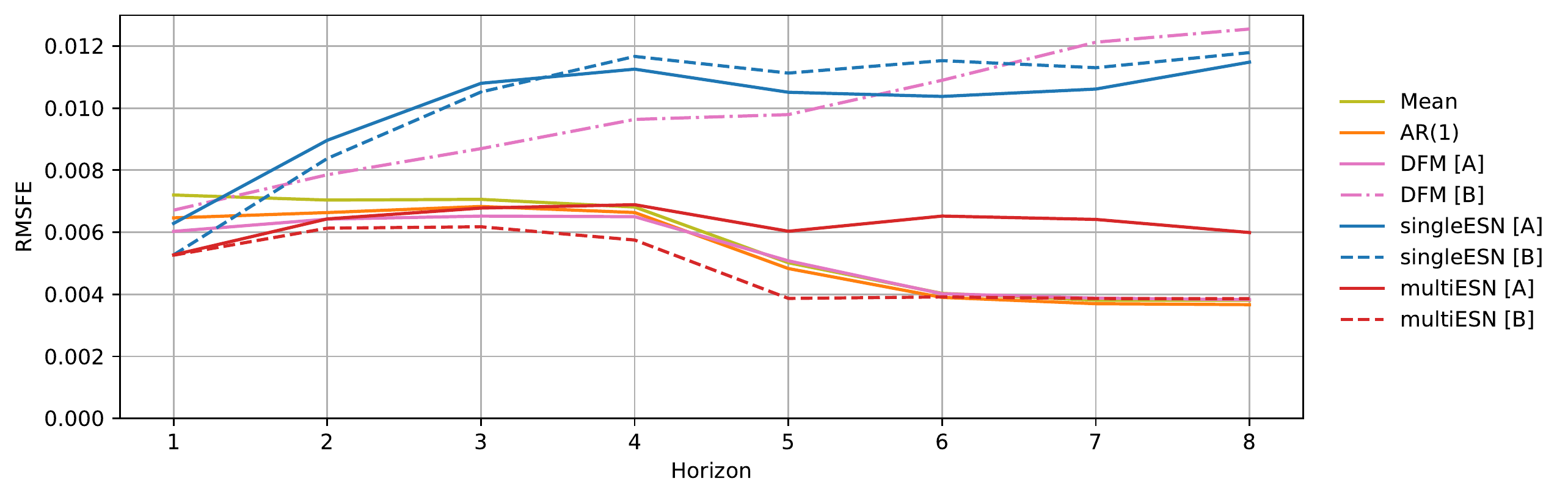}
    }
 \end{figure}
 
  \begin{figure}[p]
    \centering
    \caption{Multistep-ahead GDP Forecasting, RMSFE -- 2011 Sample -- Medium-MD Dataset}
    \label{fig:msa_2011_GDP_mediumMD}
    \vspace{1em}
    %
    \subfloat[Fixed]{%
      \includegraphics[width=1\linewidth]{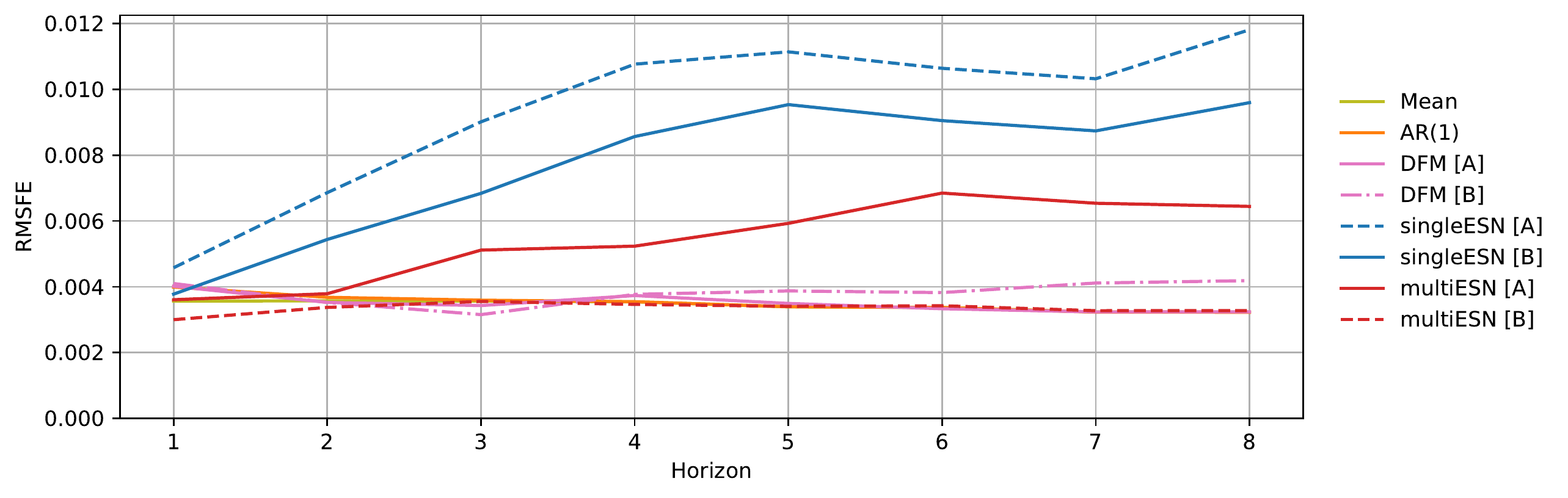}
    }
    \vspace{2em}
    \\
    \subfloat[Expanding]{%
      \includegraphics[width=1\linewidth]{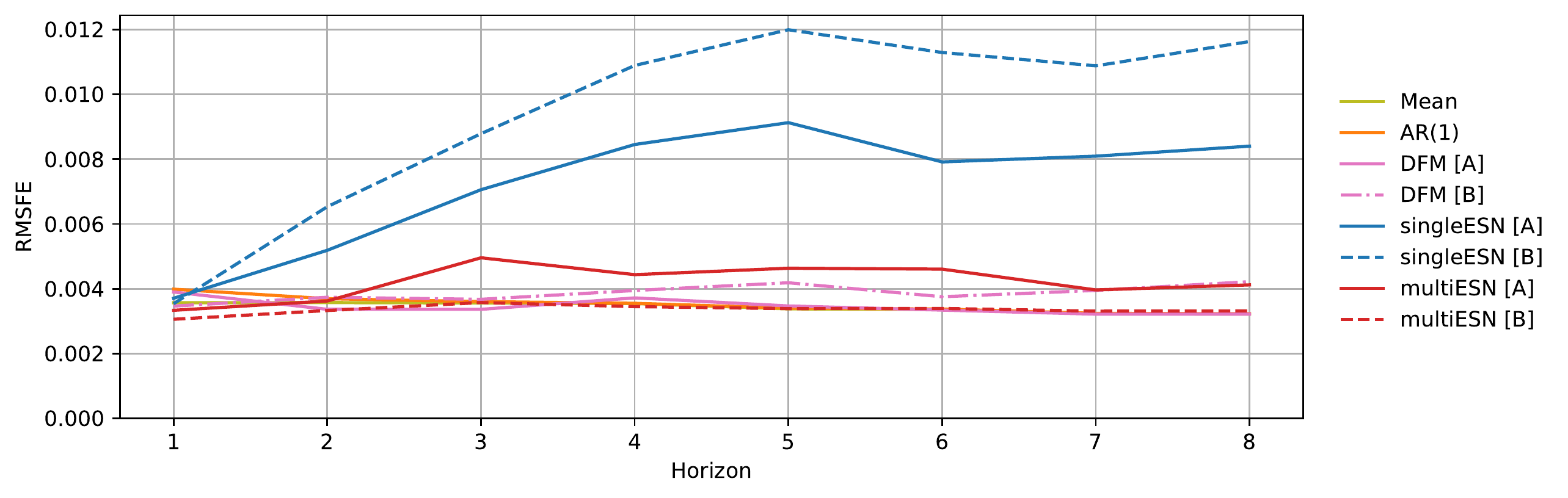}
    }
    \vspace{2em}
    \\
    \subfloat[Rolling]{%
      \includegraphics[width=1\linewidth]{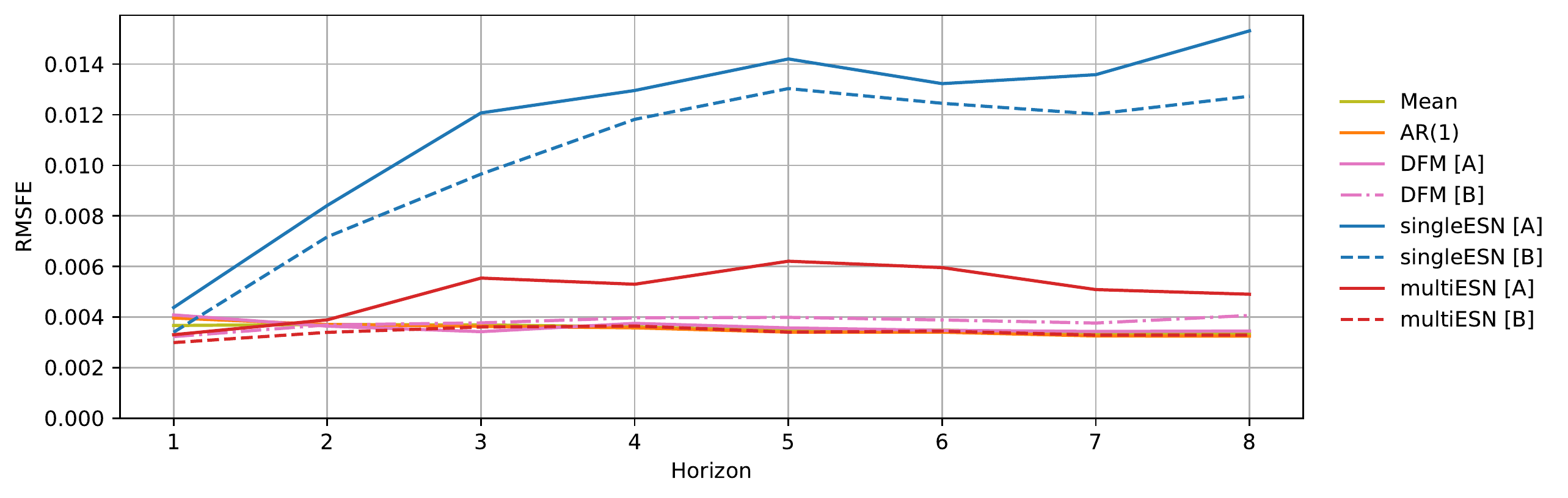}
    }
 \end{figure}

\section{Forecasting Schemes}
\label{Forecasting Schemes}

\begin{figure}[t]
	\centering
	\includegraphics[width=0.9\linewidth]{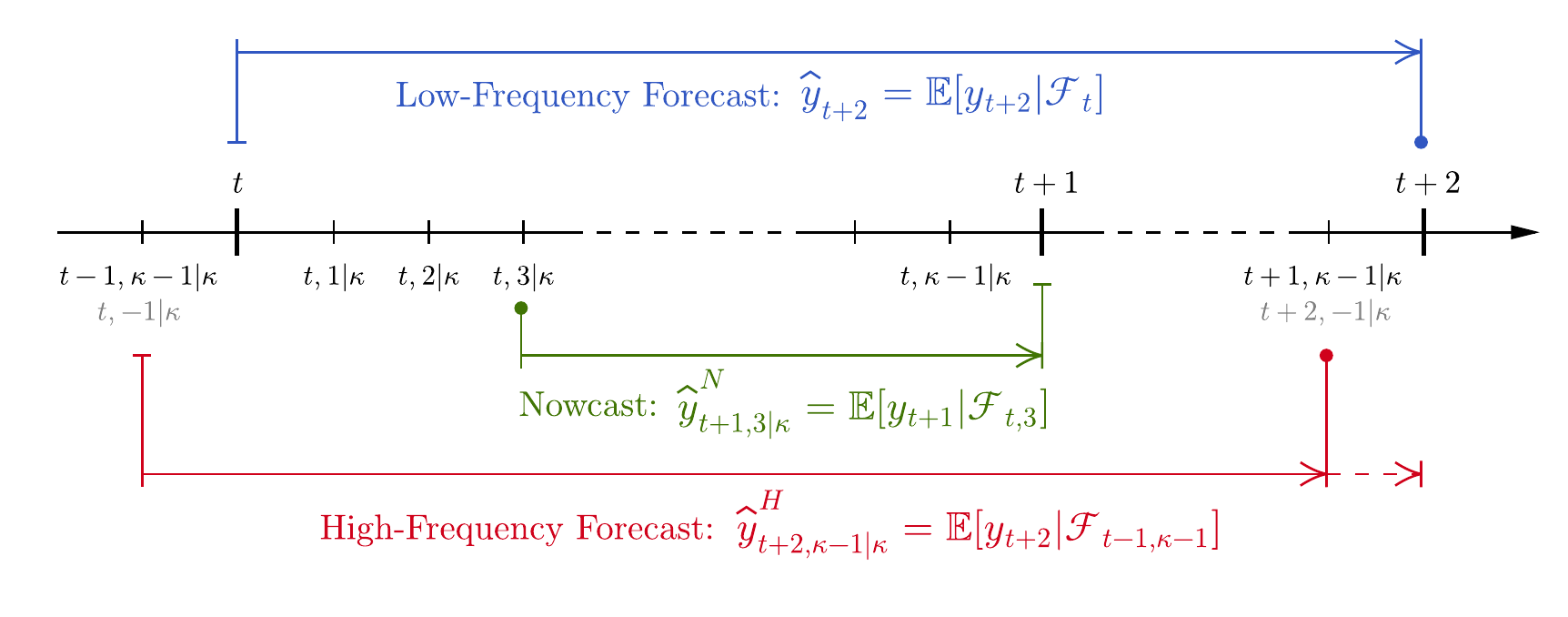}
	\caption{Diagram of the low-/high-frequency forecasting and nowcasting schemes in tempo notation. Arrows point to time indices of the forecast target, solid dots indicate the high-frequency time placeholder for the constructed high-frequency forecasts.}
	\label{fig:multicasting_diagram}
\end{figure}

To clarify the design of the forecasting experiments conducted in this paper, we present two different types of prediction illustrated in Figure~\ref{fig:multicasting_diagram}.

Let $t$ denote time in the reference frequency of the target series $(y_t)$ and suppose a regressor $(z_r)$ of frequency $\kappa$ is included in the forecasting model. The notation can be readily extended to include multiple regressors. Let $h \geq 0$ be a \textit{low-frequency} prediction horizon counted from the last available observation of  $(y_t)$. Let $l \geq 0$ be a \textit{high-frequency} horizon with respect to frequency $\kappa$. 

\paragraph{Low-frequency forecasting.}
We call an $h$-steps ahead forecast \textit{low-frequency} when predictions for the target variable are constructed only at the end of the low-frequency periods. The information set which is used at the time of $h$-steps ahead low-frequency forecasting at $t$  is the $\sigma$-algebra defined as 
\begin{equation}
    \label{eq:sigma-algebra_fixed_forecast}
    \mathcal{F}_{t} = \sigma \left(\left\{ y_{t}, y_{t-1}, y_{t-2}, \ldots, z_{t,\tempo{0}{\kappa}},  z_{t,\tempo{-1}{\kappa}}, z_{t,\tempo{-2}{\kappa}}, \ldots \right\}\right)
\end{equation}
and, when using the mean square error as a loss, the optimal forecast is given by
\begin{equation}
    \label{eq:fixed_forecast}
    \widehat{y}_{t + h} = \mathbb{E}\left[ y_{t + h} \vert \mathcal{F}_{t} \right].
\end{equation}

\paragraph{High-frequency forecasting.}
In this forecasting scheme, one may also use high-frequency regressors to produce additional high-frequency forecasts of the low-frequency target variable. For example, in the case of a target released at the end of each year and having monthly quoted covariates, the low-frequency forecasting scheme would correspond to constructing forecasts always at the end of the last month of the year (December). At the same time,  with all the information collected up to the end of December, there are other possibilities to construct forecasts. In particular, the forecaster could consider placing herself at the end of any other month of the year instead and construct predictions for the monthly proxy of the yearly variable for the next $h$th year. 

In this scheme, one often artificially \textit{reduces} the information set. Although not all the available information is exploited, this procedure has its benefits: first, it renders high-frequency forecast instances; second, it allows taking into account misspecification due to a seasonal response of $(y_t)$ to $(z_r)$. This is especially important whenever multiple time series with different sampling frequencies are combined in one model and seasonality effects are either difficult to detect or impossible to avoid.  In the context of macroeconomic forecasting, we again refer the reader to \cite{Clements2008,Clements2009,Chen2010} and \cite{Jardet2020}, where these questions are carefully discussed.

Let the forecaster place herself at time $t$: she wishes to construct a high-frequency forecast for some  $t,\tempo{l}{\kappa}$ with $l \in \mathbb{N}$. The maximal information set available at $t$ is  $\mathcal{F}_{t}$ as in \eqref{eq:sigma-algebra_fixed_forecast}. However, if she uses $\mathcal{F}_{t}$ then the forecast for $t,\tempo{l}{\kappa}$ coincides with the low-frequency forecast and is given by \eqref{eq:fixed_forecast} for any $l$. Notice that the forecasts can be constructed using the reduced information sets instead. Let $h = \lceil l / \kappa \rceil$, $\ell = l \mod \kappa$, and $m =  h  - \lfloor l / \kappa \rfloor$, and define
\begin{align*}
    \mathcal{F}_{t-m,{\ell}} & = \sigma \left(\left\{ y_{t - m}, y_{t -1 - m}, \ldots, z_{t - m,\tempo{\ell}{\kappa}}, \ z_{t - m,\tempo{(\ell - 1)}{\kappa}}, \ z_{t - m,(\tempo{\ell - 2)}{\kappa}}, \ldots \right\}\right)  \\
    & = \sigma \left(\left\{ y_{t - m}, y_{t - 1 - m}, \ldots, z_{t + 1 - m,\tempo{-(\kappa -\ell)}{\kappa}},  z_{t + 1 - m,\tempo{-(\kappa -\ell) + 1}{\kappa}}, z_{t + 1 - m,\tempo{-(\kappa -\ell  + 2)}{\kappa}}, \ldots \right\}\right).
\end{align*}
The high-frequency forecast information sets nest the low-frequency forecasting setup since $\mathcal{F}_{t-m,\ell} \equiv \mathcal{F}_{t}$ if $l = \kappa h$ for $h \in \nset$ and the forecast for the high-frequency proxy constructed for the moments $t, \tempo{l}{\kappa}$ for the low-frequency variable  is provided by the conditional expectation
\begin{equation*}
    \widehat{y}_{t+h, \tempo{\ell}{\kappa}}^H=\mathbb{E}\left[ y_{t+h} \vert \mathcal{F}_{t-m,\ell} \right].
\end{equation*}
It is easy to see that if the forecaster is interested in nowcasting, it can be readily obtained by taking $m = 0$ and writing for all $0<\ell\leq \kappa-1$:
 \begin{equation*}
    \widehat{y}_{t+1, \tempo{\ell}{\kappa}}^N=\mathbb{E}\left[ y_{t+1} \vert \mathcal{F}_{t,\ell} \right].
\end{equation*}

\paragraph{Nowcasting.} We call \textit{nowcasting} the setup in which one constructs a high-frequency proxy for a yet-unobserved target which will be available at the end of the \textit{current} low-frequency period. As such, we construct a nowcast only for horizons $0 < l \leq \kappa-1$; notice that $l= \kappa$ yields a contemporaneous regression at $t+1$, while $l = 0$ falls into the category of low-frequency forecasting considered in Section~\ref{Forecasting Schemes}, hence both these two cases are excluded. The  $\sigma$-algebras that are used in order to construct nowcasts $\widehat{y}_{t + 1, \tempo{\ell}{\kappa}}$ are given by
		\begin{align*}
		\mathcal{F}_{t,\tempo{\ell}{\kappa}} & = \sigma \left(\left\{ y_{t}, y_{t -1}, \ldots, z_{t,\tempo{\ell}{\kappa}}, \ z_{t,\tempo{(\ell - 1)}{\kappa}}, \ z_{t,(\tempo{\ell - 2)}{\kappa}}, \ldots \right\}\right)  \\
		& = \sigma \left(\left\{ y_{t}, y_{t - 1}, \ldots, z_{t + 1,\tempo{-(\kappa -\ell)}{\kappa}},  z_{t + 1,\tempo{-(\kappa -\ell) + 1}{\kappa}}, z_{t + 1,\tempo{-(\kappa -\ell  + 2)}{\kappa}}, \ldots \right\}\right).
	\end{align*}
	The $l$-steps nowcast for the high-frequency proxy constructed at moments $t, \tempo{\ell}{\kappa}$ of the current period for the low-frequency variable which becomes available at $t+1, \tempo{0}{\kappa} \equiv t+1$ is provided by the conditional expectation
	 \begin{equation*}
		\widehat{y}_{t+1, \tempo{\ell}{\kappa}}^N=\mathbb{E}\left[ y_{t+1} \vert \mathcal{F}_{t,\ell} \right].
	\end{equation*}
	
\paragraph{Multicasting.} One always aims to construct one-step and multistep forecasts by using all the available information at a given point in time. It is, therefore, natural to compare models by constructing high-frequency nowcasts for the target variable to be released at the end of the current period and its high-frequency proxy forecasts for the next periods. To avoid confusion, we refer to this situation as {\it multicasting}. More explicitly, provided that the forecaster finds herself at time index $t,\tempo{s}{\kappa}$ and is interested in all the forecasts up to some maximal low-frequency horizon $H \geq 1$, for each $1 \leq l \leq H \kappa$ the multicasting scheme yields the following combination:
	\begin{enumerate}
		\item[{(a)}] \textit{Nowcasting} when $0 < l \leq \kappa - 1$ and $\ell = l$: $\widehat{y}_{t+1, \tempo{\ell}{\kappa}}^N=\mathbb{E}\left[ y_{t+1} \vert \mathcal{F}_{t,\ell} \right]$.
		\item[{(b)}] Forecasting when $l > \kappa - 1$:
		\begin{itemize}
			\item \textit{Low-frequency forecasting} if $l$ satisfies $l \mod \kappa = 0$: $\widehat{y}_{t+h}=\mathbb{E}\left[ y_{t+h} \vert \mathcal{F}_{t} \right]$.
			\item \textit{High-frequency forecasting} if $l \mod \kappa \not= 0$: $\mathcal{F}_{t,\ell}$: $\widehat{y}_{t+h, \tempo{\ell}{\kappa}}^H=\mathbb{E}\left[ y_{t+h} \vert \mathcal{F}_{t,\ell} \right]$.
		\end{itemize}
	\end{enumerate}
	%


\newpage
\setcounter{page}{1}

\begin{center}
    {\huge Supplementary Appendix}\\
\end{center}

\section{ESN Implementation}\label{sec:ESN_implementation}

\subsection{Fixed, Expanding and Rolling Window Estimation}
\label{sec:estimation_setup}

Model parameter stability is an important and well-studied question in linear time series analysis. Indeed, identifying and explaining structural breaks play a key role in macroeconomic modeling. To account for this possibility, we compare multiple estimation setups which may reflect possible changes in model parameters.

Suppose again that a sample $Y = (\boldsymbol{y}_2, \boldsymbol{y}_3, \ldots, \boldsymbol{y}_{T})^\top \in \mathbb{M}_{T-1, J}$ of targets is available, an initial state $\boldsymbol{x}_0$ is given and regressors $Z = (\boldsymbol{z}_1, \boldsymbol{z}_2, \ldots, \boldsymbol{z}_{T-1})^\top \in \mathbb{M}_{T-1, K}$ are observed. Additionally, the researcher has available an out-of-sample dataset, $Y^\dag = (\boldsymbol{y}_{T+1}, \boldsymbol{y}_{T+2}, \ldots, \boldsymbol{y}_{T+S})^\top \in \mathbb{M}_{S, J}$, $Z^\dag = (\boldsymbol{z}_{T}, \boldsymbol{z}_{T+1}, \ldots, \boldsymbol{z}_{T+S-1})^\top \in \mathbb{M}_{S, K}$ for $S \geq 1$. We now define the estimation setups which can be used for subsequent forecasting for $h \in \mathbb{N}^+$ steps ahead and can be adjusted for the multi-frequency setup. We consider the following estimation strategies:
\begin{enumerate}
	\item[(i)] \textbf{Fixed parameters}: An estimator $\widehat{W}$ is computed strictly over sample observations $Y$ and $Z$ with some penalty $\lambda$ chosen with data available up to time $T$. Model parameters are kept fixed when the estimated model is applied to construct out-of-sample forecasts $\widehat{\boldsymbol{y}}_{T+1}, \widehat{\boldsymbol{y}}_{T+2}, \ldots, \widehat{\boldsymbol{y}}_{T+S}$ as out-of-sample regressors $\boldsymbol{z}_{T}, \boldsymbol{z}_{T+1}, \ldots, \boldsymbol{z}_{T+S-1}$ are added to the information set.
	\item[(ii)] \textbf{Expanding window:} For each out-of-sample time step $s = 0, \ldots, S$, define $\widehat{W}^{\text{EW}}_s$ as the estimate computed by ``expanding'' the sample window up to time $T+s$, given by $Y^{\text{EW}}_s := (\boldsymbol{y}_2, \boldsymbol{y}_3, \ldots, \boldsymbol{y}_{T}, \allowbreak \boldsymbol{y}_{T+1}, \ldots, \boldsymbol{y}_{T+s})^\top$ and $Z^{\text{EW}}_s := (\boldsymbol{z}_{1}, \boldsymbol{z}_{2}, \ldots, \boldsymbol{z}_{T-1}, \allowbreak \boldsymbol{z}_{T}, \ldots, \boldsymbol{z}_{T+s-1})^\top$. Coefficients $\widehat{W}^{\text{EW}}_s$ are re-estimated and penalty strength $\lambda$ is re-validated over windows $Y^{\text{EW}}_s$, $Z^{\text{EW}}_s$.
	\item[(iii)] \textbf{Rolling window:} In this setup the within-window sample size is kept fixed across windows -- that is, the sample window ``rolls'' over the data -- by defining $\widehat{W}^{\text{RW}}_s$ as the estimate over $Y^{\text{RW}}_s:= (\boldsymbol{y}_{2+s}, \boldsymbol{y}_{3+s}, \allowbreak \ldots, \boldsymbol{y}_{T+s-1}, \boldsymbol{y}_{T+s})^\top$ and $Z^{\text{RW}}_s:= (\boldsymbol{z}_{1+s}, \boldsymbol{z}_{2+s}, \allowbreak \ldots, \boldsymbol{z}_{T+s-2}, \boldsymbol{z}_{T+s-1})^\top$ for $s = 0, \allowbreak \ldots, S$. Coefficients $\widehat{W}^{\text{RW}}_s$ are re-estimated and penalty strength $\lambda$ is re-validated over windows $Y^{\text{RW}}_s$, $Z^{\text{RW}}_s$.
\end{enumerate}

In all three strategies, hyperparameters $\boldsymbol{\varphi}:= (\alpha, \rho, \gamma, \omega)$ could also be re-tuned on corresponding windows as in Supplementary Appendix~\ref{Hyperparameter Tuning}. The fixed-parameter setup is the most rigid one. It builds upon the idea that the initial sample contains sufficient information for correct model estimation and forecasting and that the model parameters are constant. Its theoretical analysis is relatively easy as there is no need to discuss the stability of the penalty and the hyperparameters across sample windows. An expanding window setup is based on the belief that newly available data contains key information to produce forecasts and, therefore, must be continuously incorporated. In essence, forecasters do this when they re-estimate a model at each data release cycle. In the case of a rolling window estimation strategy, one can theoretically handle model changes. Although proper structural break modeling would require a consistent identification of breakpoints, rolling window estimation can potentially accommodate slow drifts in model parameters over time by directly discarding old data, unlike with an expanding window. We do not explore the selection of an optimal window size, which in rolling window estimation has been shown to improve forecasting performance (\cite{inoueRollingWindowSelection2017}).

\subsection{Hyperparameter Tuning}
\label{Hyperparameter Tuning}

We now propose a general scheme for selection of hyperparameters $\boldsymbol{\varphi}:= (\alpha, \rho, \gamma, \omega)$ in \eqref{eq:esn_hyperparams} for a model of the form \eqref{eq:esn_model_states}-\eqref{eq:esn_model_output}. Our approach builds on the idea of leave-one-out cross-validation for time series models. Using a fixed, expanding, or rolling window over the training data, one can always compute the one-step forecasting errors committed by the ESN, given fixed normalized model matrices $(\overline{A}, \overline{C}, \overline{\boldsymbol{\zeta}})$ and a hyperparameter vector $\boldsymbol{\varphi}$. By choosing an appropriate loss function $\boldsymbol{\ell} : \rset^J\times\rset^J \to \rset_+$, $J\in \mathbb{N}^+$, we can thus compute the empirical ESN forecasting error
\begin{equation*}
	\mathcal{L}_T(\boldsymbol{\varphi}) := \sum_{t=T_0}^{T-1} \boldsymbol{\ell}(\boldsymbol{y}_{t+1},  \widehat{W}_t(\boldsymbol{\varphi})^\top \boldsymbol{x}_t ),
\end{equation*}
where $\widehat{W}_t(\boldsymbol{\varphi})$ is the readout coefficients estimator involving data available up to time $t$ and $1 < T_0 < T-1$ is the minimum number of observations used for fitting. Notice that if $\boldsymbol{\ell}(\boldsymbol{u}, \boldsymbol{v}) = \norm{\boldsymbol{u} -  \boldsymbol{v}}_2^2$, $\boldsymbol{u}, \boldsymbol{v}\in \rset^J$, then $\mathcal{L}_T(\boldsymbol{\varphi})$ is the cumulative squared error that is minimized in training (modulo a ridge penalty term). Here, however, the interest is not in estimating $W$, which minimizes $\mathcal{L}_T$, but rather finding the optimal hyperparameter vector
\begin{equation*}
	\boldsymbol{\varphi}^* \in \argmin_{\boldsymbol{\varphi} \in [0,1) \times [0,\overline{\rho}] \times [0, \overline{\gamma}] \times [0, \overline{\omega}]} \mathcal{L}_T(\boldsymbol{\varphi}),
\end{equation*}
where upper bounds $\overline{\rho}$, $\overline{\gamma}$, and $\overline{\omega}$ can be appropriately chosen (in our empirical exercises we use $10$ and verify that solutions are never on the boundary).
We highlight that to tune $\boldsymbol{\varphi}$ one may choose $\boldsymbol{\ell}$ that is different from the one used in the estimation of the readout coefficients $W$.

We present the entire hyperparameter optimization routine in Algorithm \ref{algorithm:hyperpar_tuning}.
Note that step (i) might entail re-normalizing inputs and targets at each window $t$. This setup is general and allows applying any global optimization routine to minimize $\mathcal{L}_T(\boldsymbol{\varphi})$. We construct the loss $\mathcal{L}_T(\boldsymbol{\varphi}_j)$ sequentially, that is by summing squared residuals of the model estimated in step (i) of Algorithm \ref{algorithm:hyperpar_tuning} when $\boldsymbol{\ell}$ is a quadratic loss. One can program $\mathcal{L}_T(\boldsymbol{\varphi}_j)$ via TensorFlow so that the gradient can be evaluated by backpropagation in Algorithm \ref{algorithm:hyperpar_tuning} (iii). Since there is no guarantee that the objective function is convex or even everywhere smooth, we suggest applying optimizers known to explore the parameter space efficiently. We emphasize that the lack of convexity guarantees is much more consequential for the other benchmarks, in particular for the MIDAS model (see Supplementary Appendix~\ref{sec:midas_robustness} for more details).


\begin{algorithm}[t]
	\caption{Hyperparameter tuning}\label{algorithm:hyperpar_tuning}
	\KwData{Sample $\boldsymbol{y}_{2:T} = \{\boldsymbol{y}_2, \boldsymbol{y}_3, \allowbreak \ldots, \boldsymbol{y}_{T}\}$, $\boldsymbol{z}_{1:T-1}  = \{\boldsymbol{z}_1, \boldsymbol{z}_2, \allowbreak \ldots, \boldsymbol{z}_{T-1}\}$, initial state $\boldsymbol{x}_0$, initial guess $\boldsymbol{\varphi}_0$, convergence criterion $\text{Crit}$, maximal algorithm iterations $\text{MaxIter}$. If ridge regression is used to estimate $W$, fixed regularization strength $\lambda > 0$.}
	\KwResult{$\boldsymbol{\varphi}^*$}
	Fix $T$ and determine the model fit windows for $t = T_0, \ldots, T-1$. Choose whether the ESN model is estimated with a fixed or rolling window\;
	$j = 0$\;
	\While{$(\textbf{not }\textnormal{Crit})$ \textbf{and}  $(j < \textnormal{MaxIter})$}{
		(i) Given $\boldsymbol{\varphi}_j$, estimate coefficient matrices 
		$$ \left(\widehat{W}_t(\boldsymbol{\varphi}_j)\right)_{T_0:T-1}, $$ 
		where possibly $\widehat{W}_t(\boldsymbol{\varphi}_j)$ does not depend on $t$, e.g. in the fixed estimation setup\;
		(ii) Compute
		$$ \mathcal{L}_T(\boldsymbol{\varphi}_j) := \sum_{t=T_0}^{T-1} \boldsymbol{\ell}(\boldsymbol{y}_{t+1}, \widehat{W}_t(\boldsymbol{\varphi}_j) ^\top \boldsymbol{x}_t ), $$ 
		the cumulative one-step-ahead forecasting loss\;
		(iii) Update $\boldsymbol{\varphi}_{j+1} \leftarrow \boldsymbol{\varphi}_j $ with an appropriate rule (for example, the  gradient descent of $\mathcal{L}_T$ in the direction of $\boldsymbol{\varphi}_j$; in our applications, we use variants L-BFGS-B and pattern search)\;
		(iv) $j \leftarrow j + 1$, update $\textnormal{Crit}$\;
	}
\end{algorithm}

One issue with the state formulation in \eqref{eq:esn_model_states} and thus with the hyperparameter optimization routine in Algorithm~\ref{algorithm:hyperpar_tuning}, is that $\boldsymbol{\varphi}$ can not always be point identified. For example, if one considers identity activation $\sigma$ and lets $\alpha = \omega = 0$, it is obvious that the ESN model is system isomorphic \cite{RC16} to $\boldsymbol{x}^*_t = d \rho \overline{A} \boldsymbol{x}^*_{t-1} + d\gamma \overline{C} \boldsymbol{z}_t$, $\boldsymbol{y}_t = d^{-1} W \boldsymbol{x}^*_t + \boldsymbol{\epsilon}_t$ for all $d \not= 0$. This issue also arises in nonlinear models, for example when $\sigma$ is taken as a hyperbolic tangent and $\gamma$ is sufficiently small. Parameter identification in nonlinear models has been extensively studied in semi- and nonparametric cross-sectional regressions. For instance, it is known that in certain setups, point identification requires a proper {normalization} to be imposed. The interested reader can refer to Section 6.3 of \cite{Horowitz2009} for a discussion in a similar vein regarding nonparametric transformation models. Since often $\omega = 0$ is used, hyperparameter identification can be a significant issue when attempting model tuning. Whenever $\omega = 0$ we propose a helpful reparametrization given by
\begin{equation*}
	\boldsymbol{x}_t =  \alpha \boldsymbol{x}_{t-1} + (1-\alpha) \sigma\left(\psi \overline{A} \boldsymbol{x}_{t-1} + \overline{C} \boldsymbol{z}_t \right),
\end{equation*}
where $\psi=\rho/\gamma$. This prescription allows decoupling $\rho$ and $\gamma$ at the cost of the constant input scaling, which may be undesirable whenever one wants to attenuate the nonlinearity induced by the sigmoid map without also reducing the spectral radius.\footnote{One can fix $\overline{C}$ to have a different scaling before optimizing the hyperparameter $\psi$. However, this amounts to one more \textit{ex ante} model tuning step.} It is immediate to modify the optimization scheme to deal with the case $\widetilde{\boldsymbol{\varphi}} = (\alpha, \psi)$. In the sequel, we assume that the ESN models are estimated using the approaches proposed in this subsection and use the conventional ESN specification as in \eqref{eq:esn_model_states}-\eqref{eq:esn_model_output} to discuss the forecasting strategy.

\subsection{Cross-validation}\label{sec:cross_validation}

Because the initial cross-validation of $\lambda$ uses an extended sample to try and improve generalization -- specifically, our concern is for the fixed estimation setups -- we use two slightly different approaches:
\begin{itemize}
	\item In all setups -- fixed, expanding, rolling -- the \textit{initial} ridge penalty cross-validation is done on the extended sample (starting January 1st, 1975 instead of January 1st, 1990). We construct 10 folds with 5 out-of-sample observations starting from the end of the sample. Each fold and out-of-sample observation set is re-normalized.
	\item Only in the expanding and rolling setups, for each subsequent window (the ones that now include at least one testing observation), we use the true sample (starting January 1st, 1990) and construct 5 folds, again with 5 out-of-sample observations. This is done to keep cross-validation computational complexity low and avoid making some folds too small, which could hurt larger MFESN models.
\end{itemize}
In practice, simple experiments show that there is not much difference between using 5 or 10 folds in the initial cross-validation.

\section{Performance measures}\label{Performance measures}

In this section we define the performance measures used throughout the paper to quantify the quality of forecasts produced by competing models. Suppose that a given model is used to produce a collection of forecasts $\{ \widehat{\boldsymbol{y}}_s \}_{s \in S}$, $\widehat{\boldsymbol{y}}_s \in \mathbb{R}^J$. The ordered index set $S = \{s_1, \ldots, s_{|S|}\}$, where $|S|$ is the number of indices in $S$, can change depending on the setup. For example, in the case of 1-step ahead forecasting, $S = \{T+1, T+2, \ldots, T+\overline{T}\}$ where the 1-step ahead forecasts are constructed using the data up to  $T, T+1,\ldots, T+\overline{T}-1$, respectively. For $h$-step ahead forecasts, we set $S = \{T+h, T+h+1, \ldots, T+\overline{T}-H+h\}$ where $H$ is the maximal forecasting horizon. This ensures that the same number of forecasts are produced at each horizon and can be compared, for example, using the uniform Model Confidence Set (MCS) test described in Supplementary Appendix \ref{uMCS}.

\paragraph{MSFE and RMSFE.}
The \textit{root mean squared forecasting error} is given by
\begin{equation*}
    \textnormal{MSFE}(S) 
    := 
    \frac{1}{|S|} \sum_{s \in S} \| {\boldsymbol{y}}_s - \widehat{\boldsymbol{y}}_s \|_2^2 ,
\end{equation*}
while the \textit{root mean squared forecasting error} is
\begin{equation*}
    \textnormal{RMSFE}(S) := \sqrt{ \textnormal{MSFE}(S) } \: .
\end{equation*}

\paragraph{Cumulative SFE and Cumulative RMSFE.}
The \textit{cumulative squared forecasting error} is given by the cumulative sum of squared errors. We define for any forecasting index $\tau \in S$, 
\begin{equation*}
    \textnormal{CSFE}(\tau) 
    := 
    \sum_{\substack{s \in S\\ s \leq \tau}} \| {\boldsymbol{y}}_s - \widehat{\boldsymbol{y}}_s \|_2^2 .
\end{equation*}
To define the \textit{cumulative RMSFE} for any $\tau \in S$ we first define $\mathcal{T}_l(\tau) := \{s \in S : s \leq \tau\}$ and then write
\begin{equation*}
    \textnormal{CRMSFE}(\tau)
    :=
    \sqrt{ \frac{1}{|\mathcal{T}_l(\tau)|} \sum_{\substack{s\in \mathcal{T}_l(\tau)}} \| {\boldsymbol{y}}_s - \widehat{\boldsymbol{y}}_s \|_2^2 }.
\end{equation*}

\paragraph{Ahead RMSFE and 1-Year-Ahead RMSFE.}
If one wants to the evaluate performance \textit{ahead} of a certain point of time, it is also possible to define the \textit{ahead RMSFE},
\begin{equation*}
    \textnormal{AheadRMSFE}(\tau)
    :=
    \sqrt{ \frac{1}{|\mathcal{T}_u(\tau)|} \sum_{\substack{s \in \mathcal{T}_u(\tau)}} \| {\boldsymbol{y}}_s - \widehat{\boldsymbol{y}}_s \|_2^2 } ,
\end{equation*}
where we introduce $\mathcal{T}_u(\tau) := \{s \in S : s \geq \tau\}$. 

In the special case where the indices of $S$ are associated to dates, one may also compare performance after a given amount of time has passed from the current time index. For example, one may evaluate how performance degrades after model estimation when parameters are fixed and not updated. In our empirical exercises where $\{{y}_t\}_{t \in \mathbb{Z}}$, $y_t\in \mathbb{R}$, is a quarterly GDP series, we can define the \textit{1-year-ahead RMSFE} as
\begin{equation*}
    \textnormal{1YAheadRMSFE}(\tau)
    :=
    \sqrt{ \frac{1}{|\mathcal{T}_u(\tau+4)|} \sum_{\substack{s \in \mathcal{T}_u(\tau+4)}} ( {{y}}_s - \widehat{{y}}_s )^2 }.
\end{equation*}

\section{Uniform Multi-Horizon MCS}\label{uMCS}

We now give details on the implementation of the Uniform Multi-Horizon MCS test for the multi-horizon forecast comparisons in our empirical analysis. Our procedure follows closely the one originally provided by \cite{quaedvliegMultiHorizonForecastComparison2021}: we provide R code for our functions, while the author's code was originally developed in the Ox programming language. 

A main difference is that we prefer to use a Bartlett kernel to compute the sample uSPA statistic, whereas \cite{quaedvliegMultiHorizonForecastComparison2021} uses the quadratic spectral (QS) kernel of \cite{Andrew1991}. Our main reason for this choice is that the QS kernel features non-zero weights for all lags, while the Bartlett kernel has finite support. This is especially important since we only have a few forecasts in our case; thus, higher lag autocovariances between model losses can only be poorly estimated. It means our uMCS procedure implements the standard Newey-West HAC estimator.
We use $B = 100$ replications for the outer and inner bootstraps. Finally, the inner bootstrap critical value is set at $\alpha = 0.1$.

\section{MIDAS}\label{subsec_midas}

A state-of-the-art methodology for incorporating data of heterogeneous frequencies into one model is the MIDAS framework developed in \cite{Ghysels2004, Ghysels2007}. Here we present MIDAS in its dynamic form, which allows the inclusion of target series autoregressive lags. We use our temporal notation given in Definition~\ref{Temporal notation} throughout.

If the MIDAS model contains only one explanatory variable $(z_r)$ with frequency multiplier $\kappa$, then it can be written as
\begin{equation}\label{eq:MIDAS}
	y_t = \alpha_0 + \sum_{i=1}^p \alpha_i y_{t-i} + \beta \sum_{k=0}^{K} \varphi(\boldsymbol{\theta}, k) z_{t,\tempo{-k}{\kappa}} + \epsilon_t,
\end{equation}
where $\alpha_0$ is a constant term, $\{\alpha_i\}_{i=1}^p$ are the autoregressive parameters, $\beta$ is a scaling parameter, $\{\varphi(\boldsymbol{\theta},k)\}_{k=0}^K$ are the MIDAS weights given as a parametric function of lag $k$ and underlying parameter vector $\boldsymbol{\theta}\in \mathbb{R}^q$, and $(\epsilon_t)$ is a martingale difference process relative to the filtration $\{\mathcal{F}_{t}\}$ generated by $\{y_{t-1-j},x_{t-j}, \ldots, x_{t-j,\tempo{-K}{\kappa}}, \epsilon_{t-1 - j} \mid j\geq 0\}$ and such that $\mathbb{E}[\epsilon_t^2] = \sigma_{\epsilon}^2<\infty$.

The MIDAS weighting scheme is the core innovation of the model. It borrows parsimony from distributed lag models in the sense that, even if $K$ is large, the vector $\boldsymbol{\theta}\in \mathbb{R}^q$  is usually restricted to contain only a handful of parameters. This greatly reduces the number of coefficients that need to be estimated, and a nonlinear least-squares estimator $\widehat{\boldsymbol{\theta}}$ can be readily implemented. There are alternative formulations of the MIDAS framework where $\varphi(\boldsymbol{\theta}, k) = \theta_k$ so that the above reduces to a full linear model, the so-called unrestricted MIDAS or U-MIDAS \citep{Foroni2011}. 

We follow the literature and use the most commonly applied weighting scheme that is based on the exponential Almon weighting polynomial map $ \varphi: \mathbb{R}^q \times \mathbb{N}^+ \longrightarrow \mathbb{R}^+$ (see \cite{almon1965} for more details). In particular, for the case of $q=2$, the two-parameter Almon weighting polynomial is given by
\begin{equation*}
	\varphi(\boldsymbol{\theta}, k) = \varphi((\theta_1, \theta_2), k) = \exp(\theta_1 k + \theta_2 k^2), \enspace k\in \mathbb{N}^+. 
\end{equation*}
Since Almon weights need not sum up to a given constant for different values of $\theta_1$ and $\theta_2$, it is often common to consider the normalized Almon scheme
\begin{equation}
	\overline{\varphi}(\boldsymbol{\theta}, k) = \frac{\exp(\theta_1 k + \theta_2 k^2)}{\sum_{k=0}^K \exp(\theta_1 k + \theta_2 k^2)}, \label{eq:almon_weights}
\end{equation}
which together with \eqref{eq:MIDAS} allows to treat $\beta$  as a rescaling constant.

Let us now consider a more general model suitable for situations where time series of different frequencies are available and must be integrated into the MIDAS equation. Consider the case of $L$ regressor time series. We assume that the $l$th time series is sampled at a frequency $\kappa_l$ and contains observations $({z}^{(l)}_{t,\tempo{s}{\kappa_l}})_{t,s}$ with ${z}^{(l)}_{t,\tempo{s}{\kappa_l}} \in \rset$ for all $t \in \mathbb{Z}$ and $s \in \{0, \ldots, \kappa_l-1\}$. It happens frequently in practice that $\kappa_l$, $l \in [L]$ takes values from a small set of integers. For example, in the case of yearly, quarterly, and monthly data $\kappa_l \in \{1, 4, 12\}$ even though $L$ could be very large (often, hundreds or thousands of series might be of interest). The MIDAS model explaining low-frequency target variable $y_t$ with $L$ regressors $({z}^{(l)}_{t,\tempo{s}{\kappa_l}})_{t,s}$, $l\in [L]$ can be written as follows
\begin{equation}\label{eq:MIDASgeneral}
	y_t = \alpha_0 + \sum_{i=1}^p \alpha_i y_{t-i} + \sum_{l=1}^L  \beta_l \sum_{k=0}^{K_l} \varphi(\boldsymbol{\theta}_l, k) z^{(l)}_{t,\tempo{-k}{\kappa_l}}  + \epsilon_t,
\end{equation}
where the martingale difference process $(\epsilon_t)$ is relative to the filtration generated by sets as in \eqref{eq:MIDAS}, modified to include all the considered $L$ regressors.

The MIDAS framework produces forecasts of the chosen target variable at the low frequency of the target. Yet, due to the MIDAS multi-frequency structure, \textit{nowcasting} is also a straightforward exercise: if, for example, the high-frequency regressor is a single series $(z_r)$ with frequency multiplier $\kappa$, one can construct exactly $\kappa$ regression equations -- one for each high-frequency release within a low-frequency period -- and use these to produce high-frequency nowcasts of the target. In fact, due to the convenience of the MIDAS model, it is easy to define high-frequency regression specifications to study high-frequency forecasts and multicasts (see Appendix~\ref{Forecasting Schemes}).

In practice, implementing \eqref{eq:MIDASgeneral} demands some care. From a computational point of view, as long as the relevant regression matrices can be constructed, estimation amounts to a nonlinear least-squares problem, which can be readily solved. In Supplementary Appendices~\ref{sec:MIDAS_implementation} and \ref{sec:midas_robustness} we discuss the technical aspects of our MIDAS implementation in more detail. One of the important issues of the MIDAS estimation is the non-convexity of the nonlinear least squares loss as a function of parameters. Often, a practitioner may obtain different estimation results depending on initialization and, more importantly, those that lead to a different quality of forecasts. Other weighting schemes that allow for convex estimation problems can be used. For example, one may adopt the Almon lag polynomial parametrization \citep{Ghysels2016,pettenuzzoMIDASApproachModeling2016} using a discrete polynomial basis for the transformation of high-frequency regressors. This specification allows for standard OLS estimation but requires careful choice of the polynomial order hyperparameter.

Another crucial disadvantage of the MIDAS specification is that practical implementations can be very challenging. This is caused mainly by the ragged edges of the ``raw'' macroeconomic data, incomplete observations, and uneven sampling frequencies. The relative inflexibility of MIDAS regression lag specifications makes integrating daily and weekly data at true calendar frequencies (i.e. without interpolation or aggregation) very complex.\footnote{One could set up a MIDAS regression with the full yearly calendar of weeks and working days as lags. However, ragged edges arising from holidays, leap years, etc. would still be non-trivial to handle coherently without resorting to downsampling, data re-alignment, or interpolation.} State-space models effectively mitigate these issues.

Finally, as shown in \cite{Bai2010}, exponential Almon MIDAS regressions have inherent connections to dynamic factor models, which we discuss in the next section. When the factor structure is not trivial, MIDAS can, however, only yield a finite-order approximation to a DFM data-generating process. Furthermore, \cite{Bai2010} prove that in well-identified setups the mapping between exponential Almon and factor model coefficients is highly nonlinear. Given the robustness evaluations in Supplementary Appendix~\ref{sec:midas_robustness}, in practice, it appears hard to formally relate MIDAS and DFM forecasting performance.


\subsection{MIDAS Implementation}\label{sec:MIDAS_implementation}

While the MIDAS regression framework is straightforward to discuss in terms of equations, some care must be taken when implementing it computationally. A key assumption that can be imposed is that the integer frequencies $\boldsymbol{\kappa}:=\{\kappa_1, \ldots, \kappa_L\}$ of $L$ regressors are such that $\kappa_{\max}:= \max(\boldsymbol{\kappa})$ is a multiple of each of the $\kappa_l$, $l\in [L]$. In this case, MIDAS parameter estimation can be written in matrix form, which allows for efficient numerical implementation, which we spell out in the following paragraphs.

Let $q_l = \kappa_{\max} / \kappa_l$, $l\in [L]$ denote the frequency ratios and define $\boldsymbol{y} := (y_1, y_2, \ldots, y_T)^\top$ the vector of target observations, where $T$ is the sample length in reference time scale. Additionally, let $\boldsymbol{z}^{(l)} := (z^{(l)}_{1}, z^{(l)}_{2}, \ldots, z^{(l)}_{T_l})^\top$ be $T_l = T \cdot \kappa_l$ long vector which consists of observations of the $l$th covariate $z^{(l)}$ released with frequency $\kappa_l$. For the parameters of the MIDAS model in \eqref{eq:MIDASgeneral} to be identifiable, we assume that 
\begin{equation*}
    T > 1 + p + \sum_{l=1}^L \left\lceil \frac{K_l}{\kappa_l} \right\rceil.
\end{equation*}
Since $\kappa_{\max}$ is a multiple of each of the $L$ frequencies, for each series we introduce 
\begin{equation*}
    \mathbf{Y} = \boldsymbol{y} \otimes \boldsymbol{i}_{\kappa_{\max}} , \quad \mathbf{Z}^{(l)} = \boldsymbol{z}^{(l)} \otimes \boldsymbol{i}_{q_l},
\end{equation*}
where $\boldsymbol{i}_{q_l}$ and $\boldsymbol{i}_{\kappa_{\max}}$ are vectors of ones of lengths ${q_l}$ and $\kappa_{\max}$, respectively. In the absence of missing observations, we have that $\mathbf{Y}, \mathbf{Z}^{(l)} \in \mathbb{R}^{T_{\max}}$ with $T_{\max} = T \cdot \kappa_{\max}$ observations. We now construct preliminary regression matrices such that their maximal rows number is $T_{\max}$ without accounting for the lags structure of both the target (autoregressive lags) and regressors (MIDAS lags) and we introduce zeros where no observations are available.\footnote{At the time of implementation of this procedure in any convenient coding environment it is more natural to introduce placeholders instead and to perform the subsequently discussed truncation via matrix manipulation rather than by using matrix multiplication.} Define for $p \geq 1$ and for $ K_l \geq 0$
\begin{equation*}
    Y_p = \left(\begin{array}{cccc}
        0 & 0 & \cdots & 0 \\
        y_1 & 0 & \cdots & 0 \\
        y_2 & y_1 & \cdots & 0 \\
        \vdots & \vdots & \vdots & \vdots\\
        y_{T-2} & y_{T-3}  &  \cdots & y_{T-p-1}\\
        y_{T-1} & y_{T-2}  &  \cdots & y_{T-p}
    \end{array}\right)
    \otimes \boldsymbol{i}_{\kappa_{\max}} \quad {\rm and}\quad
    Z_{K_l} =  \left(\begin{array}{cccc}
        z^{(l)}_{1} & 0 & \cdots & 0 \\
        z^{(l)}_{2} & z^{(l)}_{1} & \cdots & 0 \\
        z^{(l)}_{3} & z^{(l)}_{2} & \cdots & 0 \\
        \vdots & \vdots & \vdots & \vdots\\
        z^{(l)}_{T_l-1} & z^{(l)}_{T_l-2}  &  \cdots & z^{(l)}_{T_l-K_l-1}\\
        z^{(l)}_{T_l} & z^{(l)}_{T_l-1}  &  \cdots & z^{(l)}_{T_l-K_l}
    \end{array}\right)
    \otimes \boldsymbol{i}_{q_l}.
\end{equation*}
 In the special case $p = 0$ (the MIDAS model, in this case, is called \textit{static}, since it does not contain an autoregressive term) we take $Y_p$ as empty. We now follow by noticing that one should not use $Y_p\in \mathbb{M}_{T_{\max},p}$ and $Z_{K_l}\in \mathbb{M}_{T_{\max},K_l+1}$ as autoregressive and mixed-frequency regression matrices, respectively, since some observations are missing. To overcome this we introduce 
\begin{equation}
\label{s index}
    n:= \max\left\{ p, \left\lceil \frac{K_1}{q_{1}} \right\rceil, \ldots, \left\lceil \frac{K_L}{q_{L}} \right\rceil \right\} \cdot \kappa_{\max}
\end{equation}
and the so-called upper truncation (selection) matrix
\begin{align*}
   U & = \left(\begin{array}{cc} \mathbb{O}_{T_{\max}- n + 1, n - 1} & \mathbb{I}_{T_{\max}-n + 1, T_{\max}-n + 1}
    \end{array}\right)
\end{align*}
 with which we obtain the following required response vector and regression matrices
\begin{align*}
    \mathbf{Y}^{\text{resp}} & = U \mathbf{Y}\in \mathbb{R}^{T_{\max}- n + 1}, \nonumber \\[3pt]
    Y^{\text{reg}}_p & = U Y_p \in \mathbb{M}_{T_{\max}- n + 1,p},  \nonumber \\[3pt]
    Z^{\text{reg}}_{K_l} & = U Z_{K_l}\in \mathbb{M}_{T_{\max}- n + 1, K_l + 1}, \nonumber \\[3pt]
    \mathbf{Z}^{\text{reg}} & = \left(\begin{array}{cccc} Y^{\text{reg}}_p  &  Z^{\text{reg}}_{K_1}& \cdots& Z^{\text{reg}}_{K_L} \end{array}\right)\in \mathbb{M}_{T_{\max}- n + 1, \sum_{l=1}^LK_l + L},
\end{align*}
where $Y^{\text{reg}}_p$ is empty whenever $p=0$.

We can now observe that $\mathbf{Y}^{\text{resp}}$ and $\mathbf{Z}^{\text{reg}}$ are sufficient to construct all MIDAS forecasting and nowcasting regressions. In practice, some care needs to be taken to make sure that data is correctly aligned: for example, in the case of nowcasting exercise regressors in $\mathbf{Z}^{\text{reg}}$ and targets in $\mathbf{Y}^{\text{resp}}$ have to be aligned differently than in the case of forecasting exercises. Provided the aligned data is executed correctly, the estimation of MIDAS parameters can be carried out efficiently. An important thing to mention is that the truncation with the help of $s$ in \eqref{s index} may be too restrictive, as it may lead to excluding up to $K_{\max}-1$ rows from $\mathbf{Z}^{\text{reg}}$ that could be used for estimation. This can be avoided at the time of implementation. In our repository available at [the address removed for anonymous submission] we consider this detail and exclude from the final regression matrices only those rows which cannot be used due to the lag requirements in the model. We warn the reader that this comes at a cost, namely the codes are lengthier and less elegant.

\section{Mixed-frequency DFM}
\label{subsec_dfm}

Macroeconomic modeling based on dynamic factor models has been popular since their introduction in \cite{geweke1977dynamic} and \cite{sargent1977business}. The proposition of DFMs is that a low-dimensional latent factor $(\boldsymbol{f}_t)_{t \in \mathbb{Z}}$, $\boldsymbol{f}_t \in \rset^d$, drives a high-dimensional observable stochastic process $(\boldsymbol{y}_t)_{t \in \mathbb{Z}}$, $\boldsymbol{y}_t \in \rset^n$. We consider a time-inhomogeneous state-space model with dynamics
\begin{align}
	\boldsymbol{f}_{t+1}|\boldsymbol{f}_{1:t},\boldsymbol{y}_{1:t} &\sim h_{t+1,\boldsymbol{\theta}}(\cdot | \boldsymbol{f}_t) \label{eq:state_transtion} \\
	\boldsymbol{y}_{t+1}|\boldsymbol{f}_{1:t+1},\boldsymbol{y}_{0:t} &\sim g_{t+1,\boldsymbol{\theta}}(\cdot | \boldsymbol{f}_{t+1}) \label{eq:observation_model}
\end{align}
for some time-dependent state transition kernels $h_{t,\boldsymbol{\theta}}$ and observation densities $g_{t,\boldsymbol{\theta}}$ and some parameter vector $\boldsymbol{\theta}$ in a parameter space $\Theta$. A common example in the literature (see \cite{watson1983alternative} for more details) is linear Gaussian factor models with time-inhomogeneous state transitions that can be represented as
\begin{align}
	\boldsymbol{f}_{t+1}&=A_{\boldsymbol{\theta}}\boldsymbol{f}_t+R_{\boldsymbol{\theta}}\boldsymbol{u}_{t} \label{eq:linear_state_transtion} \\
	\boldsymbol{y}_{t+1}&=\Lambda_{t+1, \boldsymbol{\theta}} \boldsymbol{f}_{t+1} +  S_{t+1,\boldsymbol{\theta}} \boldsymbol{w}_{t+1} \label{eq:linear_observation_model} 
\end{align}
with state transition matrix $A_{\boldsymbol{\theta}}\in \mathbb{M}_d$, time-dependent factor loading matrices $\Lambda_{t}\in \mathbb{M}_{n,d}$, and where $\boldsymbol{u}_t$ and $\boldsymbol{w}_t$ are independent Gaussian vectors with zero mean and identity covariance matrix of dimension $p$ and $n$, respectively, and $R_{\boldsymbol{\theta}}$ and $S_{t,\boldsymbol{\theta}}$ re matrices of appropriate dimensions.
It is often assumed that the dimension $p$ of the state noise vector $\boldsymbol{u}_t$ is smaller than the latent state space dimension $d$, which implies that $R_{\boldsymbol{\theta}}R_{\boldsymbol{\theta}}^\top$ is rank deficient, such as for AR$(p)$ factor dynamics \citep{stock2016dynamic, forni2005generalized, doz2011two}. In this case, $d=kp$ for some $k\in \nset^+$,
\begin{equation}
	A_{\boldsymbol{\theta}} = 
	\left(\begin{array}{ccccc}
		A_{\boldsymbol{\theta}}^{(1)} & A_{\boldsymbol{\theta}}^{(2)} &  \cdots & A_{\boldsymbol{\theta}}^{(p-1)} & A_{\boldsymbol{\theta}}^{(p)} \\
		\mathbb{I}_k & \mathbb{O}_k & \cdots & \mathbb{O}_k & \mathbb{O}_k\\
		\mathbb{O}_k & \mathbb{I}_k &  \cdots &\mathbb{O}_k &\mathbb{O}_k \\
		\vdots & &  \ddots & & \vdots \\
		\mathbb{O}_k & \mathbb{O}_k & \cdots & \mathbb{I}_k & \mathbb{O}_k
	\end{array}\right) , \quad
	\Lambda_{t,\boldsymbol{\theta}} =\left(\begin{array}{cccc} \Lambda_{t,\boldsymbol{\theta}}^{(1)} & \Lambda_{t,\boldsymbol{\theta}}^{(2)} & \cdots & \Lambda_{t,\boldsymbol{\theta}}^{(p)}\end{array}\right) 
	\label{eq:ar_factor}
\end{equation}
with $A_{\boldsymbol{\theta}}^{(j)} \in \mathbb{M}_{k}$ and $\Lambda_{t,\boldsymbol{\theta}}^{(j)} \in \mathbb{M}_{n,k}$. Setting $\boldsymbol{f}_t=(\boldsymbol{v}_t^\top, \boldsymbol{v}_{t-1}^\top, \ldots, \boldsymbol{v}_{t-p+1}^\top)^\top$ implies that $(\boldsymbol{v}_t)_{t \in \mathbb{Z}}$ is a $k$-dimensional AR($p$) process and it is commonly assumed that $\Lambda_{t,\boldsymbol{\theta}}^{(j)}=\mathbb{O}_{n,k}$ for $j>1$.
Let the initial state $\boldsymbol{f}_0$ be distributed according to $\nu$. The joint density of the latent path $\boldsymbol{f}_{0:T}$ and observations $\boldsymbol{y}_{0:T}$ is then
\begin{equation*}
	p_{\boldsymbol{\theta},\nu}(\boldsymbol{f}_{0:T},\boldsymbol{y}_{0:T}) =\nu(\boldsymbol{f}_0) g_{0,\boldsymbol{\theta}}(\boldsymbol{y}_0|\boldsymbol{f}_0) 
	\prod_{t=1}^T h_{t, \boldsymbol{\theta}}(\boldsymbol{f}_t|\boldsymbol{f}_{t-1}) g_{t,\boldsymbol{\theta}}(\boldsymbol{y}_t|\boldsymbol{f}_t),
\end{equation*}
while the marginal likelihood of $\boldsymbol{y}_{0:T}$ is $p_{\boldsymbol{\theta},\nu}(\boldsymbol{y}_{0:T})=\int p_{\boldsymbol{\theta},\nu}(\boldsymbol{f}_{0:T},\boldsymbol{y}_{0:T}) \rmd \boldsymbol{f}_{0:T}$.
Popular procedures for learning the static parameters $\boldsymbol{\theta} \in \Theta$ are based on gradient descent 
of the negative log-likelihood function $\ell_T \colon \Theta \to \rset, \boldsymbol{\theta} \mapsto - \log p_{\boldsymbol{\theta},\nu}(\boldsymbol{y}_{0:T})$ or on the Expectation Maximization (EM) algorithm introduced in \cite{dempster1977maximum}. We consider here gradient descent algorithmsbased on a sequence of step sizes $\gamma_k>0$, that update the model parameters based on iterations of the form
\begin{equation*}
	\boldsymbol{\theta}_{k+1} = \boldsymbol{\theta}_k - \gamma_{k+1} \nabla_{\boldsymbol{\theta}} \ell_T(\boldsymbol{\theta})|_{\boldsymbol{\theta}=\boldsymbol{\theta}_k},
\end{equation*}
for $k\in \nset^+$.\footnote{Consistency of the maximum log-likelihood estimate for the dynamics \eqref{eq:linear_state_transtion}-\eqref{eq:linear_observation_model} in the time-homogeneous case has been established for instance in \cite{douc2011consistency} under regularity assumptions, including, for instance, the full-rank of the noise covariance matrix $S_{\boldsymbol{\theta}}$, of the controllability matrix $C_{\boldsymbol{\theta}}= \left( R_{\boldsymbol{\theta}} | A_{\boldsymbol{\theta}}R_{\boldsymbol{\theta}} | \cdots | A_{\boldsymbol{\theta}}^{d-1}R_{\boldsymbol{\theta}} \right)$, and of the observability matrix $O_{\boldsymbol{\theta}}= \left(\Lambda_{\boldsymbol{\theta}}^\top | (\Lambda_{\boldsymbol{\theta}}A_{\boldsymbol{\theta}})^\top | \cdots | (\Lambda_{\boldsymbol{\theta}}A_{\boldsymbol{\theta}}^{d-1})^\top \right)^\top$. It is also possible to consider an online learning setting using a recursive decomposition of the score function as in \cite{legland1997recursive}. For general latent state dynamics \eqref{eq:state_transtion} and observation densities \eqref{eq:observation_model} that can be non-linear with non-Gaussian noise, particle filtering algorithms are often utilized that make use of particle approximations in gradient-descent or EM learning approaches, see for instance \cite{kantas2015particle}.} 


Assuming a linear Gaussian setting where the transition density of the latent factor process is given by \eqref{eq:ar_factor} to yield an AR($p$) process $(\boldsymbol{v}_t)_{t \in \mathbb{Z}}$, $\boldsymbol{v}_t=(v_{1,t}, \ldots, v_{k,t})^\top$, there remains some flexibility as to how the linear mappings 
$$\mathsf{Agg}_{\boldsymbol{\theta},L} \colon \mathbb{M}_{k,p} \to \rset, ~~(\boldsymbol{v}_{t-p + 1} \ldots, \boldsymbol{v}_t) \mapsto \left(\Lambda_{t,\boldsymbol{\theta}}\boldsymbol{f}_t\right)_i = \left(\Lambda_{t,\boldsymbol{\theta}} \right)_{i, \cdot} \begin{bmatrix} \boldsymbol{v}_t\\  \boldsymbol{v}_{t-1} \\ \vdots \\ \boldsymbol{v}_{t-p+1}  \end{bmatrix}$$ 
for some lag parameter $L\leq p$ are chosen for each dimension $i\in [n]$.\footnote{The Markovian representation \eqref{eq:state_transtion}-\eqref{eq:observation_model}, that is, the companion form, is based on the autoregressive order $p$, however, one can set $A_{\boldsymbol{\theta}}^{(\ell)}=\mathbb{O}_k$ for $\ell > p$. }  
We call this linear mapping $\mathsf{Agg}_{\boldsymbol{\theta},L}$  an \emph{aggregation function} 
and consider specific examples below that yield different models for the factor loadings matrices $\Lambda_{t, \boldsymbol{\theta}}$. Notice that our aggregation functions are linear with respect to the latent factors in contrast to the non-linear approaches introduced in \citet{proietti2006dynamic} that require approximations, such as resorting to extended Kalman filtering techniques.


\begin{example}[Stock aggregation]\label{ex:stock}
	\normalfont
	For $i \in [n]$, let $\boldsymbol{\beta}_i=(\beta_{i1}, \ldots, \beta_{ik}) \in \rset^k$ and consider
	\[ \mathsf{Agg}^{\textnormal{S}}_{\boldsymbol{\theta},1}(\boldsymbol{v}_{t-p + 1} \ldots, \boldsymbol{v}_t)_i = \sum_{m=1}^k \beta_{im}  {v}_{m,t},\]
	with $\boldsymbol{\theta}=\boldsymbol{\beta}_i$. 
\end{example}

\begin{example}[Almon-Lag aggregation]\label{ex:almon}
	\normalfont
	For $i \in [n]$, let $\boldsymbol{\beta}_i \in \rset^k$, $ \boldsymbol{\psi}_i \in \rset^{2k}$ and consider
	\[ \mathsf{Agg}^{\textnormal{AL}}_{\boldsymbol{\theta},L}(\boldsymbol{v}_{t-p + 1} \ldots, \boldsymbol{v}_t)_i = \sum_{m=1}^k \beta_{im} \sum_{\ell=0}^{L-1} \overline{\varphi}({\psi}_{im}, \ell) {v}_{m,t-\ell},\]
	with $\boldsymbol{\theta}=\left(\boldsymbol{\beta}_i, \boldsymbol{\psi}_i, \boldsymbol{\beta}_i, \boldsymbol{\psi}_i \right)$ and Almon-Lag weights $\overline{\varphi}$ given in \eqref{eq:almon_weights}. 
\end{example}

\begin{example}[Trigonometric aggregation]\label{ex:sin}
	\normalfont
	For $i \in [n]$, let $\boldsymbol{\beta}_i \in \rset^{k}$, and for $K \in \nset$, let $\boldsymbol{\lambda} \in \rset_+^K$, $\boldsymbol{\omega}\in [0,1]^K$, $\boldsymbol{\gamma}\in [-\pi,\pi]^K$ and $\tau \in \rset_+$. Define
	\[ \mathsf{Agg}^{\sin}_{\boldsymbol{\theta},L}(\boldsymbol{v}_{t-p + 1} \ldots, \boldsymbol{v}_t)_i = \sum_{m=1}^k \beta_{im} \sum_{\ell=0}^{L-1} \overline{a}_p(\boldsymbol{\lambda}, \boldsymbol{\omega}, \boldsymbol{\gamma}, \tau, \ell) {v}_{m,t-\ell},\]
	with $\boldsymbol{\theta}=\left(\boldsymbol{\beta}_i, \boldsymbol{\lambda}, \boldsymbol{\omega}, \boldsymbol{\gamma}, \tau \right)$ and
	\[ \overline{a}_p(\boldsymbol{\lambda}, \boldsymbol{\omega}, \boldsymbol{\gamma}, \tau, \ell) = \frac{ \exp\left( \frac{1}{\tau} \sum_{j=1}^K \lambda_j^2 \cos(2 \pi \omega_j \ell + \gamma_j) \right)}{\sum_{\ell'=0}^{p-1}  \exp\left( \frac{1}{\tau} \sum_{j=1}^K \lambda_j^2 \cos(2 \pi \omega_j \ell' + \gamma_j) \right)}.\]
	This aggregation scheme is motivated by self-attention models (we refer the reader to \cite{bahdanau2014neural,Vaswani2017} for more details), but to retain linearity only considers a relative positional encoding with a Toeplitz structure. Observe that the aggregation parameters are shared across all $n$ dimensions in contrast to the Almon lag scheme in Example \ref{ex:almon}.
\end{example}

Some authors (see for example \cite{mariano2003new,banbura2014maximum}) have imposed different restrictions on the form of the factor loadings matrices or aggregation function, particularly for one-dimensional mixed-frequency factor models of quarterly GDP growth rates and monthly covariates, which are motivated by approximations of growth rates. We do not pursue this additional restriction in this work. 

Kalman filtering techniques have been used routinely for handling missing observations in multi-frequency DFMs, see \cite{harvey1998messy}. In this work, we leverage modern auto-differentiation libraries \citep{Abadi2016, dillon2017tensorflow} to compute the gradient of the log-likelihood based on Kalman filtering formulae and estimate the static parameters $\boldsymbol{\theta}$ by gradient ascent of the log-likelihood. For alternative estimation approaches using EM that could be extended to this setting, we refer the reader to \cite{banbura2014maximum}. Nonlinear or non-Gaussian dynamic factor models in a mixed frequency setting have been considered in \cite{gagliardini2017indirect, leippold2019particle} that rely on particle filtering methods in conjunction with backward simulation algorithms as in \cite{godsill2004monte}, while \cite{schorfheide2018identifying} consider a Bayesian approach using particle MCMC (see \cite{Andrieu2010}). Such approaches can become computationally expensive and are not considered for benchmarking purposes.

While previous mixed-frequency DFMs (see \cite{mariano2003new,banbura2014maximum} for a more thorough discussion) often consider time series which are sampled at two frequencies, we introduce here a flexible mixed-frequency DFM that describes $L\in \nset^+$ collections of distinct time series sampled at frequencies $\{\kappa_1, \ldots, \kappa_L\}$ and each consisting of $\{n_1, \ldots, n_L\}$ series, respectively. In the same setting as in Section~\ref{Multi-Frequency Echo State Models}, each group of $n_l$, $l \in [L]$, time series sampled at frequency $\kappa_l$ contains observations $(\boldsymbol{y}^{(l)}_{t,\tempo{s}{\kappa_l}})$ with $\boldsymbol{y}^{(l)}_{t,\tempo{s}{\kappa_l}} \in \rset^{n_l}$ for all $t \in \mathbb{Z}$ and $s \in \{0, \ldots, \kappa_l-1\}$. Let $\kappa_{\max} = \max_l \kappa_l$. Suppose that the latent factor dynamics are updated at the highest sampling frequency based on the linear transition
\begin{equation}
	\boldsymbol{f}_{t,s+1|\kappa_{\max}} = A_{\boldsymbol{\theta}}\boldsymbol{f}_{t,s|\kappa_{\max}} + R_{\boldsymbol{\theta}} \boldsymbol{u}_{t,s+1|\kappa_{\max}}, \label{eq:factor_dynamics}
\end{equation}
where
\begin{equation*}
	\boldsymbol{f}_{t,s|\kappa_{\max}} = \left(\boldsymbol{v}_{t,s|\kappa_{\max}}^\top, \ldots,\boldsymbol{v}_{t,s-p+1|\kappa_{\max}}^\top\right)^\top, 
\end{equation*}
with $A_{\boldsymbol{\theta}}$ given in \eqref{eq:ar_factor} for the special case where $A_{\boldsymbol{\theta}}^{(\ell)} = \mathbb{O}_{k}$ for $\ell \geq 2$, $p = \kappa_{\max}$ and 
\begin{equation*}
	A_{\boldsymbol{\theta}}^{(1)}= \bar{A} \, \frac{\rho}{\max \left\{ \rho, |\lambda_1(\bar{A})| \right\} } 
\end{equation*}
with parameters $\rho \in (0,1)$, $\bar{A}\in \mathbb{M}_{k}$ and with $\lambda_1(\bar{A})$ denoting the largest  eigenvalue of $\bar{A}$. In the simplified scenario of first-order autoregressive dynamics, we parameterize $R_{\boldsymbol{\theta}}\in \mathbb{M}_k$ to be positive definite and diagonal and $\boldsymbol{u}_{t,s+1|\kappa_{\max}}$ are a sequence of IID $k$-dimensional standard Gaussian variables.

Notice that Kalman filtering formulas yield the first moment 
$$ \widehat{\boldsymbol{f}}_{t,s|\kappa_{\max}}=\E \left[ \boldsymbol{f}_{t,s|\kappa} \big| \boldsymbol{y}_{1,0|\kappa_{\max}}, \ldots, \boldsymbol{y}_{t,s|\kappa_{\max}} \right]$$
recursively online, see for example Supplementary Appendix \ref{sec:DFM_implementation} for details in the general time-inhomogeneous case.
Due to the linearity in \eqref{eq:factor_dynamics}, for any $h \in \nset$,
$$ \widehat{\boldsymbol{f}}_{t,s+h|\kappa_{\max}}=\E \left[\boldsymbol{f}_{t,s+h|\kappa_{\max}} \big| \boldsymbol{y}_{1,0|\kappa_{\max}}, \ldots, \boldsymbol{y}_{t,s|\kappa_{\max}}\right] = A_{\boldsymbol{\theta}}^h  \widehat{\boldsymbol{f}}_{t,s|\kappa_{\max}}.$$
Furthermore, from the linearity of the aggregation scheme, we obtain the forecasts for any $s,h\in \nset$,
\begin{equation} 
	\E\left[\boldsymbol{y}^{(l)}_{t,s+h|\kappa_{l}} \big| \boldsymbol{y}_{1,0|\kappa_{l}}, \ldots, \boldsymbol{y}_{t,s|\kappa_{l}}\right] 
    = 
    \mathsf{Agg}_{\boldsymbol{\theta^{(l)}}} \left( \widehat{\boldsymbol{f}}_{t,(s+h) q_l |\kappa_{\max}} \right) ,
 \label{eq:dfm_forecasts}
\end{equation}
where $q_l = \kappa_{\max} / \kappa_l$ (c.f. Section \ref{subsec:Single-reservoir MFESN}) and $\mathsf{Agg}_{\boldsymbol{\theta^{(l)}}}$ is the aggregation scheme for frequency $l$.
%
We observe that there is a single latent factor process that describes the observations at all frequencies, in contrast, for instance, to hierarchical Hidden Markov Models (HMM) \citep{hihi1995hierarchical} where the latent variables evolve a priori at different time-scales. This time evolution of states is similar to the SMFESN models also developed in this paper.

It is possible to write the following mixed-frequency DFM model in Example \ref{ex:DFM} as a general time-inhomogeneous state-space system \eqref{eq:state_transtion}-\eqref{eq:observation_model} by suitably parameterizing the time dependencies in the aggregation matrices. We provide more details on implementing our mixed frequency DFM in Supplementary Appendix \ref{sec:DFM_implementation} below. The standard Kalman filtering recursions utilized therein for parameter estimation have a cubic complexity in the dimension $d$ or $n$ of the Markovian factor process $\boldsymbol{f}$ or the observation process $\boldsymbol{y}$, respectively, at every time step. The marginal log-likelihood is optimized based on stochastic gradient methods with adaptive step sizes \citep{kingma2014adam} and is generally not a concave function of the parameter values.\footnote{We compute gradients of the marginal log-likelihood using a Kalman filter implementation for a time-inhomogeneous linear Gaussian state space model in TensorFlow Probability \citep{dillon2017tensorflow}.}

\begin{example}[Quarterly-Monthly-Daily DFM Model]
\normalfont
\label{ex:DFM}
We consider $n_{(\textnormal{\texttt{6d}})}$ time series that result from averaging daily time series over 6 days, yieling 12 observations per quarter that are denoted as $\boldsymbol{y}^{ \dsf}$. Furthermore, we consider $n_{(\textnormal{\texttt{m}})}$ monthly $\boldsymbol{y}^{\mf}$ as well as $n_{(\textnormal{\texttt{q}})}$ quarterly time series $\boldsymbol{y}^{\qf}$. We let $\kappa_{\max} = 72/6=12$ and update the $k$-dimensional latent factor process every $6$ days in sync with $\boldsymbol{y}^{ \dsf}$. We aggregate 6 days to significantly decrease the computational cost of the factor model. The latent factors are assumed to have the VAR(1) dynamics,\footnote{Because of the AR(1) dynamics, we do not write it in the companion form of the latent factor. However, unless one uses the stock aggregation scheme, one still needs to keep track of the past factor values for modeling monthly or quarterly observables.}
$$ \boldsymbol{v}_{t,s+1|12}=A^{(1)} \boldsymbol{v}_{t,s|12}+ R \boldsymbol{u}_{t,s+1|12}, $$ 
for any $s,t \in \nset$, $A^{(1)} \in \mathbb{M}_{k,k}$ $R \in \mathbb{M}_k$ and IID $k$-dimensional standard Gaussian variables $\boldsymbol{u}_{t,s|12}$.
The averaged daily data is described by
$$ \boldsymbol{y}^{\dsf}_{t,s|12}=\beta^{\dsf}  \boldsymbol{v}_{t,s|12} + S^{\dsf} \boldsymbol{w}_{t,s|12}^{\dsf} $$
for any $s,t \in \nset$, $\beta^{\dsf} \in \mathbb{M}_{n_{(\textnormal{\texttt{6d}})},k}$, $S^{\dsf} \in \mathbb{M}_{n_{(\textnormal{\texttt{6d}})}}$ and IID $n_{(\textnormal{\texttt{6d}})}$-dimensional standard Gaussian variables $\boldsymbol{w}^{\dsf}_{t,s|12}$.
The monthly data in the stock aggregation scheme is modeled as
$$\boldsymbol{y}^{\mf}_{t,s|3}=\beta^{\mf}  \boldsymbol{v}_{t,4 s|12} + S^{\mf} \boldsymbol{w}^{\mf}_{t,s|3} , $$
with $\beta^{\mf} \in \mathbb{M}_{n_{(\textnormal{\texttt{m}})},k}$, $S^{\mf} \in \mathbb{M}_{n_{(\textnormal{\texttt{m}})}}$ and IID $n_{(\textnormal{\texttt{m}})}$-dimensional standard Gaussian variables $\boldsymbol{w}^{\mf}_{t,s|3}$.
Alternatively, an Almon aggregation scheme yields the model
$$\boldsymbol{y}^{\mf}_{t,s|3}=\beta^{\mf} \sum_{\ell=0}^3 \boldsymbol{\overline{\varphi}}(\boldsymbol{\psi^{\mf}}, \ell)  \odot \boldsymbol{v}_{t,(4 s - \ell)|12} + S^{\mf} \boldsymbol{w}^{\mf}_{t,s|3} , $$
with $\beta^{\mf} \in \mathbb{M}_{n_{(\textnormal{\texttt{m}})},k}$, $S^{\mf} \in \mathbb{M}_{n_{(\textnormal{\texttt{m}})}}$, IID $n_{(\textnormal{\texttt{m}})}$-dimensional standard Gaussian variables $\boldsymbol{w}^{\mf}_{t,s|3}$ and $\boldsymbol{\overline{\varphi}}(\boldsymbol{\psi^{\mf}}, \ell) = \left( \overline{\varphi}({\psi^{\mf}}_1, \ell), \ldots, \overline{\varphi}({\psi^{\mf}}_k, \ell)\right)^\top \in \rset^k$. The symbol $\odot$ stands for the Hadamard or componentwise matrix product.

The quarterly components can be analogously described as
$$\boldsymbol{y}^{\qf}_{t}=\beta^{\qf}  \boldsymbol{v}_{t,0|12} + S^{\qf} \boldsymbol{w}^{\qf}_{t}$$
for a stock aggregation scheme, while the Almon scheme writes as
$$\boldsymbol{y}^{\qf}_{t}=\beta^{\qf} \sum_{\ell=0}^{11} \boldsymbol{\overline{\varphi}}(\boldsymbol{\psi^{\qf}}, \ell)  \odot \boldsymbol{v}_{t,- \ell|12} + S^{\qf} \boldsymbol{w}^{\qf}_{t} , $$
with $\beta^{\qf} \in \mathbb{M}_{n_{(\textnormal{\texttt{q}})},k}$, $S^{\mf} \in \mathbb{M}_{n_{(\textnormal{\texttt{q}})}}$, IID $n_{(\textnormal{\texttt{q}})}$-dimensional standard Gaussian variables $\boldsymbol{w}^{\mf}_{t}$ and $\boldsymbol{\overline{\varphi}}(\boldsymbol{\psi^{\qf}}, \ell) = \left( \overline{\varphi}({\psi^{\qf}}_1, \ell), \ldots, \overline{\varphi}({\psi^{\qf}}_k, \ell)\right)^\top \in \rset^k$.
\end{example}

\subsection{Mixed-frequency DFM Implementation}\label{sec:DFM_implementation}

This section gives additional details on implementing non-homogeneous dynamic factor models, such as the mixed frequency model introduced in the main text. We notice that the conditioning notation in this section should not be confused with our temporal notation in Definition~\ref{Temporal notation}.

\paragraph{Kalman filtering and computational complexity.}
The sufficient statistics of the posterior distribution of the latent factor $\boldsymbol{f}_t|\boldsymbol{y}_{0:t}$ can be updated recursively by the Kalman filter updates in the linear Gaussian setting. First, propagate the prior
\begin{align*}
    \boldsymbol{\widehat{f}}_{t+1|t,\boldsymbol{\theta}}&=A_{\boldsymbol{\theta}} \boldsymbol{\widehat{f}}_{t|t,\boldsymbol{\theta}} \\
    \widehat{\Sigma}_{t+1|t,\boldsymbol{\theta}}&=A_{\boldsymbol{\theta}}\widehat{\Sigma}_{t+1|t,\boldsymbol{\theta}}A_{\boldsymbol{\theta}}^\top + S_{t+1,\boldsymbol{\theta}}S_{t+1,\boldsymbol{\theta}}^\top.
\end{align*}
Compute the innovation covariance
$$ \Gamma_{t+1,\boldsymbol{\theta}}= \Lambda_{t+1} \widehat{\Sigma}_{t+1|t,\boldsymbol{\theta}} \Lambda_{t+1}^\top + R_{\boldsymbol{\theta}}R_{\boldsymbol{\theta}}^\top$$
and the Kalman gain
$$K_{t+1,\boldsymbol{\theta}}=\widehat{\Sigma}_{t+1|t,\boldsymbol{\theta}} \Lambda_{t+1,\boldsymbol{\theta}}^\top \Gamma_{t+1,\boldsymbol{\theta}}^{-1}.$$
Then, update the statistics with the new information $y_{t+1}$,
\begin{align*}
    \boldsymbol{\widehat{f}}_{t+1|t+1,\boldsymbol{\theta}}&=\boldsymbol{\widehat{f}}_{t+1|t,\boldsymbol{\theta}}-K_{t+1,\boldsymbol{\theta}} \left(y_{t+1}-\Lambda_{t+1,\boldsymbol{\theta}} \boldsymbol{\widehat{f}}_{t+1|t,\boldsymbol{\theta}}\right) \\
    \widehat{\Sigma}_{t+1|t+1,\boldsymbol{\theta}}&=\left(\mathbb{I} - K_{t+1,\boldsymbol{\theta}} \Lambda_{t+1,\boldsymbol{\theta}} \right) \widehat{\Sigma}_{t+1|t,\boldsymbol{\theta}}.
\end{align*}

Notice that the inverse of the log-determinant of the innovation matrices $\Gamma_{t,\boldsymbol{\theta}}$ are required for computing the Kalman gains and the marginal log-likelihood, respectively, which yield a cubic computational complexity in the dimension of the observation process. Alternatively, one can apply matrix inversion or determinant lemmas to obtain a computational complexity that is cubic in the dimension of the Markovian factor process $\boldsymbol{f}_t$. For an alternative approach in high-dimensions that imposes a dynamic factor structure after a projection of the observations onto a low-dimensional space, see \citet{jungbacker2015likelihood}, and \citet{brauning2014forecasting} for a collapsed mixed-frequency DFM.

\paragraph{Model selection.}
The model parameters $\boldsymbol{\theta}$ are learned to jointly maximize the log-likelihood of the observed data for all frequencies. 
This is in contrast to the parameter estimation approach for MIDAS, which minimizes the MSE of low-frequency predictions conditional on observing the high-frequency series.
We remark that a different log-likelihood weighting for the different frequencies in DFMs has been suggested in \cite{Blasques2016}, but requires cross-validation to optimize such weightings. Nevertheless, the introduced DFM contains several hyperparameters that need to be chosen, such as the latent state space dimension $k$ or the order $p$ of the latent Markov process. One possibility is to select such hyperparameters by evaluating the low-frequency predictions
on a validation set. Approaches for choosing the dimensions of the latent factor process have been under-explored in the mixed-frequency setting, but see \citet{bai2007determining, hallin2007determining} for possible criteria in general dynamic factor models. In our implementation, we choose $p=1$, as this allows for a differentiable model parametrization with stationary factor dynamics. We set $k=5$ for the small dataset and $k=10$ for the medium dataset.

\paragraph{Parameter estimation and forecasting.}

Based on the results from the Kalman filtering recursions, the model parameters $\boldsymbol{\theta}$ are learned by maximizing the marginal log-likelihood using
$\ell_t(\boldsymbol{\theta})=-\log p_{\boldsymbol{\theta}}(\boldsymbol{y}_{0:t})=-\sum_{s=0}^t \log p_{\boldsymbol{\theta}}(\boldsymbol{y}_s|\boldsymbol{y}_{0:s-1})$
where $p_{\boldsymbol{\theta}}(\boldsymbol{y}_s|\boldsymbol{y}_{0:s-1})$ is Gaussian with mean $\Lambda_{s,\boldsymbol{\theta}} \boldsymbol{\widehat{f}}_{s|s-1,\boldsymbol{\theta}}$ and covariance $\Gamma_{s,\boldsymbol{\theta}}$. Gradients of $\ell_t(\boldsymbol{\theta})$ can be computed using algorithmic differentiation.

For fixed $\boldsymbol{\theta} \in \Theta$ and $h \in \nset$, let 
\[ \mu_{t+h|t,\boldsymbol{\theta}}(\boldsymbol{y}_{t+h}|\boldsymbol{y}_{0:t})=\int g_{t+h,\boldsymbol{\theta}}(\boldsymbol{y}_{t+h}|\boldsymbol{f}_{t+h}) \prod_{\ell=1}^h h_{t+\ell, \boldsymbol{\theta}}(\boldsymbol{f}_{t+\ell}|\boldsymbol{f}_{t+\ell-1}) \rmd \boldsymbol{f}_{t+\ell} \pi_{t|t,\boldsymbol{\theta}}(\boldsymbol{f}_t|\boldsymbol{y}_{0:t}) \rmd \boldsymbol{f}_t
\]
be the $h$-step predictive distribution of the data, while $\pi_{t|t,\boldsymbol{\theta}}(\boldsymbol{f}_t|\boldsymbol{y}_{0:t})$ is the filtering distribution of the latent state $\boldsymbol{f}_t|\boldsymbol{y}_{0:t}$. The mean of $\mu_{t+h|t,\boldsymbol{\theta}}(\cdot |\boldsymbol{y}_{0:t})$ is $\boldsymbol{\widehat{y}}_{t + h|t,\boldsymbol{\theta}} = \mathbb{E}_{\boldsymbol{\theta}}\left[ \boldsymbol{y}_{t + h} \vert \boldsymbol{y}_{0:t}\right]$. 
For some $t,\tau \geq 0$, let us write $\widehat{\boldsymbol{f}}_{t+\tau|t,\boldsymbol{\theta}}=\E_{\boldsymbol{\theta}}[\boldsymbol{f}_{t+\tau}|\boldsymbol{y}_{0:t}]$ and $\Sigma_{t+\tau|t,\boldsymbol{\theta}}=\text{Cov}_{\boldsymbol{\theta}}[\boldsymbol{f}_{t+\tau}-\boldsymbol{\widehat{f}}_{t+\tau|t,\boldsymbol{\theta}}|\boldsymbol{y}_{0:t}]$ for the mean and covariance of the latent process, respectively.
For linear Gaussian dynamics, Kalman filtering allows for computing the filtered mean $\boldsymbol{\widehat{f}}_{t|t,\boldsymbol{\theta}}$ and covariance matrices $\widehat{\Sigma}_{t|t,\boldsymbol{\theta}}$ analytically.

For fixed $\boldsymbol{\theta}$, the $\tau$-step ahead prediction function $H_{t,\boldsymbol{\theta}}^{\tau}(\boldsymbol{y}_{0:t})=\boldsymbol{\widehat{y}}_{t+\tau|t,\boldsymbol{\theta}}=\Lambda_{t+\tau,\boldsymbol{\theta}}\boldsymbol{\widehat{f}}_{t+\tau|t,\boldsymbol{\theta}}$ is linear due to the Kalman filtering recursion. 
For $s \leq t$, consider also the prediction $H_{s,t}^{\star \tau}(\boldsymbol{y}_{0:t})=\E_{\boldsymbol{\theta}^{\star}(\boldsymbol{y}_{0:s})}[\boldsymbol{y}_{t+\tau}|\boldsymbol{y}_{0:t}]$ that is based on the sample $\boldsymbol{y}_{0:t}$, but where $\boldsymbol{\theta}^{\star}(\boldsymbol{y}_{0:s})=\argmin_{\boldsymbol{\theta}} \ell_s(\boldsymbol{\theta})$ maximizes the marginal likelihood of data $\boldsymbol{y}_{0:s}$ only. This setting allows to implement different parameter estimation setups from Section \ref{sec:estimation_setup}. For instance, the fixed parameter setup corresponds to fixing $s$, which yields a fixed training set $\boldsymbol{y}_{0:s}$ to estimate $\boldsymbol{\theta}$. In the expanding window setup, both $s$ and $t$ are expanded, while a rolling window setting updates the dataset $\boldsymbol{y}_{0:s}$ by rolling over the data.

\section{High-Frequency Forecasts}

To better understand how the use of high-frequency data impacts forecasting, as an additional empirical experiment we investigate high-frequency (HF) forecasts of all models in the Small-MD dataset. We restrict our analysis to this dataset because the computational burden to construct HF forecasts can be high: when using daily data and using our suggested 24 days-per-month interpolation, one quarter amounts to 72 daily frequency observations, which means HF forecasts can involve thousands of data points, and for DFM and M-MFESN models this setup can be quite computationally onerous. 

Constructing HF forecasts with MIDAS is trivial once the aggregation weights have been estimated, even though a practical implementation requires care in constructing the appropriate lag matrices. Recall for Section \ref{subsec_midas} that the MIDAS equation with $L$ regressors $({z}^{(l)}_{t,\tempo{s}{\kappa_l}})_{t,s}$ with ${z}^{(l)}_{t,\tempo{s}{\kappa_l}} \in \rset$, $l\in[L]$ for all $t \in \mathbb{Z}$ and $s \in \{0, \ldots, \kappa_l-1\}$ can be written as 
\begin{equation*}
     y_t = \alpha_0 + \sum_{i=1}^p \alpha_i y_{t-i} + \sum_{l=1}^L  \beta_l \sum_{k=0}^{K_l} \varphi(\boldsymbol{\theta}_l, k) z^{(l)}_{t,\tempo{-k}{\kappa_l}}  + \epsilon_t.
\end{equation*}
For clarity, we suppress the dynamic autoregressive component, as it has the same frequency as the target. Now assume that we include $n_{(\textnormal{\texttt{m}})}$ monthly and $n_{(\textnormal{\texttt{d}})}$ daily frequency regressors in the model that are sampled $\kappa_{(\textnormal{\texttt{m}})} = 3$ and $\kappa_{(\textnormal{\texttt{d}})}=72$ times per quarter and hence $\kappa_{\max}=72$. Therefore we can partition the regression above in the following way
\begin{equation*}
     y_t = 
      \alpha_0 + \sum_{i=1}^p \alpha_i y_{t-i} +  \sum_{l=1}^{n_{(\textnormal{\texttt{m}})}}  \beta_l \sum_{k=0}^{K_l} \varphi(\boldsymbol{\theta}_l, k) z^{(l)}_{t,\tempo{-k}{3}} + 
     \sum_{l=n_{(\textnormal{\texttt{m}})}+1}^{L}  \beta_l \sum_{k=0}^{K_l} \varphi(\boldsymbol{\theta}_l, k) z^{(l)}_{t,\tempo{-k}{72}} + 
     \epsilon_t
\end{equation*}
with $L=n_{(\textnormal{\texttt{m}})} + n_{(\textnormal{\texttt{d}})}$.

Assuming parameter estimates $\widehat{\alpha}_0, \widehat{\alpha}_1, \ldots, \widehat{\alpha}_p$ and $\{(\widehat{\beta}_l, \widehat{\boldsymbol{\theta}}_l)\}_{l=1}^L$ are available, the HF forecast $\widehat{y}_{t+1,\tempo{s}{72}}$ is given by
\begin{equation*}
     \widehat{y}_{t+1,\tempo{s}{72}} = \widehat{\alpha}_0 + \sum_{i=1}^p \widehat{\alpha}_i y_{t-i} + 
     \sum_{l=1}^{{n_{(\textnormal{\texttt{m}})}}}  \widehat{\beta}_l \sum_{k=0}^{K_l} \varphi(\widehat{\boldsymbol{\theta}}_l, k) z^{(l)}_{t,\tempo{\lfloor s/24\rfloor-k}{3}} + 
     \sum_{l=n_{(\textnormal{\texttt{m}})}+1}^{L}  \widehat{\beta}_l \sum_{k=0}^{K_l} \varphi(\widehat{\boldsymbol{\theta}}_l, k) z^{(l)}_{t,\tempo{s-k}{72}}.
\end{equation*}

For DFMs, high-frequency forecasts can be constructed using \eqref{eq:factor_dynamics} and \eqref{eq:dfm_forecasts} to iterate factors forward in time and then aggregate them according to estimated loadings or a weighting scheme.

Multi-frequency ESN models are also able to yield high-frequency forecasts in a straightforward manner. For simplicity, let us consider the case as in Example \ref{ex:s-mfesn} of an aligned S-MFESN model that has been fit to a quarterly target with monthly and daily input data. The reservoir is run in high-frequency, $\kappa_{\max}$ steps per quarter, according to state equation 
\begin{equation*}
    \boldsymbol{x}^{(\textnormal{\texttt{m,d}})}_{t,\tempo{s}{72}} = \alpha \boldsymbol{x}^{(\textnormal{\texttt{m,d}})}_{t,\tempo{s-1}{72}} + (1-\alpha) \sigma(A \boldsymbol{x}^{(\textnormal{\texttt{m,d}})}_{t,\tempo{s-1}{72}} + C \boldsymbol{z}^{(\textnormal{\texttt{m,d}})}_{t,\tempo{s}{72}} + \boldsymbol{\zeta}).
\end{equation*}
Suppose a coefficient matrix $\widehat{W}$ has been estimated. Then, as states between low-frequency periods $t$ and $t+1$ are collected, we can immediately construct the high-frequency one-step-ahead forecasts
\begin{equation*}
    \widetilde{y}_{t+1,\tempo{s}{72}} = \widehat{W}^\top \boldsymbol{x}^{(\textnormal{\texttt{m,d}})}_{t,\tempo{s}{72}}.
\end{equation*}

For M-MFESN models HF forecasts require slightly more care. For example, when dealing with the multi-reservoir MFESN model of Example \ref{ex:m-mfesn}, we must repeat the most recent monthly state at daily frequency correctly.

\section{Robustness Analysis}

\subsection{MIDAS}
\label{sec:midas_robustness}

As we discuss briefly in the main text, parameter optimization is a principal problem in implementing any MIDAS model. Even though explicit formulas exist for both gradient and Hessian of the MIDAS loss objective when an Almon weighting scheme is used (see \cite{Kostrov2021essays}), there is no known guarantee that the loss itself is convex or even locally convex. In practice, for a given starting point (or point set) a numerical optimizer might only converge to a local minimum. 

\begin{figure}[p]
    \centering
    \caption{MIDAS Robustness Plots -- 2007 Sample -- Small-MD Dataset}
    \label{fig:midas_robustness_smallMD}
    \subfloat[MIDAS Loss t-SNE Embedding: Gradient Norm]{%
      \includegraphics[width=0.85\textwidth]{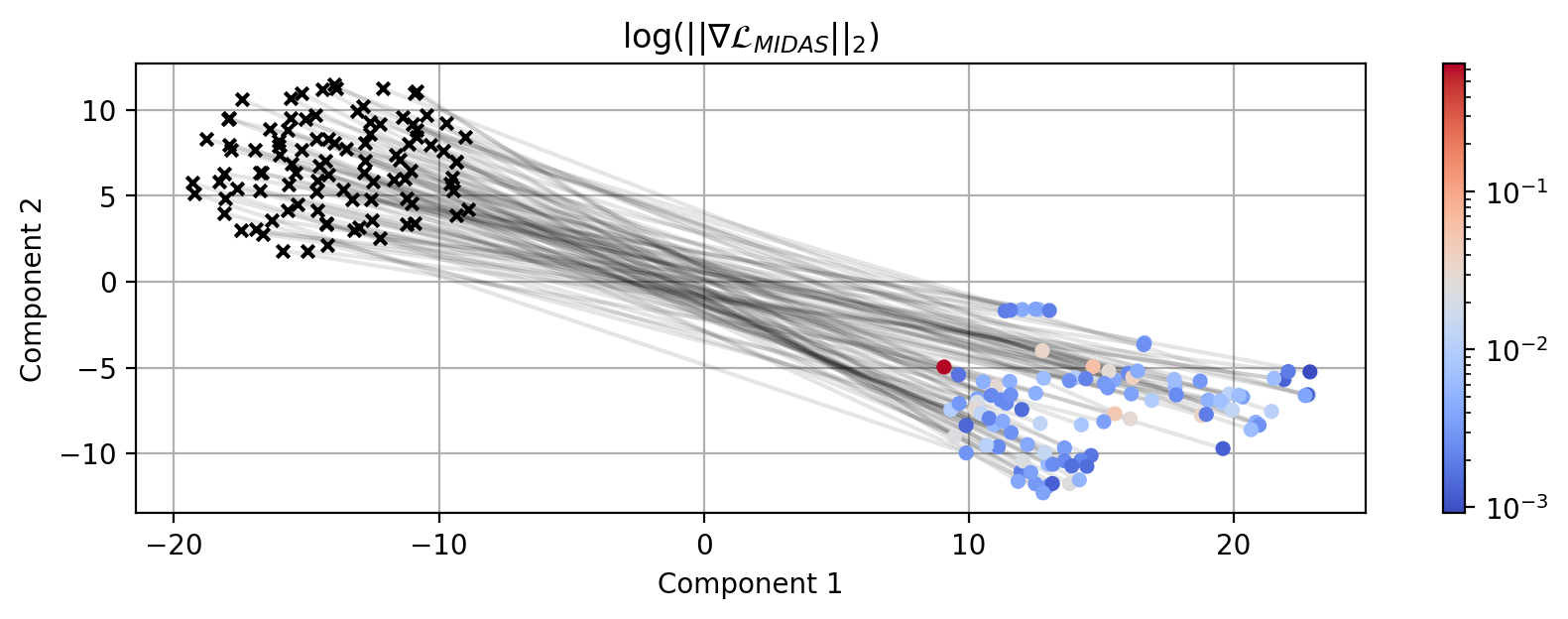}
    }
     \vspace{1em}
    \\
    \subfloat[MIDAS Loss t-SNE Embedding: Loss Norm]{%
      \includegraphics[width=0.85\textwidth]{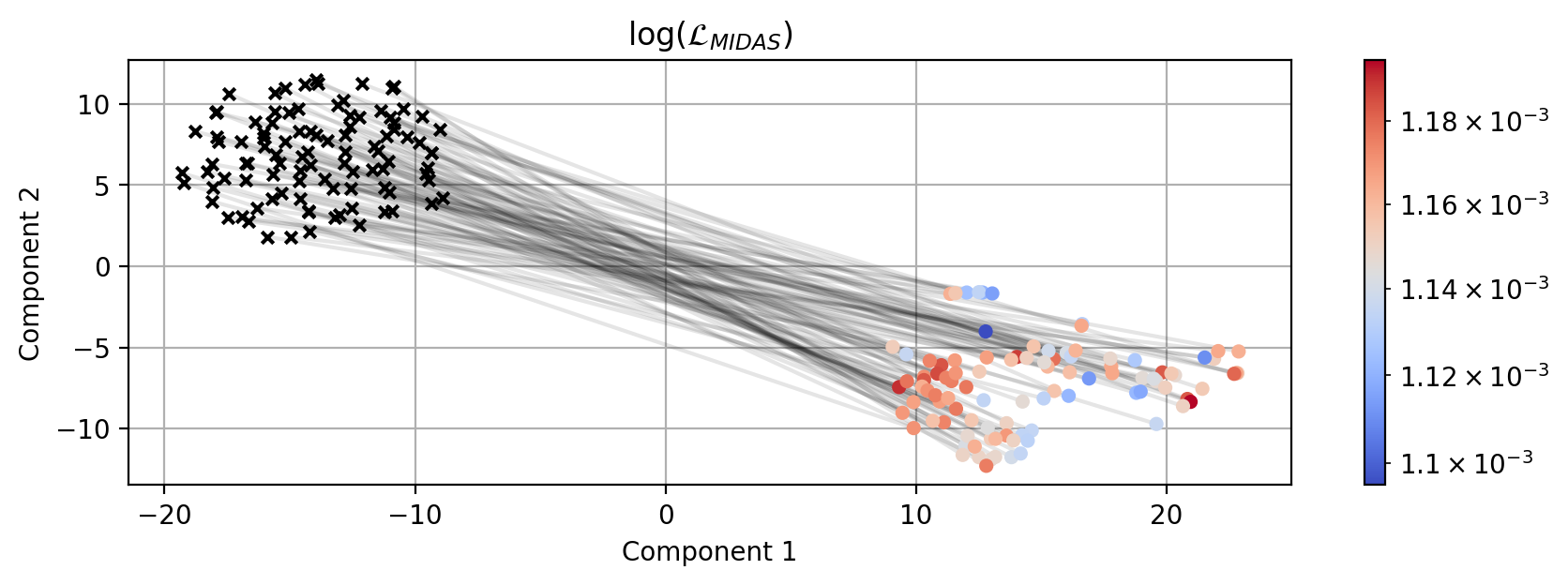}
    }
 \end{figure}

We observe this in practice, and we explore its effects on the robustness of MIDAS forecasts. We report summary results for our simulations in Figure \ref{fig:midas_robustness_smallMD}. Our proposal is, given a MIDAS model specification and a set of starting points for evaluating the loss, to run an optimizer (for example, L-BFGS-B with explicit gradient) and select the smallest local minimum. By repeating this procedure multiple times, we collect a set of MIDAS parameters and study both the variation between the parameter vectors and the implied one-step ahead forecasts.

To be precise, our procedure is as follows:
\begin{enumerate}
    \item For a total of $B$ repetitions:
    \begin{enumerate}
        \item Choose $M$ initialization points for the optimizer. We draw 64 points inside the hypercube of edge length 0.025 using a low-discrepancy Sobol sequence. The choice of a down-scaled hypercube as a domain comes from the empirical fact that the Almon exponential scheme may produce extremely large values even for relatively small coefficients. A straightforward way to see this is to notice that given any arbitrary small value for $\theta_1$ and $\theta_2$ in \eqref{eq:almon_weights}, for lag index $k$ sufficiently large weight $\exp(\theta_1 k + \theta_2 k^2)$ will overflow at any given numerical precision. This means that one should adjust the MIDAS optimization domain based on the number of lags in the model. 
        \item For each initialization point, run the optimizer of choice.
        \item Among the resulting $M$ (local) loss minimizers, select and store the one with the lowest loss value.
    \end{enumerate}
    \item With the resulting $B$ stored minimizer:
    \begin{itemize}
        \item Construct a low-dimensional projection of the high-dimensional minima to see their relative location in the parameter space and to compare their gradient and loss values, see Figure \ref{fig:midas_robustness_smallMD} (a)-(b).
        \item Use each minimizer to produce MIDAS one-step ahead forecasts and plot quantile frequency plots of the forecast variation due to initialization; see Figure \ref{fig:midas_robustness_smallMD} (c).
    \end{itemize}
\end{enumerate}

Figure \ref{fig:midas_robustness_smallMD} shows that the best minimizers among initial Sobol sets are clustered together. To construct this 2D projection of the high-dimensional Almon coefficient space (including autoregressive lags and intercept), we use the well-known t-SNE procedure developed in \cite{maatenLearningParametricEmbedding2009}, which is an unsupervised dimensionality reduction algorithm capable of preserving the latent high-dimensional structure. This approach naturally implies that the Euclidean distances in the plot are suggestive of ``clustering" rather than the actual latent distance between points. In Figures \ref{fig:midas_robustness_smallMD}~(a)-(b), we see that the L-BFGS-B optimizer with explicit gradient achieves good convergence in terms of gradient norm and also that the resulting cluster of minimizers has close loss values. However, one can see that there is no single loss minimum: Figure \ref{fig:midas_robustness_smallMD} instead suggests that the local structure of the MIDAS loss function is very uneven, and therefore many distinct local minima can be achieved even when choosing a large number of initialization values for the optimizer. This means that the ``multi-start" strategy suggested in \cite{Kostrov2021essays} to alleviate issues in MIDAS model estimation is insufficient.

The effects of non-negligible variation in parameter values on forecasts appear to be significant. Looking at Figure~\ref{fig:midas_robustness_smallMD} (c), we can see wide frequency bands for the one-step ahead forecasts constructed using the Small-MD dataset and fixed parameter values. In particular, the Financial Crisis period seems to induce larger deviations in forecasts, consistent with the intuition that data with larger variation causes stronger model sensitivity when making forecasts.

\subsection{MFESN}
\label{sec:mfesn_robustness_resample}

Since ESN models, and thus MFESN models, require random sampling of parameter matrices, the size of which is often large, there is inherent variability in any ESN model forecast. In theory, because all MFESN state parameters $(\widetilde{A}, \widetilde{C}, \widetilde{\boldsymbol{\zeta}})$ are drawn independently of each other, one could try to decompose the variance of any MFESN into the share due to parameter sampling and the share due to data sampling. Unfortunately, in practice, such decomposition is hard to derive. Cross-validation of ridge penalties and rolling and expanding window estimation are non-trivial data-dependent operations that complicate inference. In this work, we limit ourselves to numerically evaluating the effect of reservoir coefficient sampling on MFESN forecast variance.

\begin{figure}[p]
    \centering
    \caption{ESN Robustness Plots -- 2007 Sample -- Small-MD Dataset}
    \label{fig:singleesn_a_robustness_smallMD}
    \subfloat[Fixed Parameters]{%
      \includegraphics[width=0.85\textwidth]{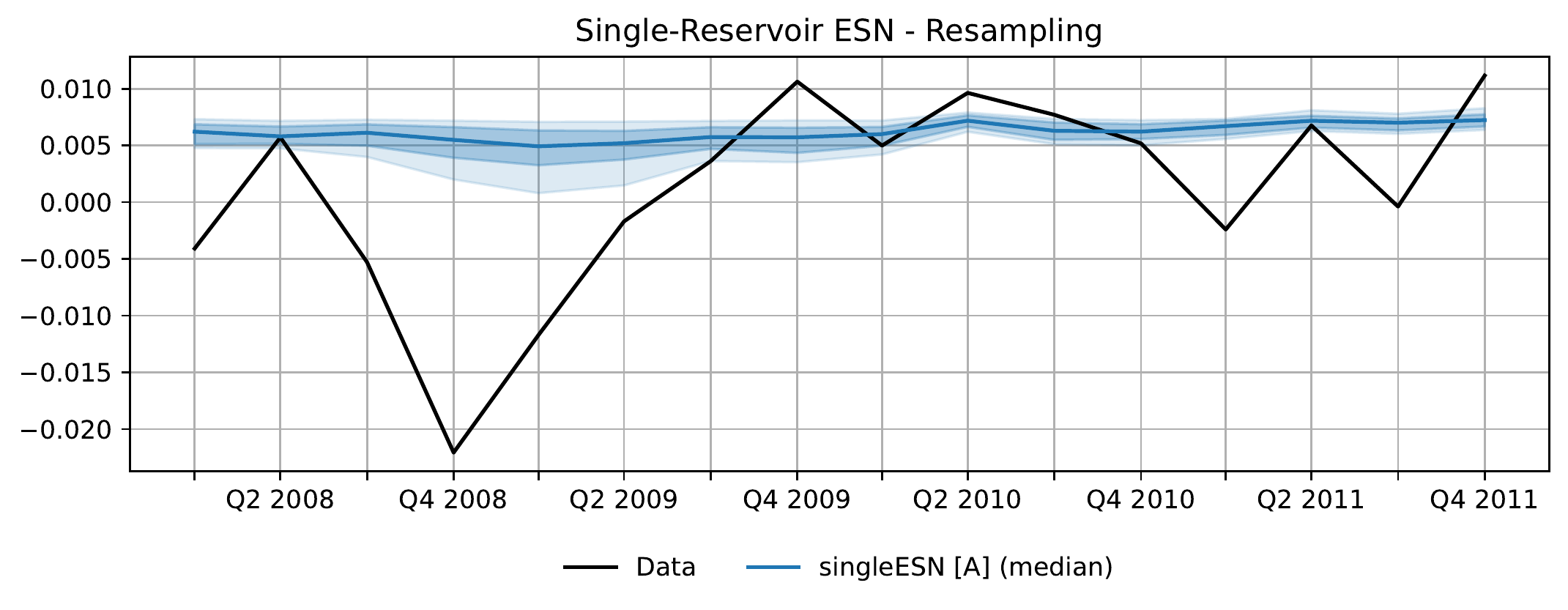}
    }
     \vspace{1em}
    \\
    \subfloat[Expanding Window]{%
      \includegraphics[width=0.85\textwidth]{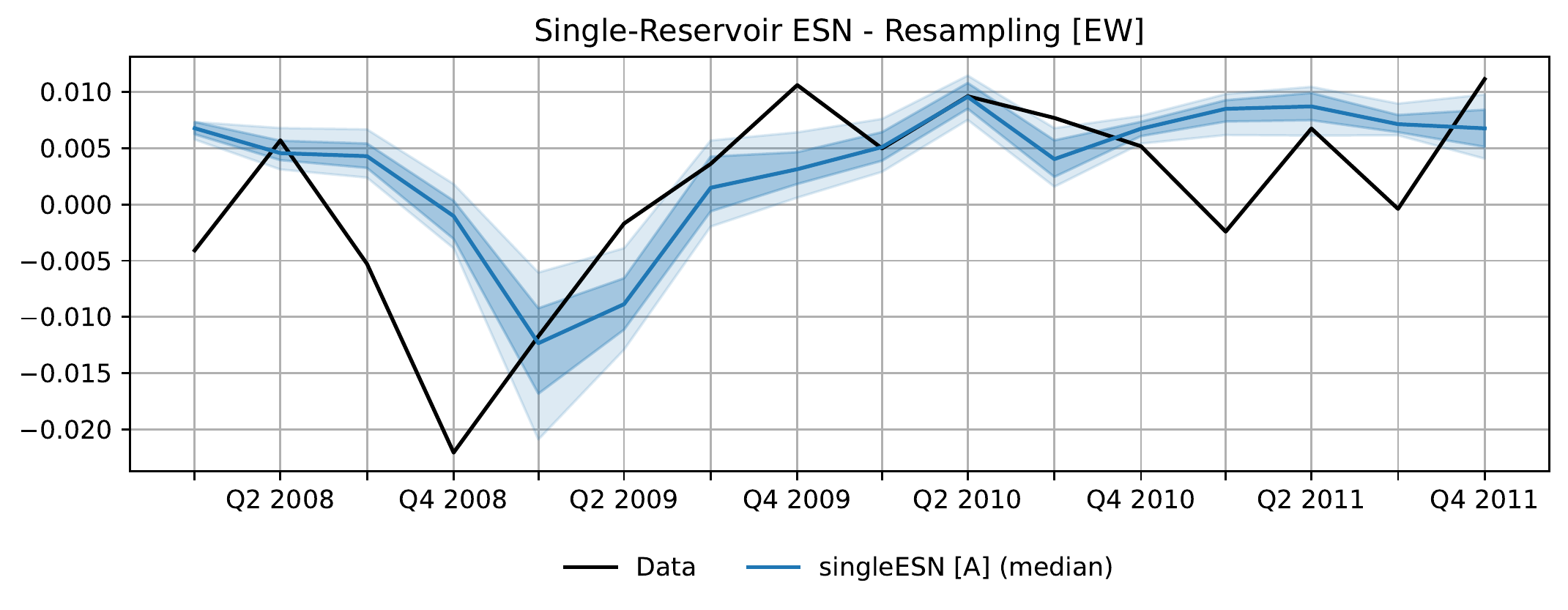}
    }
     \vspace{1em}
    \\
    \subfloat[Rolling Window]{%
      \includegraphics[width=0.85\textwidth]{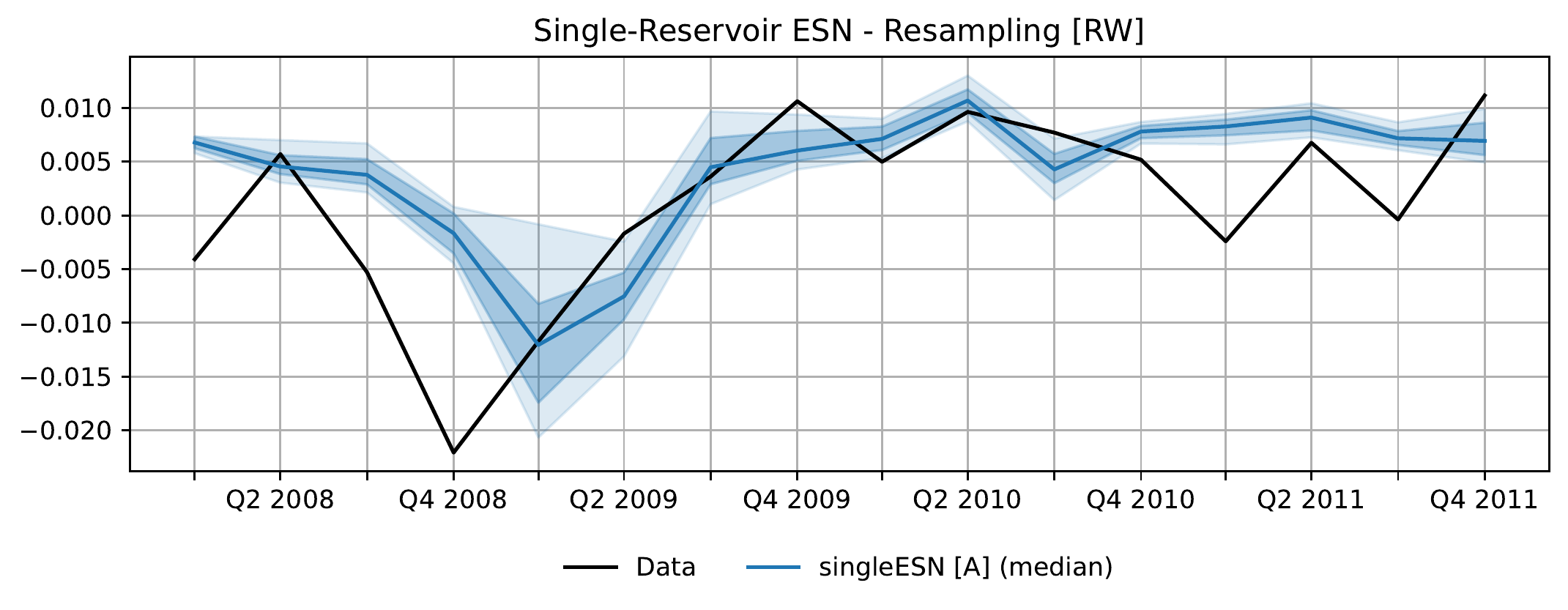}
    }
 \end{figure}

\begin{figure}[p]
    \centering
    \caption{ESN Robustness Plots -- 2007 Sample -- Small-MD Dataset}
    \label{fig:multiesn_a_robustness_smallMD}
    \subfloat[Fixed Parameters]{%
      \includegraphics[width=0.85\textwidth]{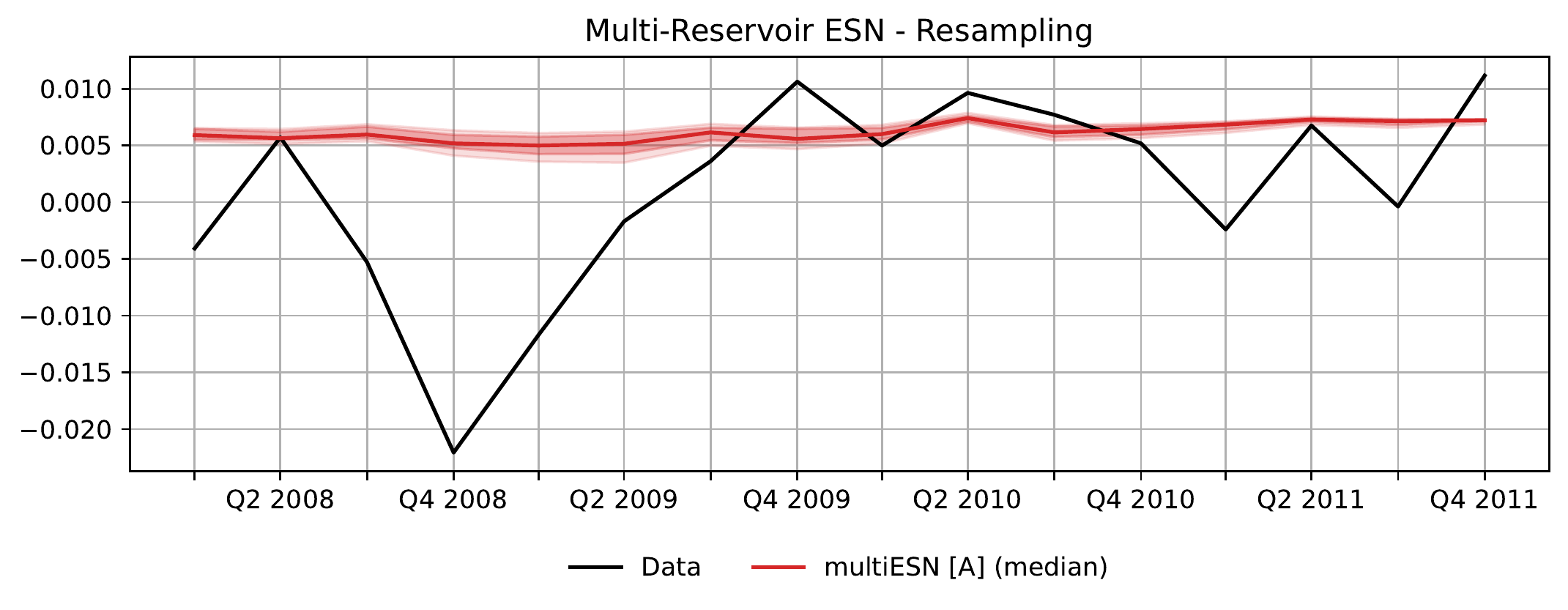}
    }
     \vspace{1em}
    \\
    \subfloat[Expanding Window]{%
      \includegraphics[width=0.85\textwidth]{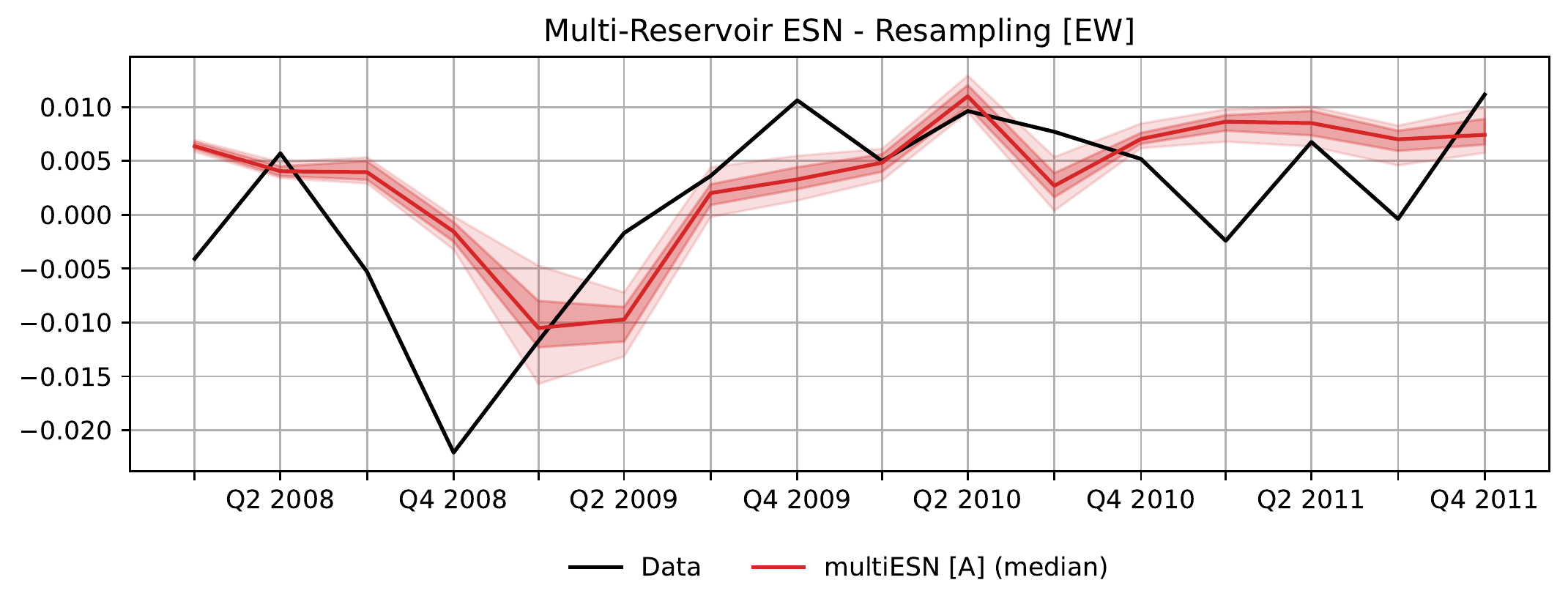}
    }
     \vspace{1em}
    \\
    \subfloat[Rolling Window]{%
      \includegraphics[width=0.85\textwidth]{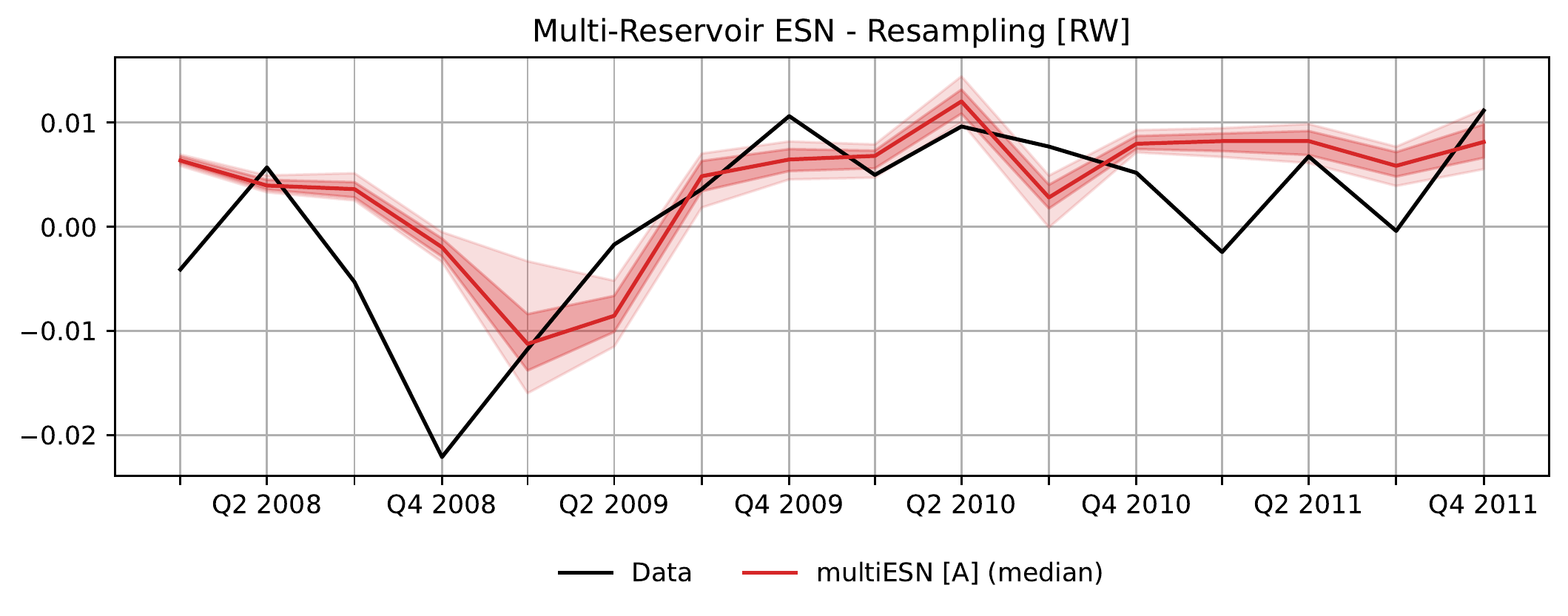}
    }
 \end{figure}

Our approach is straightforward: given an MFESN model specification, c.f. Table \ref{table:models_empirical}, and a forecasting setup (fixed parameters, expanding or rolling window), we resample the reservoir state matrix parameters, perform cross-validation and possibly train-test sample windowing, and finally construct pointwise forecasts. Once a sufficiently large set of resampling forecasts has been computed, we plot frequency intervals in Figures \ref{fig:singleesn_a_robustness_smallMD} and \ref{fig:multiesn_a_robustness_smallMD}. From Figure \ref{fig:singleesn_a_robustness_smallMD}, we can see that the single-reservoir MFESN model with reservoir size 30 produces forecasts with a meaningful amount of variability induced by parameter resampling. Forecasts exhibit more variation when using an expanding or rolling window estimation strategy, even though the overall forecasts align with the GDP realizations. A similar discussion to that of MIDAS applies here: forecast sensitivity increases with underlying data variation, exacerbated in periods of systemic economic crisis.

Figure \ref{fig:multiesn_a_robustness_smallMD} suggests that larger MFESN models produce significantly more stable forecasts regarding model resampling. Note that the M-MFESN model [A] has a monthly frequency reservoir that is approximately 3 times the size of the S-MFESN model [A]. This stability is preserved even in expanding or rolling window settings, even though a slightly higher variation is apparent at the height of the 2008 Financial Crisis. We hypothesize that this reduction in variance due to model parameter sampling is due to the concentration of measure phenomena that prevail in high-dimensional spaces. 
Figure \ref{fig:multiesn_a_robustness_smallMD} suggests that larger MFESN models produce significantly more stable forecasts regarding model resampling. Note that the M-MFESN model [A] has a monthly frequency reservoir that is approximately 3 times the size of the S-MFESN model [A]. This stability is preserved even in expanding or rolling window settings, even though a slightly higher variation is apparent at the height of the 2008 Financial Crisis. We hypothesize that this reduction in variance due to model parameter sampling is due to the concentration of measure phenomena that prevail in high-dimensional spaces.

\newpage

\section{Additional Figures}
\label{subsec:additional_figures}


\begin{figure}[h!]
	\centering
	\includegraphics[width=0.97\linewidth]{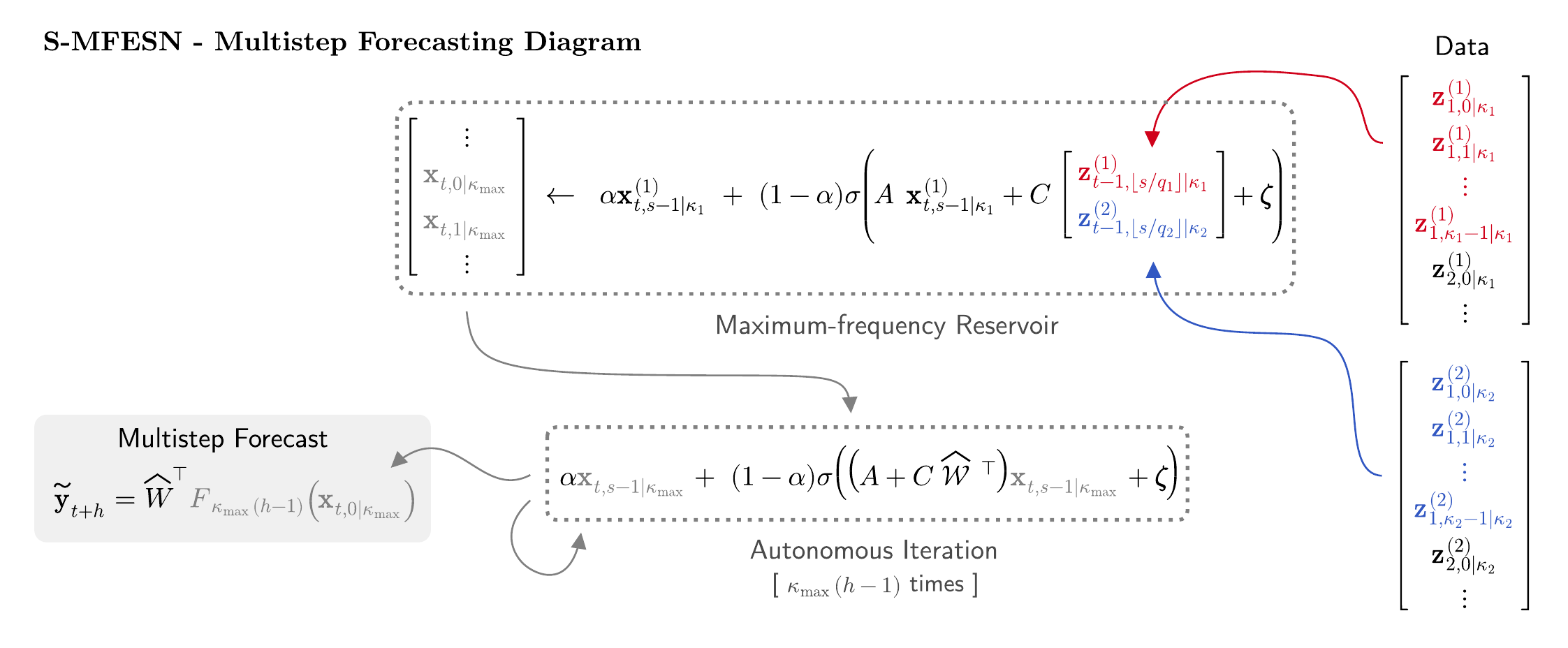}
	\caption{}
	\label{fig:single_reservoir_sfesn_multistep_diagram}
\end{figure}

\begin{figure}[h!]
	\centering
	\includegraphics[width=0.97\linewidth]{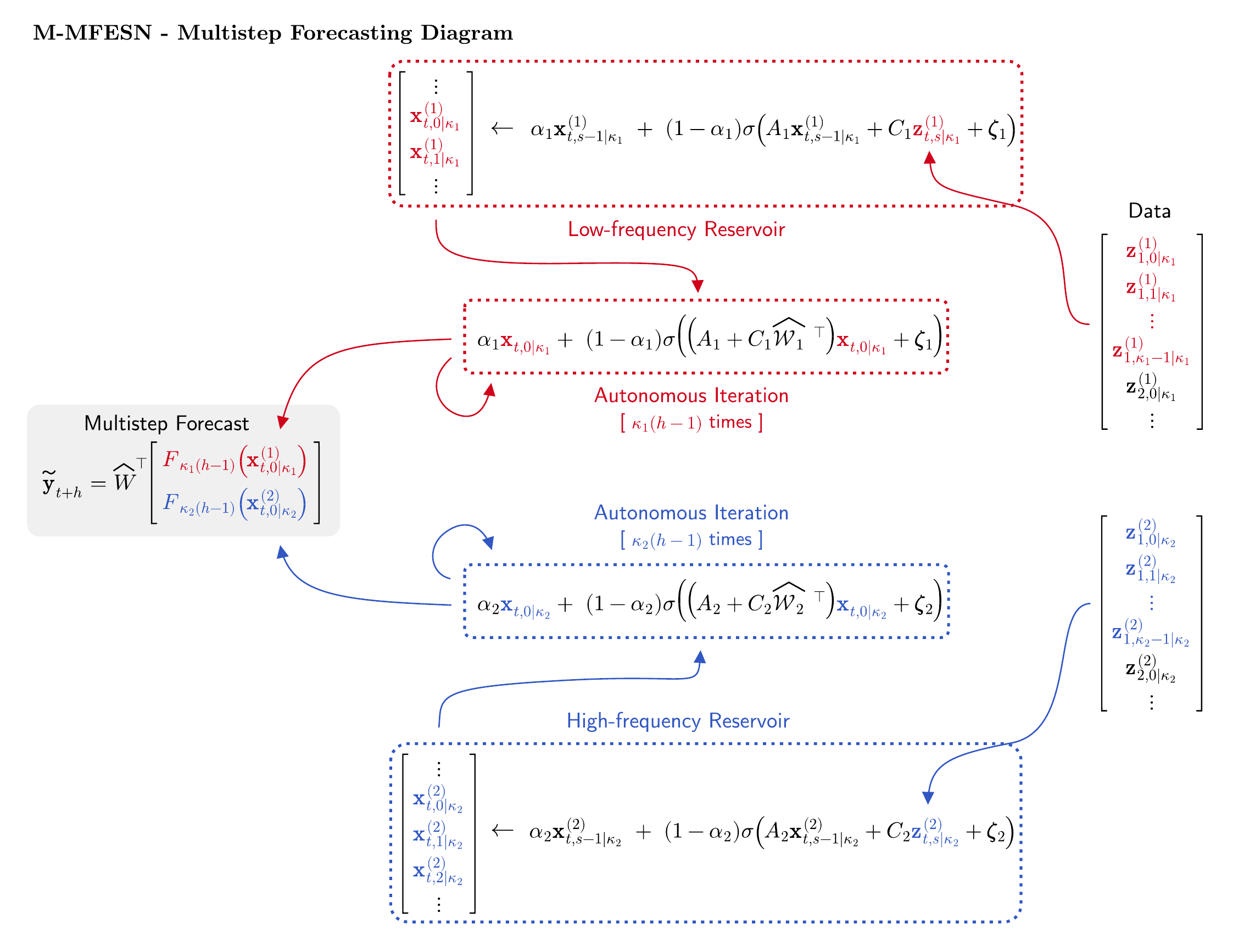}
	\caption{}
	\label{fig:multi_reservoir_mfesn_multistep_diagram}
\end{figure}

\begin{figure}[p]
	\caption{1-Step-ahead GDP Forecasting -- Modified Diebold-Mariano -- Small-MD Dataset}
	\label{fig:1sa_GDP_smallMD_diebold_mariano}
	\subfloat[Fixed 2007]{%
		\includegraphics[width=0.48\textwidth, trim=30 0 0 25, clip]{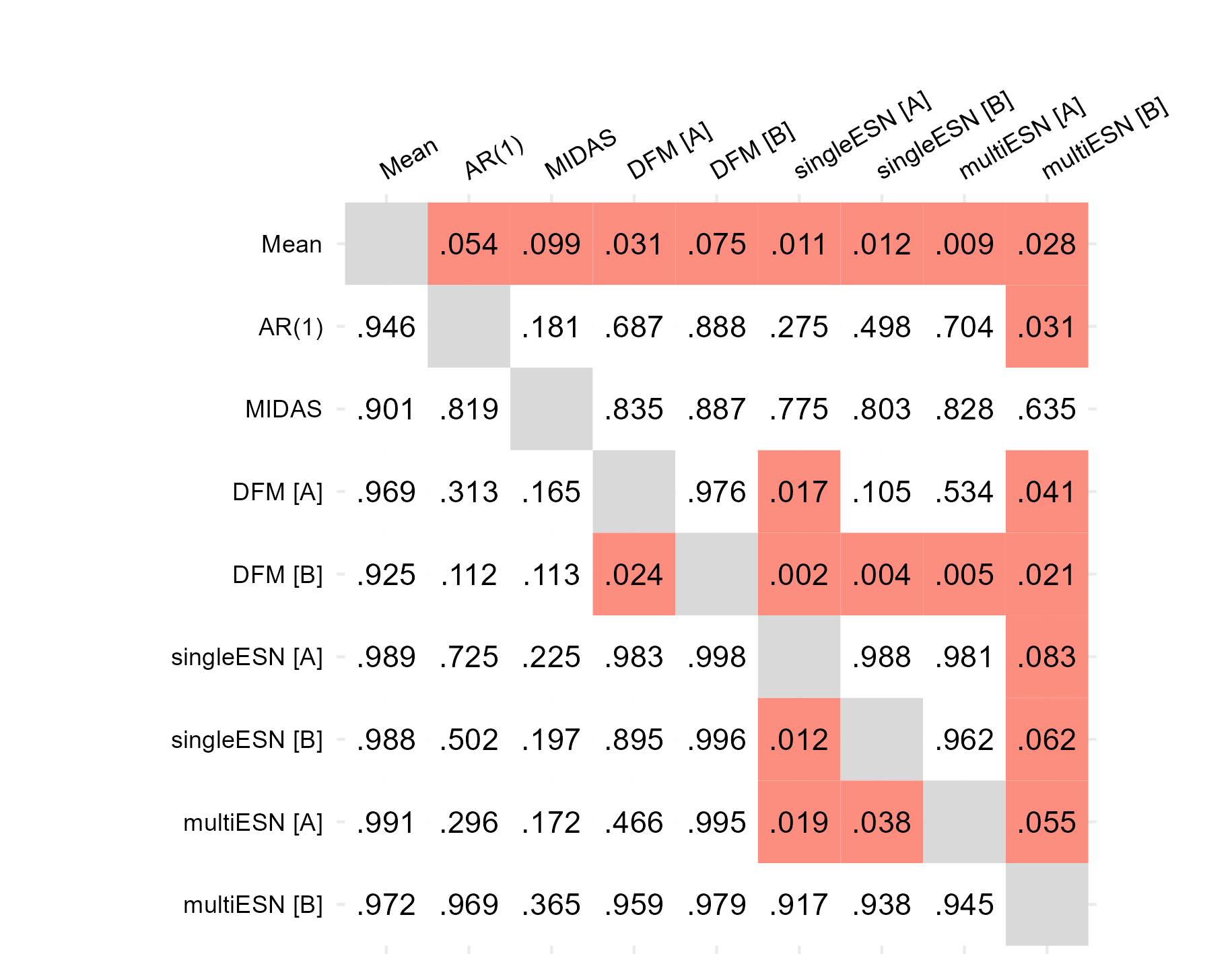}
	}
	\hfill
	\subfloat[Fixed 2011]{%
		\includegraphics[width=0.48\textwidth, trim=30 0 0 25, clip]{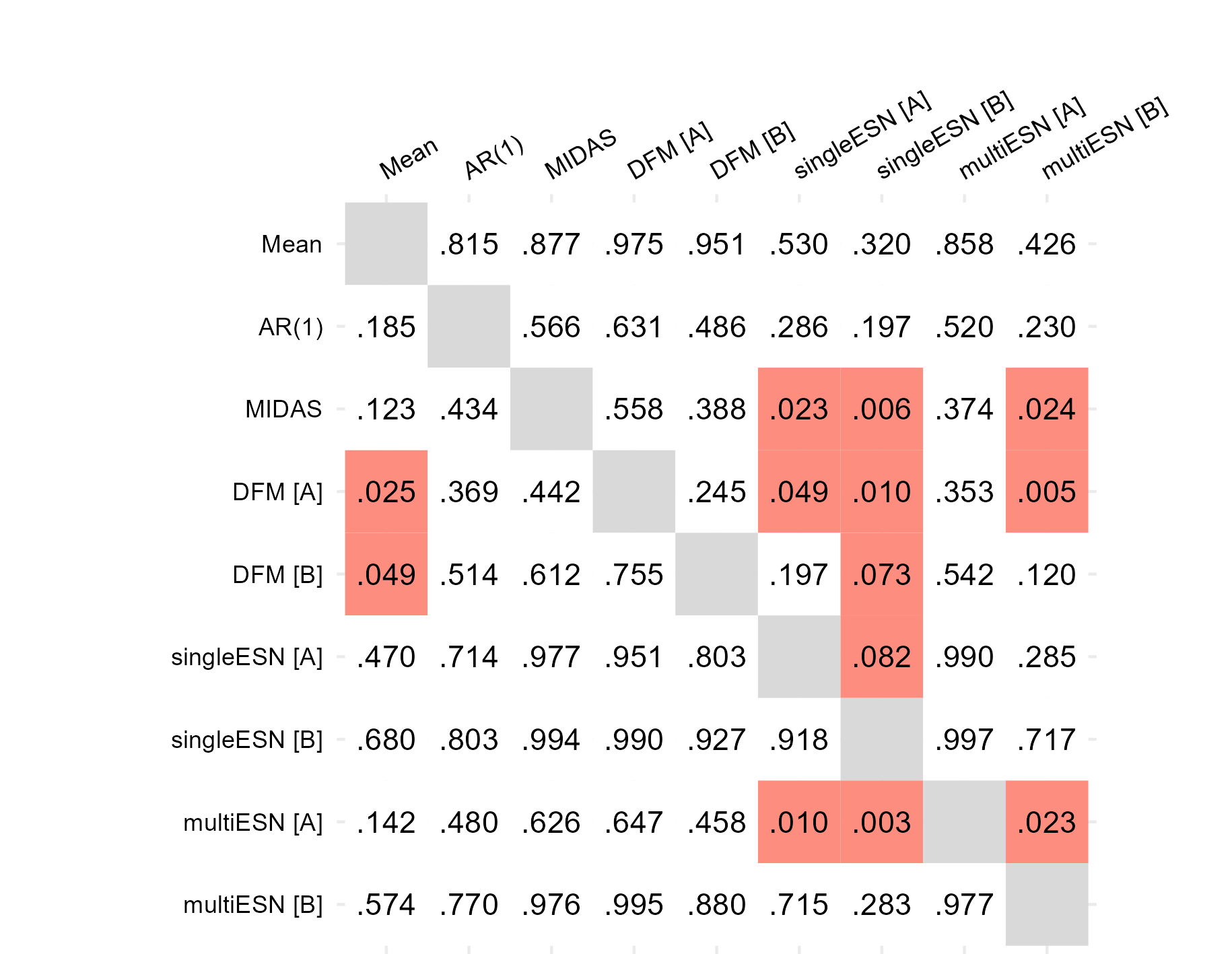}
	}
	\\
	\subfloat[Expanding 2007]{%
		\includegraphics[width=0.48\textwidth, trim=30 0 0 25, clip]{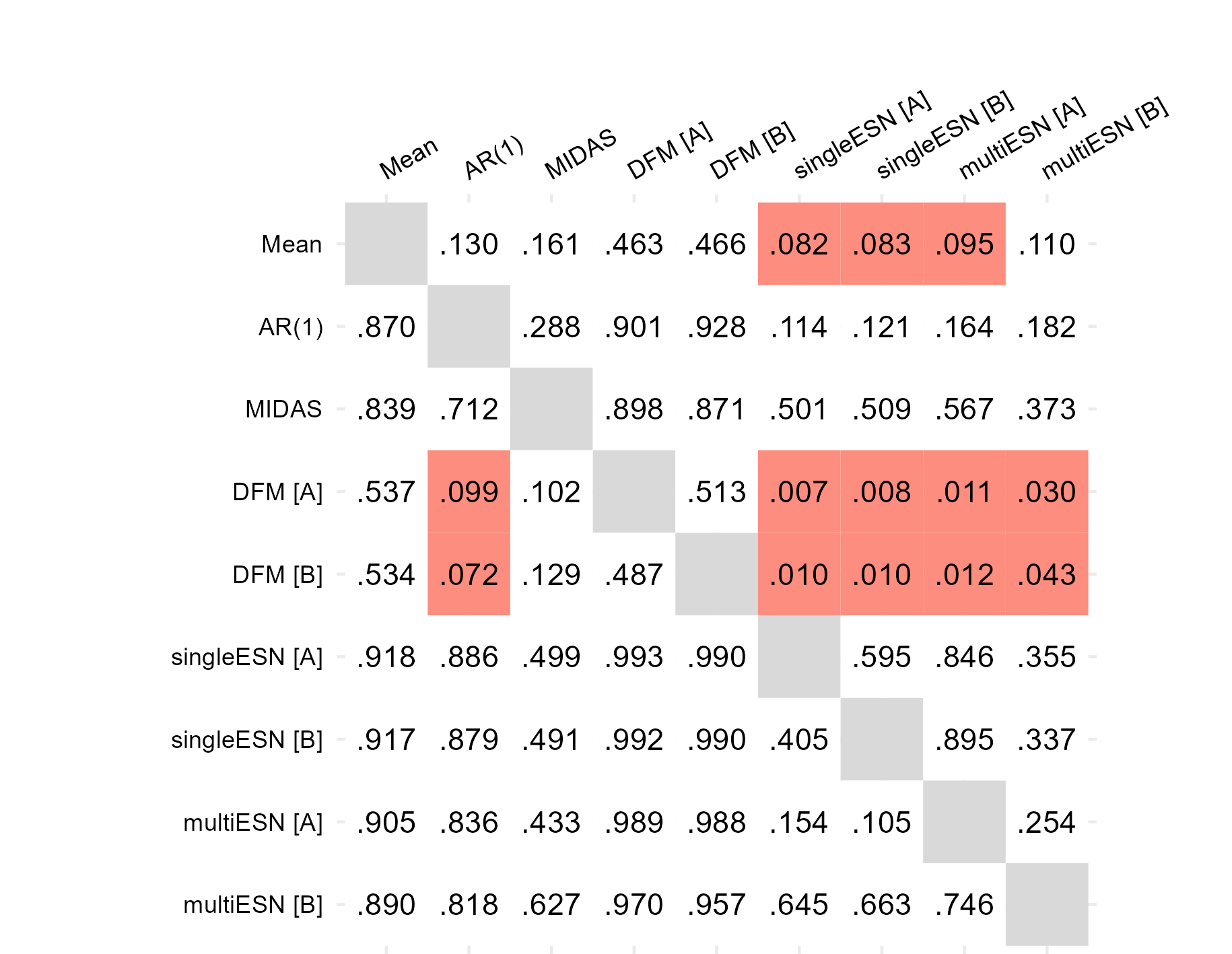}
	}
	\hfill
	\subfloat[Expanding 2011]{%
		\includegraphics[width=0.48\textwidth, trim=30 0 0 25, clip]{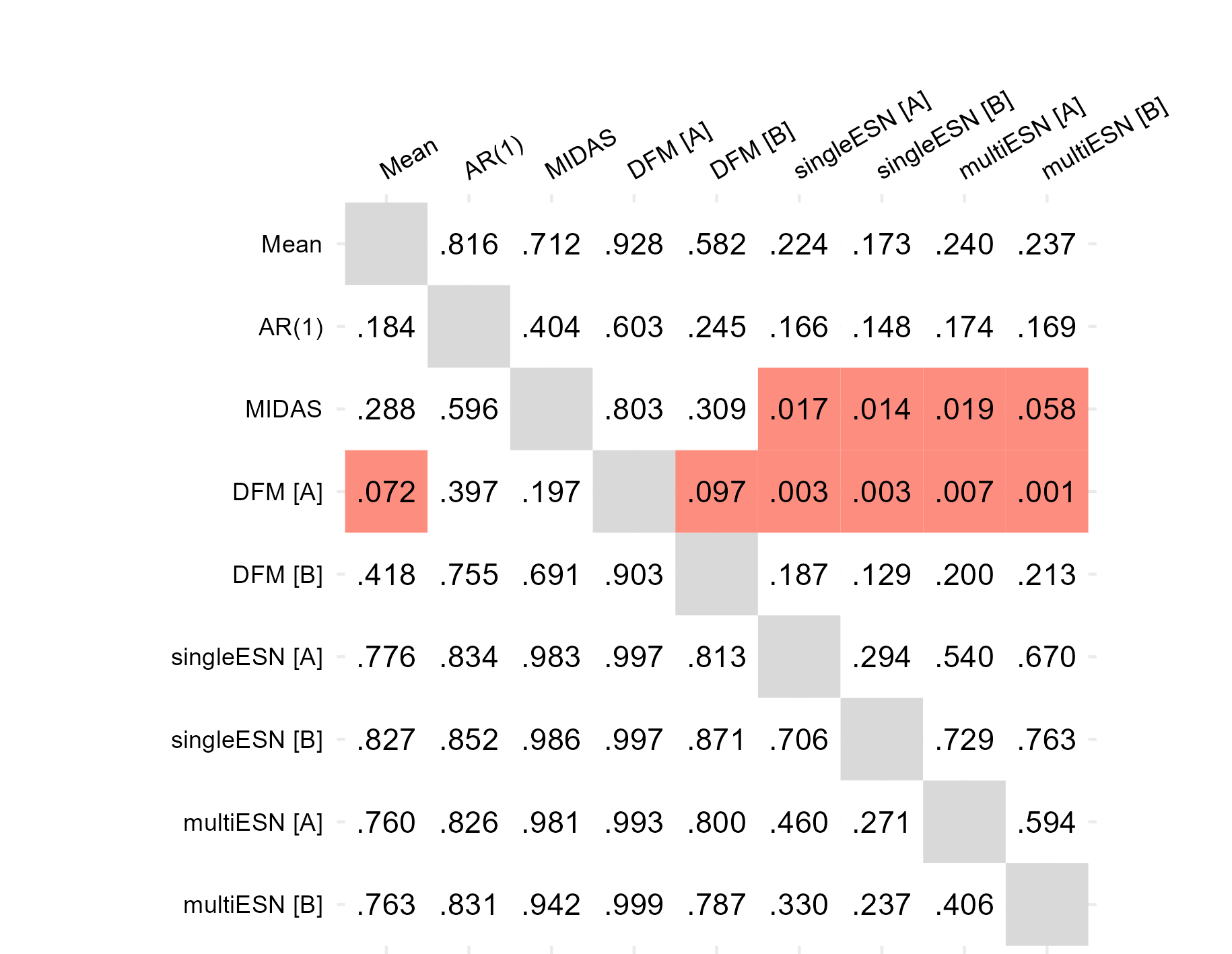}
	}
	\\
	\subfloat[Rolling 2007]{%
		\includegraphics[width=0.48\textwidth, trim=30 0 0 25, clip]{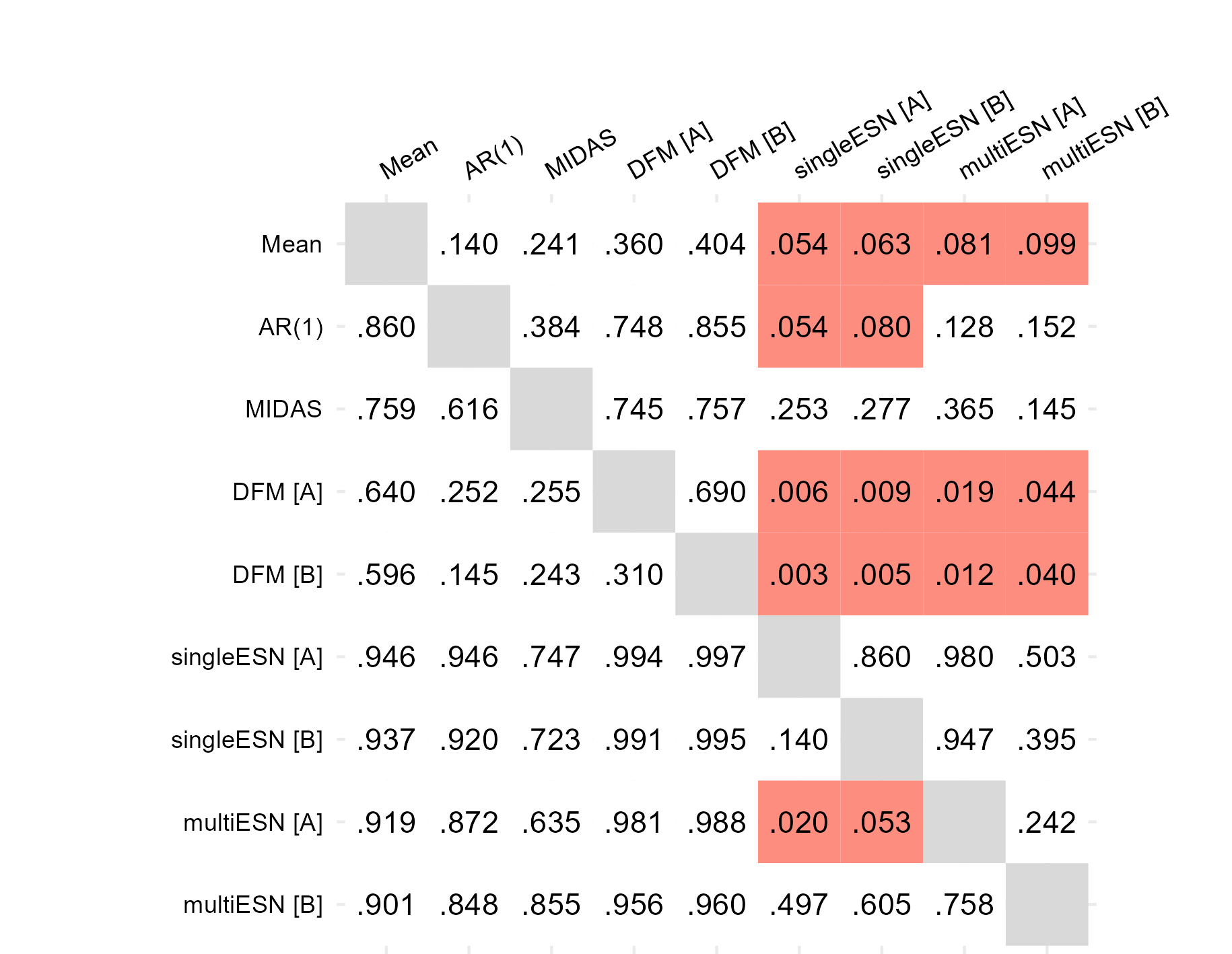}
	}
	\hfill
	\subfloat[Rolling 2011]{%
		\includegraphics[width=0.48\textwidth, trim=30 0 0 25, clip]{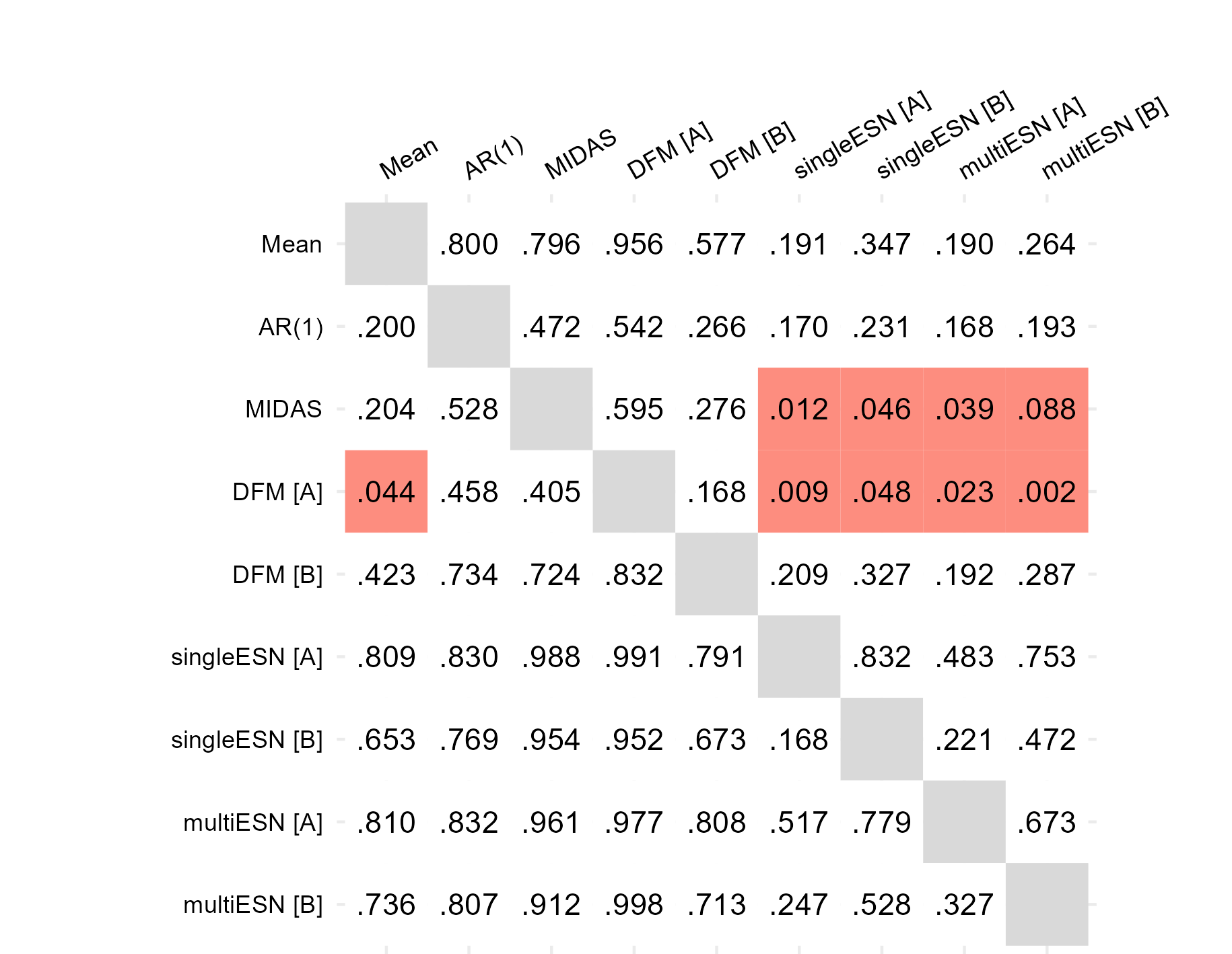}
	}
	\vspace{1em}
	\caption{p-values of the pairwise Modified Diebold-Mariano tests between models of Table \ref{table:1sa_GDP_smallMD}. Tests are one-sided and carried out row-wise: the null hypothesis for row $i$ and column $j$ reads as ``the $i$th-row model forecasts \textit{more accurately} than the $j$th-column model''. Rejections at the 10\% level are highlighted in red.}
\end{figure}

\begin{figure}[p]
	\caption{1-Step-ahead GDP Forecasting -- Modified Diebold-Mariano -- Medium-MD Dataset}
	\label{fig:1sa_GDP_mediumMD_diebold_mariano}
	\subfloat[Fixed 2007]{%
		\includegraphics[width=0.48\textwidth, trim=30 0 0 25, clip]{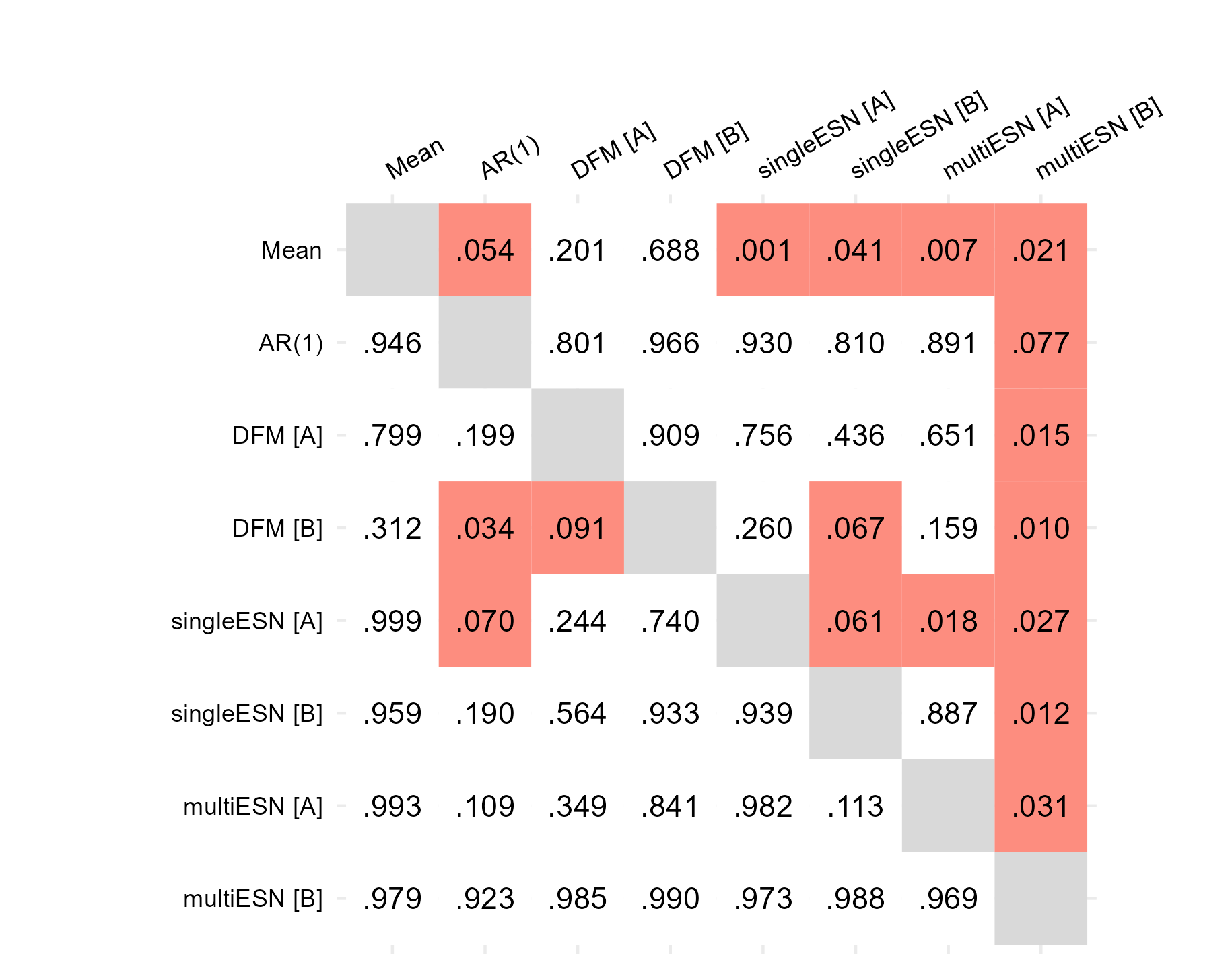}
	}
	\hfill
	\subfloat[Fixed 2011]{%
		\includegraphics[width=0.48\textwidth, trim=30 0 0 25, clip]{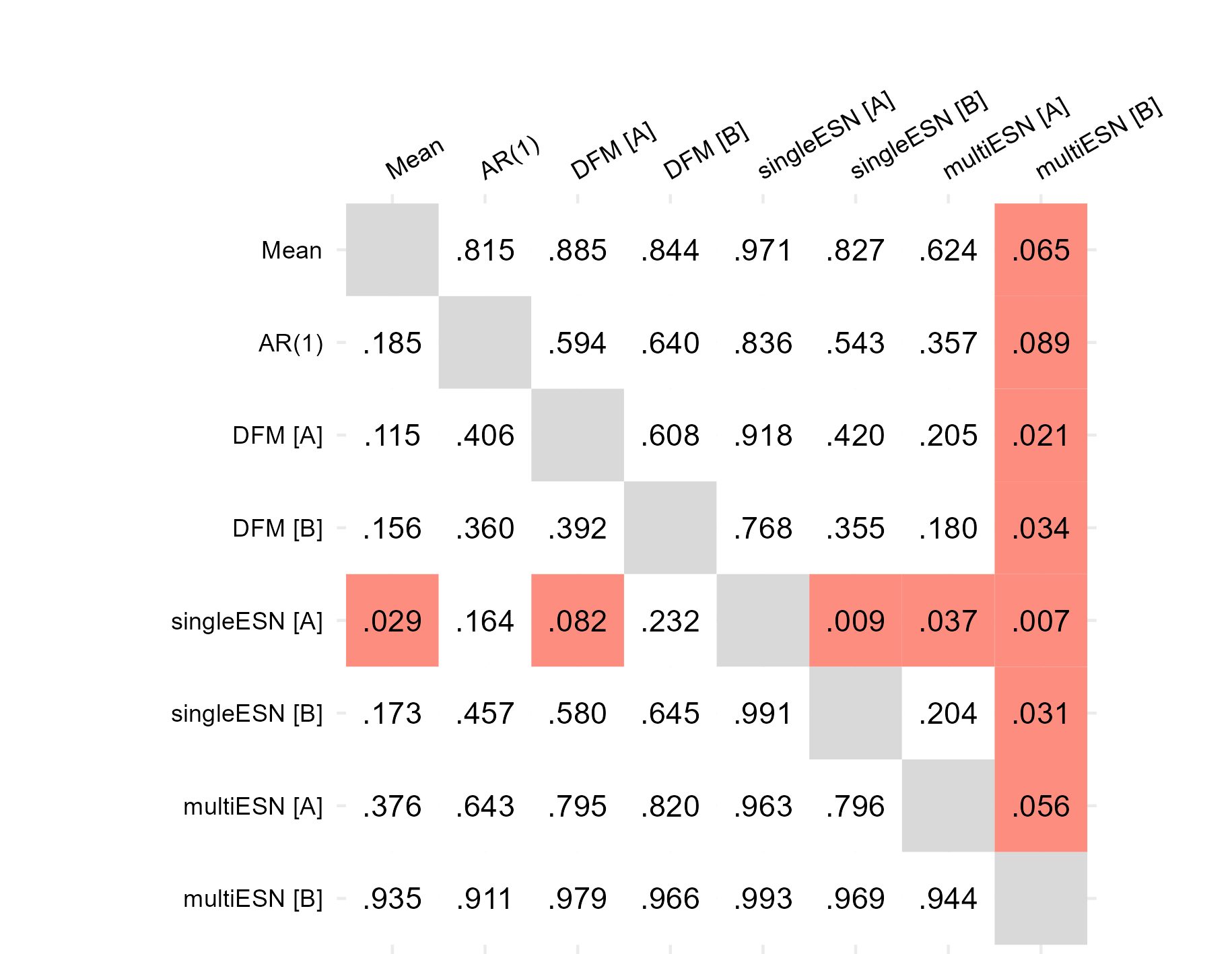}
	}
	\\
	\subfloat[Expanding 2007]{%
		\includegraphics[width=0.48\textwidth, trim=30 0 0 25, clip]{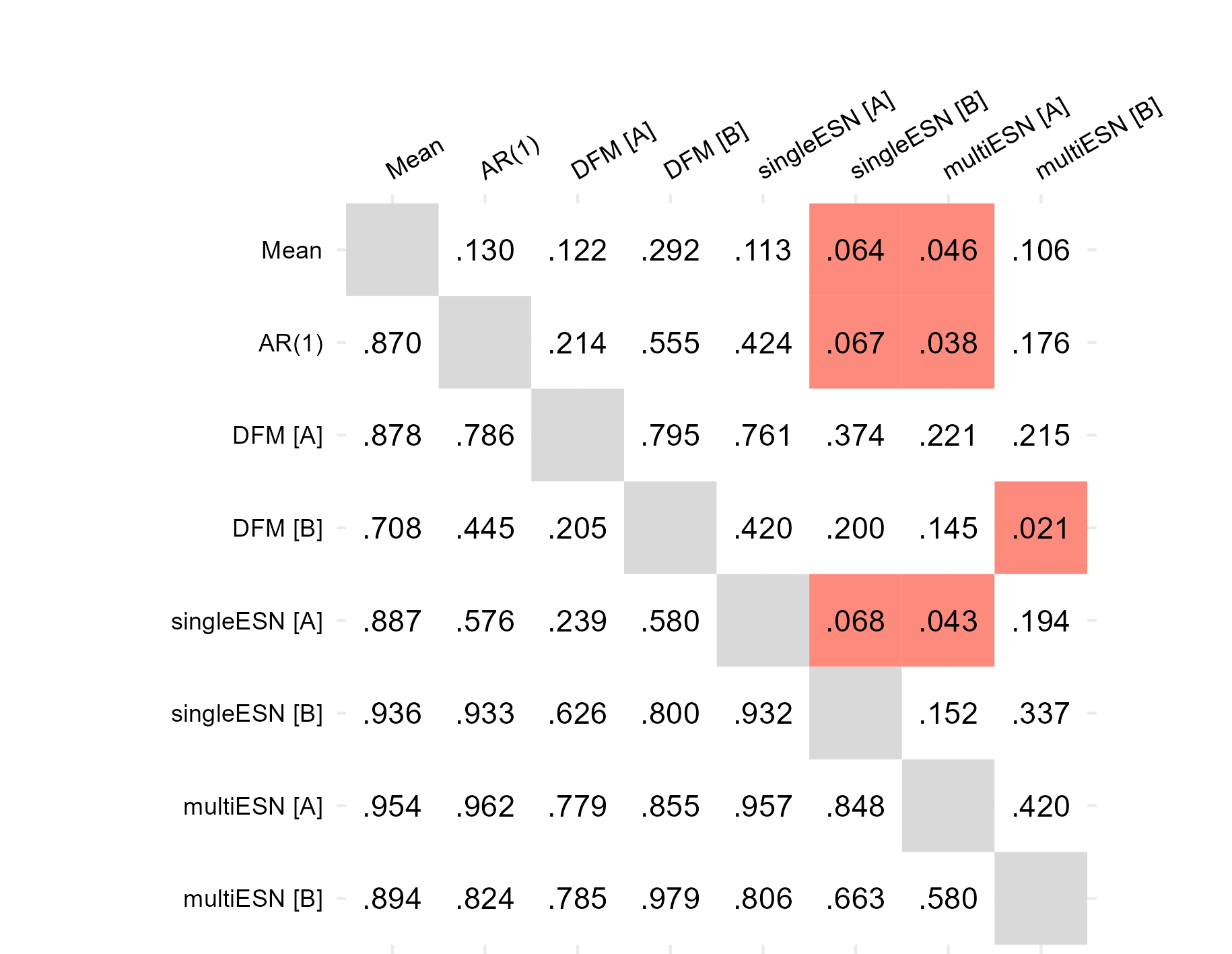}
	}
	\hfill
	\subfloat[Expanding 2011]{%
		\includegraphics[width=0.48\textwidth, trim=30 0 0 25, clip]{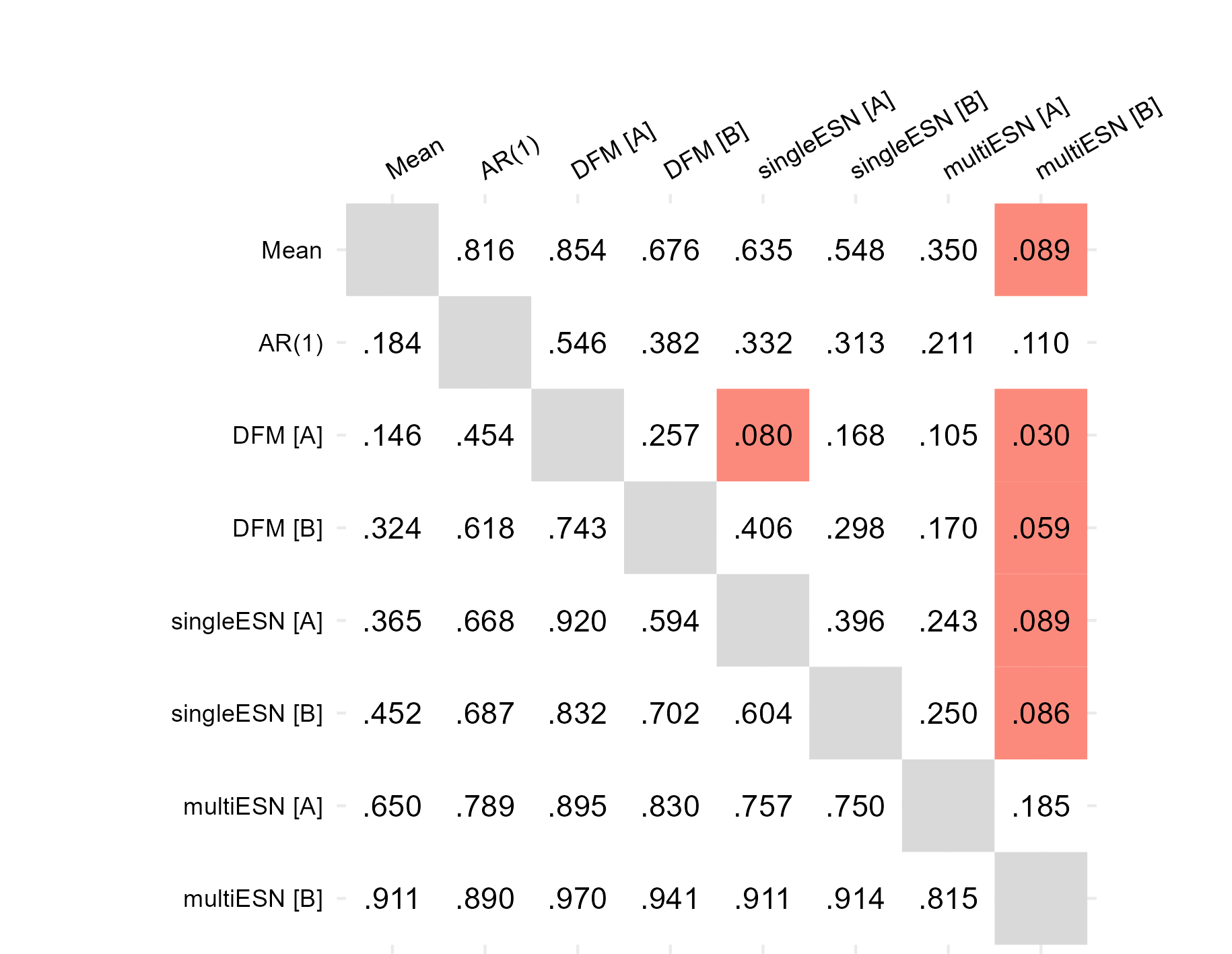}
	}
	\\
	\subfloat[Rolling 2007]{%
		\includegraphics[width=0.48\textwidth, trim=30 0 0 25, clip]{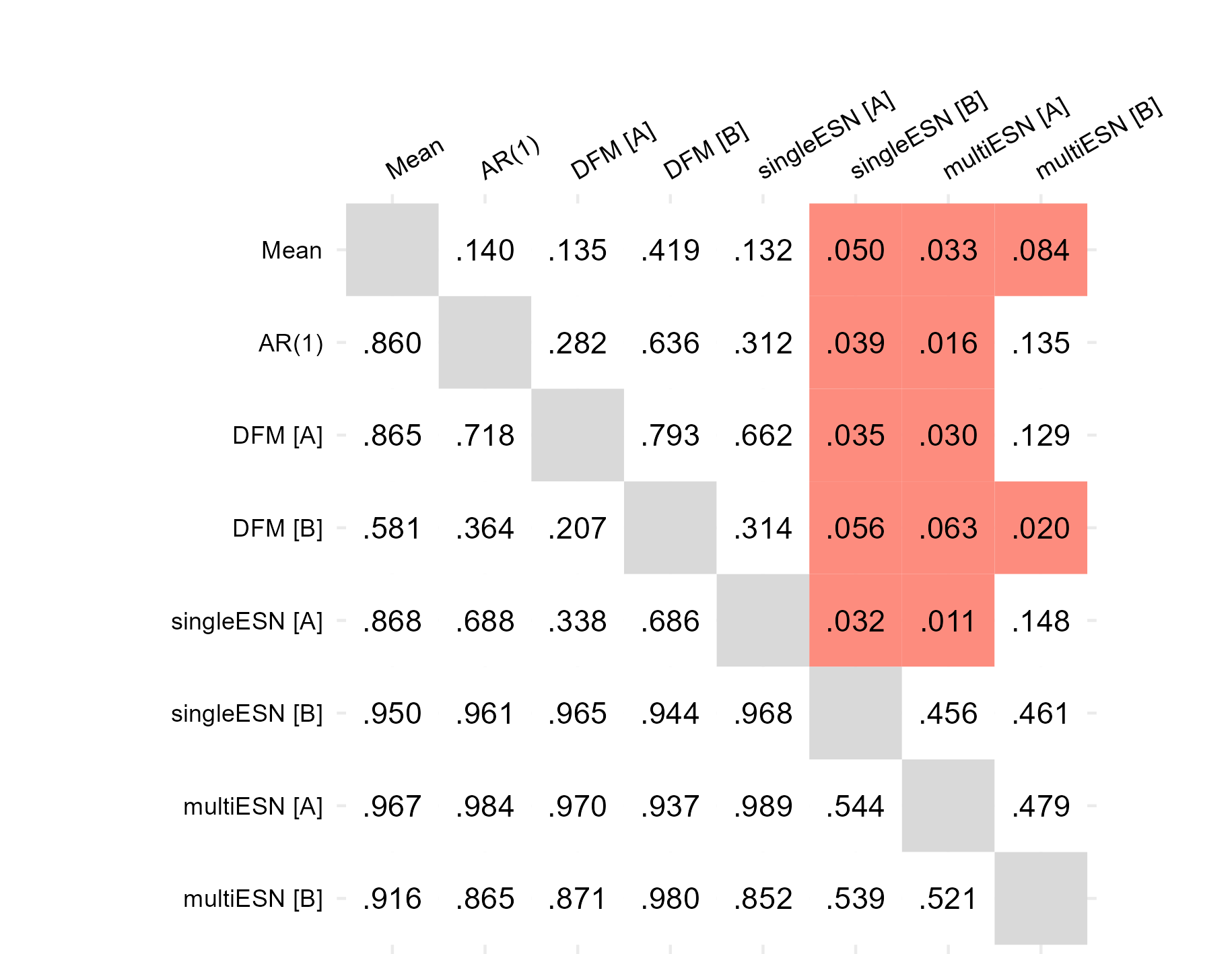}
	}
	\hfill
	\subfloat[Rolling 2011]{%
		\includegraphics[width=0.48\textwidth, trim=30 0 0 25, clip]{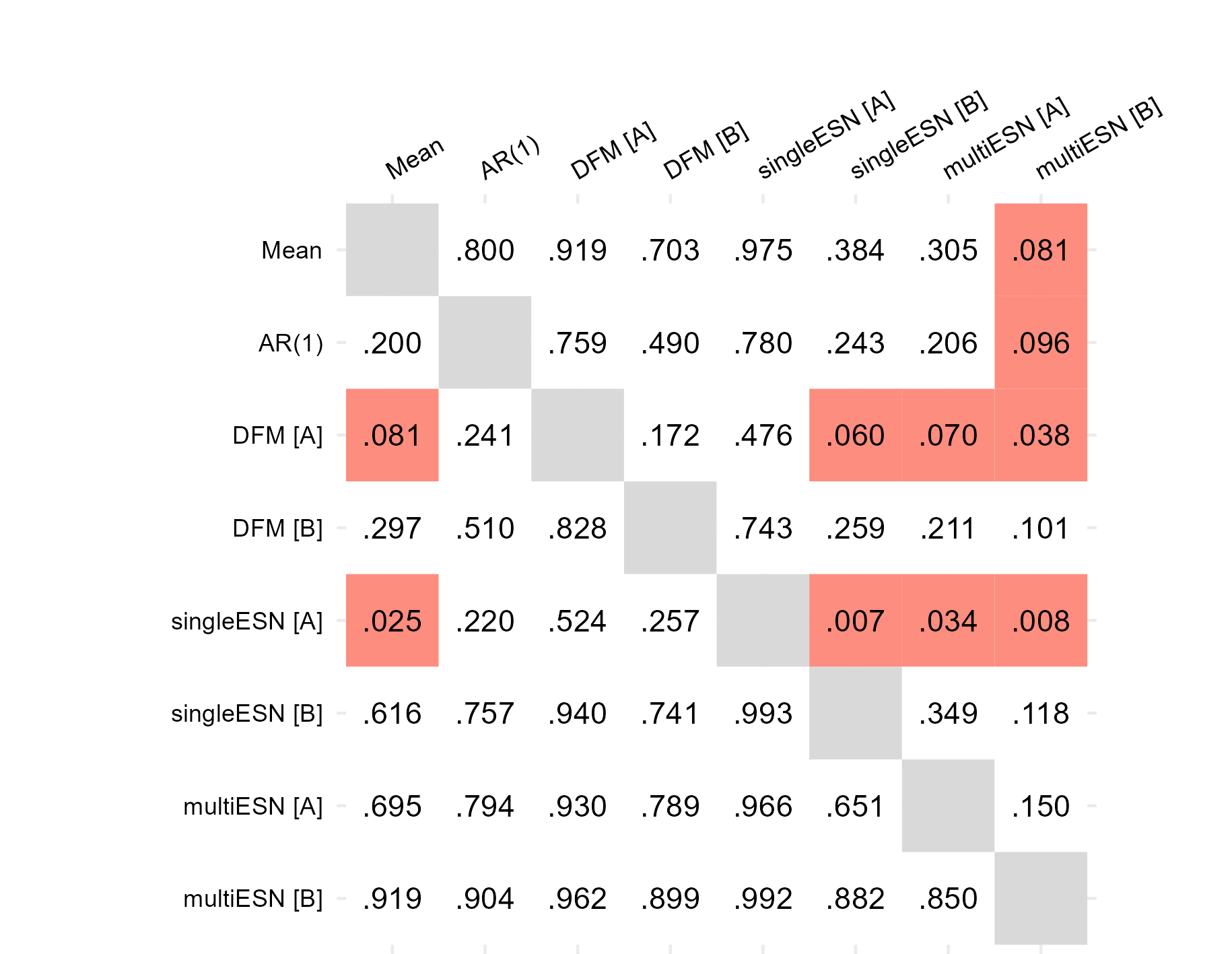}
	}
	\vspace{1em}
	\caption{p-values of pairwise Modified Diebold-Mariano tests between models of Table \ref{table:1sa_GDP_smallMD}. Tests are one-sided and carried out row-wise: the null hypothesis for row $i$ and column $j$ reads as ``the $i$th-row model forecasts \textit{more accurately} than the $j$th-column model''. Rejections at the 10\% level are highlighted in red.}
\end{figure}

\begin{figure}[h!]
    \centering
    \caption{1-Step-ahead {High-Frequency} GDP Forecasting -- 2007 Sample -- Small-MD Dataset}
    \label{fig:1sa_highfreq_forecasts_2007_GDP_smallMD}
    \subfloat[Fixed]{%
      \includegraphics[width=0.8\textwidth]{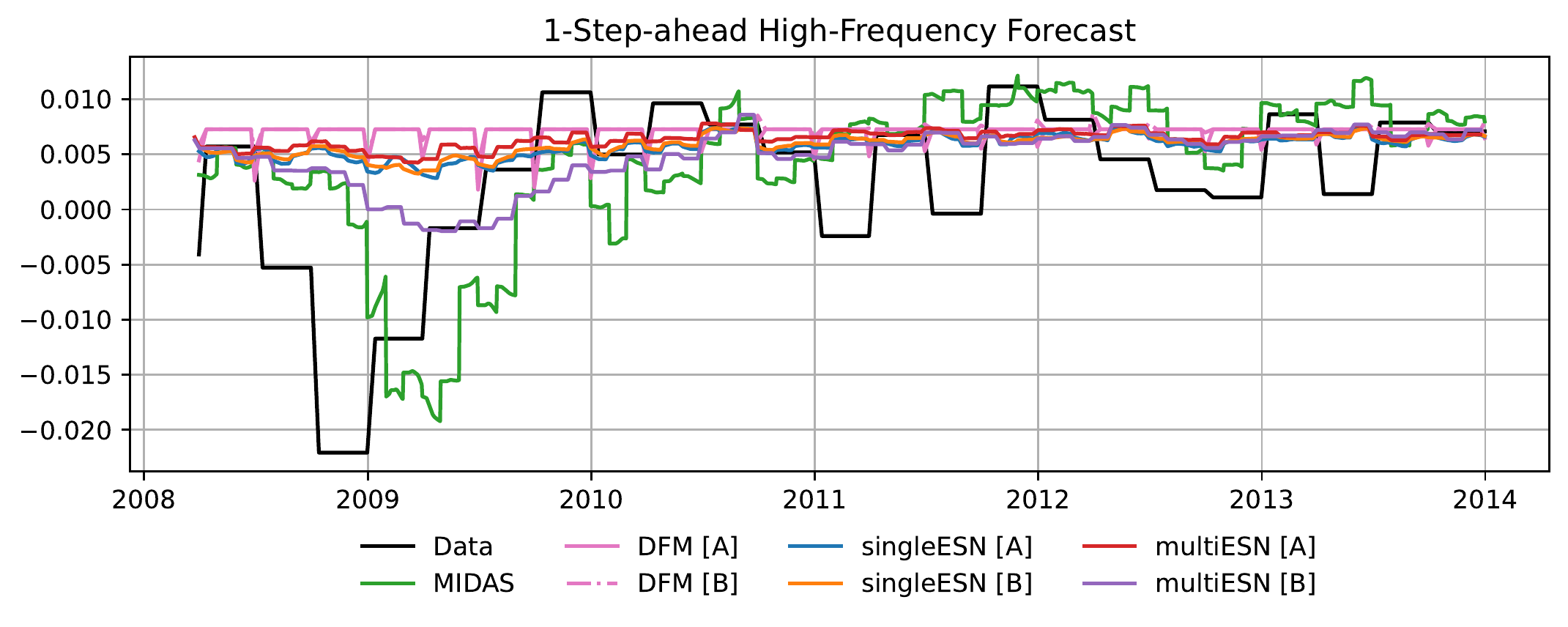}
    }
    \vspace{1em}
    \\
    \subfloat[Expanding]{%
      \includegraphics[width=0.8\textwidth]{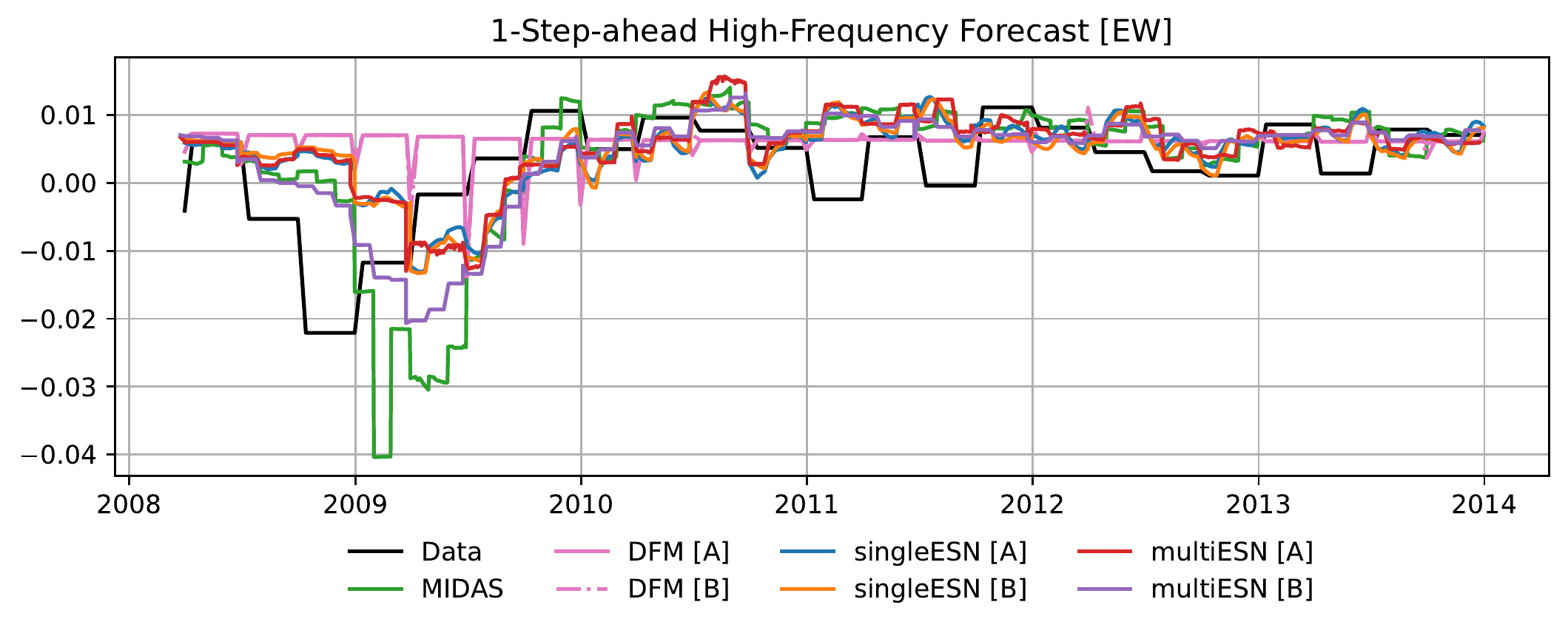}
    }
    \vspace{1em}
    \\
    \subfloat[Rolling]{%
      \includegraphics[width=0.8\textwidth]{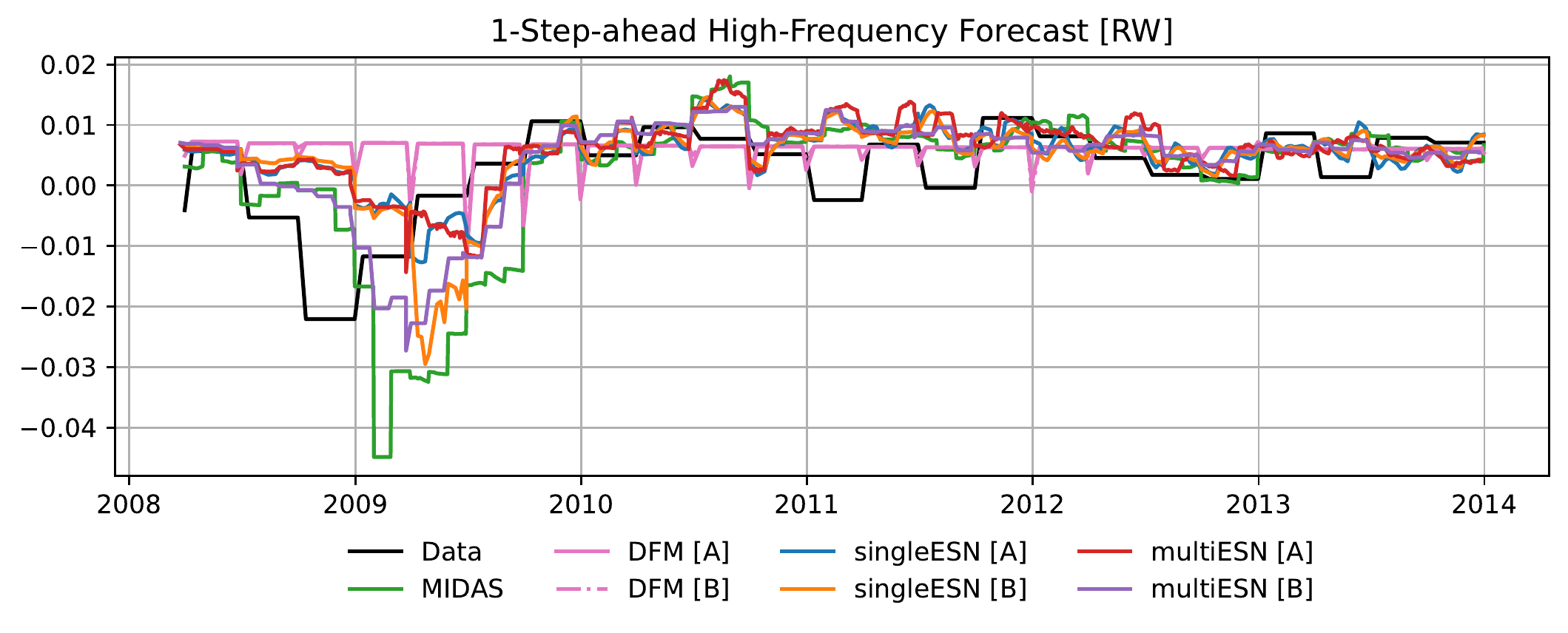}
    }
    \\
    \vspace{2em}
    \begin{tablenotes}[para,flushleft]
      \small
      Note: Forecasts for the 2007 sample are presented up to Q4 2013 to better display the high-frequency behavior of models during the Financial Crisis period.  
    \end{tablenotes}
 \end{figure}
 
 \begin{figure}[p]
    \centering
    \caption{1-Step-ahead {High-Frequency} GDP Forecasting -- 2011 Sample -- Small-MD Dataset}
    \label{fig:1sa_highfreq_forecasts_2011_GDP_smallMD}
    \subfloat[Fixed]{%
      \includegraphics[width=0.8\textwidth]{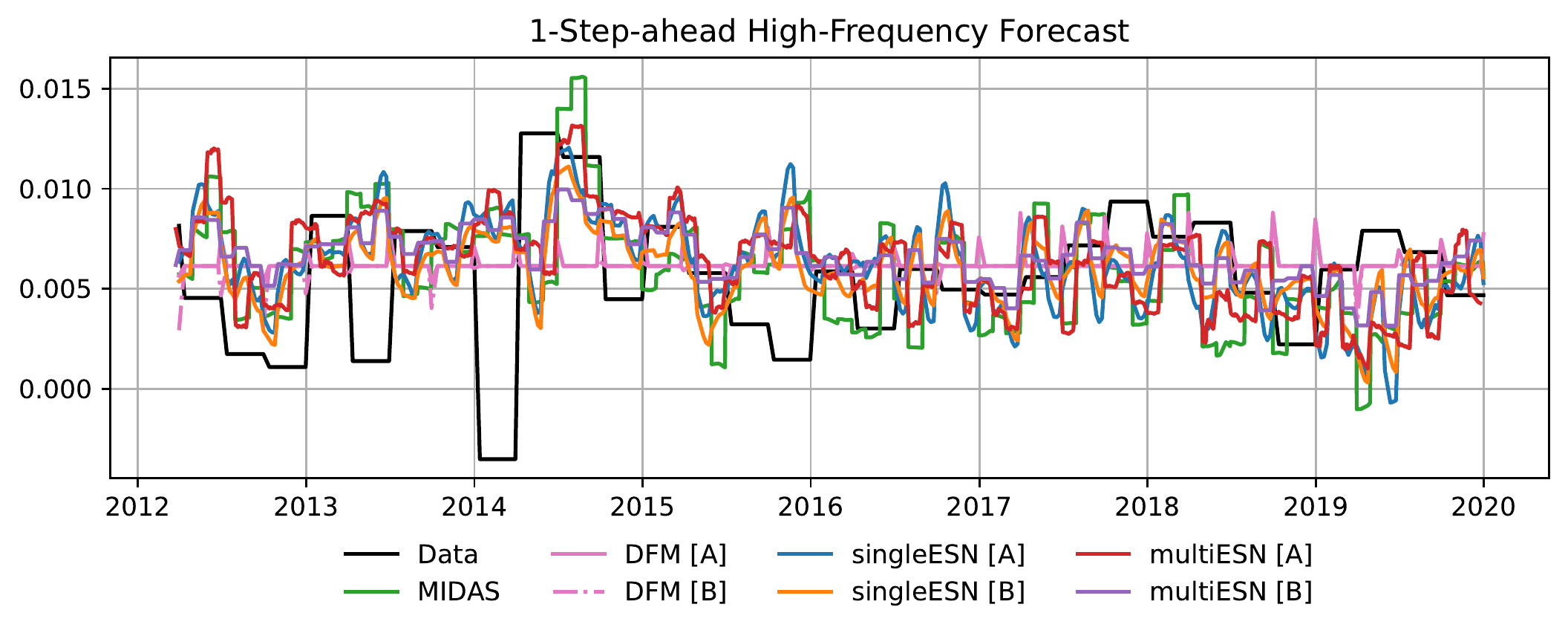}
    }
    \vspace{1em}
    \\
    \subfloat[Expanding]{%
      \includegraphics[width=0.8\textwidth]{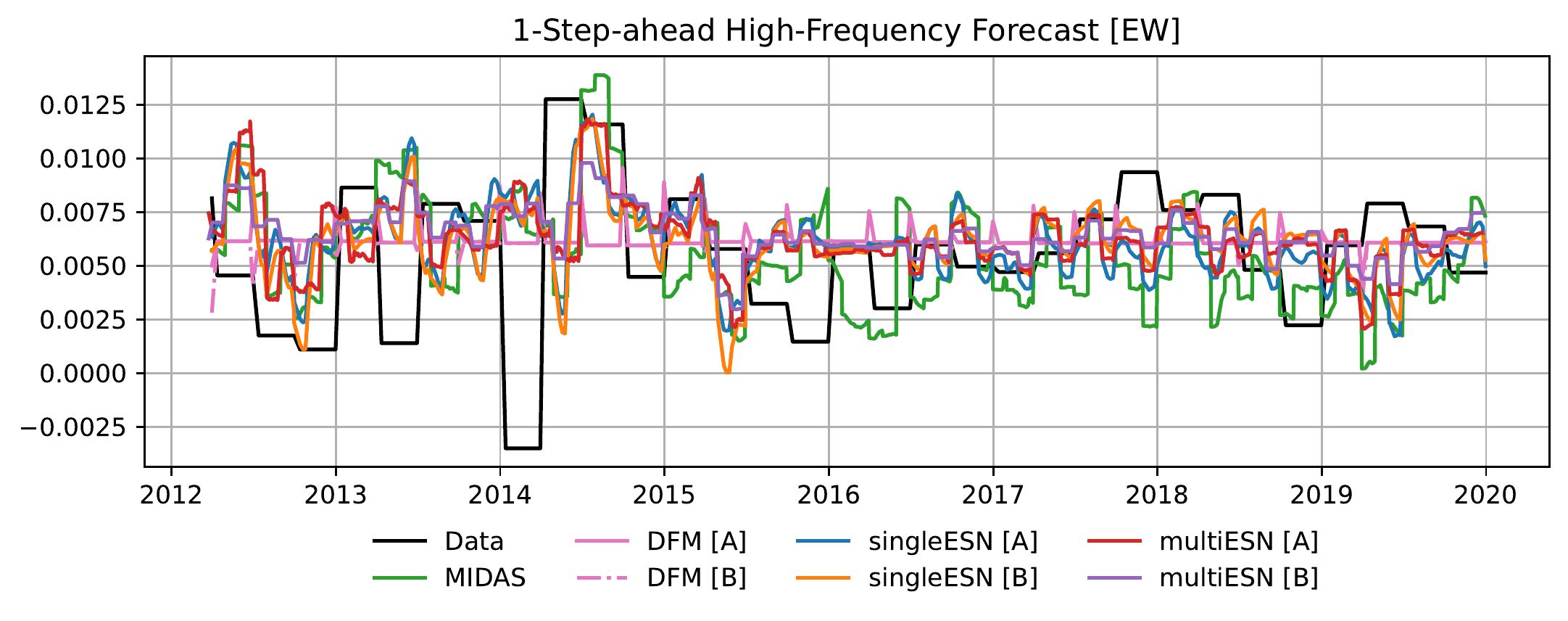}
    }
    \vspace{1em}
    \\
    \subfloat[Rolling]{%
      \includegraphics[width=0.8\textwidth]{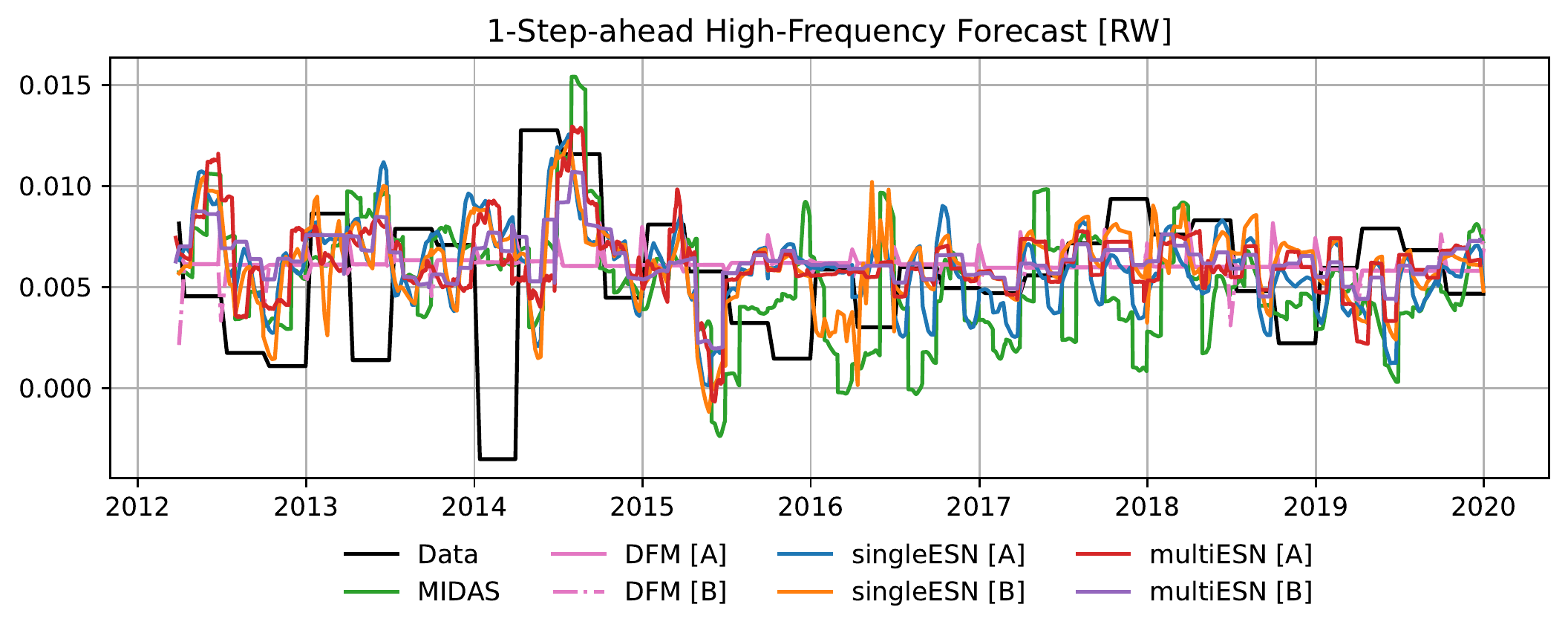}
    }
 \end{figure}
 
 \begin{figure}[p]
    \centering
    \caption{MIDAS Robustness Plots -- 2007 Sample -- Small-MD Dataset}
    \label{fig:}
    \subfloat[Fixed Parameters]{%
      \includegraphics[width=0.85\textwidth]{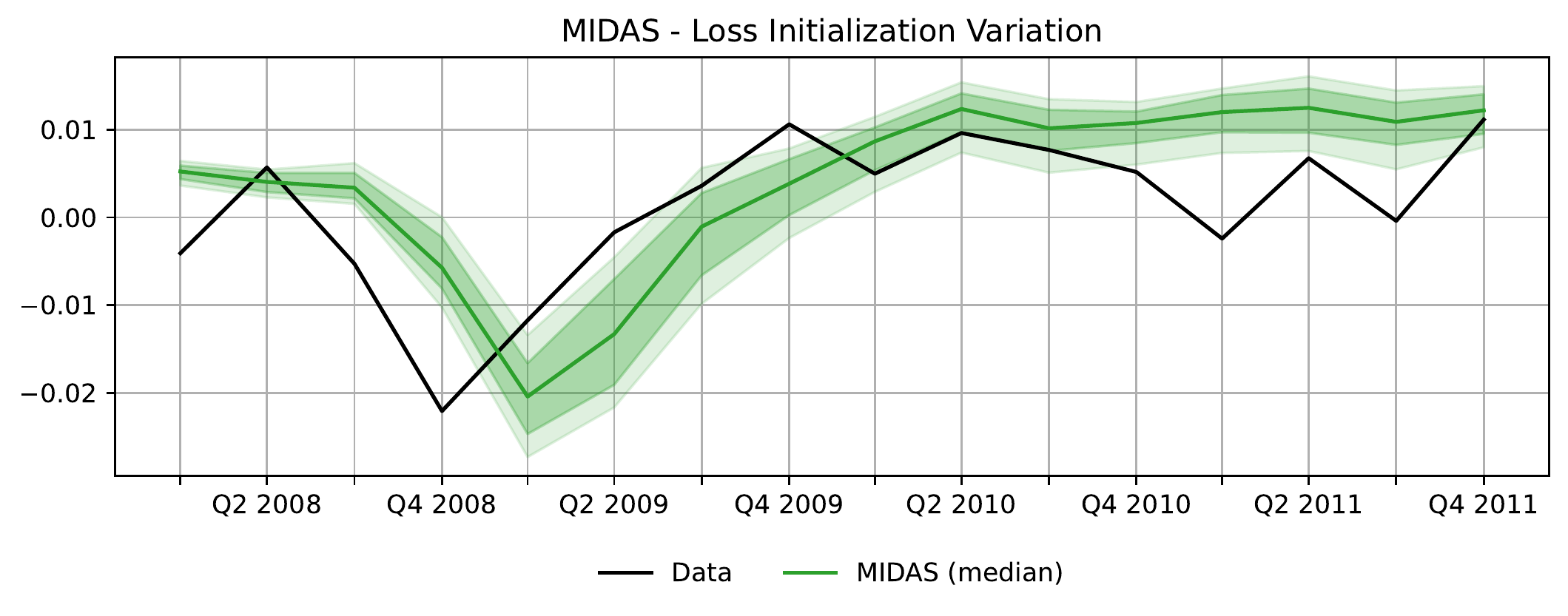}
    }
     \vspace{1em}
    \\
    \subfloat[Expanding]{%
      \includegraphics[width=0.85\textwidth]{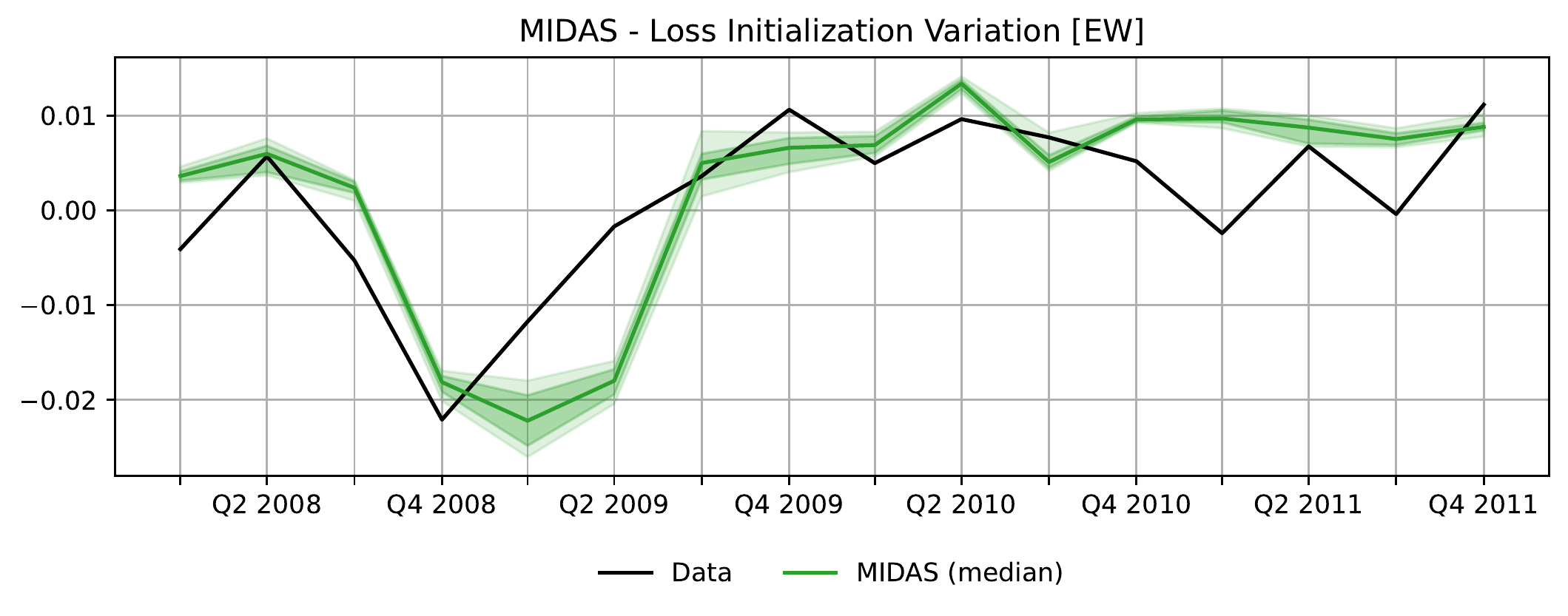}
    }
     \vspace{1em}
    \\
    \subfloat[Rolling]{%
      \includegraphics[width=0.85\textwidth]{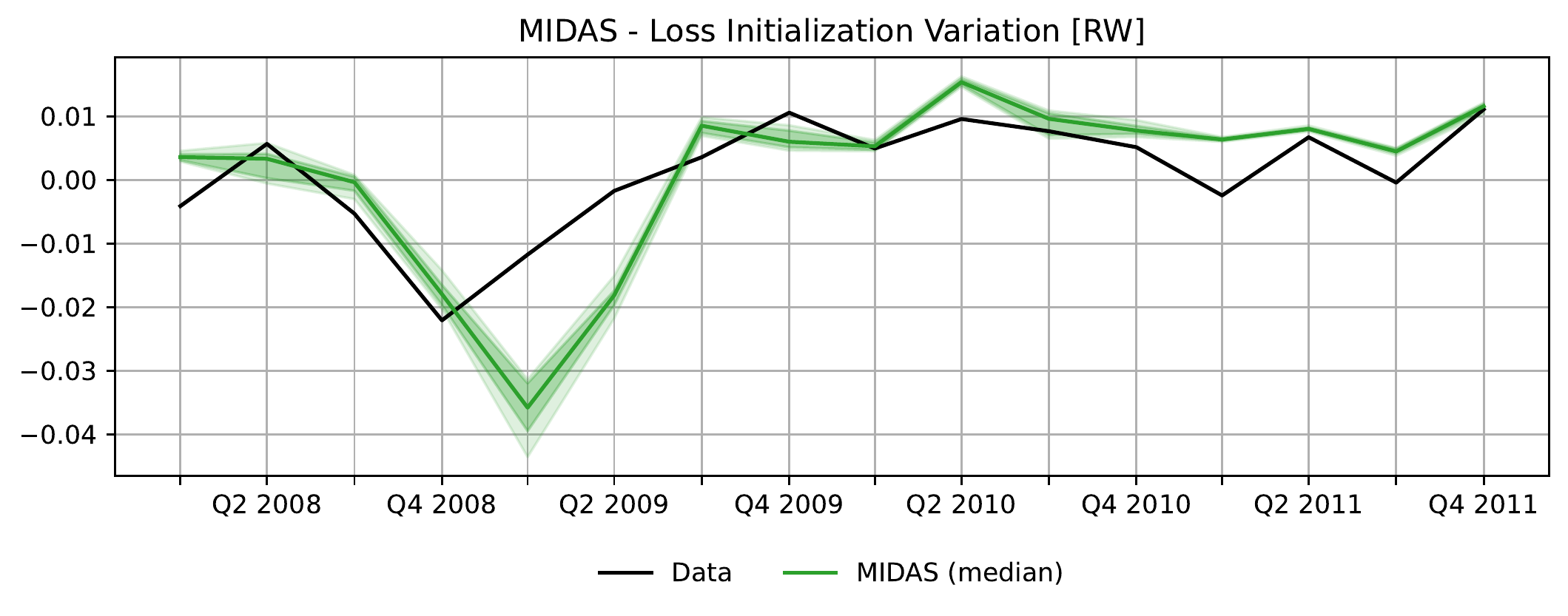}
    }
 \end{figure}
 
 \begin{figure}[p]
    \centering
    \caption{ESN Robustness Plots -- 2007 Sample -- Small-MD Dataset}
    \label{fig:singleesn_b_robustness_smallMD}
    \subfloat[Fixed Parameters]{%
      \includegraphics[width=0.85\textwidth]{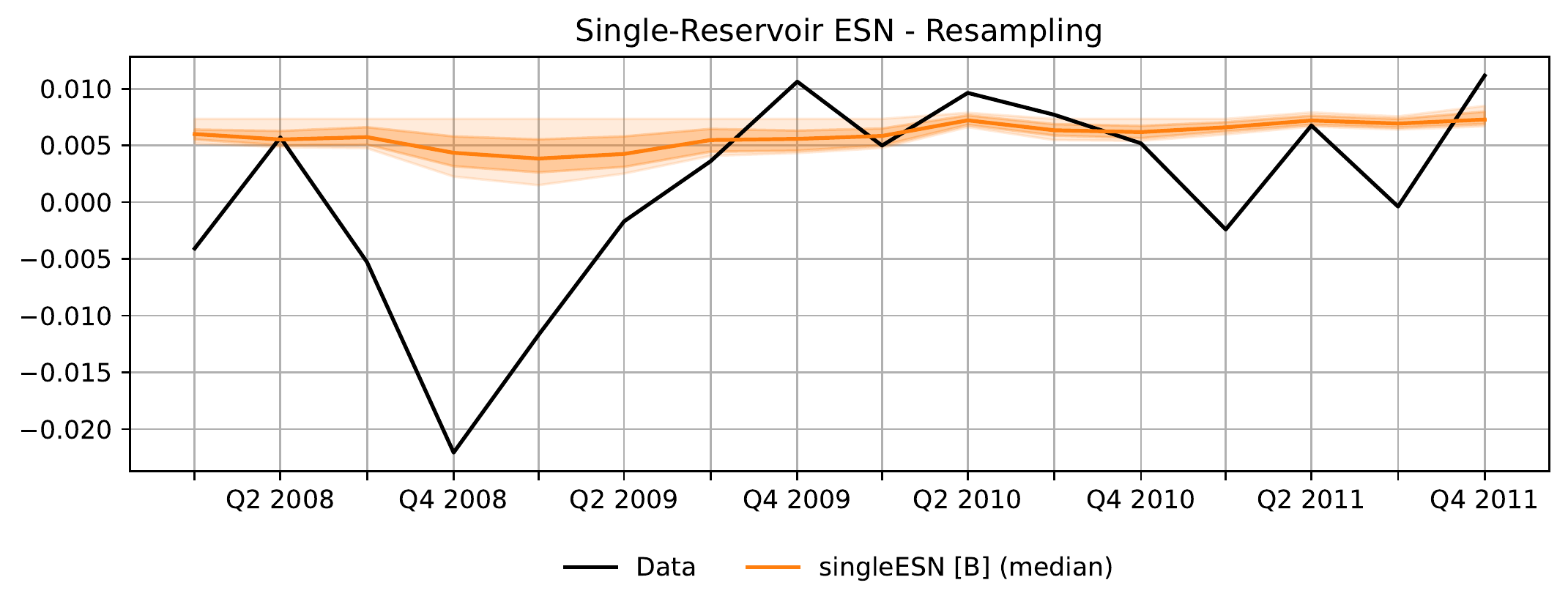}
    }
     \vspace{1em}
    \\
    \subfloat[Expanding Window]{%
      \includegraphics[width=0.85\textwidth]{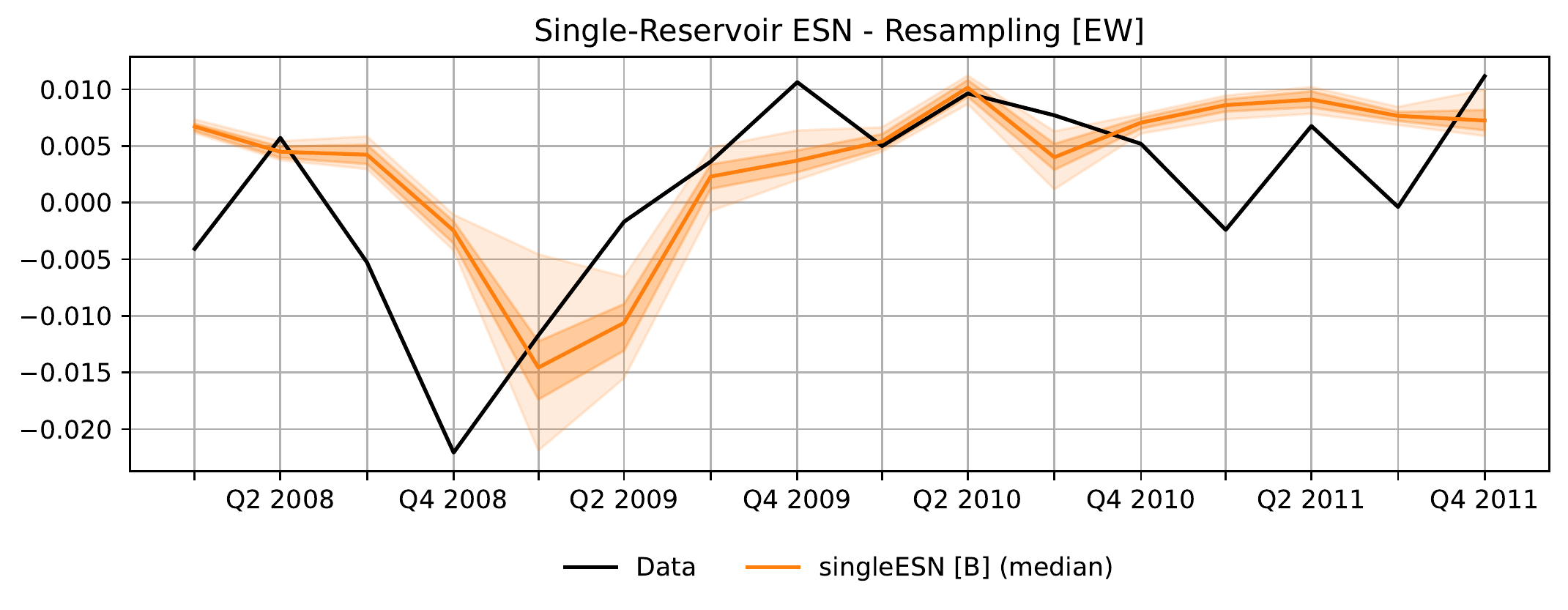}
    }
     \vspace{1em}
    \\
    \subfloat[Rolling Window]{%
      \includegraphics[width=0.85\textwidth]{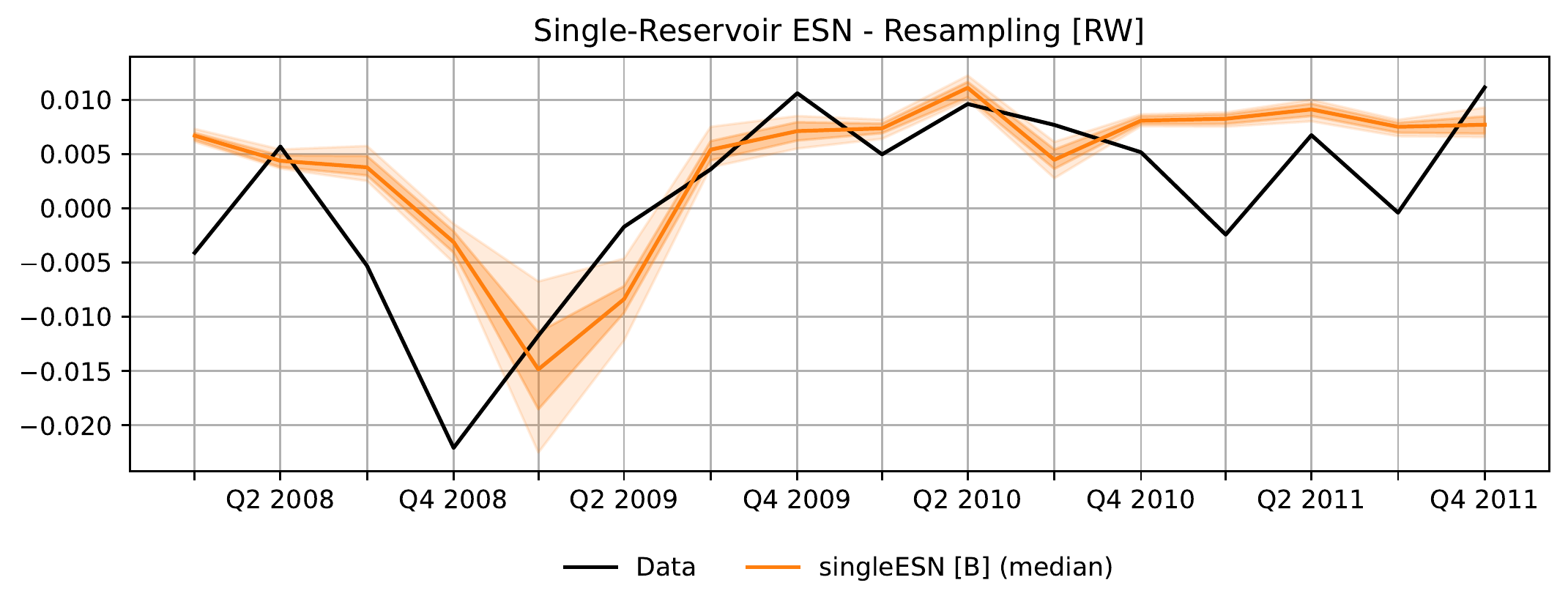}
    }
 \end{figure}
 
 \begin{figure}[p]
    \centering
    \caption{ESN Robustness Plots -- 2007 Sample -- Small-MD Dataset}
    \label{fig:multiesn_b_robustness_smallMD}
    \subfloat[Fixed Parameters]{%
      \includegraphics[width=0.85\textwidth]{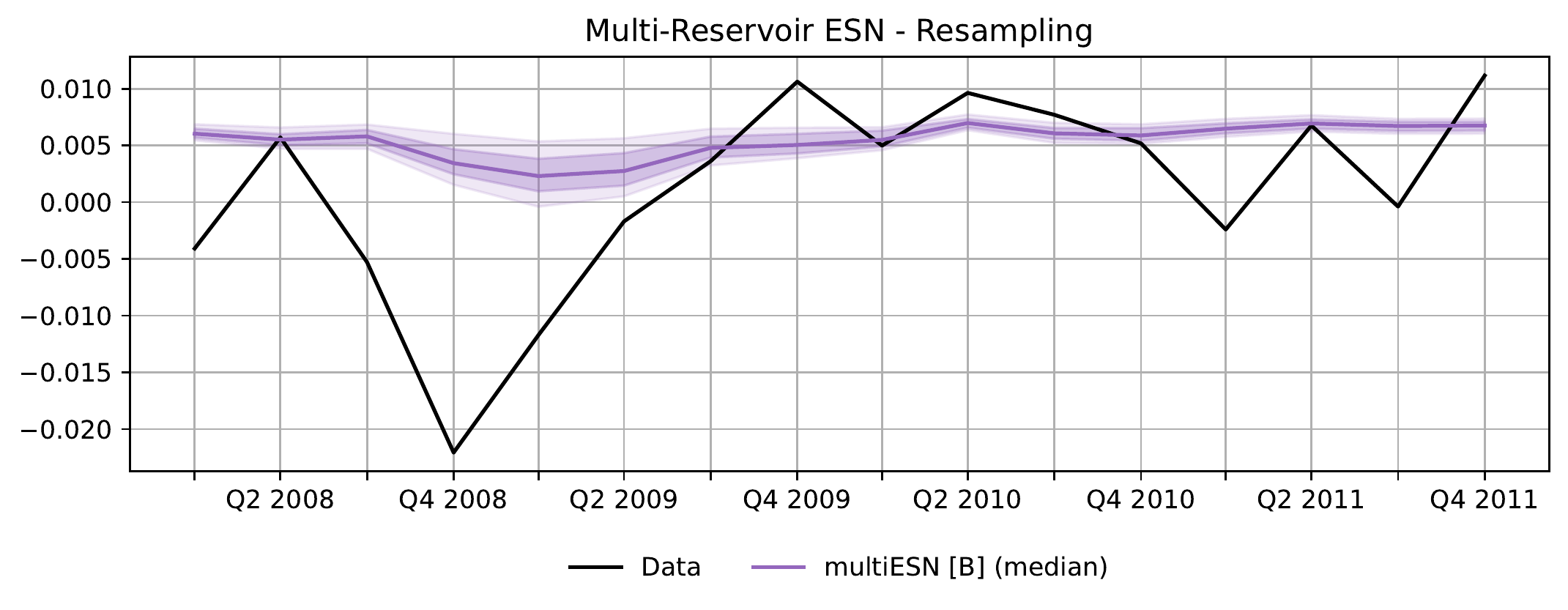}
    }
     \vspace{1em}
    \\
    \subfloat[Expanding Window]{%
      \includegraphics[width=0.85\textwidth]{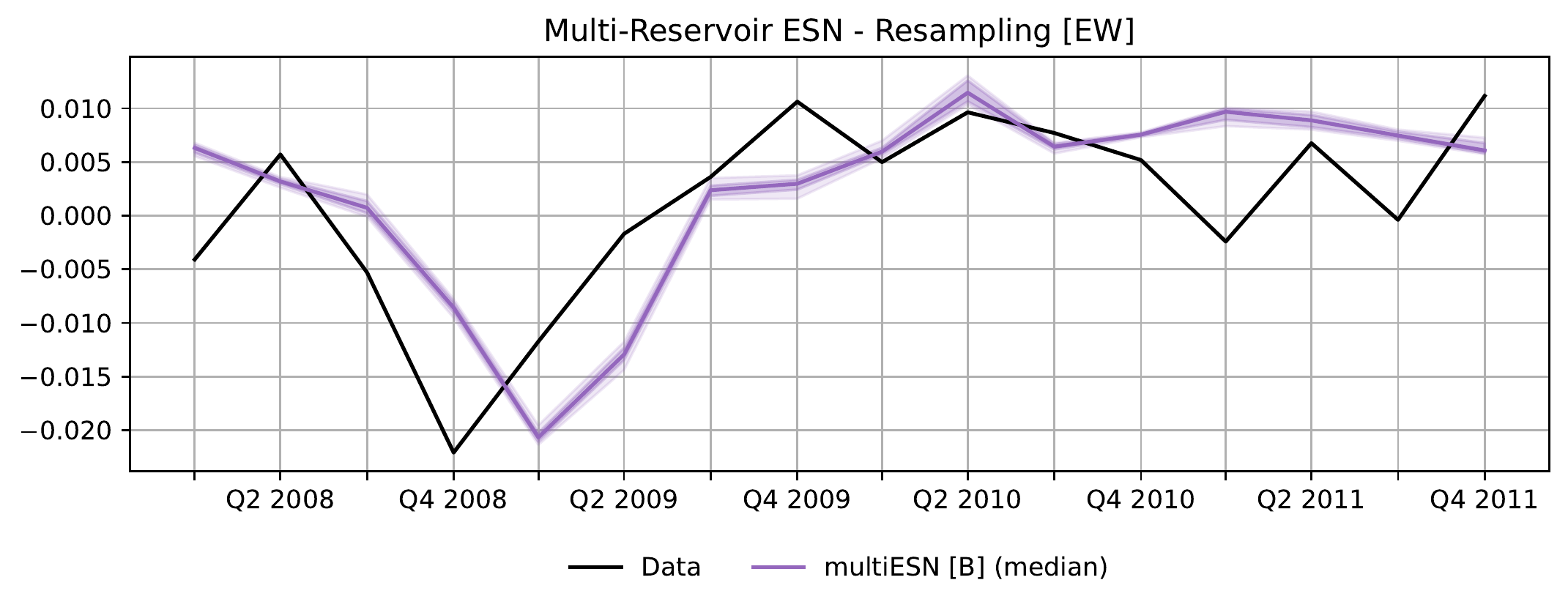}
    }
     \vspace{1em}
    \\
    \subfloat[Rolling Window]{%
      \includegraphics[width=0.85\textwidth]{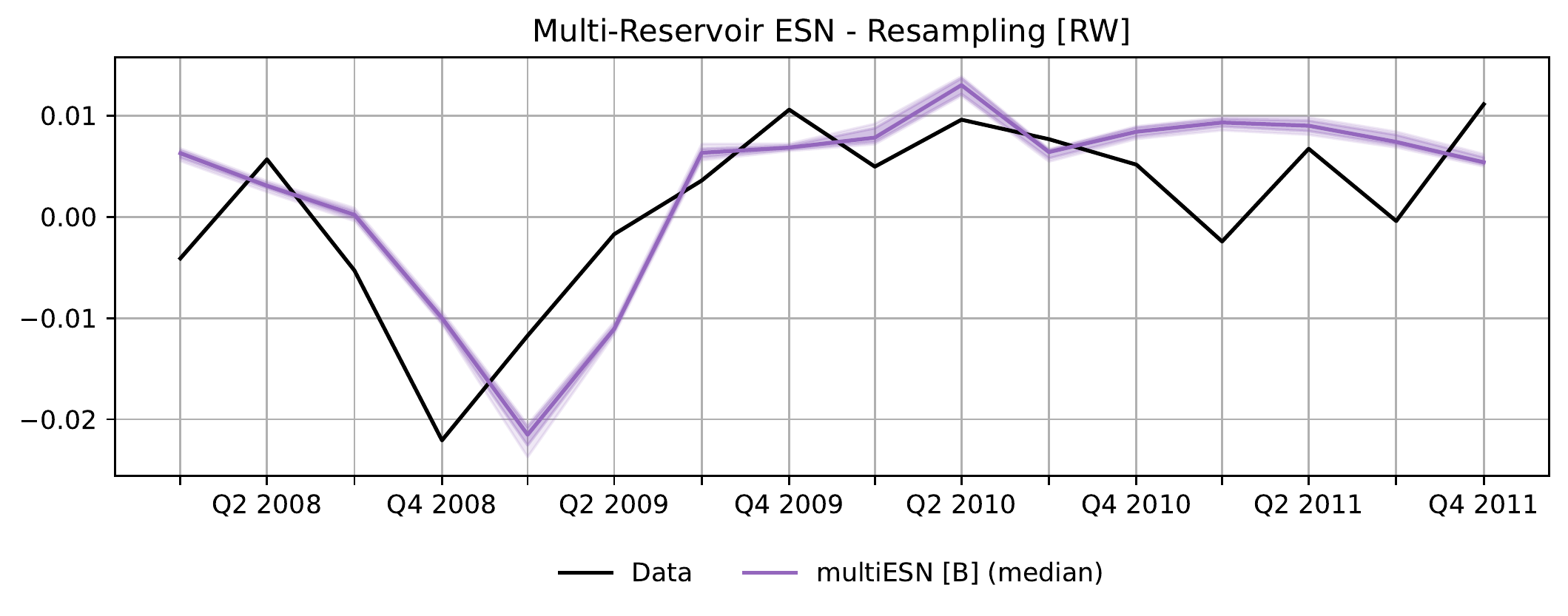}
    }
 \end{figure}

\begin{figure}[ht]
    \caption{1-Step-ahead GDP Forecasting, Fixed Parameters - Small-MD Dataset}
    \label{fig:1sa_GDP_fixed_smallMD}
    %
    \subfloat[Pre-crisis model]{%
      \includegraphics[width=0.48\textwidth]{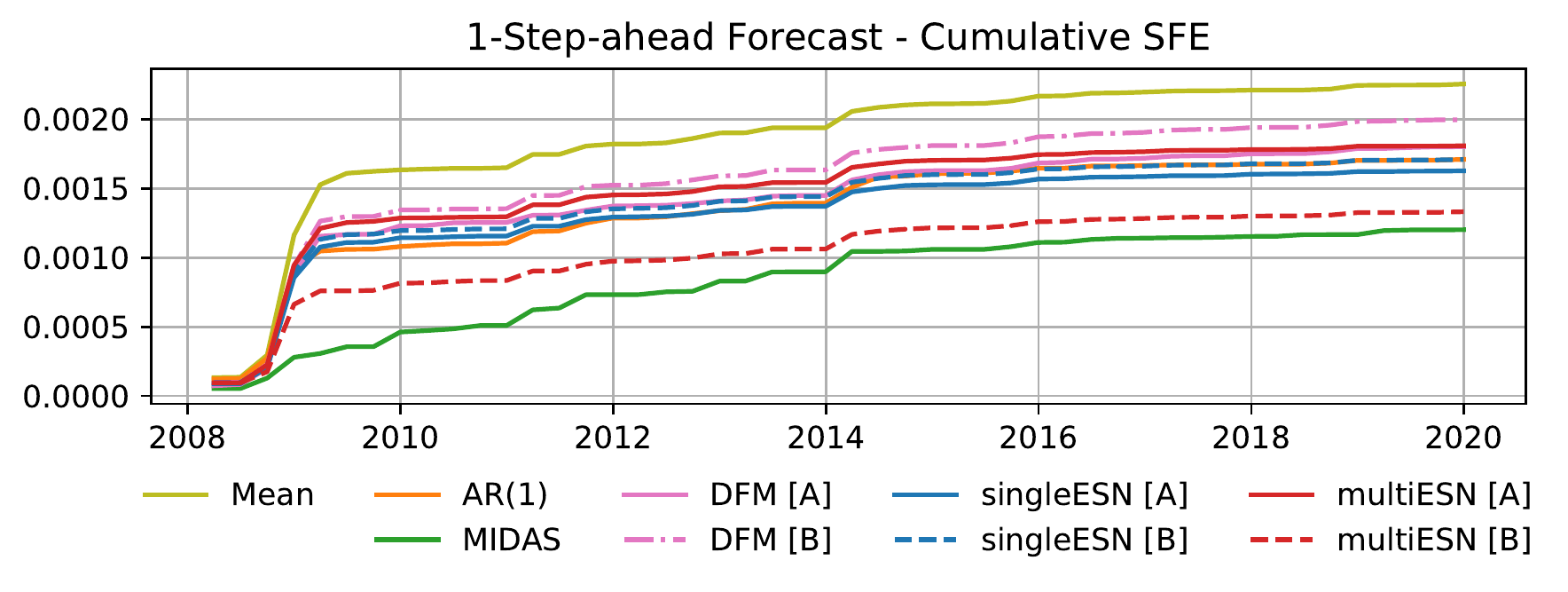}
    }
    \hfill
    \subfloat[Post-crisis model]{%
      \includegraphics[width=0.48\textwidth]{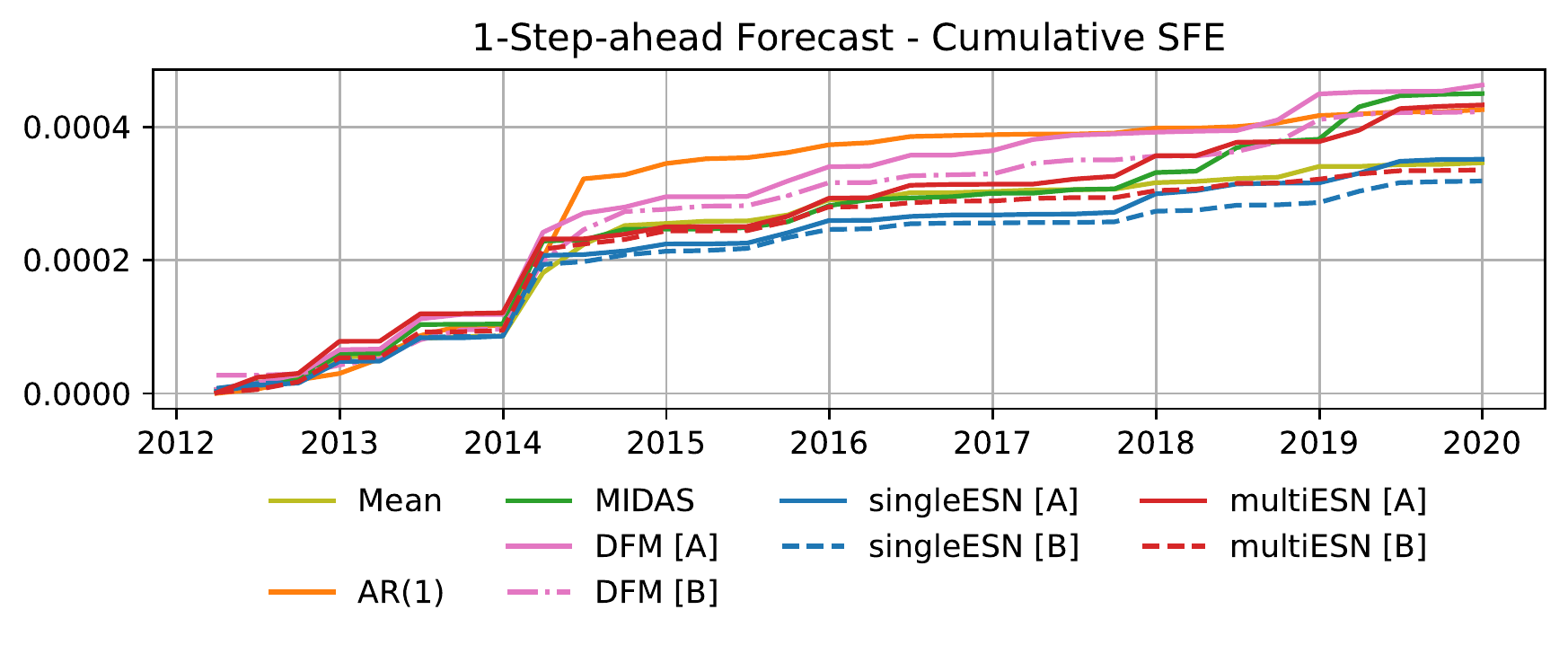}
    }
    \\
    \subfloat[]{%
      \includegraphics[width=0.48\textwidth]{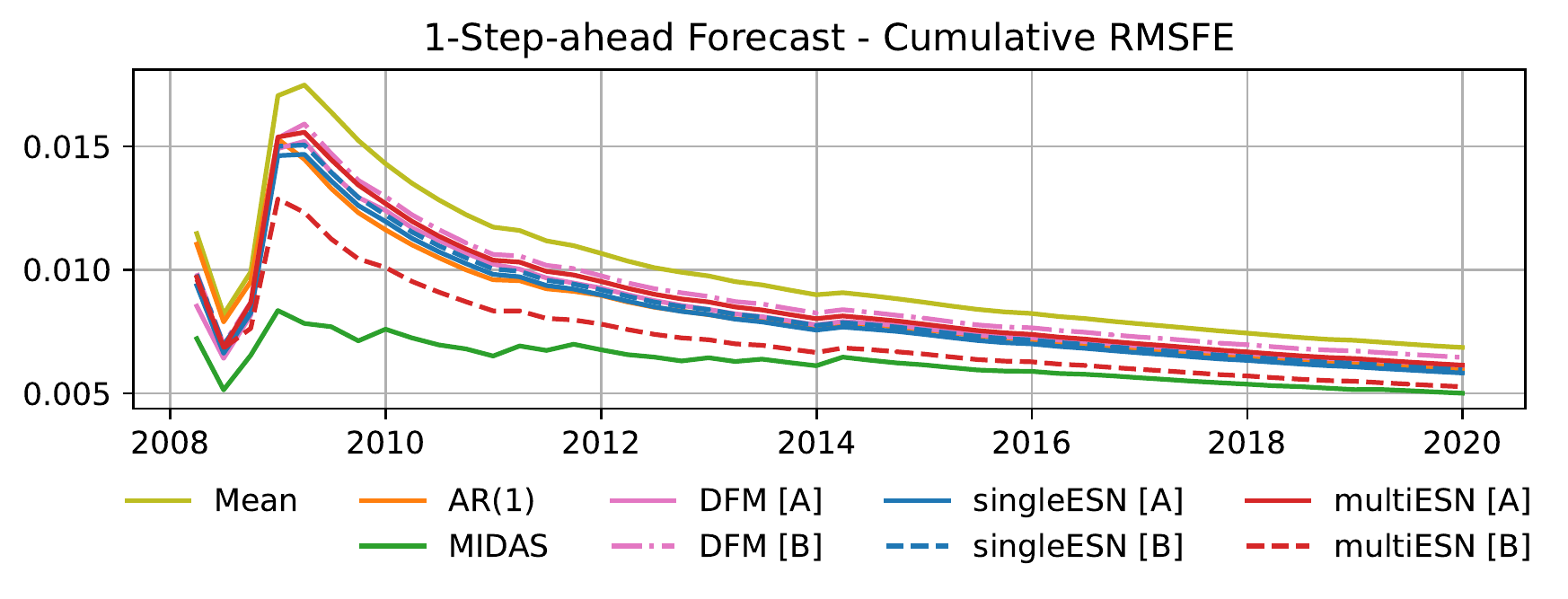}
    }
    \hfill
    \subfloat[]{%
      \includegraphics[width=0.48\textwidth]{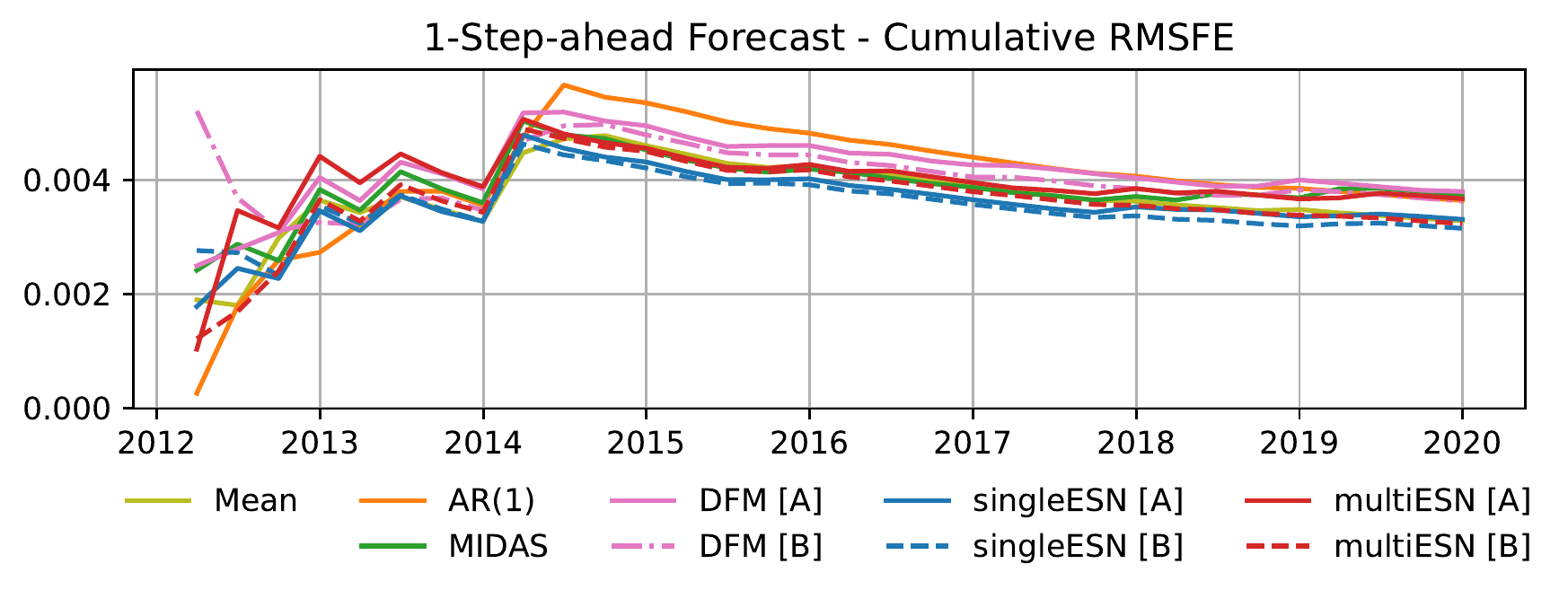}
    }
    \\
    \subfloat[]{%
      \includegraphics[width=0.48\textwidth]{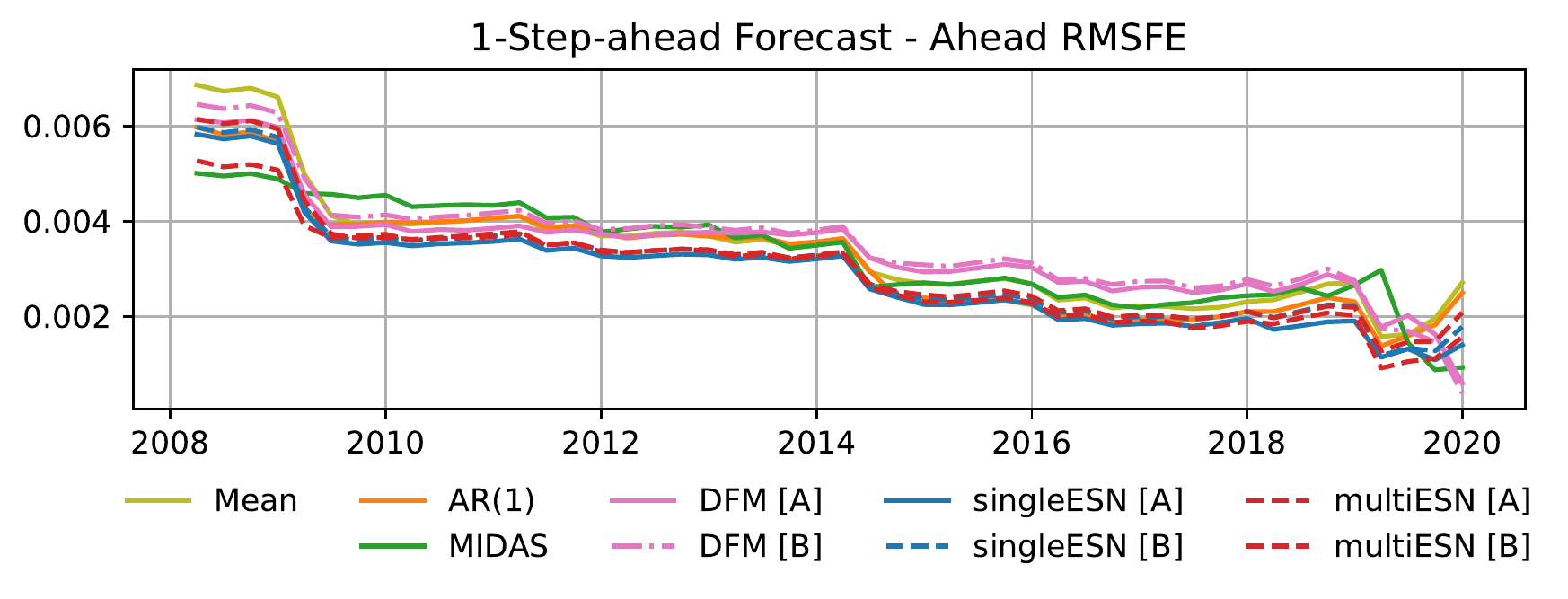}
    }
    \hfill
    \subfloat[]{%
      \includegraphics[width=0.48\textwidth]{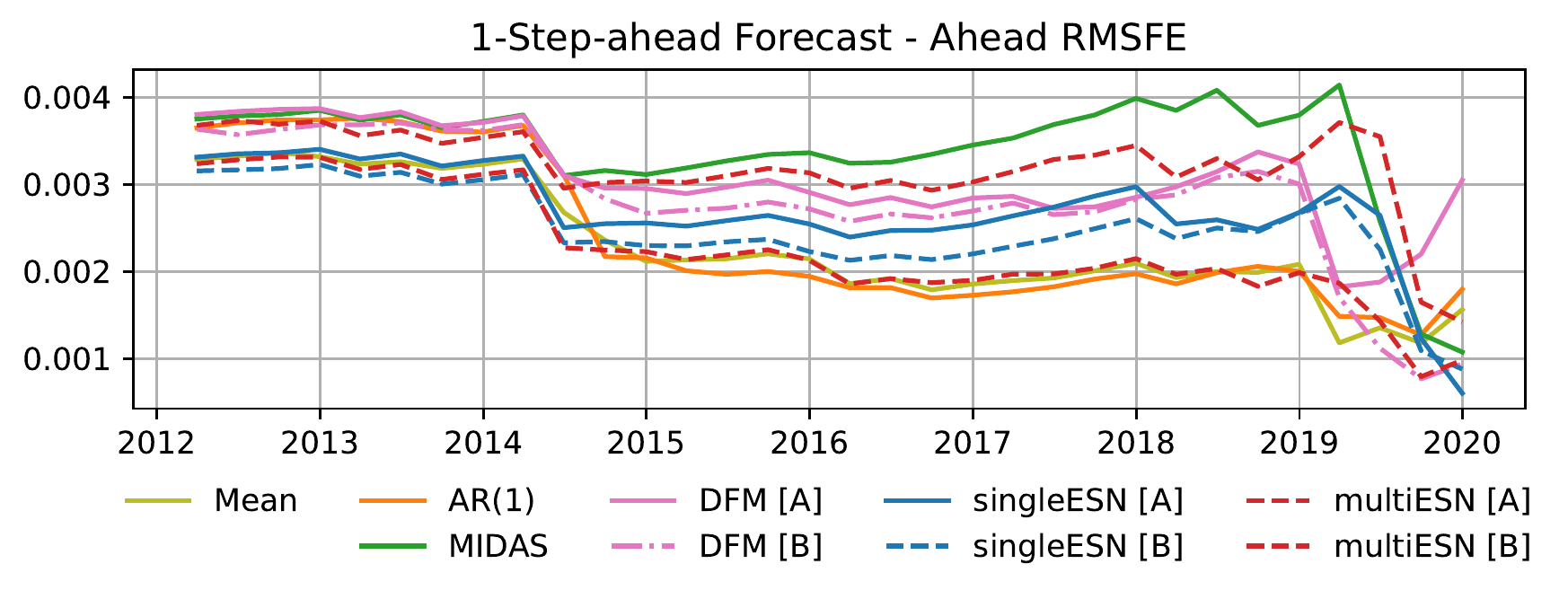}
    }
    \\
    \subfloat[]{%
      \includegraphics[width=0.48\textwidth]{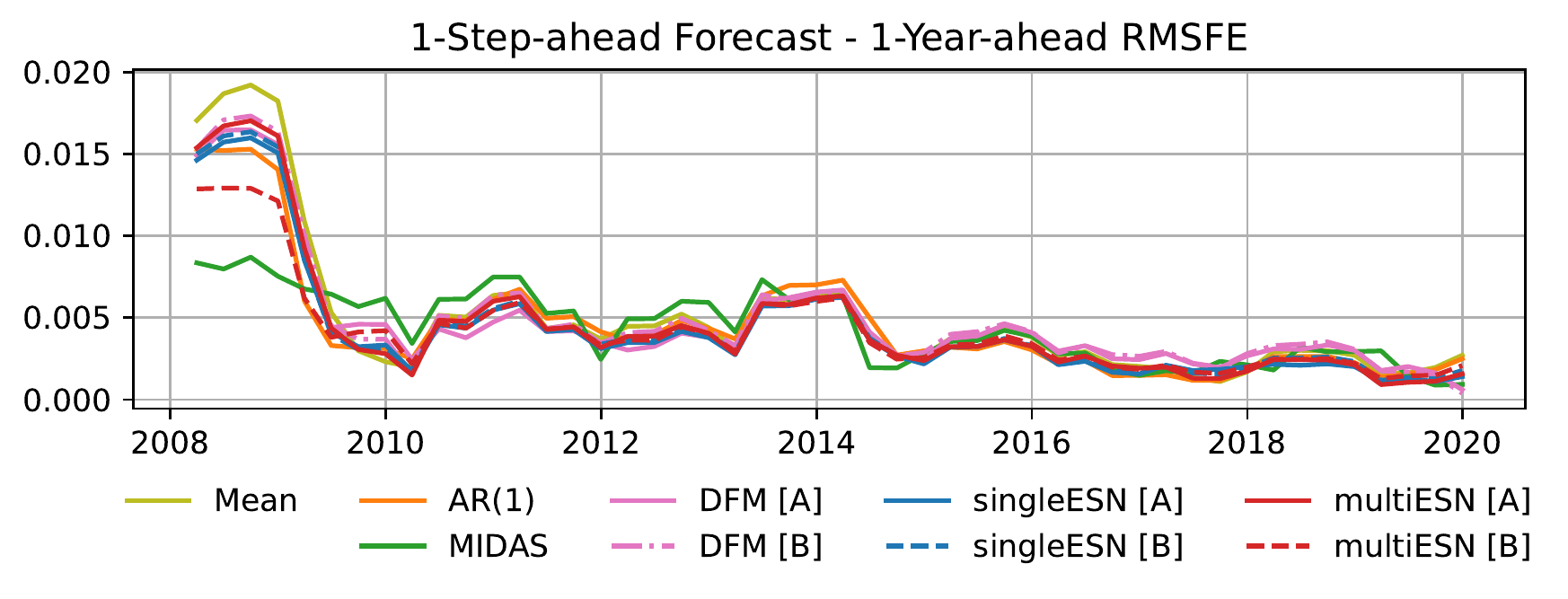}
    }
    \hfill
    \subfloat[]{%
      \includegraphics[width=0.48\textwidth]{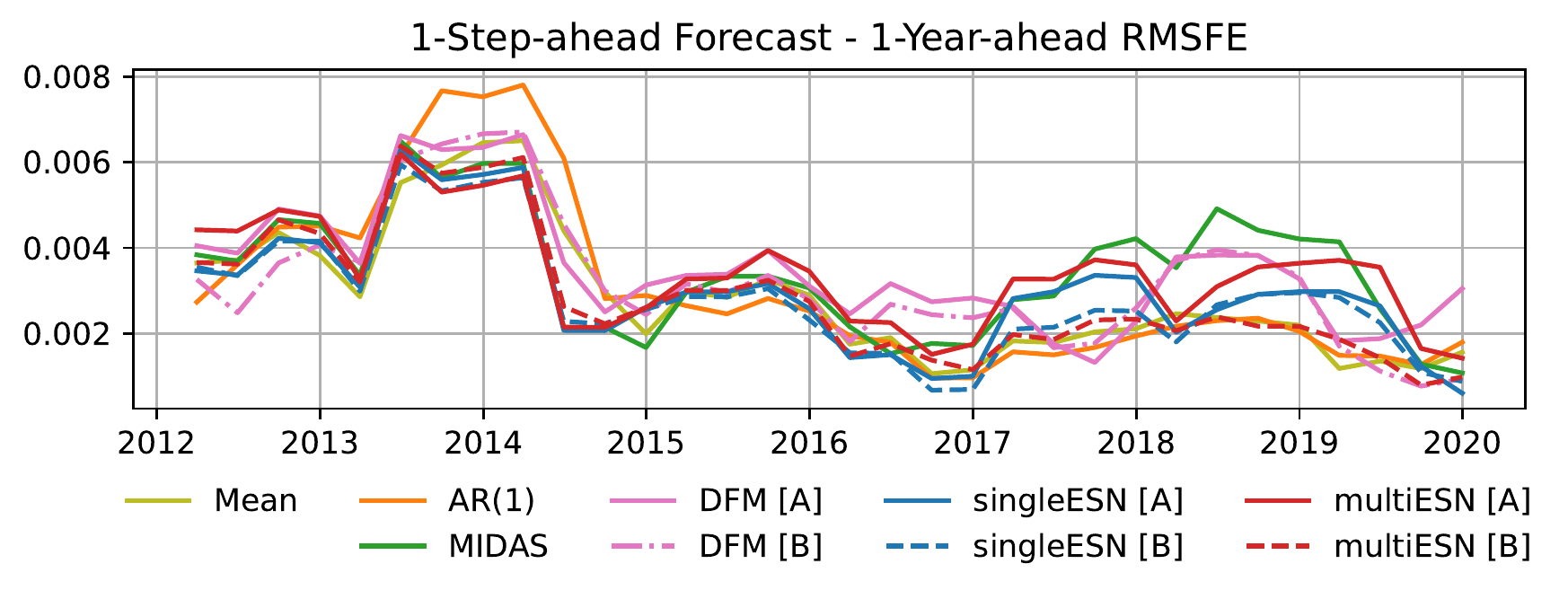}
    }
 \end{figure}
 
 \newpage

\begin{figure}[ht]
    \caption{1-Step-ahead GDP Forecasting, Expanding Window - Small-MD Dataset}
    \label{fig:1sa_GDP_expanding_smallMD}
    %
    \subfloat[Pre-crisis model]{%
      \includegraphics[width=0.48\textwidth]{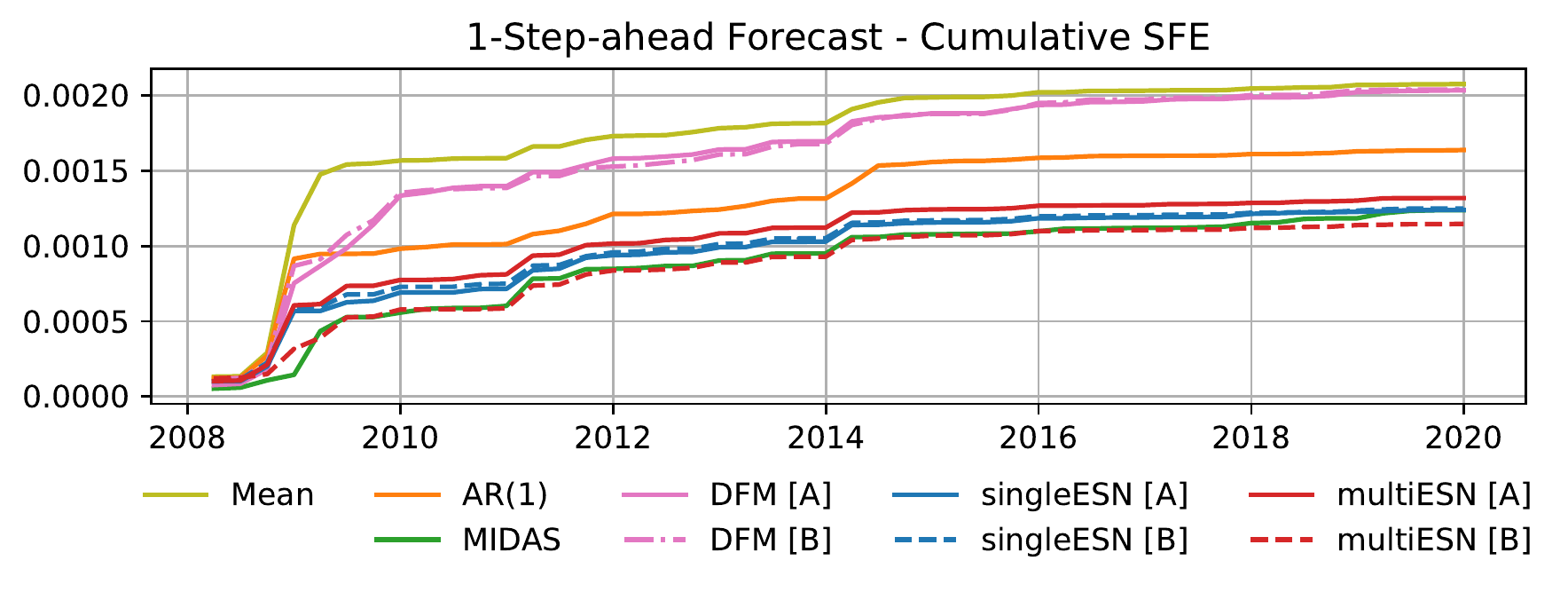}
    }
    \hfill
    \subfloat[Post-crisis model]{%
      \includegraphics[width=0.48\textwidth]{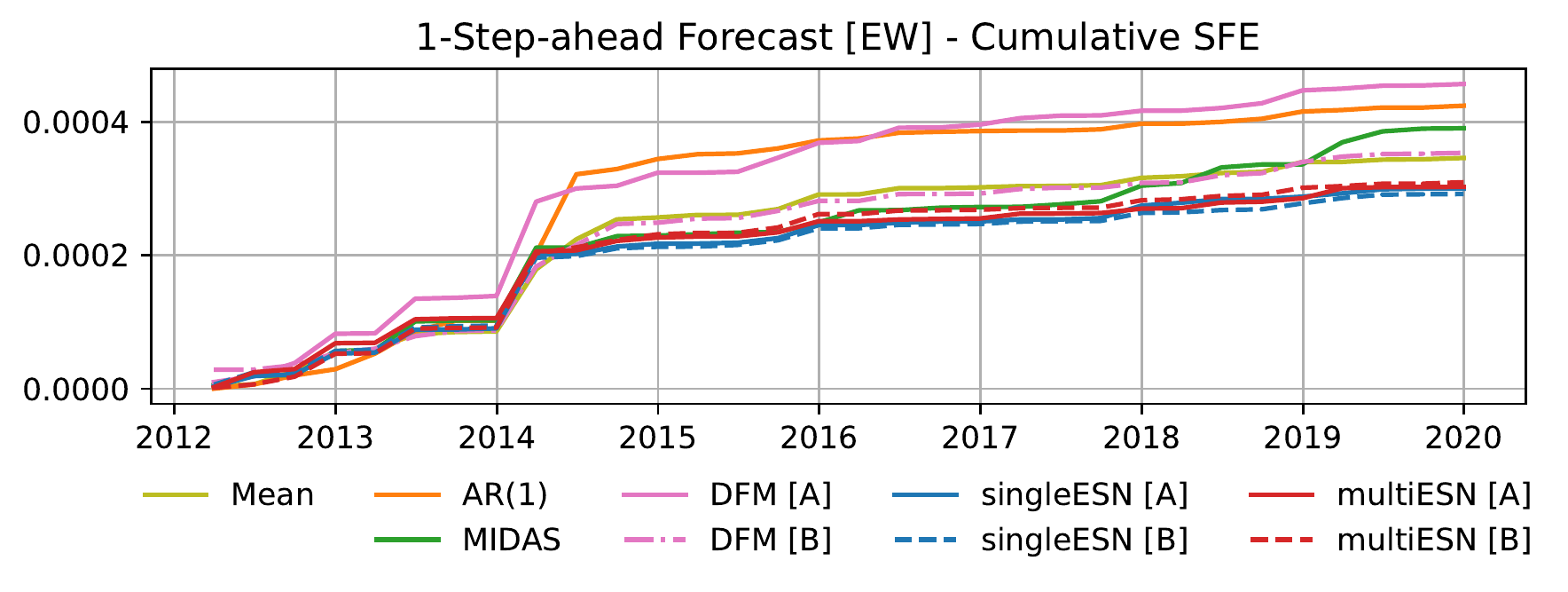}
    }
    \\
    \subfloat[]{%
      \includegraphics[width=0.48\textwidth]{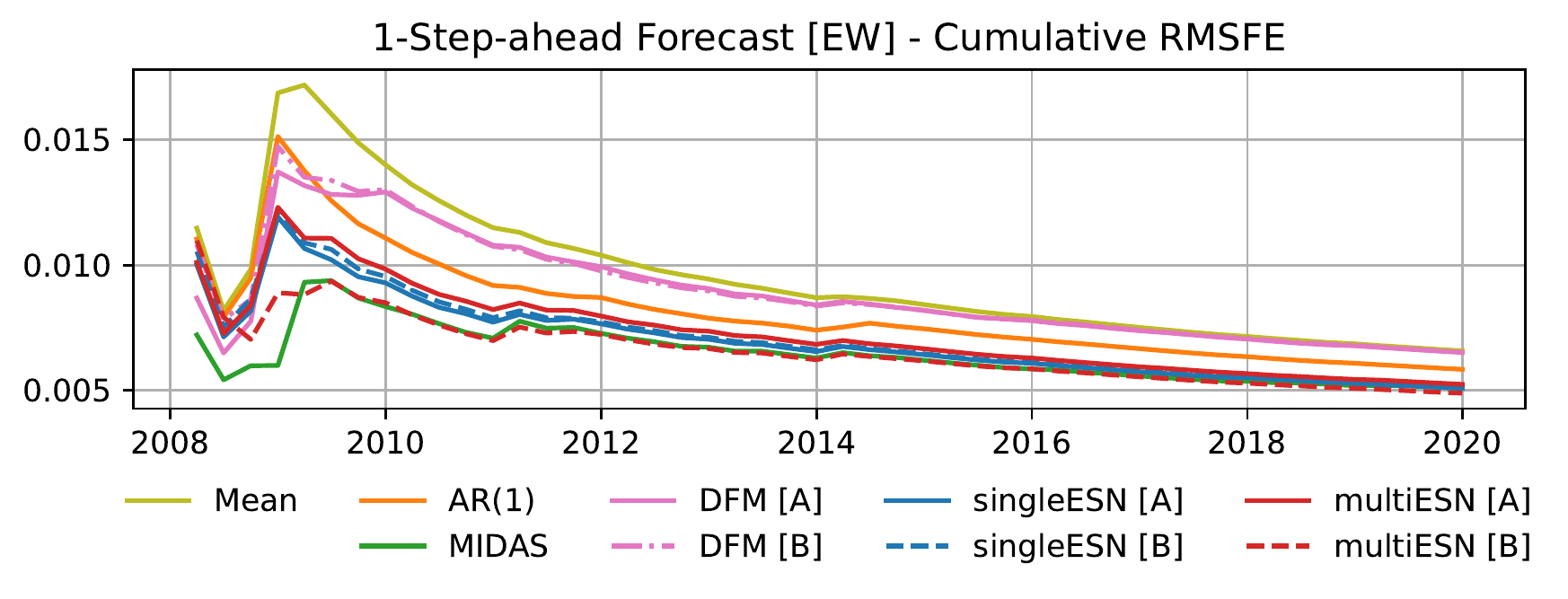}
    }
    \hfill
    \subfloat[]{%
      \includegraphics[width=0.48\textwidth]{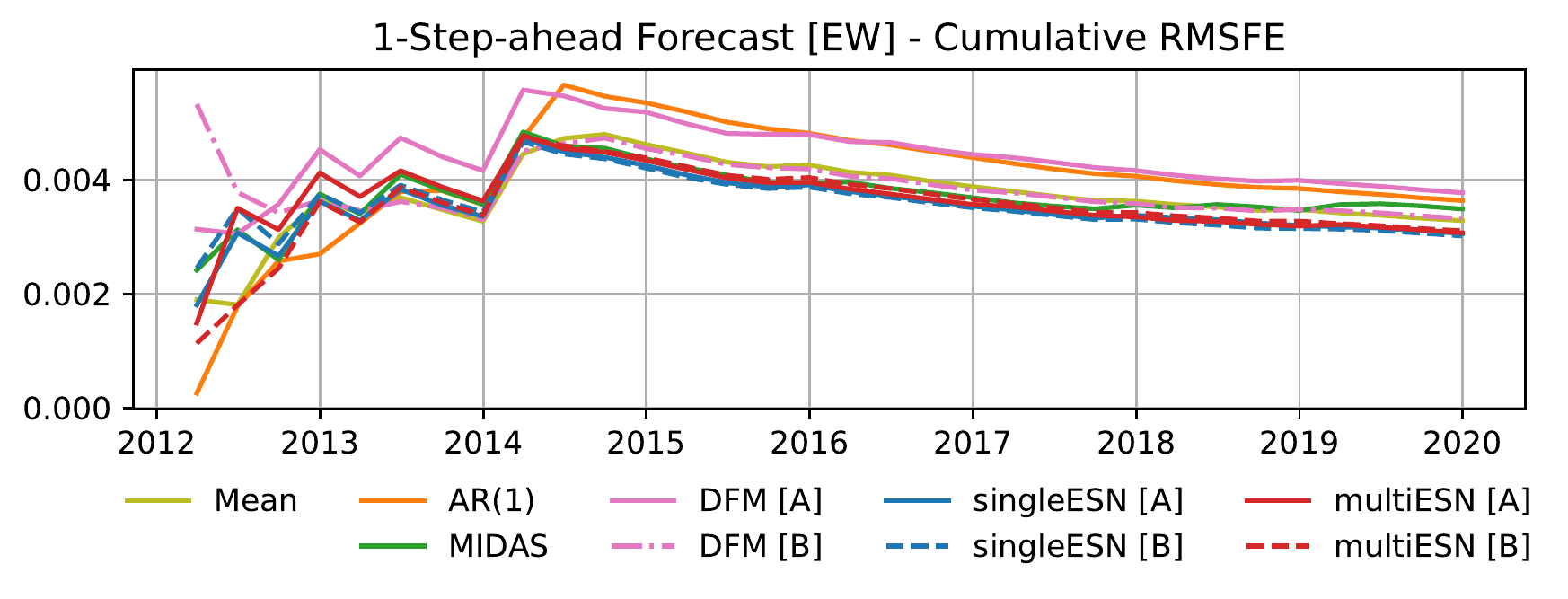}
    }
    \\
    \subfloat[]{%
      \includegraphics[width=0.48\textwidth]{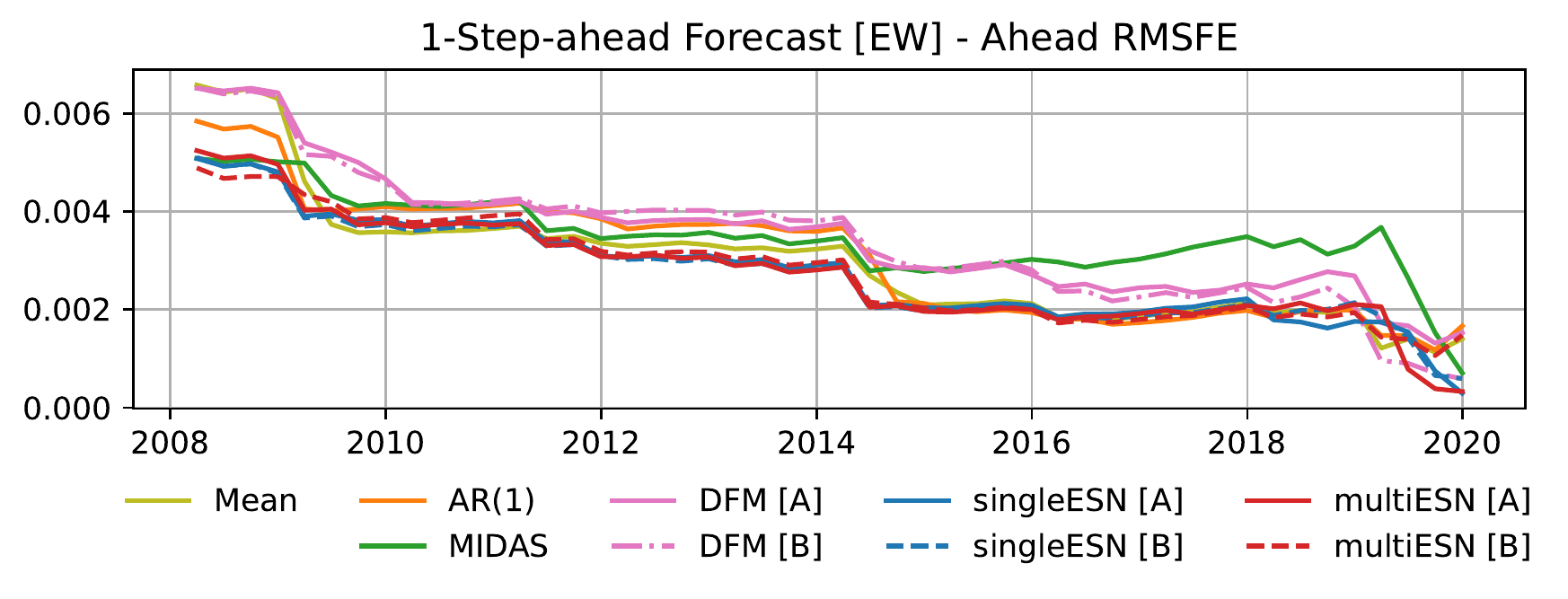}
    }
    \hfill
    \subfloat[]{%
      \includegraphics[width=0.48\textwidth]{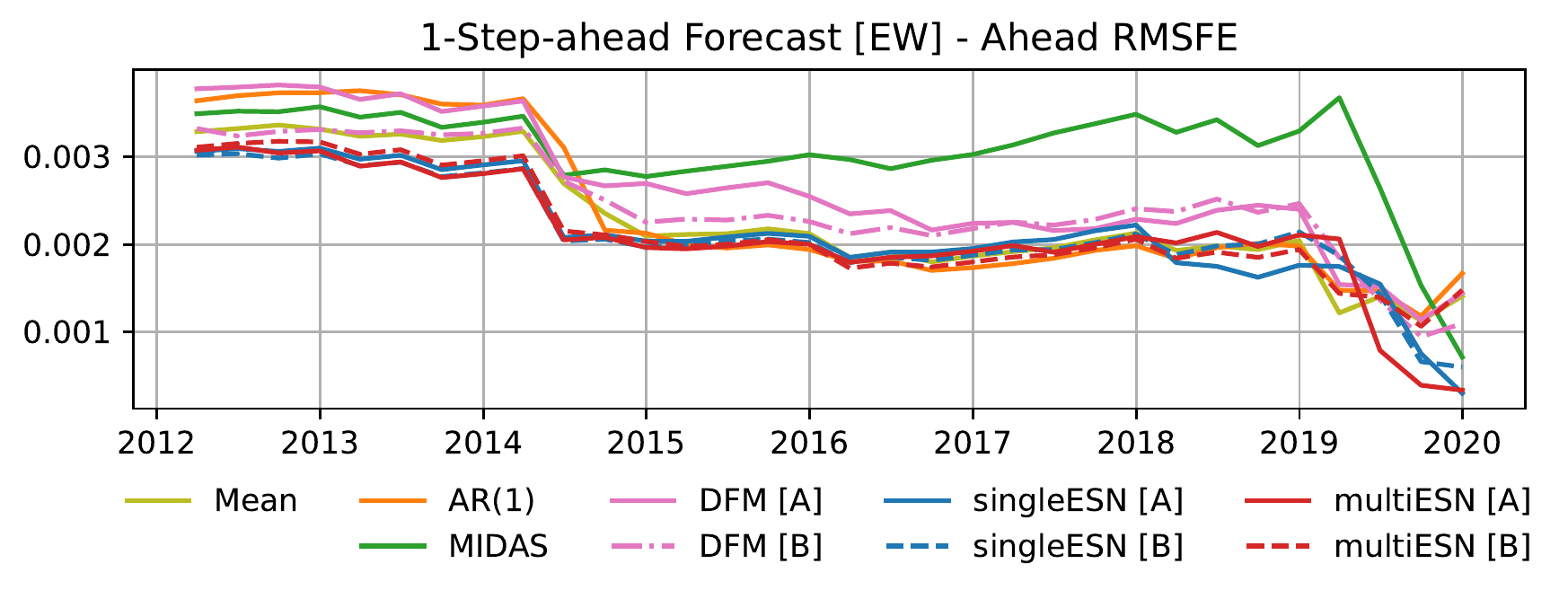}
    }
    \\
    \subfloat[]{%
      \includegraphics[width=0.48\textwidth]{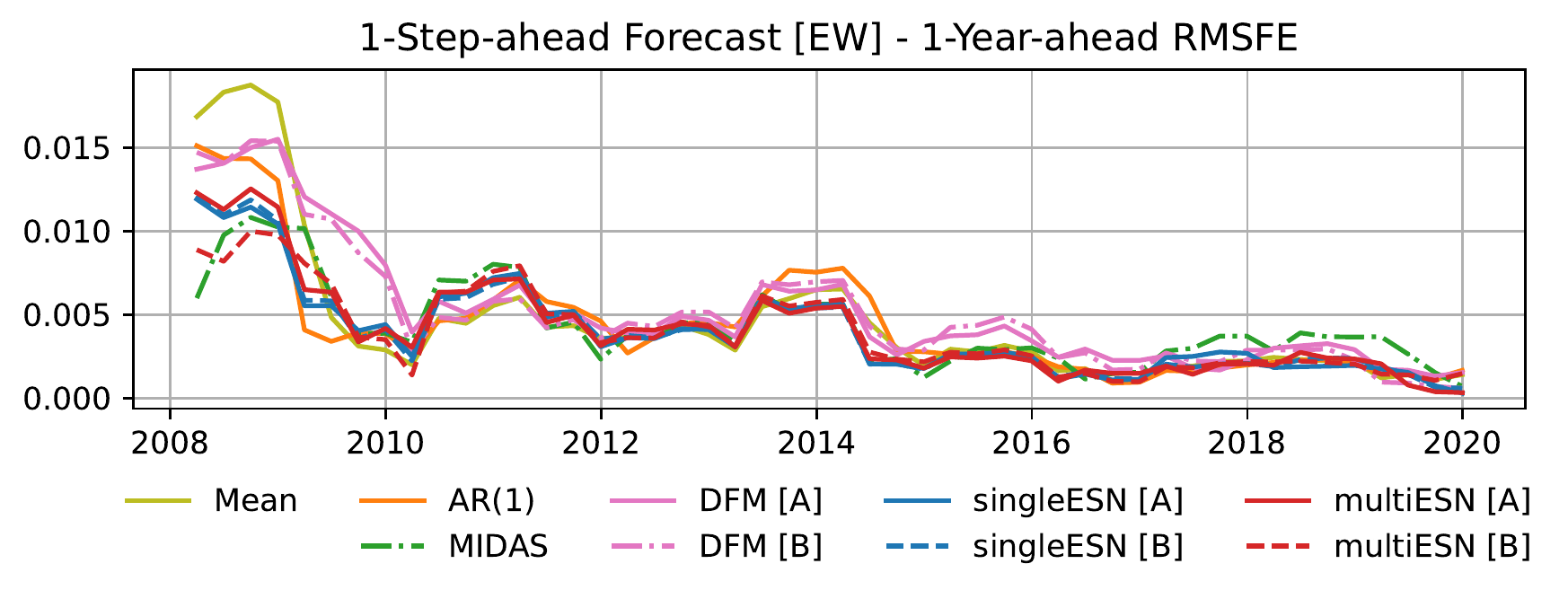}
    }
    \hfill
    \subfloat[]{%
      \includegraphics[width=0.48\textwidth]{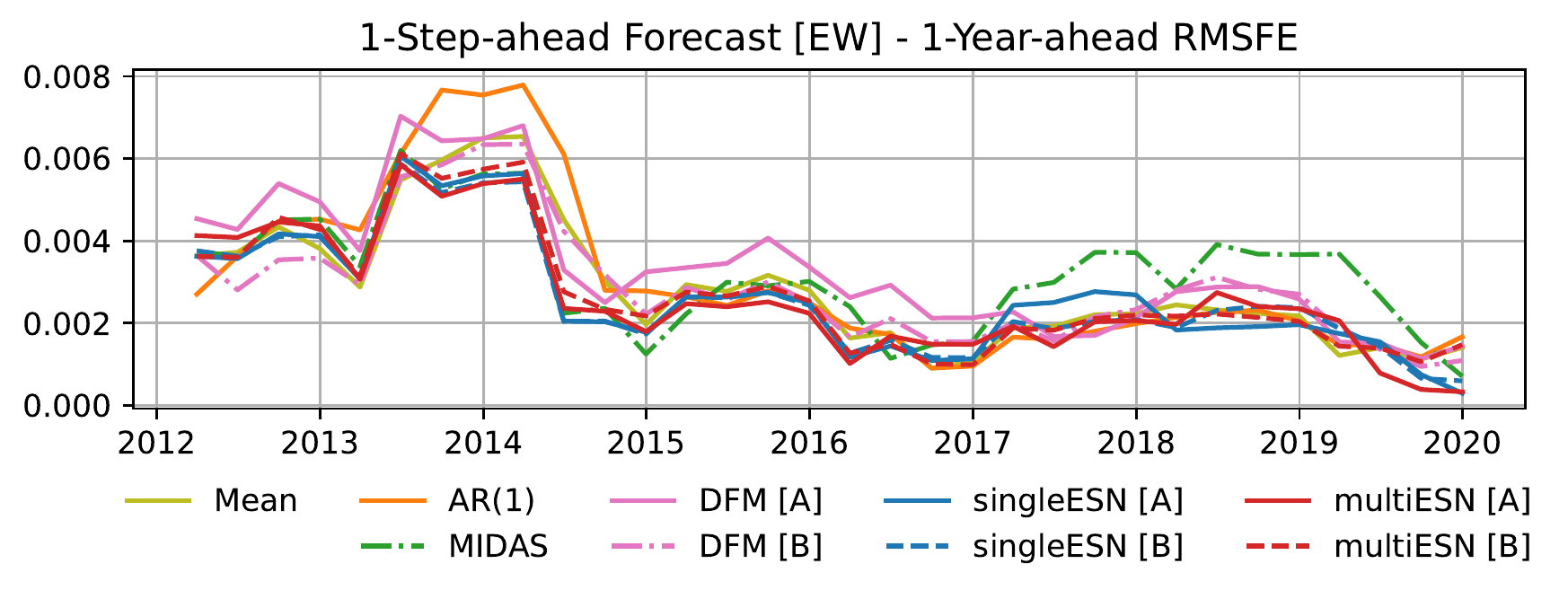}
    }
 \end{figure}
 
 \newpage
 
 \begin{figure}[ht]
    \caption{1-Step-ahead GDP Forecasting, Rolling Window - Small-MD Dataset}
    \label{fig:1sa_GDP_rolling_smallMD}
    %
    \subfloat[Pre-crisis model]{%
      \includegraphics[width=0.48\textwidth]{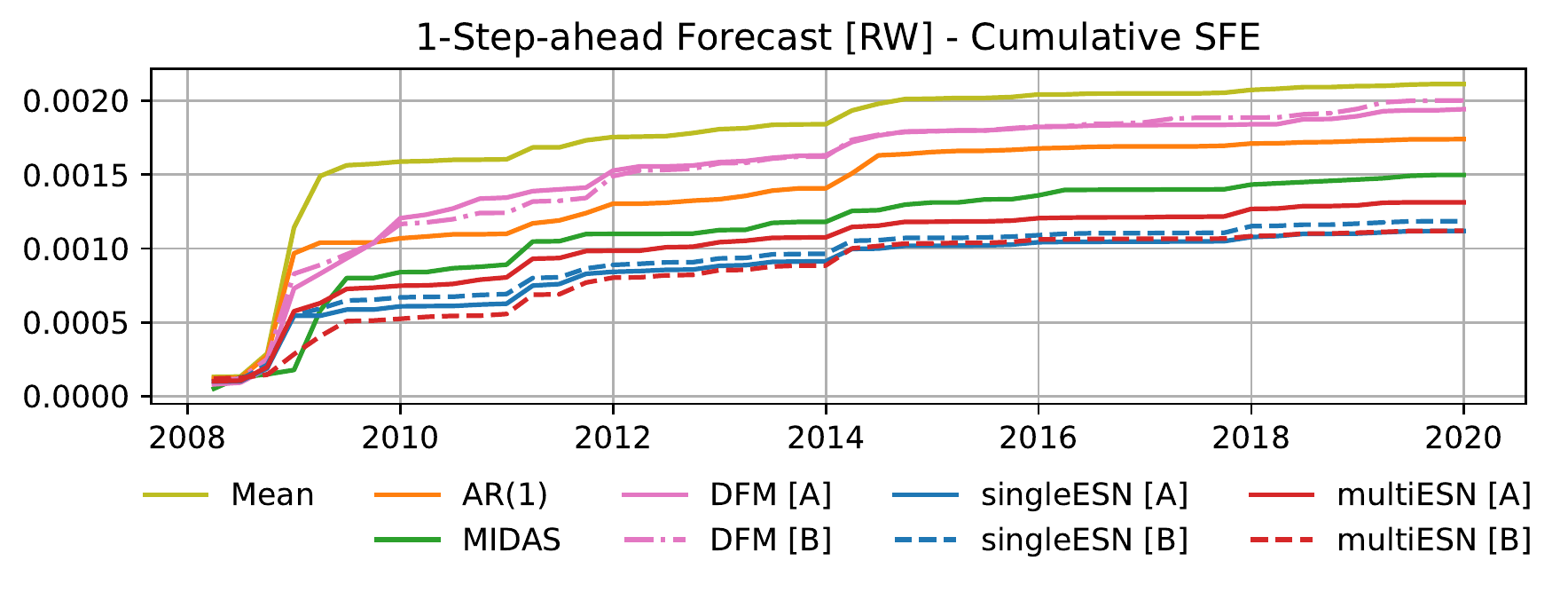}
    }
    \hfill
    \subfloat[Post-crisis model]{%
      \includegraphics[width=0.48\textwidth]{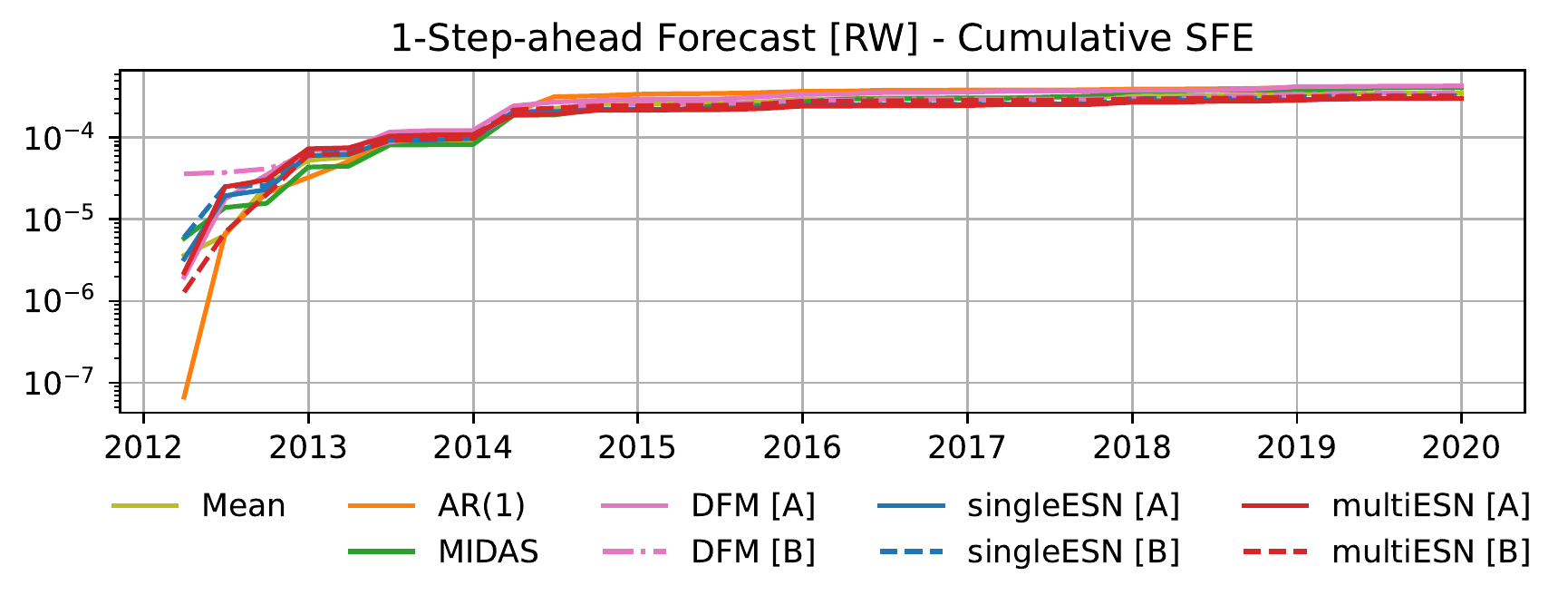}
    }
    \\
    \subfloat[]{%
      \includegraphics[width=0.48\textwidth]{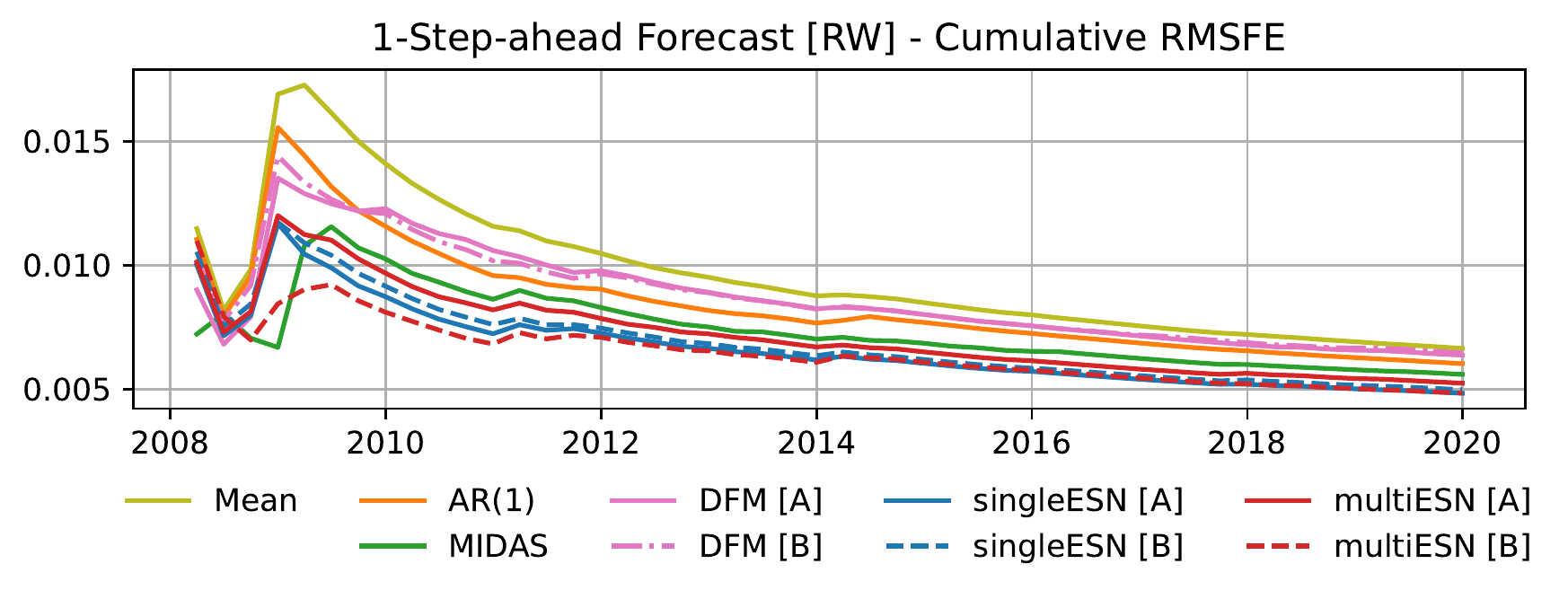}
    }
    \hfill
    \subfloat[]{%
      \includegraphics[width=0.48\textwidth]{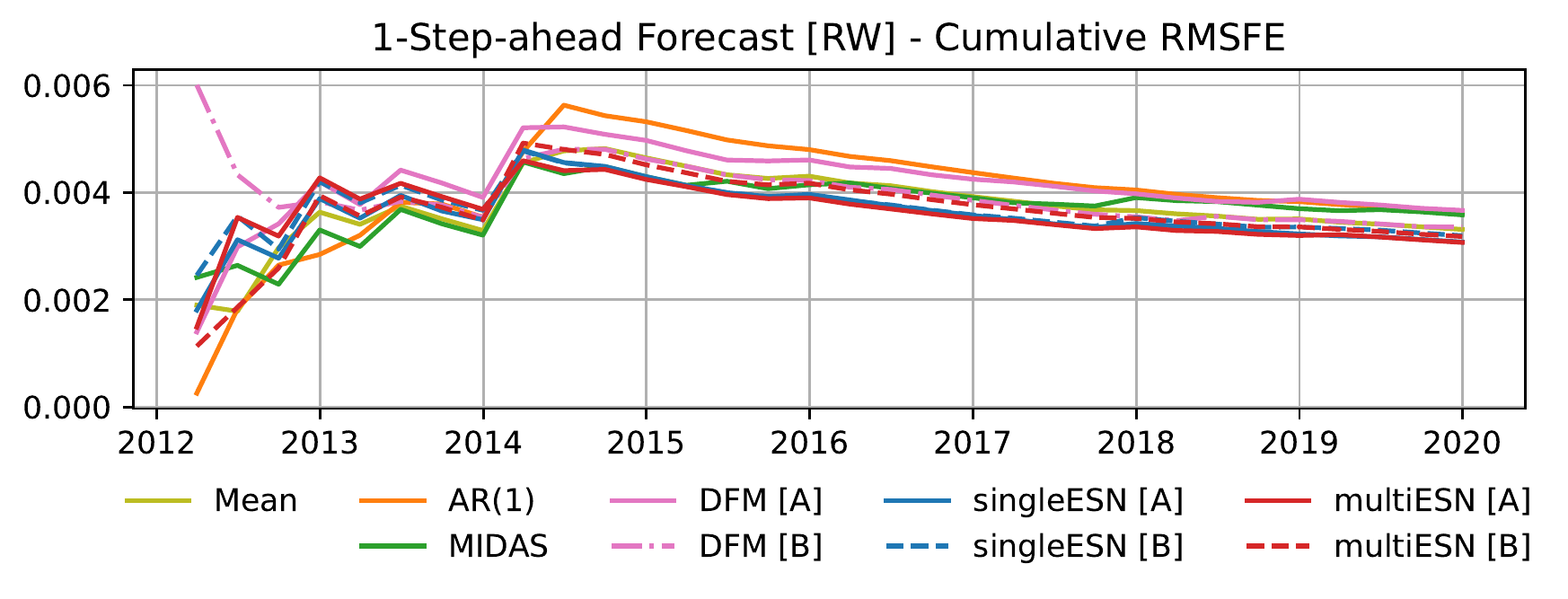}
    }
    \\
    \subfloat[]{%
      \includegraphics[width=0.48\textwidth]{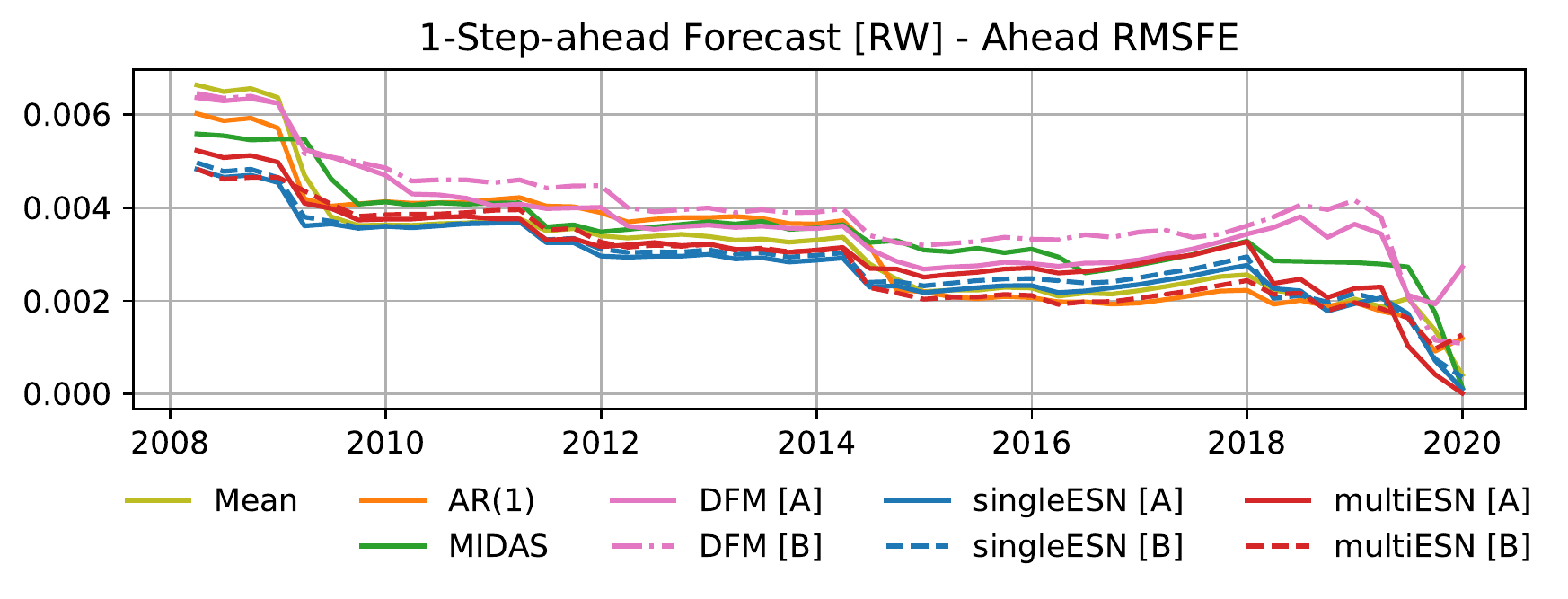}
    }
    \hfill
    \subfloat[]{%
      \includegraphics[width=0.48\textwidth]{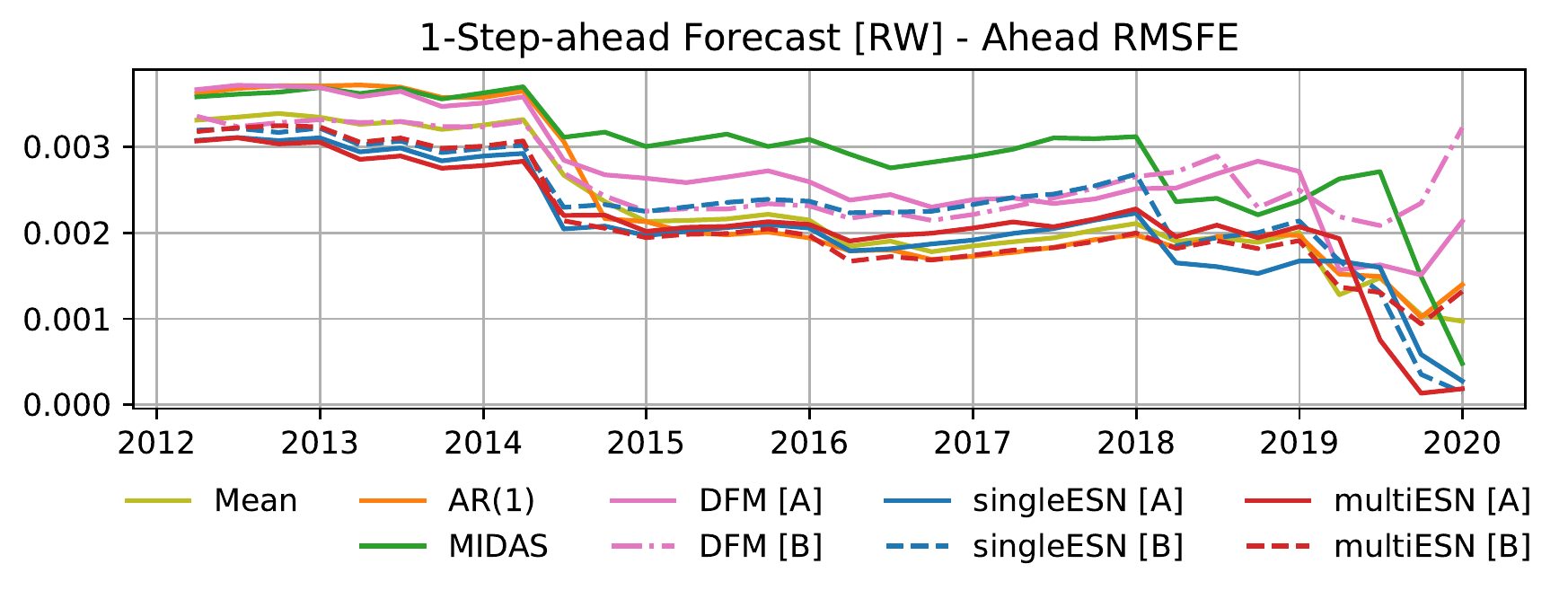}
    }
    \\
    \subfloat[]{%
      \includegraphics[width=0.48\textwidth]{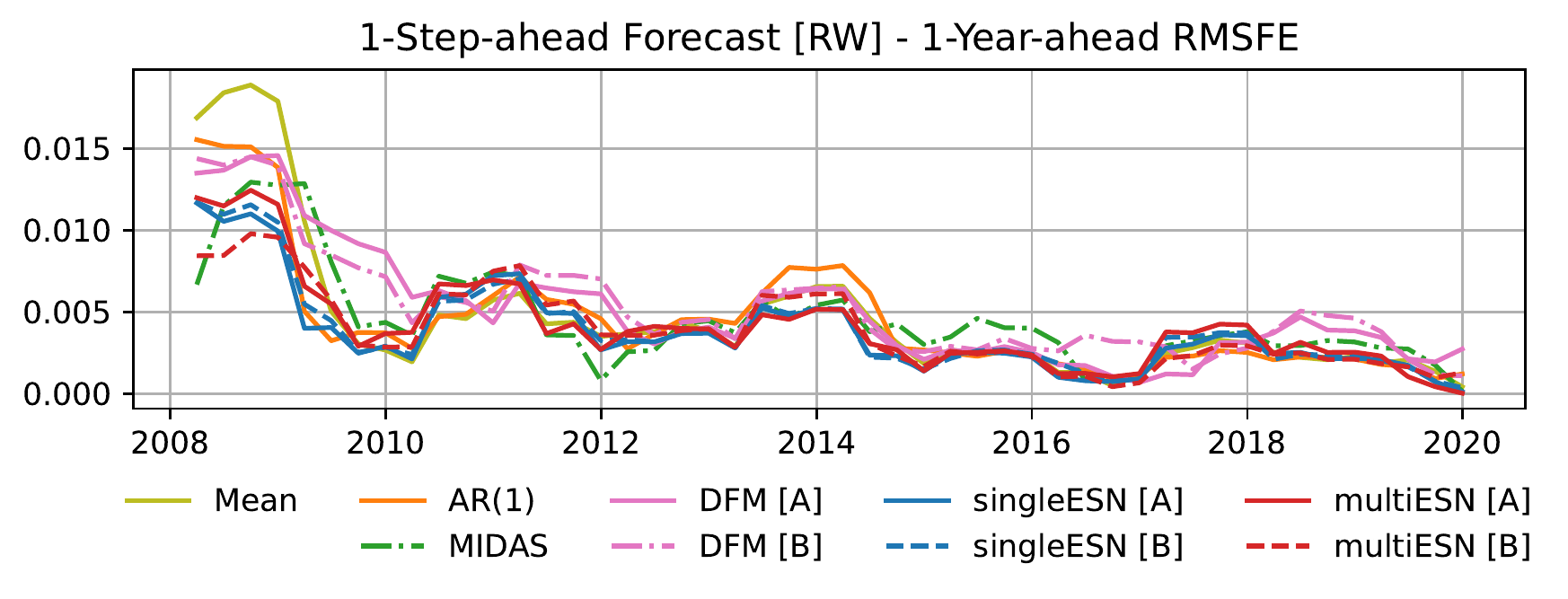}
    }
    \hfill
    \subfloat[]{%
      \includegraphics[width=0.48\textwidth]{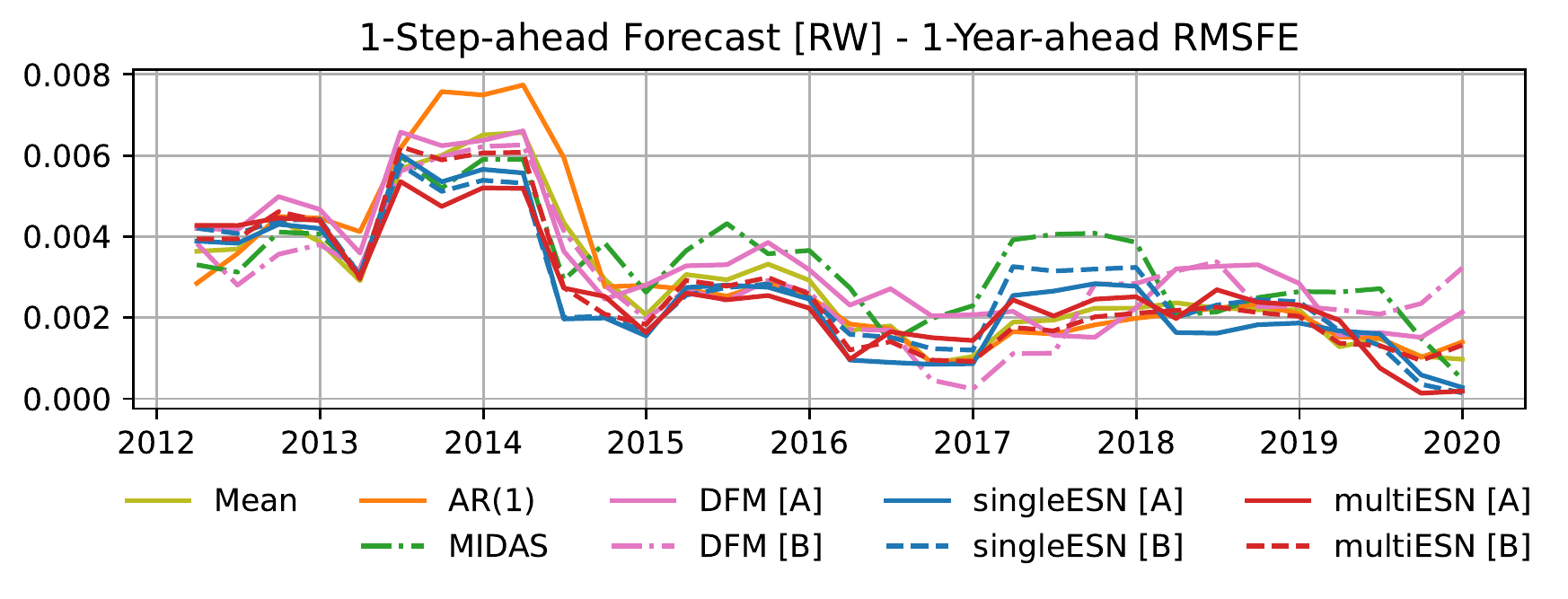}
    }
 \end{figure}

\end{document}